\documentclass[english,nofootinbib]{revtex4-1}
\usepackage{lmodern}

\usepackage{color}
\usepackage{babel}
\usepackage{array}
\usepackage{units}
\usepackage{mathtools}
\usepackage{multirow}
\usepackage{amsmath}
\usepackage{amssymb}
\usepackage{graphicx}
\usepackage[unicode=true,pdfusetitle,
 bookmarks=true,bookmarksnumbered=true,bookmarksopen=true,bookmarksopenlevel=1,
 breaklinks=false,pdfborder={0 0 1},backref=false,colorlinks=true]
 {hyperref}

\usepackage{mathrsfs}

\hypersetup{
 pdfborderstyle=,linkcolor=blue,citecolor=blue,urlcolor={}}

\makeatletter

\@ifundefined{showcaptionsetup}{}{
\PassOptionsToPackage{caption=false}{subfig}}
\usepackage{subfig}
\makeatother

\begin{document}

\title{All fundamental electrically charged thin shells
in general relativity: From star shells to tension shell black holes
and regular black holes and beyond}







\author{Jos\'{e} P. S. Lemos}
\email{joselemos@ist.utl.pt}

\affiliation{Centro de Astrof\'{\i}sica e Gravita\c{c}\~{a}o - CENTRA,
Departamento
de F\'{\i}sica, Instituto Superior T\'{e}cnico - IST, Universidade
de Lisboa - UL, Av. Rovisco Pais 1, 1049-001 Lisboa, Portugal}
\author{Paulo Luz}
\email{paulo.luz@ist.utl.pt}
\affiliation{Centro de Astrof\'{\i}sica e Gravita\c{c}\~{a}o - CENTRA,
Departamento
de F\'{\i}sica, Instituto Superior T\'{e}cnico - IST, Universidade
de Lisboa - UL, Av. Rovisco Pais 1, 1049-001 Lisboa, Portugal,}
\affiliation{Center for Mathematical Analysis, Geometry and Dynamical
Systems, Instituto Superior T\'{e}cnico - IST, Universidade de Lisboa
- UL, Avenida Rovisco Pais 1, 1049 Lisboa, Portugal,}
\affiliation{
Departamento de Matem\'atica, ISCTE - Instituto Universit\'ario de Lisboa, Portugal}

\begin{abstract}

We classify all fundamental electrically charged thin shells in
general relativity, i.e., static spherically symmetric perfect fluid
thin shells with a Minkowski spacetime interior and a
Reissner-Nordstr\"om spacetime exterior, characterized by the
spacetime mass $M$, which we assume positive, and the electric charge
$Q$, which without loss of generality in our analysis can always be
assumed as being the modulus of the electric charge, be it positive or
negative.  The fundamental shell can exist in three states, namely,
nonextremal when $\frac{Q}{M}<1$, which includes the Schwarzschild
$\frac{Q}{M}=0$ state, extremal when $\frac{Q}{M}=1$, and overcharged
when $\frac{Q}{M}>1$.  The nonextremal state, $\frac{Q}{M}<1$, allows
the shell to be located in such a way that the radius $R$ of the shell
can be outside its own gravitational radius $r_+$, i.e., $R>r_+$,
where $r_+$ is given in terms of $M$ and $Q$ by
$r_+=M+\sqrt{M^2-Q^2}$, or can be inside its own Cauchy radius $r_-$,
i.e., $R<r_-$, where $r_-$ is given in terms of $M$ and $Q$ by
$r_-=M-\sqrt{M^2-Q^2}$.  The extremal state, $\frac{Q}{M}=1$, allows
the shell to be located in such a way that the radius $R$ of the shell
can be outside its own gravitational radius $r_+$, i.e., $R>r_+$,
where now $r_+=r_-$, or can be inside its own gravitational radius,
i.e., $R<r_+$, or can be at its own gravitational radius $r_+$, i.e.,
$R=r_+$.  The overcharged state, $\frac{Q}{M}>1$, allows the shell to
be located anywhere $R\geq0$.  There is yet a further division,
indeed, one has still to specify the orientation of the shell, i.e.,
whether the normal out of the shell points toward increasing radii or
toward decreasing radii. For the shell's orientation, the analysis in
the nonextremal state is readily performed using Kruskal-Szekeres
coordinates, whereas in the extremal and overcharged states the
analysis can be performed in the usual spherical coordinates. There is
still a subdivision in the extremal state $r_+=r_-$ when the shell is
at $r_+$, $R=r_+$, in that the shell can approach $r_+$ from above or
approach $r_+$ from below.
The shell is assumed to be composed of an electrically charged perfect
fluid characterized by the energy density, pressure, and electric
charge density, for which an analysis of the energy conditions, null,
weak, dominant, and strong, is performed.  In addition, the shell
spacetime has a corresponding Carter-Penrose diagram that can be built
out of the diagrams for Minkowski and Reissner-Nordstr\"om spacetimes.
Combining these two characterizations, specifically, the physical
properties and the Carter-Penrose diagrams, one finds that there are
fourteen cases that comprise a bewildering variety of shell
spacetimes, namely,
nonextremal star shells,
nonextremal tension shell black holes,
nonextremal tension shell regular and
    nonregular black holes,
nonextremal compact shell naked singularities,
Majumdar-Papapetrou star
   shells,
extremal tension shell singularities,
extremal tension shell regular and nonregular black holes,
Majumdar-Papapetrou compact shell naked singularities,
Majumdar-Papapetrou shell quasiblack holes,
extremal null shell quasinonblack holes,
extremal null shell singularities,
Majumdar-Papapetrou null shell singularities,
overcharged star shells,
and overcharged compact shell naked singularities.

\end{abstract}
\maketitle

\section{Introduction}

An important group of solutions in a theory of gravitation, in
particular solutions to general relativity,
are those that represent stars. A
nonrotating star in general relativity is a static spherically
symmetric solution composed of some matter fields that either extend
indefinitely, with its fields decaying sufficiently fast to yield a
well-defined asymptotic infinity structure, or fill some interior part
that has a well defined boundary that in turn connects to a vacuum
exterior. A particular interesting example of this latter instance of
a general relativistic star is an interior composed of vacuum plus a
boundary made of a thin shell of matter, that is joined to a
vacuum exterior. The matter concentration on the thin shell
can produce with faithfulness several local and global properties
of the spacetime, with particular relevance it can generate
the formation of apparent and event horizons.

In general relativity the generic study of uncharged timelike
and spacelike thin
shells
was initiated by Israel \cite{Israel_1966}, followed by
Papapetrou and Hamoui \cite{papapetrouhamoui} and Taub \cite{taub},
with lightlike thin shells being treated by Barrab\`es and Israel
\cite{Barrabes_Israel_1991}.  In the instance that the
uncharged thin shell is spherically symmetric, the interior spacetime
can be of any type, in many situations Minkowski can be used, and the
exterior spacetime of pure vacuum is the Schwarzschild spacetime
according to 
Birkoff's theorem.  There are many applications of spherically
symmetric uncharged shells with a Schwarzschild exterior and we
mention a few of those.  Thin shell gravitational collapse was treated
by Israel \cite{israel2}, spacetimes with counter-rotating particles
in thin shells were studied by Evans \cite{evans} and Papapetrou and
Hamoui \cite{papapetrouhamoui2}, collisions of spherical thin shells
were analyzed by 't Hooft and Dray \cite{draythooft}, cosmic bubbles
with inflation were considered by Blau, Guendelman, and Guth
\cite{blauguendelmanguth}, properties of thin shells with a black hole
interior were investigated by Frauendiener, Hoenselaers, and Konrad
\cite{frauen} and Brady, Louko and Poisson \cite{brad}, the study of
tension shell black holes was performed by Katz and Lynden-Bell
\cite{Katz_Lynden-Bell_1991} and Comer and Katz
\cite{Comer_Katz_1994}, thin shell wormholes were constructed by
Visser \cite{visserbook} and by Lemos, Lobo, and Oliveira
\cite{Lemos_Lobo_2008},
and properties of black holes, such as
black hole entropy through thin shells, were done by Andr\'e, Lemos,
and Quinta \cite{l5} and Bergliaffa, Chiapparini, and Reyes
\cite{bergli}.

The study of spherically symmetric electrically charged thin shells
with a Reissner-Nordstr\"om spacetime exterior was dealt by de la Cruz
and Israel \cite{cruzisrael} and Kucha\v{r} \cite{Kuchar_1968},
gravitational collapse of electrically charged shells was performed by
Chase \cite{chase} and Boulware \cite{boulware}, Vilenkin and Fomin
\cite{vilenkinfomin} inspected the problem of the self-energy of the
electron through thin shells,
the topology of the collapse of charged thin shells and fluids
was analyzed by Hiscock \cite{hiscock},
interacting electric thin shells were
studied in Lemos and Zanchin, \cite{lemoszanchin2006}, Dias, Gao, and
Lemos studied electric collapsing shells in Lovelock theory \cite{lz2}
with the general relativistic counterpart and its relation to cosmic
censorship studied by Gao and Lemos \cite{gaolemos}, properties of
highly compact electric shells together with their energy conditions
where a maximum bound, of the kind of the Buchdahl bound has been
found, have been discussed by Andr\'easson \cite{andreasson2009},
the inclusion of a Vlasov fluid into a thin shell
was studied by
Andr\'easson, Eklund, and Rein \cite{andreasson2},
thin
shell electrically charged wormholes were constructed by Dias and
Lemos \cite{diaslemosworm2010}, regular
black holes with thin shells were worked out in
\cite{lemoszanchinregularbhs},
the highest compact shells were
analyzed by Lemos and Zaslavskii \cite{qbh4lemoszaslacp},
Berezin and Dokuchaev performed a
thorough study of gravitational collapse of electric thin shells
\cite{berezin1}, and properties of black holes, such as black hole
entropy, through thin shells, were analyzed by Lemos, Quinta, and
Zaslavskii \cite{l4,l6}, tension electric shells
on the other side of the Reissner-Nordst\"om universe
were studied by Luz and Lemos \cite{luzlemos},
further study on thin shells with matter obeying the 
Einstein-Maxwell equations with the inclusion
of a Vlasov fluid were analyzed in 
\cite{thaller},
and  for the highest compact electric thin shells
that form quasiblack holess
see the work of Lemos and Zaslavskii
\cite{lemoszasla2020}.

A feature of the electric thin shells is that the exterior
Reissner-Nordstr\"om spacetime can be in three different states
depending on the ratio of the charge $Q$ to spacetime mass $M$,
$\frac{Q}{M}$, noting that we use units in which the constant of
gravitation and the speed of light are equal to one.  Indeed, the
exterior Reissner-Nordstr\"om spacetime can be nonextremal when
$\frac{Q}{M}<1$, extremal when $\frac{Q}{M}=1$, and overcharged when
$\frac{Q}{M}>1$. The matter that forms the thin shells can be of
several types, one type often considered is an electric charged
perfect fluid, with energy density $\sigma$ and pressure $p$. Of
course the electric charge density $\sigma_e$ of the fluid is related
to $Q$ and the rest mass density of the fluid $\sigma$ is related to
$M$.  An interesting particular situation is when $\sigma_e=\sigma$,
for which the matter is called Majumdar-Papapetrou matter and which
gives rise to an exterior extremal Reissner-Nordstr\"om spacetime with
$\frac{Q}{M}=1$.  For all the three different states it is of interest
to test the energy conditions for the matter, the most important of
which are the null, weak, dominant and strong.  Many of the studied
shell spacetimes can only be understood through the maximal extension
of the corresponding exterior spacetime. Indeed, to appreciate a
spacetime in its totality, in particular a spherical symmetric
spacetime, one should maximally extend it and draw the corresponding
Carter-Penrose diagram by using the techniques available, see, e.g.,
\citep{Graves_Brill_1960,Penrose1964,Carter_1966_2,
Hawking_Ellis_book,MTW_Book,felicebook}.  In addition, to have a full grasp on
electric shells, certain properties of the Reissner-Nordstr\"om
spacetime have to be understood, such as the regions where electric
repulsion dominates over gravitational attraction.

The aim of the paper is to find and classify all the fundamental
electrically charged thin shells in the Einstein-Maxwell theory,
i.e., static
spherically symmetric electrically charged
general relativistic thin shells with a
Minkowski interior and a Reissner-Nordstr\"om exterior.  We thus
extend the Katz-Lynden-Bell solution and analysis done for uncharged
shells with a Schwarzschild exterior.  The fundamental electric thin
shells are assumed to have a perfect fluid stress-energy tensor
$S_{\alpha\beta}$ on the shell which is defined through the junction
of the interior and exterior spacetimes. We consider
that the shells have positive
spacetime mass $M$, in the no-shell
limit they have zero mass,
$M\geq0$, and the radius of the shell obeys
$R\geq0$.  In the nonextremal state, $\frac{Q}{M}<1$, there are two
natural intrinsic radii, the gravitational radius $r_+$ given by
$r_+=M+\sqrt{M^2-Q^2}$ and the Cauchy radius $r_-$ given by
$r_-=M-\sqrt{M^2-Q^2}$. Depending on the location radius $R$ of the
shell, $r_+$ and $r_-$ can be horizon radii.  In this nonextremal
state it is useful to define Kruskal-Szekeres coordinates for the
maximally extended spacetime, as they allow to analyze in a natural way
the physical properties including the energy conditions of the thin
shell at any allowable location radius $R$.  In the extremal state,
$\frac{Q}{M}=1$, the two intrinsic radius merge into one, $r_+=r_-$,
and it is possible to analyze the physical properties of the thin
shell including the energy conditions at any allowable location radius
$R$ simply by resorting to the usual spacetime spherical coordinates.
In the overcharged state, $\frac{Q}{M}>1$, the two intrinsic radius do
not exist, one can analyze the physical properties of the thin shell
including the energy conditions at any allowable location radius $R$
also by resorting to the usual spacetime spherical coordinates.  The
shell spacetime classification that we present can only be
fully understood
through the maximal extension of the outer Reissner-Nordstr\"om
spacetime and the drawing of the Carter-Penrose diagrams.
A bewildering variety of fourteen cases appear, namely,
nonextremal star shells,
nonextremal tension shell black holes,
nonextremal tension shell regular and
    nonregular black holes,
nonextremal compact shell naked singularities,
Majumdar-Papapetrou star
   shells,
extremal tension shell singularities,
extremal tension shell regular and nonregular black holes,
Majumdar-Papapetrou compact shell naked singularities,
Majumdar-Papapetrou shell quasiblack
     holes,
extremal null shell quasinonblack holes,
extremal null shell singularities,
Majumdar-Papapetrou null shell singularities,
overcharged star shells,
and overcharged compact shell naked singularities.
In contrast to the Schwarzschild shell analyzed by Lynden-Bell and
Katz which has only two cases, here we have indeed a
wealth of cases.
To all the fourteen cases, a physical interpretation can be given with
the help of the two main features that we mentioned, namely, the
matter properties in conjunction with the energy conditions, and the
causal and global structure based on the Carter-Penrose diagrams.
Which cases are familiar and which cases are peculiar, or
even strange, depends on the
analysis one makes.  Some cases have the energy conditions verified
and the geometrical setup seems to be physically reasonable, other
cases have the energy conditions verified and the resulting spacetime
is rather peculiar, and yet other cases have the energy conditions
violated with a physically reasonable geometrical setup.

The article is organized as follows.
In Section~\ref{Sec:Junction_formalism},
we set the framework
and devise the manner
to study the
physical properties of the fundamental electric thin shells
through the formalism of 
junction conditions in general relativity, joining a
Minkowski interior to a
Reissner-Nordstr\"om exterior. We  present
the main features of the Minkowski
and the
Reissner-Nordstr\"om 
spacetimes together with their
Carter-Penrose diagrams. We also establish the classification
scheme and the nomenclature we use.
In Section~\ref{Sec:Nonextremal-thin-shells_outside_event_horizon}, we
study nonextremal electric thin shells outside the gravitational
radius and show there are two types, namely, star shells and tension
shell black holes.
In Section~\ref{insidecauchy}, we study nonextremal electric thin shells
inside  the Cauchy radius and show there are two types, namely,
tension shell regular and nonregular black holes and compact shell
naked singularities.
In Section~\ref{Sec:Extremal-thin-shells-outside}, we study extremal
electric thin shells outside the gravitational radius and show there
are two types, namely, Majumdar-Papapetrou star shells and tension
shell black holes.
In Section~\ref{Sec:Extremal-thin-shells-inside12}, we study extremal
electric thin shells inside the gravitational radius and show there
are two types, namely, tension shell regular and nonregular black
holes and
Majumdar-Papapetrou compact shell naked singularities.
In Section~\ref{Sec:Horizon_thin_shells}, we study extremal electric
thin shells at the gravitational radius and show there are four types,
namely, Majumdar-Papapetrou shell quasiblack holes,
extremal null shell quasinonblack holes,
extremal null shell singularities, and
Majumdar-Papapetrou null shell singularities.
In Section~\ref{Sec:Overcharged_thin_shells}, we study overcharged thin
shells and show there are two types, namely, star shells and compact
shell naked singularities.
In Section~\ref{Sec:SinopECandCP}, we study the
weak, null, dominant, and strong energy conditions
for all the fundamental electric thin
shells 
and present a chart with all Carter-Penrose diagrams for the
shells.  In Section~\ref{Sec:Conclusions}, we conclude.
In Appendix~\ref{Appendix_sec:Kruskal-Szekeres_coordinates_RN}, we
present the maximal extension of the nonextremal 
Reissner-Nordstr\"om spacetime
through Kruskal-Szekeres coordinates important to deal with the
shells in a nonextremal state.  In
Appendix~\ref{Appendix_sec:Extrinsic_curvature}, we present the
calculation of the shell's extrinsic curvature in a 
nonextremal 
Reissner-Nordstr\"om exterior
spacetime
also important to the whole development of the paper.

\newpage

\section{Preliminaries:
Physical properties of  fundamental electric thin shells
through 
junction conditions,
Minkowski interior and 
Reissner-Nordstr\"om exterior and their
Carter-Penrose diagrams, and classification
scheme and nomenclature}
\label{Sec:Junction_formalism}

\global\long\def\ps#1#2{\prescript{#1}{}{#2}}

\subsection{Physical properties of fundamental
electric thin shells
through 
junction conditions}
\label{Sec:Junction_formalismproper}


We work with general relativity coupled to electric matter, so the 
appropriate equations are the Einstein-Maxwell-charged matter
field equations, i.e.,
\begin{align}
R_{\alpha\beta}-\frac{1}{2}g_{\alpha\beta}R & =8\pi
T_{\alpha\beta}\,,\label{eq:EFE1}\\
\nabla_{\beta}F^{\alpha\beta} & =4\pi J^{\alpha}\,,
\label{eq:Maxwell_FE}
\end{align}
where $R_{\alpha\beta}$ is the Ricci tensor, $R$ the Ricci scalar,
$T_{\alpha\beta}$ the stress-energy tensor, $F_{\alpha\beta}$ the
Faraday-Maxwell tensor, $J^{\alpha}$ is the electromagnetic 4-current,
and $\alpha,\beta=0,1,2,3$ are
the usual spacetime indices.  The other Maxwell
equations,
$\nabla_{\left[\alpha\right.}F_{\left.\beta\gamma\right]}=0$, where
square brackets represent antisymmetrization in the delimited indices,
are automatically satisfied for a properly defined $F_{\alpha\beta}$.
We use units in which the constant of gravitation and the speed of
light are equal to one, and assume the metric signature
$\left(-+++\right)$.

We consider a general
relativistic spacetime that is built from an
interior
$\mathcal{M}_{\rm i}$ with metric $g_{\rm i}$,
an exterior 
$\mathcal{M}_{\rm e}$ with metric $g_{\rm e}$,
glued together at a common hypersurface $\cal S$.
We will assume $\mathcal{M}_{\rm i}$
to be described by the Minkowski
solution and $\mathcal{M}_{\rm e}$ to be the
Reissner-Nordstr\"om spacetime, but for the time being we
can keep the analysis quite general. 
In joining  $\mathcal{M}_{\rm i}$ with $\mathcal{M}_{\rm e}$
the whole spacetime solution $\mathcal{M}$ has still to obey
the Einstein-Maxwell field equations,
Eqs.~(\ref{eq:EFE1}) and (\ref{eq:Maxwell_FE}).
The hypersurface $\mathcal{S}$ can be timelike,
lightlike, or spacelike.  Here, we  work with the timelike and
possible spacelike situations, 
and we will
revise briefly the Darmois-Israel junction
formalism for these type of hypersurfaces in the theory of general
relativity~\citep{Israel_1966}, see
also~\citep{Kuchar_1968}
for the inclusion of the electromagnetic field.
The lightlike case can sometimes
be dealt with by extension of
these two situations  or generically within an appropriate
formalism~\citep{Barrabes_Israel_1991}.

To start, we assume that it is possible to define
a common coordinate system
$\left\{ x^{\alpha}\right\} $
on both sides of the hypersurface
$\mathcal{S}$,
where the index $\alpha$
runs from 0 to 3, for the time and
the three space components, respectively.
We also assume
the existence of a vector field $n$, well defined on
both sides of $\mathcal{S}$, to be orthogonal at each point to the
matching surface. We choose $n$, the normal to $\mathcal{S}$, to point
from $\mathcal{M}_{\rm i}$ to $\mathcal{M}_{\rm e}$ and without loss
of generality
\begin{equation}
n^{\alpha}n_{\alpha}=\varepsilon\,,\label{eq:normal_normalized}
\end{equation}
where $n^{\alpha}$ are the components of $n$ in the coordinate system
$\left\{ x^{\alpha}\right\} $ and $\varepsilon$ is $\pm1$ depending on
$n$ being spacelike or timelike, respectively, the null case having
$\varepsilon=0$ but will not be treated here. The normal vector
field $n$ is such that it is spacelike or timelike if the hypersurface
is timelike or spacelike, respectively.
We denote by  $\left\{ y^{a}\right\}$
a local coordinate system on
$\mathcal{S}$, where the index $a$ has three components only,
which depending on the character of $\mathcal{S}$,
can be one for the time
and two for the other
space coordinates, or three for the space coordinates.
Now,
the normal vector field $n$ must be orthogonal, at each
point, to the tangent vectors to the hypersurface $\mathcal{S}$,
$e_{a}\equiv
\frac{\partial\;\;\;}{\partial y^{a}}$, such that
\begin{equation}
e_{a}^{\alpha}n_{\alpha}=0\,,\label{eq:normal_orthogonal}
\end{equation}
with $e_{a}^{\alpha}\equiv\frac{\partial x^{\alpha}}{\partial
y^{a}}$. The
induced metric on $\mathcal{S}$ as seen from each region
$\mathcal{M}_{\rm i}$ and $\mathcal{M}_{\rm e}$, is 
\begin{equation}
h_{{\rm i}\,ab}\equiv
g_{{\rm i}\, \alpha\beta}e_{a}^{\alpha}e_{b}^{\beta}\,,\quad\quad
h_{{\rm e}\, ab}\equiv
g_{{\rm e}\, \alpha\beta}e_{a}^{\alpha}e_{b}^{\beta}\,,
\label{eq:inducedh}
\end{equation}
respectively,
where $g_{{\rm i}\, \alpha\beta}$
and $g_{{\rm e}\, \alpha\beta}$
are the components of
the interior and exterior metrics 
in the coordinate system
$\left\{ x^{\alpha}\right\}$, respectively.
Notice that in general, the induced metric on $\mathcal{S}$ by each
metric
$g_{{\rm i}\, \alpha\beta}$
or
$g_{{\rm e}\, \alpha\beta}$
may not coincide, hence we use the
notation
$h_{{\rm i}\, \alpha\beta}$
and
$h_{{\rm e}\, \alpha\beta}$
to refer to the metric induced
by each spacetime $\mathcal{M}_{{\rm i}}$ and
$\mathcal{M}_{{\rm e}}$, respectively. The
extrinsic curvature
$K_{{\rm i}\, ab}$ or $K_{{\rm e}\, ab}$
of $\mathcal{S}$, as an embedded manifold in
$\mathcal{M}_{{\rm i}}$ or $\mathcal{M}_{{\rm e}}$, respectively,
is defined as 
\begin{equation}
{K_{{\rm i}\, ab}}\equiv e_{a}^{\alpha}e_{b}^{\beta}
\nabla_{{\rm i}\,\alpha}n_{\beta}\,,\quad\quad
{K_{{\rm e}\, ab}}\equiv e_{a}^{\alpha}e_{b}^{\beta}
\nabla_{{\rm e}\,\alpha}n_{\beta}
\,,\label{eq:extrinsic1}
\end{equation}
where $\nabla_{{\rm i}}$ and
$\nabla_{{\rm e}}$ are the covariant derivatives with respect to
$g_{{\rm i}\, \alpha\beta}$
and 
$g_{{\rm e}\, \alpha\beta}$, respectively.
Their traces are 
\begin{equation}
K_{\rm i}\equiv {{h_{\rm i}}^{ab}}
{K_{{\rm i}\, ab}}\,,\quad\quad
K_{\rm e}\equiv {{h_{\rm e}}^{ab}}
{K_{{\rm e}\, ab}}
\,,
\end{equation}
respectively.

We need to find the conditions under which the matching of the two
spacetimes $\mathcal{M}_{{\rm i}}$ and $\mathcal{M}_{{\rm e}}$
form a valid solution of the field
equations, namely, of Eqs.~(\ref{eq:EFE1}) and~(\ref{eq:Maxwell_FE}).
So, following the Darmois-Israel
formalism~\citep{Israel_1966}
to join the two regions $\mathcal{M}_{\rm i}$ and $\mathcal{M}_{\rm e}$
at $\mathcal{S}$, such that the union of
$g_{{\rm i}\, \alpha\beta}$ and
$g_{{\rm e}\, \alpha\beta}$ forms a valid solution to the Einstein
field equations~(\ref{eq:EFE1}), two junction conditions must be
verified at the matching surface $\mathcal{S}$: (i) The induced metric
as seen from each region $\mathcal{M}_{\rm i}$
and $\mathcal{M}_{\rm e}$,
see Eq.~(\ref{eq:inducedh}), must be the same, i.e., 
\begin{equation}
\left[h_{ab}\right]=0\,,
\label{eq:1st_junct_cond}
\end{equation}
where, we use $\left[\psi\right]$ to represent the
difference of a field as seen from each sub-manifold at $\mathcal{S}$,
i.e., $\left[\psi\right]\equiv
\left.
{\psi}_{\rm e}\right|_{\mathcal{S}}-\left.\
{\psi}_{\rm i}\right|_{\mathcal{S}}$,
with $\psi_{\rm i}$ and
$\psi_{\rm e}$ referring to
a field $\psi$ defined in $\mathcal{M}_{\rm i}$ or $\mathcal{M}_{\rm
e}$, respectively;
(ii) If the extrinsic curvature, see Eq.~(\ref{eq:extrinsic1}),
is not the same on both sides of the boundary $\mathcal{S}$, then
a thin shell is present at $\mathcal{S}$.
The relation between the extrinsic curvature $K_{ab}$
of $\mathcal{S}$
and
the stress-energy tensor $S_{ab}$ of the thin shell is 
given by 
\begin{equation}
-\varepsilon\big(\left[K_{ab}\right]-
h_{ab}\left[K\right]\big)=8\pi\,S_{ab}\,.
\label{eq:2nd_junct_cond}
\end{equation}
We further
assume that the stress-energy $S_{ab}$ of the thin
shell is a perfect fluid stress-energy tensor on
$\mathcal{S}$,
i.e., we assume that the total stress-energy tensor,
defined as the sum
of the matter stress-energy tensor
and the
electromagnetic stress-energy tensor, can be written as
\begin{equation}
S_{ab}=\sigma u_{a}u_{b}+p\left(h_{ab}+u_{a}u_{b}\right)\,,
\label{eq:perfect}
\end{equation}
where $\sigma$ is the energy density, $p$ is the pressure, and $u_a$
is the fluid's velocity on $\mathcal{S}$. This $u_a$
is uniquely defined from projecting the interior 
four-velocity
${u_{\rm i}}^\alpha$ onto $\mathcal{S}$ itself
as $u_a={u_{\rm i}}^\alpha e_{a}^{\alpha}$, 
or projecting the exterior
four-velocity
${u_{\rm e}}^\alpha$ onto $\mathcal{S}$ itself
as $u_a={u_{\rm e}}^\alpha e_{a}^{\alpha}$,
with both projections obviously
yielding the same  $u_a$.

In the presence of electromagnetic fields, to guarantee that the
spacetime $\mathcal{M}$ is a valid solution of the Maxwell field
equations, Eq.~(\ref{eq:Maxwell_FE}), in addition to
Eqs.~(\ref{eq:1st_junct_cond}) and (\ref{eq:2nd_junct_cond}) we must
impose that the Faraday-Maxwell tensor, $F_{\alpha\beta}$, obeys
certain conditions.  Defining the projected Faraday-Maxwell tensor at
$\mathcal{S}$ from the interior as as
$F_{{\rm i}\,ab}=F_{{\rm i}\,
\alpha\beta}e_{a}^{\alpha}e_{b}^{\beta}$
and from the exterior
as $F_{{\rm e}\,ab}=F_{{\rm e}\,
\alpha\beta}e_{a}^{\alpha}e_{b}^{\beta}$,
and the projected Faraday-Maxwell vector at $\mathcal{S}$
from the interior as 
as $F_{{\rm i}\,a}=F_{{\rm i}\,
\alpha\beta}e_{a}^{\alpha}n^{\beta}$,
and from the exterior
as
$F_{{\rm e}\,a}=F_{{\rm e}\,
\alpha\beta}e_{a}^{\alpha}n^{\beta}$,
the electromagnetic matching conditions at
$\mathcal{S}$ are then
\begin{equation}
\left[F_{ab}\right]
=0\,,
\label{eq:junct_cond_Faradaya}
\end{equation}
and
\begin{equation}
\left[F_{a}\right] = 4\pi\,s_a \,,
\label{eq:junct_cond_Faradayb}
\end{equation}
where $s_a$ is
the electromagnetic surface current at $\mathcal{S}$
given by 
\begin{equation}
s_a=\sigma_e u_a\,,
\label{eq:junct_cond_Faraday2}
\end{equation}
with $\sigma_e$ being the electric charge density
on $\mathcal{S}$.

Two notes are in order.  The first note is to mention that
an infinitesimally
thin shell is certainly an approximation
to a thick shell with very small thickness.  The second note is to
draw the attention that our assumption on the form of $S_{ab}$, see
Eq.~(\ref{eq:perfect}), is a restriction on the properties of the
fluid, namely, it imposes that the thin shell fluid has no effective
anisotropic pressure. Since under certain density regimes, realistic
matter is expected to be anisotropic, it would be interesting to
understand how the presence of anisotropic pressure affects our
results.  Surely, both assumptions, namely, infinitesimal thickness
and isotropic matter for the shell, simplify the analysis, which
nevertheless, as we will see, can be quite complex.

\subsection{Minkowski interior and Reissner-Nordstr\"om
exterior spacetimes}

\subsubsection{Minkowski interior spacetime}
\label{Sec:Mpacetime}

The interior spacetime $\mathcal{M}_{\rm i}$
that will be considered is the Minkowski spacetime.
The Minkowski spacetime is a totally empty spacetime
and is a
solution, a trivial solution, to the Einstein-Maxwell equations,
Eqs.~(\ref{eq:EFE1}) and (\ref{eq:Maxwell_FE}).
It is 
characterized in spherical coordinates
by the following line
element 
\begin{equation}
ds^{2}=-dt^{2}+dr^{2}+r^{2}d\Omega^{2}\,,
\label{eq:Mink_metric0}
\end{equation}
where $t$ and $r$ are the time and radial coordinates, respectively,
and $d\Omega^{2}\equiv d\theta^{2}+\sin^{2}\theta\,d\varphi^{2}$,
with $\theta$ and $\varphi$ being the angular coordinates.

\begin{figure}[h!]
\centering\includegraphics[height=0.29\paperheight]
{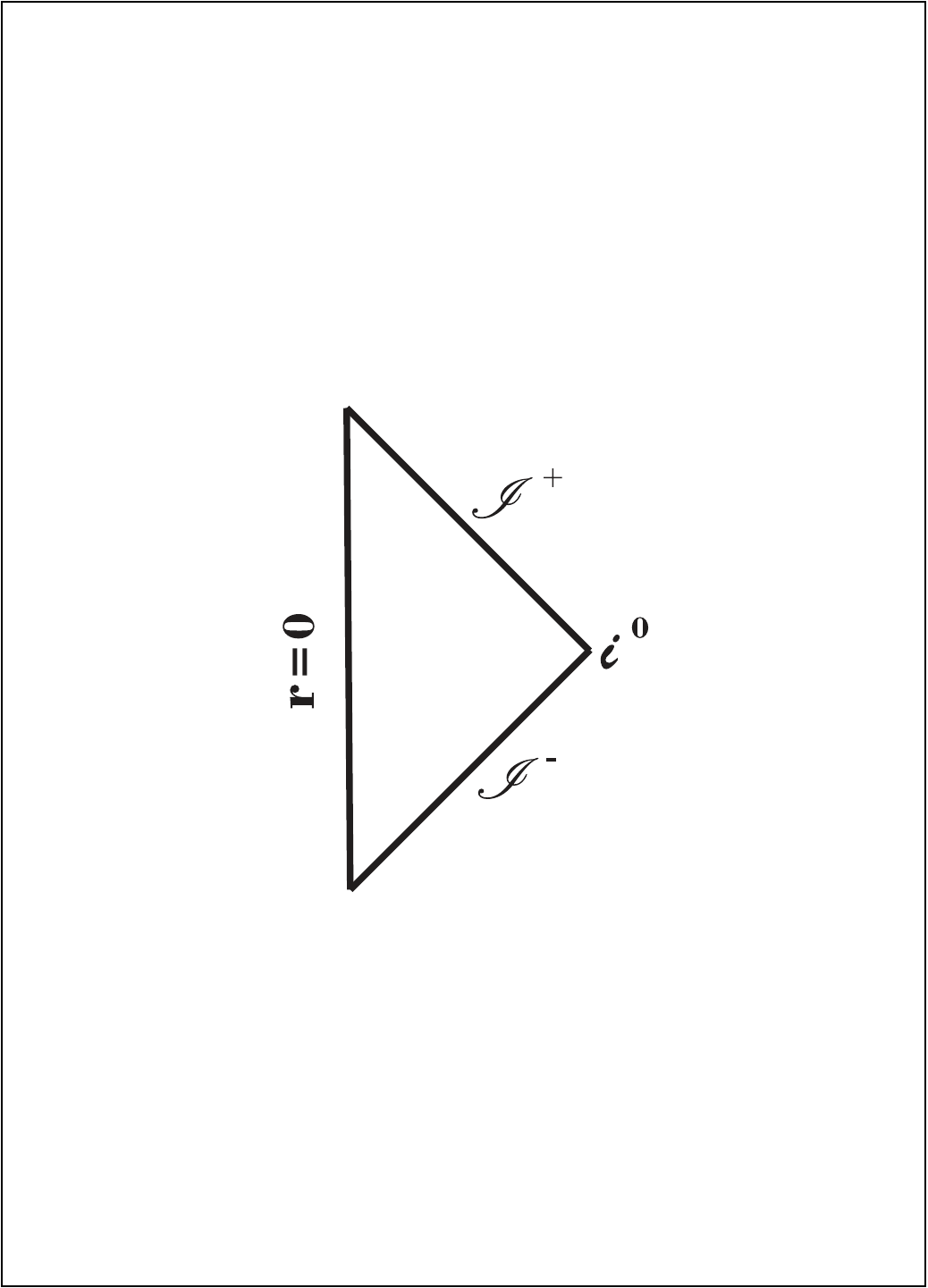}
\caption{\label{Fig:Minkowski_Carter_Penrose_Diagram}
Carter-Penrose diagram
of a spacetime described by the Minkowski solution.
See text for details.}
\end{figure}

The Carter-Penrose diagram of the Minkowski is given in
Figure~\ref{Fig:Minkowski_Carter_Penrose_Diagram}.  It is worth noting
the timelike line $r=0$ which is the origin of coordinates, the null
infinities, the past one ${\mathscr I}^-$ and the future one ${\mathscr
I}^+$, and the spacelike infinity $i_0$. There are also the timelike
infinities $i^-$ and $i^+$ at the lower and upper vertices of the
triangle that are not drawn to not overload the figure.  Each point in
the diagram represents a two-sphere of radius $r$.

\subsubsection{Reissner-Nordstr\"om exterior spacetime}
\label{Sec:RN_spacetime}

\begin{figure}[h]
{\includegraphics[height=0.29\paperheight]
{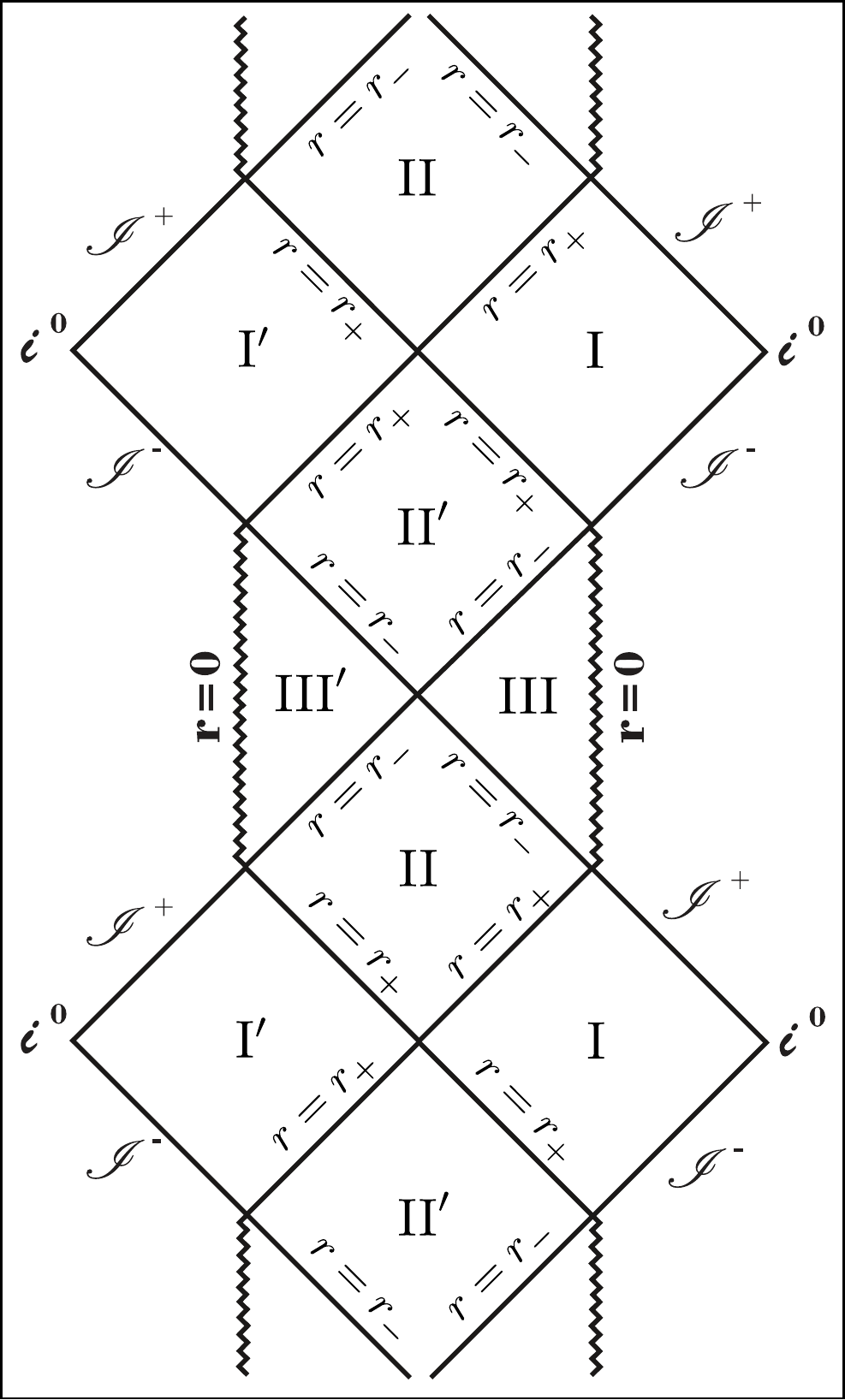}}
\caption{Carter-Penrose diagram of the nonextremal
Reissner-Nordstr\"om spacetime. See text for details.}
\label{Fig:Penrose_diagram_RN_non_extremal}
\end{figure}

\begin{figure}[h]
{\includegraphics[height=0.29\paperheight]
{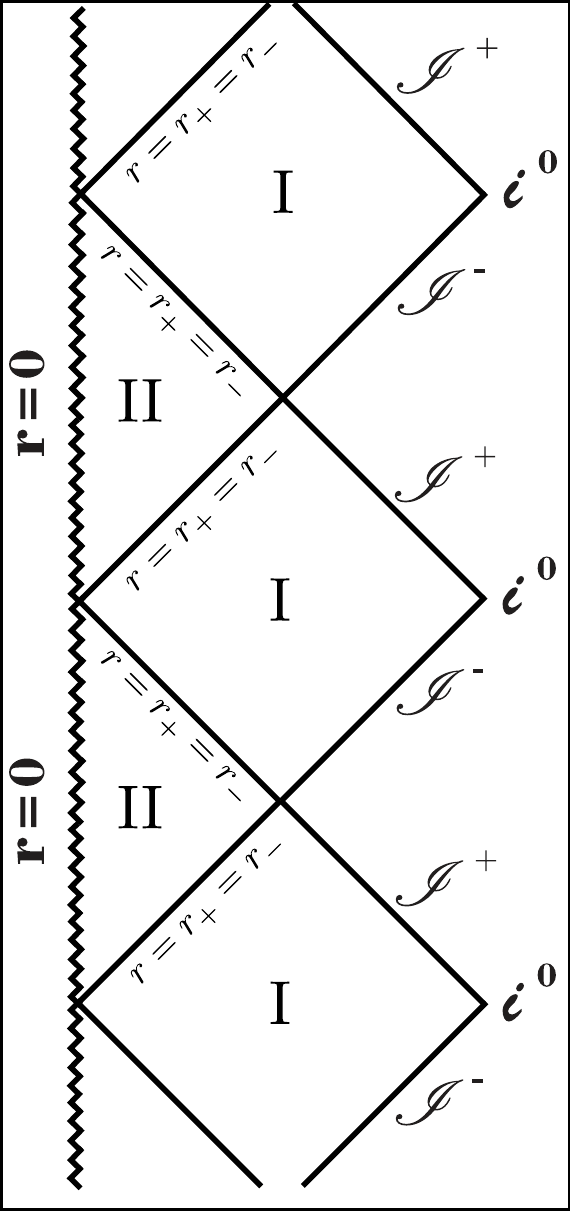}}
\caption{Carter-Penrose diagram of the
extremal Reissner-Nordstr\"om
spacetime. See text for details.}
\label{Fig:Penrose_diagram_RN_extremal}
\end{figure}

\begin{figure}[h]
{\includegraphics[height=0.29\paperheight]
{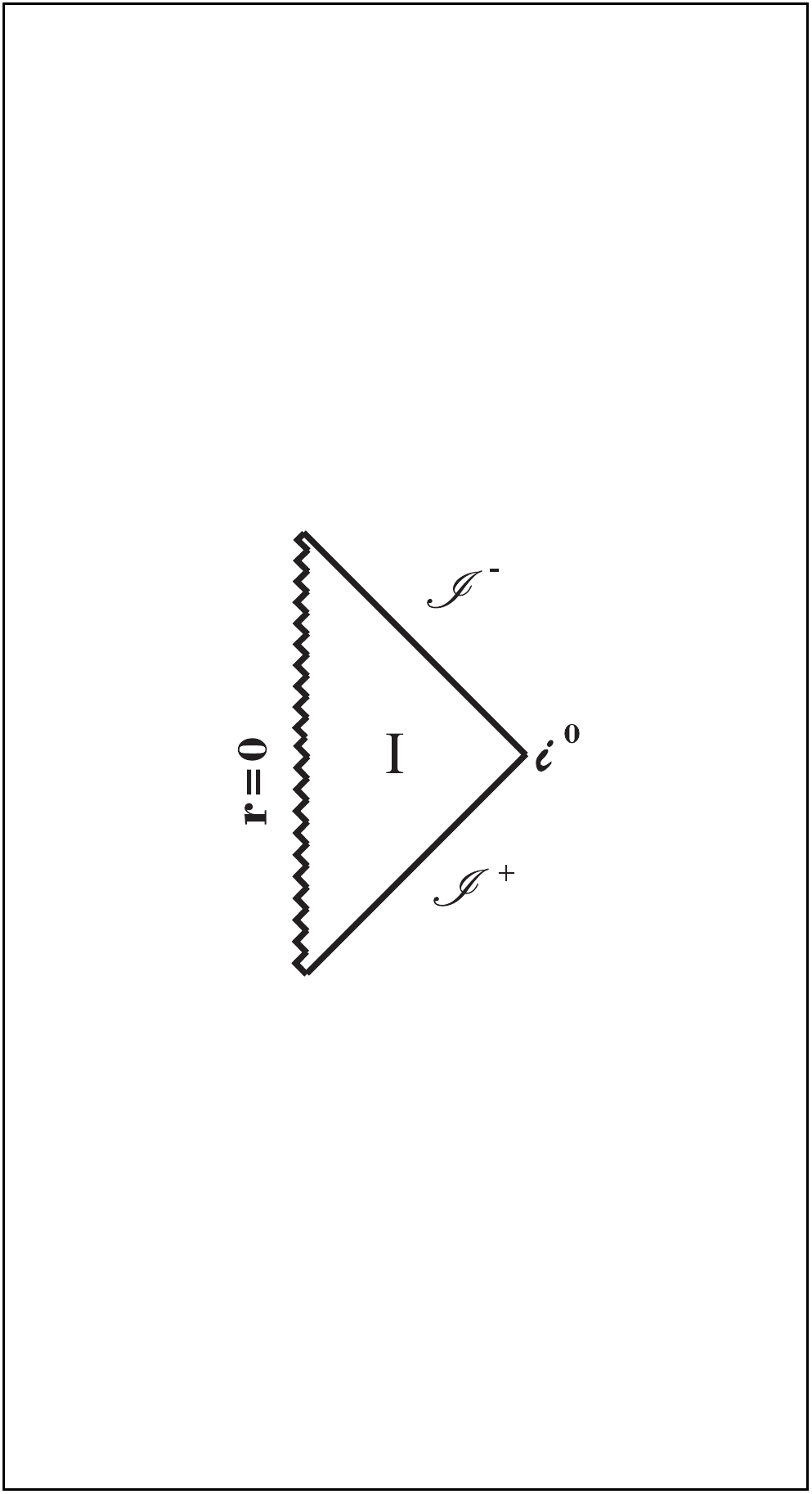}}
\caption{Carter-Penrose diagram of the
overcharged Reissner-Nordstr\"om spacetime. See text for details.}
\label{Fig:Penrose_diagram_RN_overcharged}
\end{figure}

The exterior spacetime $\mathcal{M}_{\rm e}$
that will be considered is the Reissner-Nordstr\"om spacetime.
The Reissner-Nordstr\"om spacetime is a
solution fo the Einstein-Maxwell equations,
Eqs.~(\ref{eq:EFE1}) and (\ref{eq:Maxwell_FE}),
and 
describes an empty massive electrically charged spherically
symmetric spacetime.
The full vacuum  Reissner-Nordstr\"om spacetime is the unique vacuum
spherically symmetric electric charged
solution in general relativity, following Birkhoff's
theorem.
In
what are commonly
called Schwarzschild coordinates $\left(t,r,\theta,\varphi\right)$,
the Reissner-Nordstr\"om spacetime is a solution of
the Einstein-Maxwell
field equations
characterized by the following line
element 
\begin{equation}
ds^{2}=-\left(1-\frac{2M}{r}
+\frac{Q^2}{r^2}\right)dt^{2}+
\frac{dr^{2}}{
1-\frac{2M}{r}
+\frac{Q^2}{r^2}}+r^{2}d\Omega^{2}\,.
\label{eq:RN_metric0MQ}
\end{equation}
with $M$ being the spacetime mass and $Q$ its charge.
We assume the mass to be zero or positive
$M\geq0$ and $Q$ we treat, without loss
of generality here,
as the modulus of the electric charge, be it
positive or negative. 
A function
that turns up often is the redshift function $k(r,M,Q)$ at
some radius $r$, or simply $k$, given by 
\begin{equation}
k=\sqrt{1-\frac{2M}{r}+\frac{Q^{2}}{r^{2}}}\,.
\label{eq:redshift2}
\end{equation}
The Faraday-Maxwell tensor, solution of Maxwell field equation in
Eq.~(\ref{eq:Maxwell_FE}), is in terms of $Q$ given by
\begin{equation}
F_{rt}=-F_{tr}=\frac{Q}{r^{2}}\,,
\label{eq:RN_FaradayMaxwell_valueQ}
\end{equation}
with the remaining components being identically null.
When $\frac{Q}{M}\leq1$ there are two important
characteristic radii, 
the event horizon radius $r_+$
and the Cauchy horizon $r_-$,
with $M$ and $Q$ being given in terms of these by
\begin{equation}
M=\frac{r_++r_-}{2}\,,\quad Q=\sqrt{r_+r_-}\,.
\label{eq:M_Q_radius_relations}
\end{equation}
The line element Eq.~(\ref{eq:RN_metric0MQ})
then represents a black hole solution and can written as
\begin{equation}
ds^{2}=-\left(1-\frac{r_+}{r}\right)\left(1-\frac{r_-}{r}
\right)dt^{2}+\frac{dr^{2}}{\left(1-\frac{r_+}{r}\right)
\left(1-\frac{r_-}{r}\right)}+r^{2}d\Omega^{2}\,,
\label{eq:RN_metric0}
\end{equation}
The redshift function 
of Eq.~(\ref{eq:redshift2})
at
some radius $r$, is now $k(r,r_+,r_-)$
given by
\begin{equation}
k=\sqrt{\left(1-\frac{r_+}{r}\right)
\left(1-\frac{r_-}{r}\right)}\,.
\label{eq:redshift}
\end{equation}
The Faraday-Maxwell tensor
of Eq.~(\ref{eq:RN_FaradayMaxwell_valueQ})
is in terms of $r_+$ and $r_-$  given by
\begin{equation}
F_{rt}=-F_{tr}=\frac{\sqrt{r_+r_-}}{r^{2}}\,.
\label{eq:RN_FaradayMaxwell_value}
\end{equation}
Inverting Eq.~(\ref{eq:M_Q_radius_relations}) 
one obtains $r_+$ and $r_-$
in terms of the spacetime mass
$M$ and the electric charge $Q$ as 
\begin{equation}
r_+=M+\sqrt{M^{2}-Q^{2}}\,,\quad
r_-=M-\sqrt{M^{2}-Q^{2}}\,.
\label{eq:KS_horizons_radius0}
\end{equation}
When the solution is nonextremal, one has $r_+>r_-$,
which in terms of $M$ and $Q$ is given by the condition
$M^{2}>Q^{2}$,
see Eq.~(\ref{eq:KS_horizons_radius0}),
i.e., $\frac{Q}{M}\leq1$.
For the nonextremal solution we work with the line element
as given in Eq.~(\ref{eq:RN_metric0}).
A particular
important instance here, is when $r_-=0$,
i.e., the electric charge is zero, $Q=0$, $r_+=2M$, and
the Schwarzschild solution is recovered,
$ds^{2}=-\left(1-\frac{r_+}{r}\right)dt^{2}+
\frac{dr^{2}}{1-\frac{r_+}{r}} +r^{2}d\Omega^{2}$.
For the extremal solution, one has $r_+=r_-$, which in terms of $M$
and $Q$ is given by the condition $M^{2}=Q^{2}$,
see Eq.~(\ref{eq:KS_horizons_radius0}), i.e., $\frac{Q}{M}=1$.
For the extremal solution we work with the line element
as given in Eq.~(\ref{eq:RN_metric0}) putting $r_+=r_-$.
For the overcharged
solution, $r_+$ and $r_-$ take complex values, and in terms of $M$ and
$Q$ it is given by the condition $M^{2}<Q^{2}$, 
see Eq.~(\ref{eq:KS_horizons_radius0}), i.e., $\frac{Q}{M}>1$,
so for the overcharged solution it is definitely
better to work not with the line element
of Eq.~(\ref{eq:RN_metric0})
but with the line element of Eq.~(\ref{eq:RN_metric0MQ}).
The Reissner-Nordstr\"om spacetime, given by the line element
Eq.~(\ref{eq:RN_metric0}),
or  Eq.~(\ref{eq:RN_metric0MQ}),
has a maximal extension.  The
Carter-Penrose diagrams for the three
possible solutions, namely, nonextremal,
extremal, and
overcharged~\citep{Carter_1966_2,Hawking_Ellis_book,MTW_Book},
are given in
Figures~\ref{Fig:Penrose_diagram_RN_non_extremal},
\ref{Fig:Penrose_diagram_RN_extremal}, and
\ref{Fig:Penrose_diagram_RN_overcharged}, respectively.
Looking at the Carter-Penrose diagrams it is clear that the
nonextremal, extremal, and overcharged Reissner-Nordstr\"om
spacetimes have a very distinct causal structure.
Let us look at them one at a time. 

For the nonextremal solution,
$r_+>r_-$, see Figure~\ref{Fig:Penrose_diagram_RN_non_extremal},
the Carter-Penrose
diagram shows that
there is a central block that repeats itself.
The central block is composed of regions I and I',
regions II and II', and regions III and III'.
Regions I and I' are delimited by the
two null lines
$r_+$, which are the event horizon of the spacetime,
the past
${\mathscr I}^-$ and the future ${\mathscr I}^+$,
the spacelike infinity $i_0$, and the timelike
infinities $i^-$ and $i^+$ at the lower and upper vertices of the
triangle that are not drawn to not overload the figure,
regions II and II' are delimited by $r_+$ and $r_-$, the
event and the Cauchy horizons, respectively, and regions III and
III' by $r_-$ and the singularity at $r = 0$.
From region I to I' there is an Einstein-Rosen bridge, more precisely, a
dynamic wormhole. Region II' is a white hole region, region II a black hole.
Thus, in summary the central block is composed of a white and a black
hole and two asymptotically flat regions connected by a wormhole that end in a
region that contains Cauchy horizons which in turn can
be past to the singularities at $r=0$.
A detailed analysis of the
causal structure and the
construction of the diagram is given in
Appendix~\ref{Appendix_sec:Kruskal-Szekeres_coordinates_RN} for the
mathematics and properties of the maximal analytical extension of the
vacuum  nonextremal 
Reissner-Nordstr\"om solution.
It will be seen that instead of the $(t,r)$ coordinates
of Schwarzschild it is necessary to resort to $(T,X)$
coordinates of Kruskal to have a better hold on the extension.

For the extremal solution, $r_+=r_-$,
see Figure~\ref{Fig:Penrose_diagram_RN_extremal},
 the Carter-Penrose
diagram shows that there are no mirror regions, and so there is no
Einstein-Rosen bridge, i.e., no dynamic wormhole.
Region I is the asymptotically
flat region,
and region II is a black and white hole together.

For the overcharged solution, $r_+$ and $r_-$ do
not exist, see 
Figure~\ref{Fig:Penrose_diagram_RN_overcharged}, the Carter-Penrose
diagram shows that there is only the asymptotic region with $r=0$ being
both the origin of coordinates and a timelike singularity.

\newpage
\centerline{}
\newpage

\subsection{Classification scheme and nomenclature}

\subsubsection{Classification scheme}
\label{Sec:Classification}

Armed with the necessary formalism to make a junction of an interior
Minkowski to an exterior Reissner-Nordstr\"om solution we can now
proceed to find and classify all fundamental electrically charged
static thin shells.

There are three distinct main electric states for the thin shell,
namely, nonextremal, extremal, and overcharged.  These yield an
exterior Reissner-Nordstr\"om spacetime, that is nonextremal, i.e.,
$\frac{Q}{M}<1$ or $r_+>r_-$, extremal, i.e., $\frac{Q}{M}=1$ or
$r_+=r_-$, and overcharged, i.e., $\frac{Q}{M}>1$, respectively.

For each of the three main states one has to assign a location for the
shell, assumed to have radius $R$.  In the nonextremal state,
$r_+>r_-$, there are two possible locations, the shell can be located
outside $r_+$, i.e., $R>r_+$, or inside $r_-$, i.e., $R<r_-$.  In the
extremal state, $r_+$ and $r_-$ coincide, $r_+=r_-$, there are three
possible locations, the shell can be located outside $r_+$, i.e.,
$R>r_+$, inside $r_+$, i.e., $R<r_+$, or at $r_+$, i.e., $R=r_+$.  In
the overcharged state, there are no $r_+$ and $r_-$, there is one
generic location, the shell can be located anywhere without
distinction, $R>0$.

For each of the locations one has to assign an orientation for the
normal vector to the shell, indeed as seen by an external observer the
normal vector to the shell can either point outward to increasing $r$
or point inward to decreasing $r$.
Thus, in the nonextremal state with the
shell located
outside $r_+$, one can have either a shell with the orientation
of the normal vector pointing to spatial infinity or 
one can have a shell with the orientation
of the normal vector pointing to $r_+$.
In the nonextremal state with the
shell located inside $r_-$, one can have either a shell with the
orientation
of the normal vector pointing to $r_-$ or 
one can have a shell with the orientation
of the normal vector pointing to $r=0$.
In the extremal state with the
shell located outside $r_+$, one can have either a shell with the orientation
of the normal vector pointing to  spatial infinity or 
one can have a shell with the orientation
of the normal vector pointing to $r=r_+$.
In the extremal state with the
shell located inside $r_+$, one can have  either a shell with the orientation
of the normal vector pointing to $r_+$ or 
one can have a shell with the orientation
of the normal vector pointing to $r=0$.
In the extremal state with the shell located at $r_+$, one can have
either a shell with the orientation of the normal vector pointing to
spatial infinity
or one can have a shell with the orientation of the
normal vector pointing to $r=0$, and for each of these two
orientations, there are two possible approaches, the shell approaches
$r_+$ from above, i.e., $R>r_+$ with $R\to r_+$, or 
the shell approaches
$r_+$ from below, i.e., $R<r_+$ with $R\to r_+$.
In the overcharged state,  with the
shell  located anywhere, one can have  either
a shell with the orientation
of the normal vector pointing to spatial infinity or 
one can have a shell with the orientation
of the normal vector pointing to $r=0$.

So, the classification we perform for a fundamental electric thin shell
and respective spacetime, namely, state, location, orientation, yields
fourteen different cases.  All the fourteen cases, four
for nonextremal, eight for extremal, and two for overcharged
will be analyzed, in particular, 
the physics and geometry of all the cases will be displayed.

\subsubsection{Nomenclature}
\label{Sec:Remarks}
There is a question of nomenclature that we must clarify.  Note that
the thin shell spacetime solution has a characteristic radius which is
the radius $R$ of the location of the shell.  For a shell in the
nonextremal state, $r_+>r_-$, the exterior Reissner-Nordstr\"om
exterior solution has two further characteristic radii, namely, $r_+$
and $r_-$ themselves.  The question of nomenclature is the distinction
between gravitational radius and event horizon radius on one hand, and
between Cauchy radius and Cauchy horizon radius on the other.  The
gravitational radius and Cauchy radius of a nonextremal spacetime are
a characteristic of the spacetime, more precisely, given a mass $M$
and charge $Q$ then there is a one-to-one correspondence to $r_+$ and
$r_-$.  It can happen that the nonextremal shell spacetime has no
horizons in which situation $r_+$ and $r_-$ are simply the
gravitational radius and Cauchy radius of the spacetime, respectively,
or it can happen that the nonextremal shell spacetime has horizons, in
which situation the  event
horizon radius $r_+$ is also the
gravitational radius and the Cauchy horizon
radius $r_-$ is also the  Cauchy radius.
E.g., a nonextremal star object for which $\frac{Q}{M}<1$ has
gravitational radius but no event horizon, since the spacetime in
which it is inserted has no event horizon.  For the same reason, it
also has a Cauchy radius but no Cauchy horizon.  On the other hand, a
nonextremal black hole for which also $\frac{Q}{M}<1$ has the property
that its gravitational radius is also
its event horizon radius and its Cauchy radius
is also its Cauchy horizon radius.
For a shell in the extremal state, $r_+=r_-$,
besides the radius $R$ of the shell, the exterior
Reissner-Nordstr\"om exterior solution has one characteristic
radius, namely, $r_+$, one could use $r_-$ also but it is clearly more
appropriate to use $r_+$.  As before, depending on the location of the
shell, $r_+$ can be a gravitational radius alone when the extremal
shell spacetime has no horizon, or it can be a gravitational radius
and an event horizon radius as well, when the shell spacetime has a
horizon.  For a shell in the overcharged state, $r_+$ and $r_-$ are
not defined, and so the nomenclature does not apply.

Let us see in detail this nomenclature when applied to
the fundamental electric thin shells.
First, the shell in the nonextremal state has four cases.
If the radius $R$ of the shell is greater than $r_+$ and its normal
points towards spatial infinity, so the shell is located in region I of
Figure~\ref{Fig:Penrose_diagram_RN_non_extremal}, then there is no
event horizon and no Cauchy horizon, and $r_+$ is the gravitational
radius and $r_-$ is the Cauchy radius.
If the radius $R$ of the shell is greater than $r_+$ and its normal
points towards $r_+$, so the shell is located in region I' of
Figure~\ref{Fig:Penrose_diagram_RN_non_extremal}, then there are event
and Cauchy horizons, and $r_+$ is, in addition to a gravitational
radius, an event horizon radius and $r_-$ is also the Cauchy horizon
radius.
If the radius $R$ of the shell is less than $r_-$ and its normal
points towards $r_-$, so the shell is located in region III or III' of
Figure~\ref{Fig:Penrose_diagram_RN_non_extremal}, then $r_+$ is the
event horizon radius and $r_-$ is the Cauchy horizon radius.
If the radius $R$ of the shell is less than $r_-$ and its normal
points towards the singularity $r=0$, so the shell is still
located in
region III or III' of
Figure~\ref{Fig:Penrose_diagram_RN_non_extremal}, then there is no
event horizon and no Cauchy horizon, and $r_+$ is the gravitational
radius and $r_-$ is the Cauchy radius.
Second, the shell in the extremal state has eight cases.  For the eight
cases of the extremal state one has that, since $r_+=r_-$, there is no
need for the name Cauchy which drops out, and the radius $r_+=r_-$ is
called gravitational radius and event horizon radius in the
appropriate cases. Here one follows the nonextremal nomenclature, see
also Figure~\ref{Fig:Penrose_diagram_RN_extremal} representing the
shell's exterior region.
Third, the shell in the overcharged state has two cases.  The
nomenclature in the overcharged state with
Figure~\ref{Fig:Penrose_diagram_RN_overcharged} representing the
shell's exterior region, does not apply since $r_+$ and $r_-$ do not
exist.

It is clearly convenient to use this nomenclature and to distinguish
when there no horizons from when there are horizons. We follow it
in the 
study of the fourteen different cases.

\newpage



\section{Nonextremal electric thin shells outside the gravitational
radius: Star shells and tension shell black holes}
\label{Sec:Nonextremal-thin-shells_outside_event_horizon}

\subsection{Nonextremal electric thin shells outside the
gravitational radius: Star shells}
\label{Subsec:nonextremalnormaloutside}

Here we study the case of a fundamental electric thin
shell in the
nonextremal state, i.e., $r_+>r_-$ or $M>Q$, for which the shell's 
location obeys $R>r_+$, and for which the orientation is such that the
normal to the shell points towards spatial infinity.  In this case
horizons do not exist and so, following the nomenclature, $r_+$ and
$r_-$ are the gravitational radius and Cauchy radius, respectively.
Two remarks are important. First, when we write 
$M>Q$ it is meant $M>|Q|$, but to simplify the notation,
we drop the modulus in these instances, expecting that
the context makes clear the meaning. Second, the normal to the shell
pointing towards spatial infinity means in the notation of the Kruskal
coordinate $X$ that we take $\text{sign}\left(X\right)=+1$, see the
end of this section and
Appendix~\ref{Appendix_sec:Kruskal-Szekeres_coordinates_RN} for
details.

\begin{figure}[h!]
\subfloat[]{\includegraphics[scale=0.40]{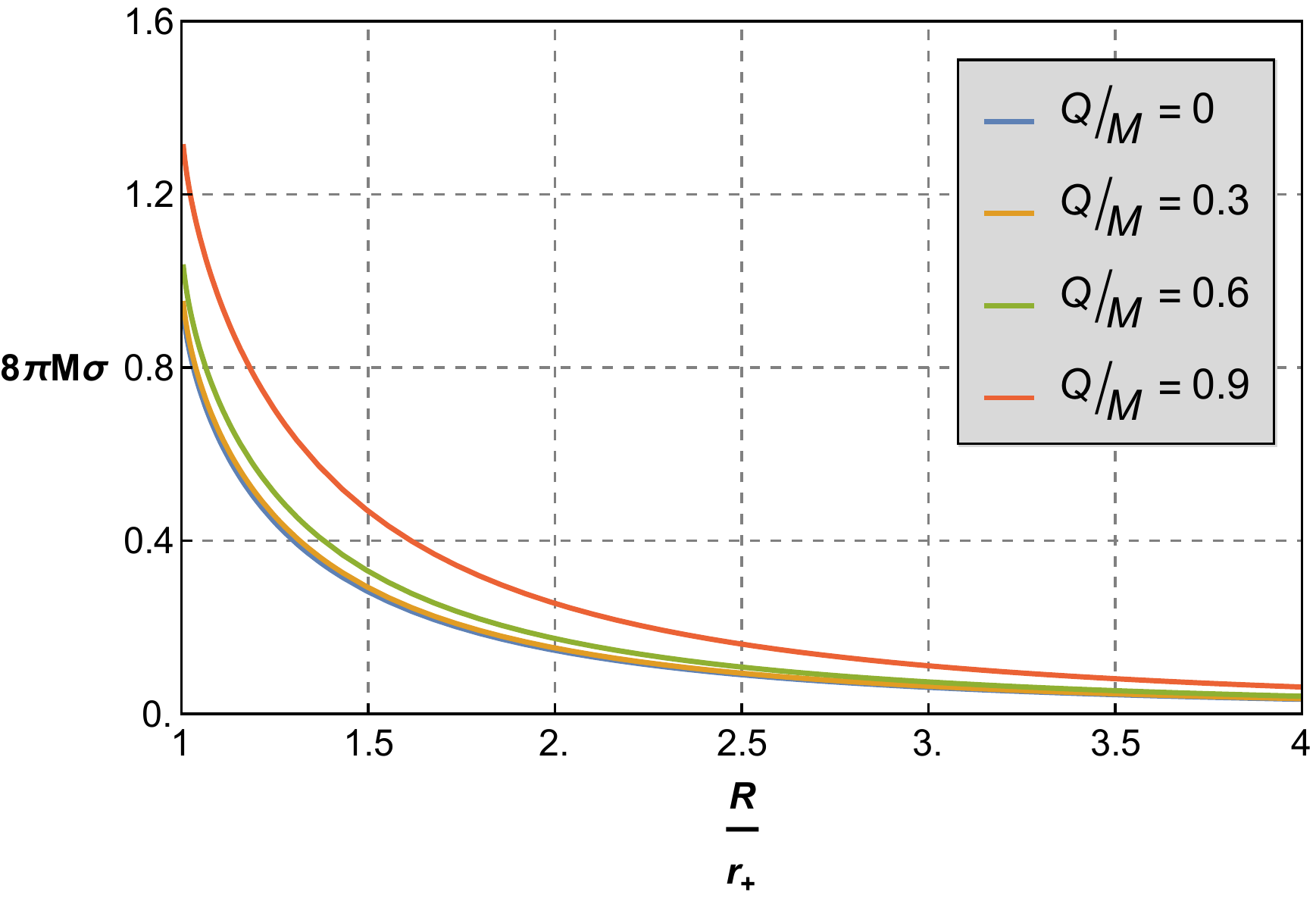}}
\hspace*{\fill}
\subfloat[\label{Fig:pressure_subregion_I}]
{\includegraphics[scale=0.40]{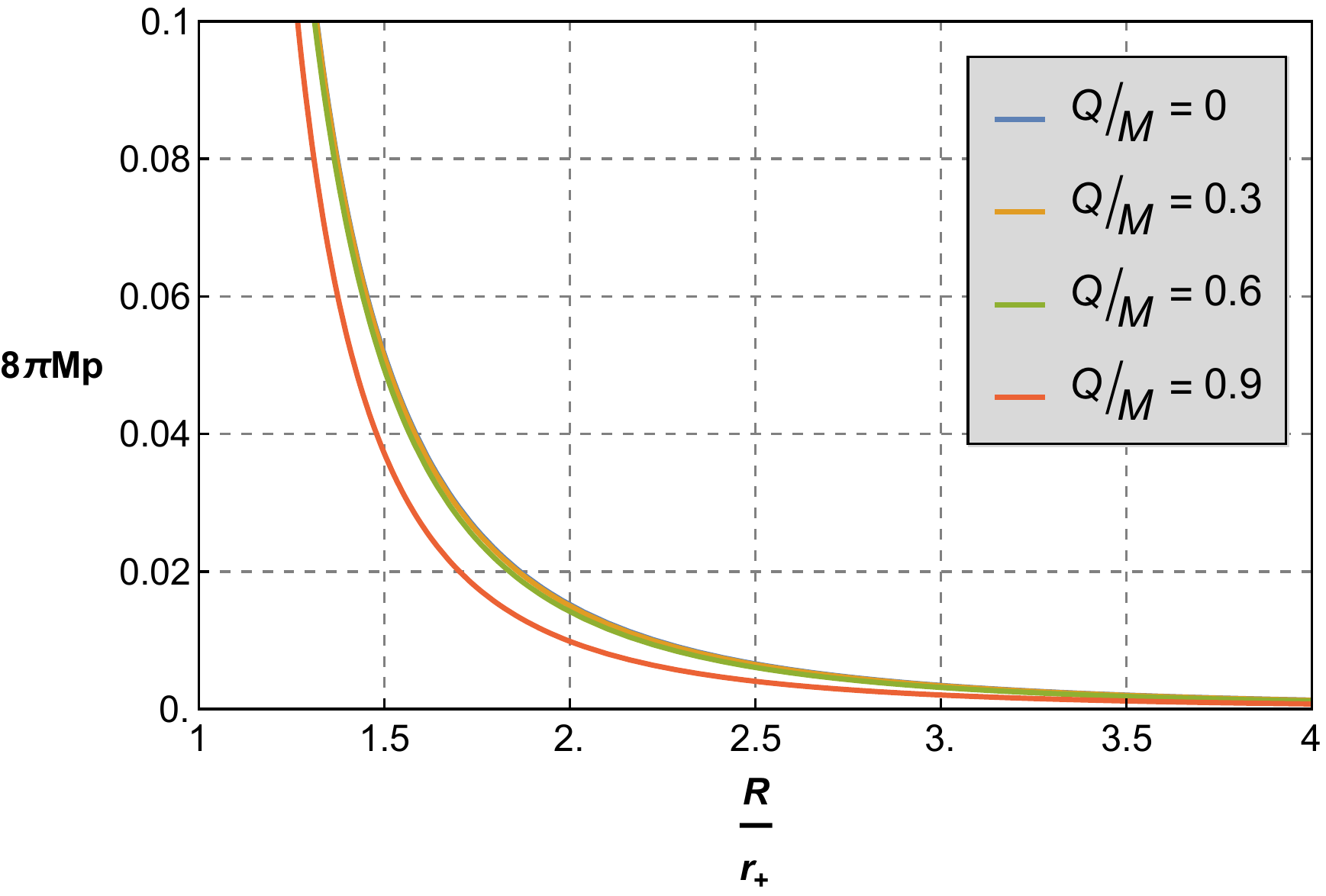}}
\caption{\label{Fig:Properties_region_I}
Physical properties of a nonextremal
star shell, i.e., an electric perfect fluid
thin shell in a nonextremal Reissner-Nordstr\"om state, in the
location $R>r_+$, i.e., located outside the gravitational radius, and
with orientation such that the normal points towards
spatial infinity.  The
interior is Minkowski and the exterior is nonextremal
Reissner-Nordstr\"om spacetime.  Panel (a) Energy density $\sigma$ of
the shell as a function of the radius $R$ of the shell for various
values of the $\frac{Q}{M}$ ratio. The energy density is
adimensionalized through the mass $M$, $8\pi M\sigma$, and the radius
is adimensionalized through the gravitational radius $r_+$,
$\frac{R}{r_+}$.  Panel (b) Pressure $p$ on the shell as a function of
the radius $R$ of the shell for various values of the $\frac{Q}{M}$
ratio. The pressure is adimensionalized through the mass $M$, $8\pi
Mp$, and the radius is adimensionalized through the gravitational
radius $r_+$, $\frac{R}{r_+}$.
}
\end{figure}

As functions of $M$, $Q$, and $R$, the shell's energy density $\sigma$
and pressure $p$, are, see the end of this section,
\begin{equation}
8\pi\sigma=\frac{2}{R}\left(1-k\right)\,,\
\label{eq:sigma_value_rplus_MQ_sign_plus}
\end{equation}
\begin{equation}
8\pi p=\frac{1}{2Rk}\left[\left(1-k\right)^{2}-
\frac{Q^{2}}{R^{2}}\right]\,,
\label{eq:pressure_value_rplus_MQ_sign_plus}
\end{equation}
with $k=\sqrt{1-\frac{2M}{R}+\frac{Q^{2}}{R^{2}}}$.
Also, the electric charge density $\sigma_{e}$
is given in terms of $M$, $Q$, and $R$, by
\begin{equation}
8\pi\sigma_{e}=\frac{2Q}{ R^{2}}\,,
\label{eq:chargedensity12}
\end{equation}
explicitly showing that the shell's electric charge density is the
source of the exterior Reissner-Nordstr\"om spacetime electric
charge. 
The behavior of $\sigma$ and $p$ as functions of the radius 
$R$ of the shell for various values of the $\frac{Q}{M}$ ratio in this
case is shown in Fig.~\ref{Fig:Properties_region_I}.
Since $M>Q$, and also $R>r_+$, and so $R>M$,
one has that $\sigma$ and $p$,
Eqs.~(\ref{eq:sigma_value_rplus_MQ_sign_plus}) and
(\ref{eq:pressure_value_rplus_MQ_sign_plus}), are always positive for
this type of shells, as can be also checked
in Fig.~\ref{Fig:Properties_region_I}.
These are the star shells.  Qualitatively, one
can understand why the pressure is positive for shells with normal
pointing towards spatial infinity, i.e.,
$\text{sign}\left(X\right)=+1$.  A free-falling particle in the region
outside the shell sees a geometry that is indistinguishable from the
nonextremal Reissner-Nordstr\"om spacetime. Therefore, a particle,
momentarily comoving with the shell but detached from it say, would
tend to fall to the inside as if an event horizon existed. Therefore,
in order to be static, a thin shell located at the junction
hypersurface with $R>r_+$ must by supported by pressure.
Notice from Fig.~\ref{Fig:Properties_region_I}
that as the charge $Q$ is increased one needs less
pressure support, as expected, the electric repulsion makes up
for the pressure. 
Notice that when
$R\to\infty$, the energy density $\sigma$, the
pressure $p$, and the charge
density $\sigma_{e}$, all tend to zero,
i.e., the shell disperses away.  Notice
also that when $R\to r_+$, the energy density is finite, but the
pressure of the shell goes to infinity,
while the electric charge
density is also finite. Indeed, for $R= r_+$ one
has a quasiblack hole. 
When $Q=0$ the outer solution is Schwarzschild.
In relation to the energy conditions of the shell one can work out and
find that the null and the weak energy conditions are verified for
$R>r_+$, the dominant energy condition for $R\geq R_{\mathrm{I}}$,
where $R_{\mathrm{I}}$
is some specific radius that we present later,
and the strong
energy condition for $R>r_+$, see a detailed presentation ahead.

\begin{figure}[h]
\includegraphics[height=0.39\paperheight]
{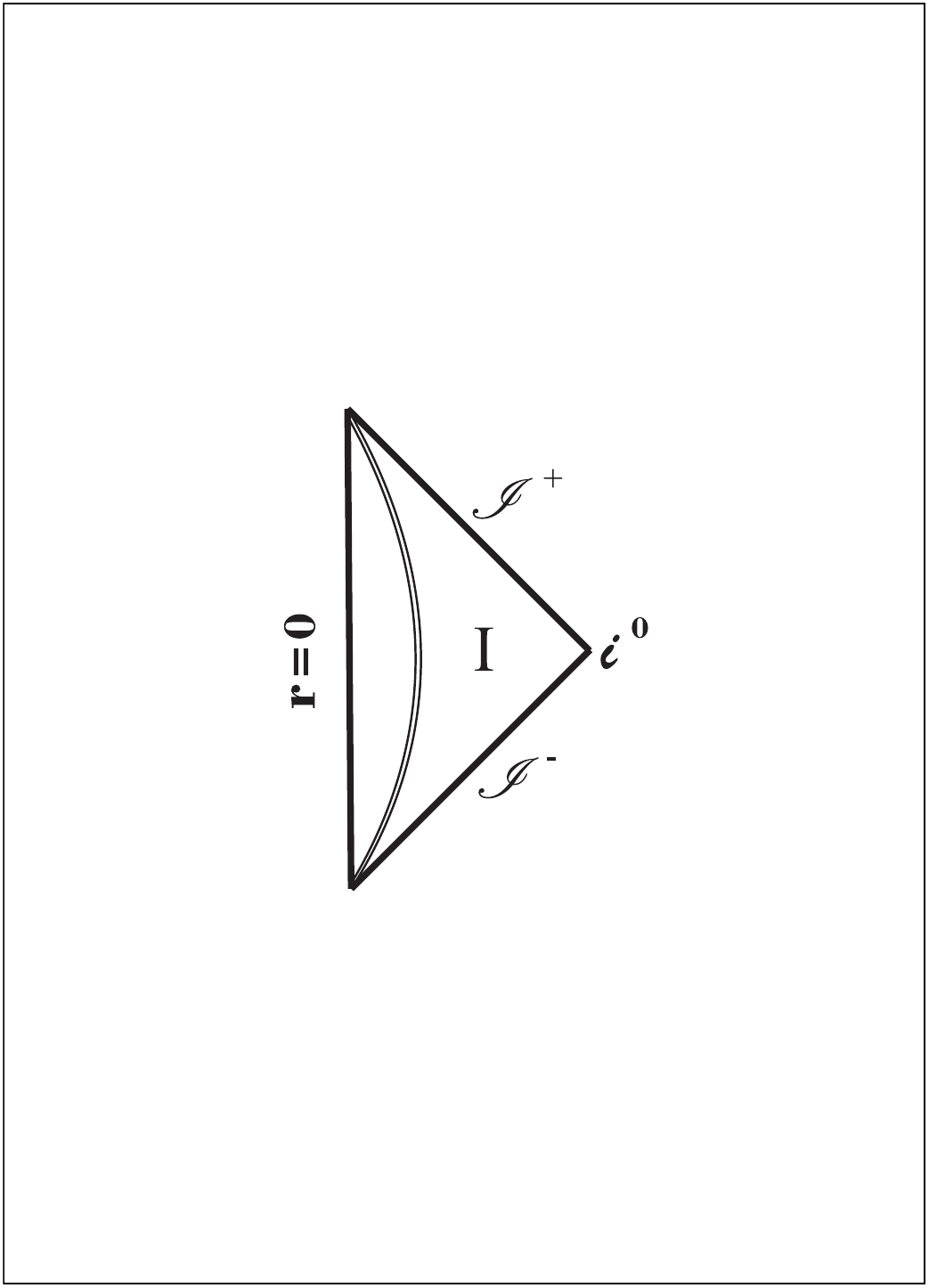}\caption{
\label{Fig:Penrose_diagram_Mink_RN_regions_I}
Carter-Penrose diagram of a star shell, i.e., a thin shell spacetime
in a nonextremal Reissner-Nordstr\"om state, in the location $R>r_+$,
i.e., located outside the gravitational radius, with orientation such
that the normal points towards spatial infinity.  The interior is
Minkowski, the exterior is nonextremal Reissner-Nordstr\"om.
For zero electric charge the exterior is
Schwarzschild, in which case the diagram looks the
same.}
\end{figure}

The Carter-Penrose diagram for this case
can be drawn directly from the building
blocks of an interior Minkowski spacetime and the
exterior asymptotic infinite region of the
nonextremal Reissner-Nordstr\"om spacetime.
In Fig.~\ref{Fig:Penrose_diagram_Mink_RN_regions_I}
the Carter-Penrose diagram of a shell spacetime
in a nonextremal Reissner-Nordstr\"om state,
which includes Schwarzschild,
in
the location $R>r_+$, with orientation such that
the normal points towards spatial infinity, i.e.,
$\text{sign}\left(X\right)=+1$, is shown.
It is clearly a star shell, a star in an asymptotically
flat spacetime.

The physical interpretation of this case is clear cut. This nonextremal
thin shell solution mimics a familiar star.  The energy density and
pressure obey the energy conditions if the radius of the shell is
sufficiently large. When this radius approaches the gravitational
radius the energy conditions are not obeyed, and at the gravitational
radius itself the solution turns into a quasiblack hole an object with
very interesting properties. The causal and global structure as
displayed by the Carter-Penrose diagram are well behaved and rather
elementary.  So, this case falls into the category of having the energy
conditions verified and the geometrical setup is physically
reasonable.

\newpage

\subsection{Nonextremal electric thin shells outside the event
horizon: Tension shell black holes}
\label{Subsec:nonextremaltensionoutside}

Here we study the case of a fundamental electric thin shell in the
nonextremal state, i.e., $r_+>r_-$ or $M>Q$, for which the shell's
location
obeys $R>r_+$, and for which the orientation is
such that the normal to the shell points towards $r_+$.  In this case
horizons do exist and so, following the nomenclature, $r_+$ is both
the gravitational and the event horizon radius, and $r_-$ is both the
Cauchy radius and the Cauchy horizon radius.  The normal to the shell
pointing towards $r_+$ means in the notation of the Kruskal coordinate
$X$ that we take $\text{sign}\left(X\right)=-1$, see the end of this
section and
Appendix~\ref{Appendix_sec:Kruskal-Szekeres_coordinates_RN} for
details.

\begin{figure}[h]
\subfloat[\label{Fig:energy_subregion_I_prime}]
{\includegraphics[scale=0.45]{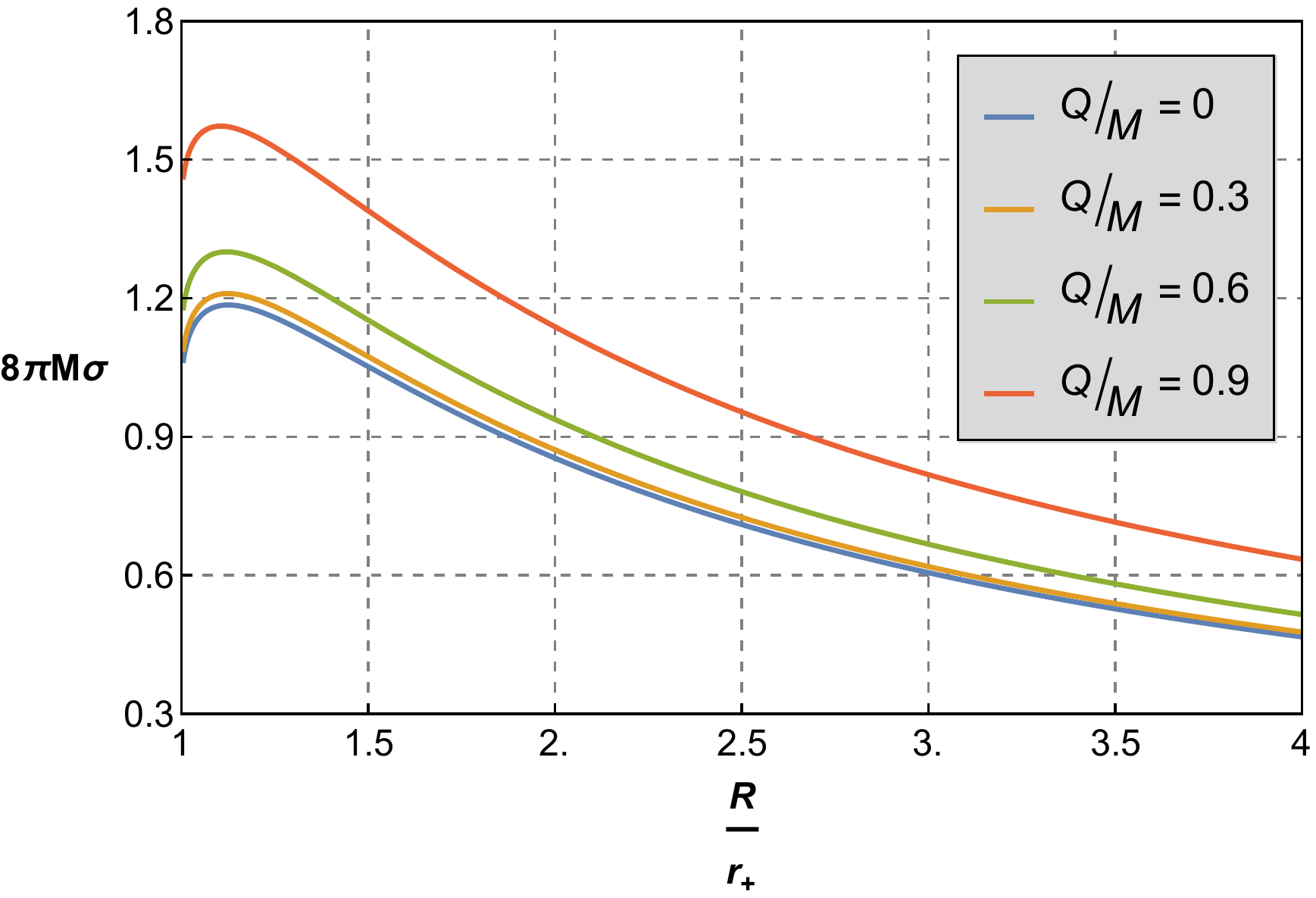}}
\hspace*{\fill}
\subfloat[\label{Fig:pressure_subregion_I_prime}]
{\includegraphics[scale=0.45]{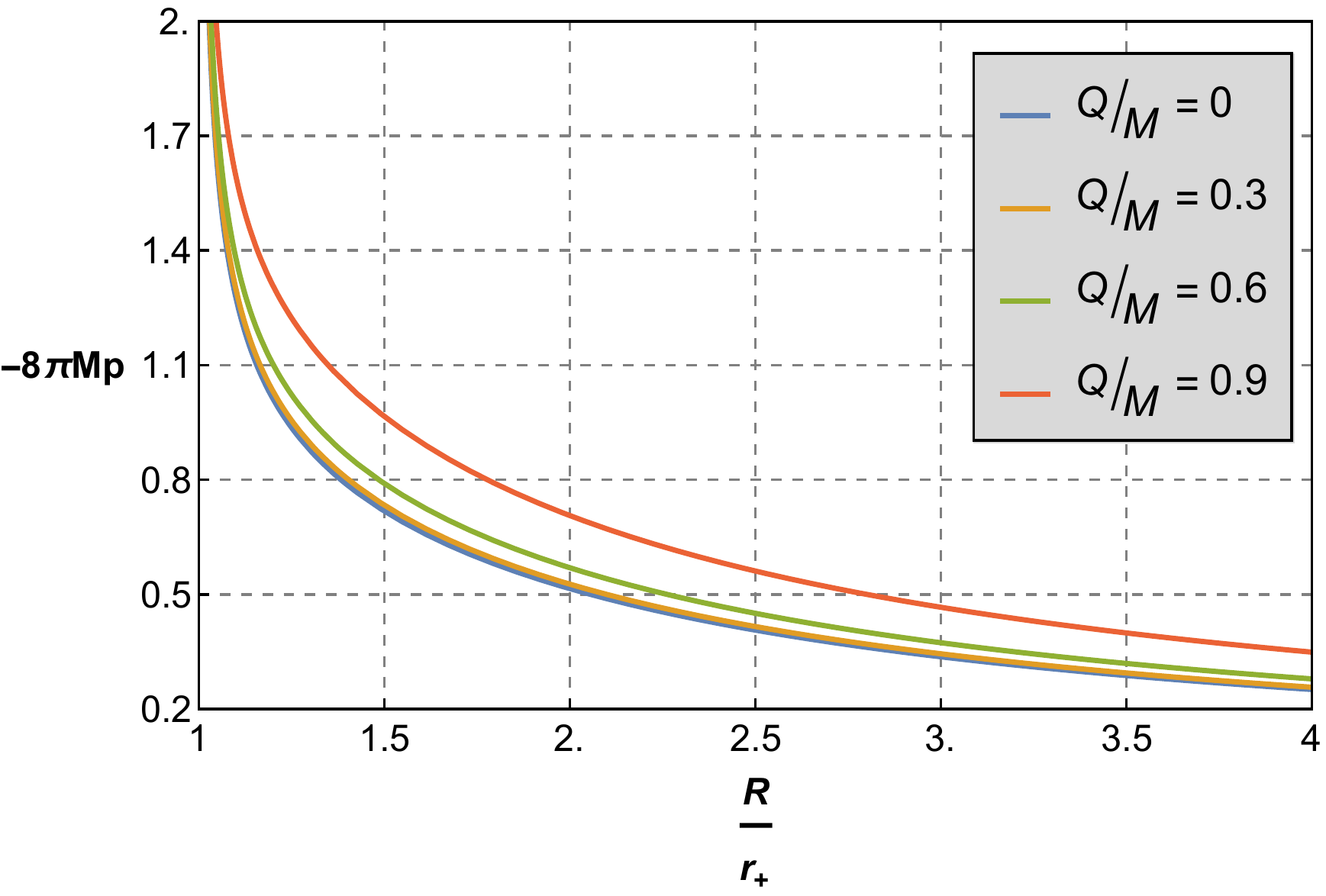}}
\caption{\label{Fig:Properties_region_I_prime}
Physical properties of a
nonextremal tension shell black hole,
i.e.,
an electric perfect fluid thin shell
in a nonextremal Reissner-Nordstr\"om state, in the
location $R>r_+$, i.e., located outside the event horizon, with
orientation such that the normal points towards $r_+$.  The interior
is Minkowski, the exterior is nonextremal Reissner-Nordstr\"om
spacetime.  Panel (a) Energy density $\sigma$ of the shell as a
function of the radius $R$ of the shell for various values of the
$\frac{Q}{M}$ ratio. The energy density is adimensionalized through
the mass $M$, $8\pi M\sigma$, and the radius is adimensionalized
through the gravitational radius $r_+$, $\frac{R}{r_+}$.  Panel (b)
Tension $-p$ on the shell as a function of the radius $R$ of the shell
for various values of the $\frac{Q}{M}$ ratio. The tension is
adimensionalized through the mass $M$, $-8\pi Mp$, and the radius is
adimensionalized through the gravitational radius $r_+$,
$\frac{R}{r_+}$.
}
\end{figure}

As functions of $M$, $Q$, and $R$,  
the shell's energy density $\sigma$ and pressure $p$,
are, see the end of this section, 
\begin{equation}
8\pi\sigma=\frac{2}{R}\left(1+k\right)\,,
\label{eq:sigma_value_rplus_MQ_sign_minus}
\end{equation}
\begin{equation}
8\pi p=-\frac{1}{2Rk}\left[\left(1+k\right)^{2}-\frac{Q^{2}}{R^{2}}
\right]\,,\label{eq:pressure_value_rplus_MQ_sign_minus}
\end{equation}
where the redshift parameter $k$ is again
$k=\sqrt{1-\frac{2M}{R}+\frac{Q^{2}}{R^{2}}}$.
The  electric charge density $\sigma_{e}$
is given in terms of $M$, $Q$, and $R$
also by Eq.~(\ref{eq:chargedensity12}).
The behavior of $\sigma$ and $p$ as functions of the radial coordinate
$R$ of the shell for various values of the $\frac{Q}{M}$ ratio in this
case is shown in 
Fig.~\ref{Fig:Properties_region_I_prime}.  We see that $\sigma$ is
always positive but $p$ is negative, it is
rather a tension. These are
the tension shells. 
Qualitatively, one can
understand why these shells,
with normal pointing to $r_+$, i.e.,
$\text{sign}\left(X\right)=-1$, must be
supported by tension, by remembering that a free-falling particle in the
region outside the event horizon will infall
towards the event horizon $r_+$
itself.  Therefore, a particle momentarily comoving with the shell but
detached from it will infall towards the black hole region of the
exterior Reissner-Nordstr\"om spacetime, hence, a perfect fluid thin
shell located at the junction hypersurface, in order to be static, must
by supported by tension.
Notice from Fig.~\ref{Fig:Properties_region_I_prime}
that as the charge $Q$ is increased one needs more
tension support, as expected, the electric repulsion obliges
an increase in the tension. 
Notice that here $R$ is finite, although it can be arbitrarily large,
in which case 
the energy density $\sigma$, the
tension $-p$, and the charge
density $\sigma_{e}$, all tend to zero.
Notice that $\sigma$ has a nonmonotonic behavior.
Notice
also that when $R\to r_+$, the energy density is finite, but the
tension of the shells goes to infinity,
while the charge
density is also finite. Indeed, for $R= r_+$ one
has a shell at the horizon with properties
similar to a 
quasiblack hole, although one with
additional structures.
When $Q=0$ the outer solution is Schwarzschild.
In relation to the energy conditions of the shell
one can work out and find
that the null, the weak, and the dominant
energy conditions are verified for $R\geq R_{I'}$, where
$R_{I'}$ is some specific radius that we present later,
and the strong energy condition is never verified,
see a detailed presentation ahead.

\newpage

\begin{figure}[h]
\subfloat[
\label{Fig:Penrose_diagram_Mink_RN_regions_Iprime1}]
{\includegraphics[height=0.31\paperheight]
{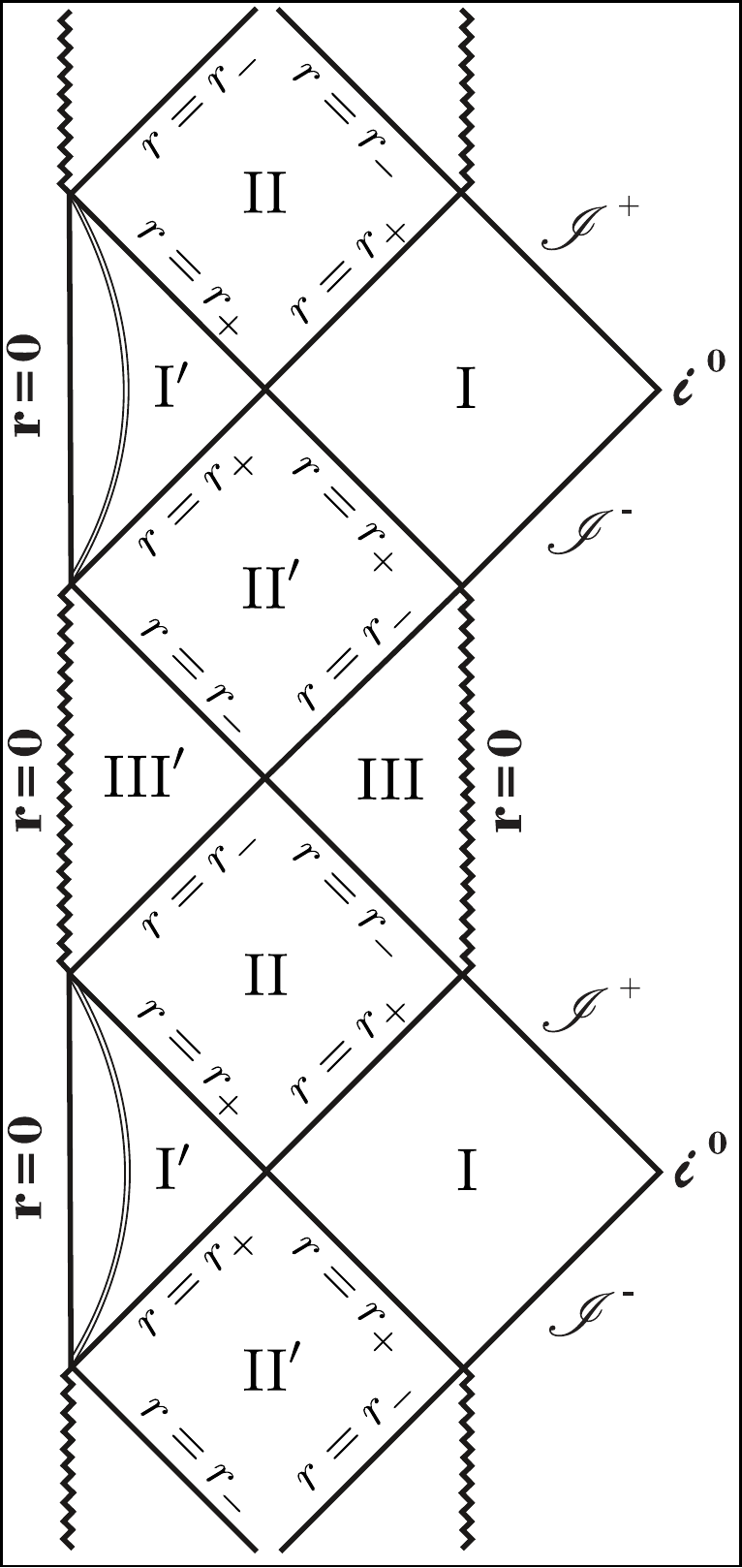}}
\hskip 2cm
\subfloat[\label{Fig:Penrose_diagram_Mink_RN_region_IIIrepeated}]
{\includegraphics[height=0.31\paperheight]
{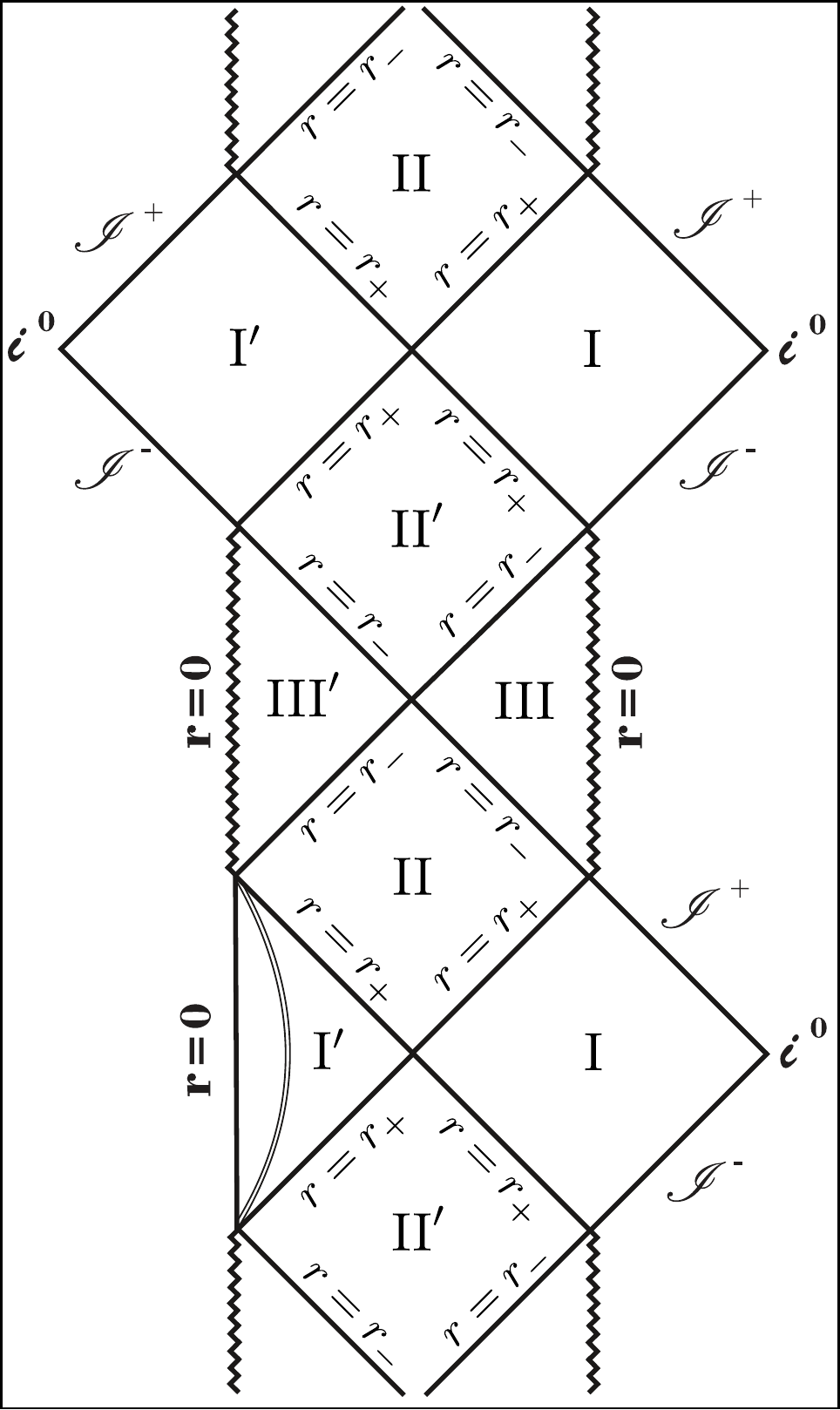}}
\caption{
\label{Fig:Penrose_diagram_Mink_RN_regions_Iprime}
Carter-Penrose diagrams of a tension shell black hole, i.e., a thin
shell spacetime in a nonextremal Reissner-Nordstr\"om state, in the
location $R>r_+$, i.e., located outside the event horizon, with
orientation such that the normal points towards $r_+$.  The interior
is Minkowski, the exterior is nonextremal Reissner-Nordstr\"om
spacetime. For zero electric charge the exterior is
Schwarzschild, in which case the timelike singularities
turn into spacelike ones.
Panel~(a) The Carter-Penrose diagram contains a shell in
the regions $\mathrm{I'}$ shown and another shell in the next region
$\mathrm{I'}$, which is repeated for all regions $\mathrm{I'}$.
Panel~(b) The Carter-Penrose diagram contains a shell in region
$\mathrm{I'}$ and an infinity in the next region $\mathrm{I'}$, which
is then repeated for all regions $\mathrm{I'}$. An infinite number of
different Carter-Penrose diagrams can be drawn, since there are an
infinite number of combinations to locate the shell and infinity.
}
\end{figure}

The Carter-Penrose diagram for this case can be drawn from the
building blocks of an interior Minkowski spacetime and the full
nonextremal Reissner-Nordstr\"om spacetime.  In
Fig.~\ref{Fig:Penrose_diagram_Mink_RN_regions_Iprime} the
Carter-Penrose diagram of a shell spacetime in a nonextremal
Reissner-Nordstr\"om state, in the location $R>r_+$, with orientation
such that the normal points towards $r_+$, i.e.,
$\text{sign}\left(X\right)=-1$, is shown.  In the diagram it is clear
that the tension shell is in the other side of the Carter-Penrose
diagram of a Reissner-Nordstr\"om spacetime.  From
Fig.~\ref{Fig:Penrose_diagram_Mink_RN_regions_Iprime} it is seen, that
it is clearly a black hole solution, not a vacuum black hole, neither
a regular black hole. The solutions represent tension shell black
holes.  Note $r_+$ and $r_-$ are the event horizon and the Cauchy
horizon radii, and there is an Einstein-Rosen bridge, provided by a
dynamic wormhole in the spacetime. Tension shell black holes were
found in~\citep{Katz_Lynden-Bell_1991} for the zero electric charge
case, i.e., for the Schwarzschild shells, in which case the
Carter-Penrose diagram is similar, only the $r=0$ singularity is
spacelike, and the diagram does not repeat itself.  In the
Reissner-Nordstr\"om spacetime, contrary to Schwarzschild, there is an
infinitude of possible diagrams.  In the diagram (a) of
Fig.~\ref{Fig:Penrose_diagram_Mink_RN_regions_Iprime} it is clear that
the tension shell is outside the event horizon in the other side of
the diagram in the region $\mathrm{I'}$ shown. One can then put
another shell in the region $\mathrm{I'}$ above and repeating the
procedure ad infinitum.  In the diagram (b) of
Fig.~\ref{Fig:Penrose_diagram_Mink_RN_regions_Iprime} the tension
shell is again outside the event horizon in the other side of the
diagram in the region $\mathrm{I'}$ shown. One can then put an
infinity region in the region $\mathrm{I'}$ above and repeating the
procedure ad infinitum.  As what one puts in the regions
$\mathrm{I'}$, either a tension shell or infinity, is not decided by
the solution, an infinite number of different Carter-Penrose diagrams
can be drawn, since there are an infinite number of combinations to
locate a shell or infinity when one goes upward or downward through the
diagram.  When $R= r_+$ the shell with its interior forms a
tension quasiblack hole with special features since
it is attached to the other regions of the
Reissner-Nordstr\"om spacetime.

\newpage

The physical interpretation of this case has some complexity.  This
nonextremal thin shell solution carries with it a white hole connected
to a black hole through a wormhole.  The energy density and pressure
obey some of the energy conditions if the radius of the shell is sufficiently
large, i.e., is sufficiently larger than the gravitational
radius. When the radius of the shell approaches the gravitational
radius, the energy conditions are not obeyed, and when the radius of
the shell is at the gravitational radius the solution turns into a
tension quasiblack hole an object with interesting properties.
The causal and global structures as displayed by the
Carter-Penrose diagram in its simplest form shows the important
spacetime regions.  We have called this solution a tension shell black
hole, but it could be called as well a tension shell nontraversable
wormhole, since there is a nontraversable wormhole that links the
white hole to the black hole region. As in Reissner-Nosdstr\"om
solution, This tension shell black hole possesses Cauchy horizons,
and, as in the vacuum Reissner-Nosdstr\"om solution, it is subject to
be destroyed by perturbations. Presumably, the perturbation would turn
the Cauchy horizon into a null or spacelike singularity, turning in
turn the nonextremal tension shell solution into a solution similar to
the electrically uncharged Lynden-Bell-Katz tension shell black hole
solution. Moreover, these solutions, in the same way as the full
Reissner-Nosdstr\"om or Schwarzschild solutions, are universes in
themselves, and, if they existed, they would have to be given directly
by mother nature, rather than appear by, say, a straight gravitational
collapse or some other process.  So, this case falls into the category
of having some of
the energy conditions verified and the geometrical setup
being physically peculiar,  although full of interest, as matter
solutions on the other side of the Carter-Penrose diagram are
rare. Moreover these solutions are familiar, in the sense that
nontraversable wormholes with white and black holes are well known.


\centerline{}
\newpage

\subsection{Formalism for nonextremal electric thin shells
outside the gravitational radius}
\label{Subsec:Induced_MinkowskiandRN_ouside_event_horizon}

\subsubsection{Preliminaries}
\label{prel1}

We now make a careful study to derive the properties of the
fundamental electric thin shell used in the two previous subsections,
i.e., the thin shell in the nonextremal state, i.e., $r_+>r_-$ or
$M>Q$, for which the shell's location obeys $R>r_+$, and
for which the orientation is such that the normal to the shell points
towards spatial infinity or towards $r_+$.
It should be read as an appendix to the previous two
subsections.  We use the formalism developed in
Sec.~\ref{Sec:Junction_formalism} and
Appendix~\ref{Appendix_sec:Kruskal-Szekeres_coordinates_RN}.

\subsubsection{Induced metric and extrinsic curvature of $\mathcal{S}$
as seen from $\mathcal{M}_{\rm i}$}
\label{asseenfrom1}

Let us start by analyzing the interior
Minkowski spacetime, $\mathcal{M}_{\rm i}$, whose line element in spherical
coordinates is given by 
\begin{equation}
ds_{\rm i}^{2}=-dt_{\rm i}^{2}+d\mathrm{r}^{2}+\mathrm{r}^{2}d\Omega^{2}\,,
\label{eq:Mink_metric_interior}
\end{equation}
where $t_{\rm i}$ and $\mathrm{r}$ are the time and radial coordinates,
respectively, and
$d\Omega^{2}\equiv d\theta^{2}+\sin^{2}\theta d\varphi^{2}$,
with $\theta$ and $\varphi$ being the angular coordinates.
The subscript $\rm i$ denotes interior or inside from now onwards.

The junction from the interior to the exterior is made through a
hypersurface $\mathcal{S}$.  We assume the hypersurface $\mathcal{S}$
to be static, i.e., static as seen from a free-falling observer in the
interior Minkowski spacetime.  In general, $\mathcal{S}$ can be either
timelike or spacelike, however, since we are considering  Minkowski
spacetime, it is not possible to have a static spacelike surface
hence, $\mathcal{S}$ must be timelike.  It is convenient to choose the
coordinates on $\mathcal{S}$ to be $\left\{ y^{a}\right\}
=\left(\tau,\theta,\varphi\right)$, where $\tau$ is the proper time
measured by an observer comoving with $\mathcal{S}$.  It follows that
denoting $u_{\rm i}$ as the 4-velocity of an observer comoving with
the shell as seen from the inside, we can define a unit vector
$e_{\tau}$ such that $e_{\tau}\equiv u_{\rm i}$.  The hypersurface
$\mathcal{S}$, as seen from the interior 
spacetime $\mathcal{M}_{\rm i}$,
is parameterized by $\tau$, such that the surface's radial
coordinate is described by a function
$\mathrm{r}\vert_\mathcal{S}\equiv R=R\left(\tau\right)$. The fact
that $\mathcal{S}$ is assumed to be static implies
$\frac{d\,R}{d\tau}=0$, from which we have that $u_{\rm
i}^{\alpha}=\left(\frac{dt_{\rm i}}{d\tau},0,0,0\right)$, where
$u_{\rm i}^{\alpha}$ represents the components of the 4-velocity
$u_{\rm i}$ as seen from the interior spacetime $\mathcal{M}_{\rm
i}$. Since $\mathcal{S}$ is a timelike hypersurface, it must verify
$u_{{\rm i}\alpha}u_{\rm i}^{\alpha}=-1$.  With these latter two
equations we find that $\frac{dt_{\rm i}}
{d\tau}=\pm1$.  Imposing that $u_{\rm
i}$ points to the future leads to the choice of the plus sign, thus
\begin{equation}
u_{\rm i}^{\alpha}=\left(1,0,0,0\right)\,.
\label{eq:Mink_vel_explicit}
\end{equation}
From Eqs.~(\ref{eq:inducedh}) and (\ref{eq:Mink_vel_explicit})
we can find the induced metric on $\mathcal{S}$ by the spacetime
$\mathcal{M}_{\rm i}$, such that 
\begin{equation}
\left.ds_{\rm i}^{2}\right|_{\mathcal{S}}
=-d\tau^{2}+R^{2}d\Omega^{2}\,.
\label{eq:induced_metric_Mink}
\end{equation}
Also, with the expression for the 4-velocity of an observer comoving
with $\mathcal{S}$, we can now use Eqs.~(\ref{eq:normal_orthogonal})
and (\ref{eq:Mink_vel_explicit}) to find the expression for the components
of the unit normal as seen from $\mathcal{M}_{\rm i}$,
$n_{\rm i}^{\alpha}$,
hence $n_{{\rm i}\alpha}=\lambda\left(0,1,0,0\right)$ where $\lambda$
is a normalization factor.
Using Eqs.~(\ref{eq:normal_normalized})
and (\ref{eq:Mink_metric_interior})
and the condition that $n$ is
spacelike, yields $\lambda=\pm1$. Since we are studying the case
where the interior Minkowski spacetime is spatially compact and enclosed
by the hypersurface $\mathcal{S}$, we must choose the plus sign,
such that, the expression for the outward pointing unit normal to
$\mathcal{S}$ is given by 
\begin{equation}
n_{{\rm i}\alpha}=\left(0,1,0,0\right)\,.
\label{eq:normal_Mink}
\end{equation}
We are now in position to compute the components of the extrinsic
curvature of $\mathcal{S}$ as seen from $\mathcal{M}_{\rm i}$,
$K_{{\rm i}\,ab}$.
In the case where the matching surface $\mathcal{S}$ is timelike,
static and spherically symmetric, the nonzero components of the extrinsic
curvature are given by $K_{\tau\tau}=-a^{\alpha}n_{\alpha}$,
$K_{\theta\theta}=\nabla_{\theta}n_{\theta}$,
$K_{\varphi\varphi}=\nabla_{\varphi}n_{\varphi}$,
where $a^{\alpha}\equiv u^{\beta}\nabla_{\beta}u^{\alpha}$,
see Appendix~\ref{Appendix_sec:Extrinsic_curvature}.
Taking into account Eqs.~(\ref{eq:normal_orthogonal}),
(\ref{eq:Mink_metric_interior}),
(\ref{eq:induced_metric_Mink}),
and~(\ref{eq:normal_Mink}),
we find that the nontrivial
components of the extrinsic curvature as seen from the interior
Minkowski spacetime, see Eq.~(\ref{eq:extrinsic1}), are given by
\begin{equation}
{K_{\rm i}}^{\tau}{}_{\tau}=0\,,\quad\quad
{K_{\rm i}}^{\theta}{}_{\theta}=
{K_{\rm i}}^{\varphi}{}_{\varphi}=\frac{1}{R}\,,
\label{eq:Extrinsic_curvature_Mink}
\end{equation}
where the induced metric taken from
Eq.~(\ref{eq:induced_metric_Mink}) was
used to raise the indices.

\subsubsection{Induced metric and extrinsic
curvature of $\mathcal{S}$ as seen
from $\mathcal{M}_{\rm e}$}
\label{induceM+1}

To proceed we have now to find the expressions for the induced metric
on $\mathcal{S}$ and the extrinsic curvature components as seen from
the exterior spacetime, $\mathcal{M}_{\rm e}$, in the nonextremal
state, i.e., $r_+>r_-$ or $M>Q$, see
Figure~\ref{Fig:Penrose_diagram_RN_non_extremal}, for which the
shell's location obeys $R>r_+$, and for which the
orientation is such that the normal to the shell points towards
increasing $r$ or towards decreasing $r$ as seen from the exterior, as
used in the two previous subsections.

For a nonextremal shell with $R>r_+$ we work
with the coordinate pathch that has no
coordinate singularity at the
gravitational radius $r=r_+$. For
the setting of coordinate patches 
in the nonextremal Reissner-Nordstr\"om spacetime see 
Appendix~\ref{Appendix_sec:Kruskal-Szekeres_coordinates_RN}, see
also \cite{Comer_Katz_1994} for
the coordinate patches of an uncharged shell
matched to the Schwarzschild spacetime.
In this region and for the chosen coordinate
patch, the line element for the
Reissner-Nordstr\"om spacetime in Kruskal-Szekeres coordinates is
given by
\begin{eqnarray}
ds_{\rm e}^{2}=4\left(\frac{r_++r_-}{r_+-r_-}\right)^{2}&&
\frac{r_+^{4}}{r^{2}}e^{-\frac{r\,\left(r_+-r_-\right)}{r_+^{2}}}
\left(\frac{r-r_-}{r_++r_-}
\right)^{1+\left(\frac{r_-}{r_+}\right)^{2}}\left(dX^{2}-dT^{2}
\right)+r^{2}\left(T,X\right)d\Omega^{2}\,,
\label{eq:metric_RN_rplus}\\
&&X^{2}-T^{2}=e^{\frac{r\,\left(r_+-r_-\right)}{r_+^{2}}}
\left(\frac{r-r_+}{r_++r_-}\right)\left(
\frac{r-r_-}{r_++r_-}\right)^{-\left(
\frac{r_-}{r_+}\right)^{2}}\,,
\nonumber
\end{eqnarray}
with $r\left(T,X\right)$ being given implicitly by the latter equation. 
The subscript $\rm e$ denotes exterior from now onwards.

The shell's radial coordinate when measured by an observer at
$\mathcal{M}_{\rm e}$ is described by a function
$r\vert_\mathcal{S}\equiv{R}=
{R}\left(\tau\right)$, where $\tau$ is the proper time of an
observer comoving with the surface $\mathcal{S}$, which, since we
assume it to be static, is such that $\frac{d{R}}{d\tau}=0$.
Strictly, $R$ should be written as another letter, say
${\cal R}$, but as we will see we can put ${\cal R}=R$ and so
we stick to the letter $R$ from the start.
Considering the second of the equations
given in Eq.~(\ref{eq:metric_RN_rplus}), $\frac{d{R}}{d\tau}=0$
implies that the $X$ and $T$ coordinates of a point on $\mathcal{S}$
must verify $X^{2}-T^{2}=\text{constant}$. Taking the derivative of
$X^{2}-T^{2}=\text{constant}$ in order to the proper time we find the
relation $\frac{\partial X}{\partial\tau}=\frac{T}{X}\frac{\partial
T}{\partial\tau}$.  In our previous analysis of the
$\mathcal{M}_{\rm i}$ spacetime, we found that the
hypersurface $\mathcal{S}$ must be
timelike, then, due to the first junction condition, $\mathcal{S}$
must also be timelike when seen from the exterior
$\mathcal{M}_{\rm e}$ spacetime. Therefore, the components
of the 4-velocity of an
observer comoving with it as seen from $\mathcal{M}_{\rm e}$ are,
$u_{\rm e}^{\alpha}=\left(\frac{\partial T}{\partial\tau},\frac{\partial
X}{\partial\tau},0,0\right)$.  Using $\frac{\partial
X}{\partial\tau}=\frac{T}{X}\frac{\partial T}{\partial\tau}$ and
$u_{{\rm e}\alpha}u_{\rm e}^{\alpha}=-1$ we find $\frac{\partial
T}{\partial\tau}=\pm\sqrt{\frac{g^{^{XX}}\,X^{2}}{X^{2}-T^{2}}}$ and
$\frac{\partial
X}{\partial\tau}=\pm\sqrt{\frac{g^{^{XX}}\,T^{2}}{X^{2}-T^{2}}}$,
so that,
\begin{equation}
u_{\rm e}^{\alpha}=\sqrt{\frac{g^{^{XX}}}{X^{2}-T^{2}}}
\left(X,T,0,0\right)\,,\label{eq:4velocity_value_rplus}
\end{equation}
where the sign was chosen in order that $u_{\rm e}$ points to the
future and $g^{^{XX}}$ is the $XX$ component of the inverse metric
associated with Eq.~(\ref{eq:metric_RN_rplus}).  Notice that the
expression found for the components of $u_{\rm e}$,
Eq.~(\ref{eq:4velocity_value_rplus}), only makes sense, physically, if
$X^{2}>T^{2}$.
Looking at
the second of the equations
given in Eq.~(\ref{eq:metric_RN_rplus}),
one has that $X^{2}>T^{2}$ implies that ${R}>r_+$ so,
either the shell is located in the region $\mathrm{I}$ or in the
region $\mathbb{\mathrm{I}}'$, see
Figure~\ref{Fig:Penrose_diagram_RN_non_extremal}. The restriction on
the allowed regions for the shell is a consequence of the shell being
assumed static, if we were to consider a dynamic shell or a different
interior spacetime, then shells in the black hole or the white hole
region
could also be treated. Note also that our choice of the plus sign in
Eq.~(\ref{eq:4velocity_value_rplus}), such that $u_{\rm e}$ points to the
future, is the correct one in both $\mathrm{I}$ or $\mathrm{I}'$
regions. Equation~(\ref{eq:4velocity_value_rplus}) can now be used to
find the induced metric on the hypersurface $\mathcal{S}$ by the
spacetime $\mathcal{M}_{\rm e}$, such that
$
\left.ds_{\rm e}^{2}\right|_{\mathcal{S}}
=-d\tau^{2}+{R}^{2}d\Omega^{2}$.
From the first junction condition, Eq.~(\ref{eq:1st_junct_cond}),
matching Eq.~(\ref{eq:induced_metric_Mink}) with this equation
for $\left.ds_{\rm e}^{2}\right|_{\mathcal{S}}$, we find
that ${R}$, the radial coordinate of $\mathcal{S}$ when measured by
an observer at $\mathcal{M}_{\rm e}$, and $R$, the radial coordinate
of $\mathcal{S}$ when measured by an observer at $\mathcal{M}_{\rm i}$,
must be indeed equal, as we have anticipated.
So, generically, $R$  describes the radial
coordinate of the shell for either the interior and exterior
spacetime, and so, the intrinsic line elements of the shell,
namely, 
$
\left.ds_{\rm i}^{2}\right|_{\mathcal{S}}
=-d\tau^{2}+R^{2}d\Omega^{2}
$
and 
$
\left.ds_{\rm e}^{2}\right|_{\mathcal{S}}
=-d\tau^{2}+{R}^{2}d\Omega^{2}$,
can be written 
uniquely as
\begin{equation}
\left.ds^{2}\right|_{\mathcal{S}}=-d\tau^{2}+R^{2}d\Omega^{2}\,.
\label{eq:induced_metric_RN}
\end{equation}
Now, using the fact that the unit normal to $\mathcal{S}$ is
spacelike, implies $n_{\rm e}^{\alpha}n_{{\rm e}\alpha}=+1$. Then,
taking into account Eqs.~(\ref{eq:normal_orthogonal}) and
(\ref{eq:4velocity_value_rplus}),
we find $n_{{\rm e}\alpha}=\pm\sqrt{
\frac{g_{_{XX}}}{X^{2}-T^{2}}}\left(-T,X,0,0\right)$.
To proceed, we must choose the sign for the normal. The choice of
the sign is related with the orientation
of the shell, i.e., the direction of the normal, and we 
impose that it points in the direction of increasing $X$ coordinate.
This implies that the choice of the sign is different if we consider
the shell to be in the region $\mathrm{I}$ or $\mathrm{I}'$, see
Figure~\ref{Fig:Penrose_diagram_RN_non_extremal} and 
also Figure~\ref{Appendix_fig:Coordinate_patch_1}
of Appendix~\ref{Appendix_subsec:Kruskal-Szekeres_coordinates}.
One of the simplifications
that the use of the Kruskal-Szekeres coordinates introduces is that
the choice of the sign can be written in a concise manner, such that
\begin{equation}
n_{{\rm e}\alpha}=\text{sign}\left(X\right)\sqrt{
\frac{g_{_{XX}}}{X^{2}-T^{2}}}\left(-T,X,0,0\right)\,,
\label{eq:normal_value_rplus}
\end{equation}
where the quantities on the right-hand side are to be evaluated at
$r=R$ and $\text{sign}\left(X\right)$ is the signum function of the
coordinate $X$ of the shell. Notice however, that the usage of this
notation is simply to treat in a concise way the two possible
directions of the normal of the shell. Physically, there is nothing
different between a shell located in either region, i.e., with
positive or negative values of $X$. Having found the normal to the
hypersurface $\mathcal{S}$ as seen from the exterior nonextremal
Reissner-Nordstr\"om spacetime, we can now compute the nonzero
components of the extrinsic curvature. Following the results in
Appendix~\ref{Appendix_subsec:Extrinsic_curvature_outside_event_horizon}
we have 
\begin{equation}
{K_{\rm e}}^{\tau}{}_{\tau}=\frac{\text{sign}
\left(X\right)}{2R^{2}k}\left(r_++r_--2
\frac{r_+r_-}{R}\right)\,,\quad\quad
{K_{\rm e}}^{\theta}{}_{\theta}=
{K_{\rm e}}^{\varphi}{}_{\varphi}=\frac{\text{sign}
\left(X\right)\left(r_+-r_-\right)}{2r_+^{2}R}
\sqrt{g_{_{XX}}\left(X^{2}-T^{2}\right)}\,,
\label{eq:Nonextremal_Extrinsic_RN_outside_event_horizon}
\end{equation}
where $k$, here, is the redshift function given in
Eq.~(\ref{eq:redshift}),
evaluated at $R$, i.e., $k(R,r_+,r_-)=
\sqrt{\left(1-\frac{r_+}{R}\right)\left(1-\frac{r_-}{R}\right)}$.

\subsubsection{Shell's energy density and pressure
\label{subSubsec:shellsenergydensityandpressure}}

We are now in position to find the properties of a perfect fluid thin
shell in a nonextremal Reissner-Nordstr\"om state, located outside the
gravitational radius or event horizon radius, depending on the case.
The shell's stress-energy tensor is given in Eq.~(\ref{eq:perfect}),
an expression containing
the energy per unit area  $\sigma$, the
tangential pressure of the fluid $p$,
the velocity $u_a$, and
the induced metric $h_{ab}$.
From our choice of coordinates on
$\mathcal{S}$ we have that  $\left\{ y^{a}\right\}
=\left(\tau,\theta,\varphi\right)$, the four-velocity $u_{\rm i}^\alpha$
is given in Eq.~(\ref{eq:Mink_vel_explicit}),
and the metric $h_{ab}$ is given through
Eq.~(\ref{eq:induced_metric_RN}). Putting
everything together we find that
$S_{\tau}^{\tau}=-\sigma$,
$S_{\theta}^{\theta}=S_{\varphi}^{\varphi}=p$.
Comparing these latter equations
with the second junction condition, Eq.~(\ref{eq:2nd_junct_cond}),
taking into account the components of the induced metric,
given through Eq.~(\ref{eq:induced_metric_RN}),
and the fact that 
$\left[K_{\theta}^{\theta}\right]=\left[K_{\varphi}^{\varphi}\right]$,
we find $\sigma=-\frac{1}{4\pi}\left[K_{\theta}^{\theta}\right]$
and $p=\frac{1}{8\pi}\left[K_{\tau}^{\tau}\right]-\frac{\sigma}{2}$.
With the components of the extrinsic curvature found in
Eqs.~(\ref{eq:Extrinsic_curvature_Mink})
and~(\ref{eq:Nonextremal_Extrinsic_RN_outside_event_horizon}) 
we then obtain
\begin{equation}
8\pi\sigma=\frac{2}{R}\left(1-\text{sign}\left(X\right)k\right)\,,
\label{eq:sigma_value_rplus}
\end{equation}
\begin{equation}
8\pi p=\frac{\text{sign}\left(X\right)}{2Rk}\left[
\left(1-\text{sign}\left(X\right)k\right)^{2}-
\frac{r_+r_-}{R^{2}}\right]\,,
\label{eq:pressure_value_rplus}
\end{equation}
where $k$ here is the redshift function given
in Eq.~(\ref{eq:redshift})
evaluated at $R$, i.e., $k(R,r_+,r_-)
=\sqrt{\left(1-\frac{r_+}{R}\right)\left(1-\frac{r_-}{R}\right)}$.
As the surface electric current density $s_a$ on the
thin shell is defined as $s_a=\sigma_{e}u_a$,
where $\sigma_{e}$ represents the
electric charge density and $u_a$ is the velocity
of the shell, and since the Minkowski spacetime has zero electric
charge,
from Eqs.~(\ref{eq:junct_cond_Faradayb})-(\ref{eq:junct_cond_Faraday2})
and~(\ref{eq:RN_FaradayMaxwell_value})
it follows that 
\begin{equation}
8\pi \sigma_{e}=2\frac{\sqrt{r_+r_-}}{ R^{2}}\,.
\label{eq:chargedensity1}
\end{equation}
In Eqs.~(\ref{eq:sigma_value_rplus}) and~(\ref{eq:pressure_value_rplus})
it is clear that it is necessary to pick the sign in $\text{sign}
\left(X\right)$.
Let us start with $\text{sign}\left(X\right)=+1$. It is useful here to
give the expressions for the shell's energy density and pressure,
$\sigma$ and $p$, in terms of $M$ and $Q$.
Using Eq.~(\ref{eq:KS_horizons_radius0}) in Eqs.~(\ref{eq:sigma_value_rplus})
and~(\ref{eq:pressure_value_rplus}) with $\text{sign}
\left(X\right)=+1$ we have
$8\pi\sigma=\frac{2}{R}\left(1-k\right)$,
$8\pi p=\frac{1}{2Rk}\left[\left(1-k\right)^{2}-
\frac{Q^{2}}{R^{2}}\right]$,
where 
$k(R,M,Q)
=\sqrt{
1-\frac{2M}{R}+
\frac{Q^2}{R^2}}$,
and 
also from Eq.~(\ref{eq:chargedensity1}) we have
$8\pi \sigma_{e}=\frac{2Q}{R^{2}}$.
Let us now take $\text{sign}\left(X\right)=-1$. It is also useful here to
give the expressions for the shell's energy density and pressure,
$\sigma$ and $p$, in terms of $M$ and $Q$.
Using Eq.~(\ref{eq:KS_horizons_radius0}) in Eqs.~(\ref{eq:sigma_value_rplus})
and~(\ref{eq:pressure_value_rplus}) with $\text{sign}
\left(X\right)=-1$ we have
$8\pi\sigma=\frac{2}{R}\left(1+k\right)$,
$8\pi p=-\frac{1}{2Rk}\left[\left(1+k\right)^{2}-\frac{Q^{2}}{R^{2}}
\right]$, where again
$k(R,M,Q)
=\sqrt{
1-\frac{2M}{R}+
\frac{Q^2}{R^2}}$, and 
also from Eq.~(\ref{eq:chargedensity1}) we have
$8\pi\sigma_{e}=\frac{2Q}{R^{2}}$.
These are the expressions used in the two previous
subsections.
Note also that when $r_-=0$, then $\sigma_{e}=0$ and the
electric charge $Q$ is zero, $Q=0$,
so the outside spacetime is
described by the Schwarzschild
solution, for which Eqs.~(\ref{eq:sigma_value_rplus})
and (\ref{eq:pressure_value_rplus})
can be written explicitly as
$8\pi\sigma\vert_{r_-=0}=\frac{2}{R}
\left(1-\text{sign}\left(X\right)\sqrt{1-
\frac{r_+}{R}}\right)$
and $8\pi p\vert_{r_-=0}=\frac{\text{sign}
\left(X\right)}{2R}\sqrt{\frac{1}{1-
\frac{r_+}{R}}}\left[\left(1-\text{sign}\left(X\right)
\sqrt{1-\frac{r_+}{R}}\right)^{2}\right]$,  and which
are the energy density and the tangential pressure
for a shell matching Minkowski to 
the Schwarzschild spacetime.

\clearpage{}

\section{Nonextremal electric thin shells inside the Cauchy radius:
Tension shell regular and nonregular
black holes and compact shell naked singularities}
\label{insidecauchy}

\subsection{Nonextremal electric thin shells inside the Cauchy horizon:
Tension shell regular and nonregular black holes}
\label{Subsec:nonextremalnormalcauchy}

Here we study the case of a fundamental electric thin shell in the
nonextremal state, i.e., $r_+>r_-$ or $M>Q$, for which the shell's
location obeys $R<r_-$, and for which the orientation is
such that the normal to the shell points towards $r_-$.  In this case
horizons do exist and so, following the nomenclature, $r_+$ is both
the gravitational and the event horizon radius, and $r_-$ is both the
Cauchy radius and the Cauchy horizon radius.  The normal to the shell
pointing towards $r_-$ means in the notation of the Kruskal
coordinate $X$ that we take $\text{sign}\left(X\right)=+1$, see the
end of this section and
Appendix~\ref{Appendix_sec:Kruskal-Szekeres_coordinates_RN} for
details.

As functions of $M $, $Q$, and $R$,  the shell's energy
density $\sigma$ and
pressure $p$, are, see the end of this section,
\begin{equation}
8\pi\sigma=\frac{2}{R}\left(1-k\right)\,,
\label{eq:sigma_value_rminus_MQ_sign_plus}
\end{equation}
\begin{equation}
8\pi p=\frac{1}{2Rk}\left[\left(1-k\right)^{2}-\frac{Q^{2}}{R^{2}}
\right]\,,
\label{eq:pressure_value_rminus_MQ_sign_plus}
\end{equation}
respectively, with $k=\sqrt{1-\frac{2M}{R}+\frac{Q^{2}}{R^{2}}}$.
\begin{figure}[h]
\subfloat[]{\includegraphics[width=0.8\textwidth]{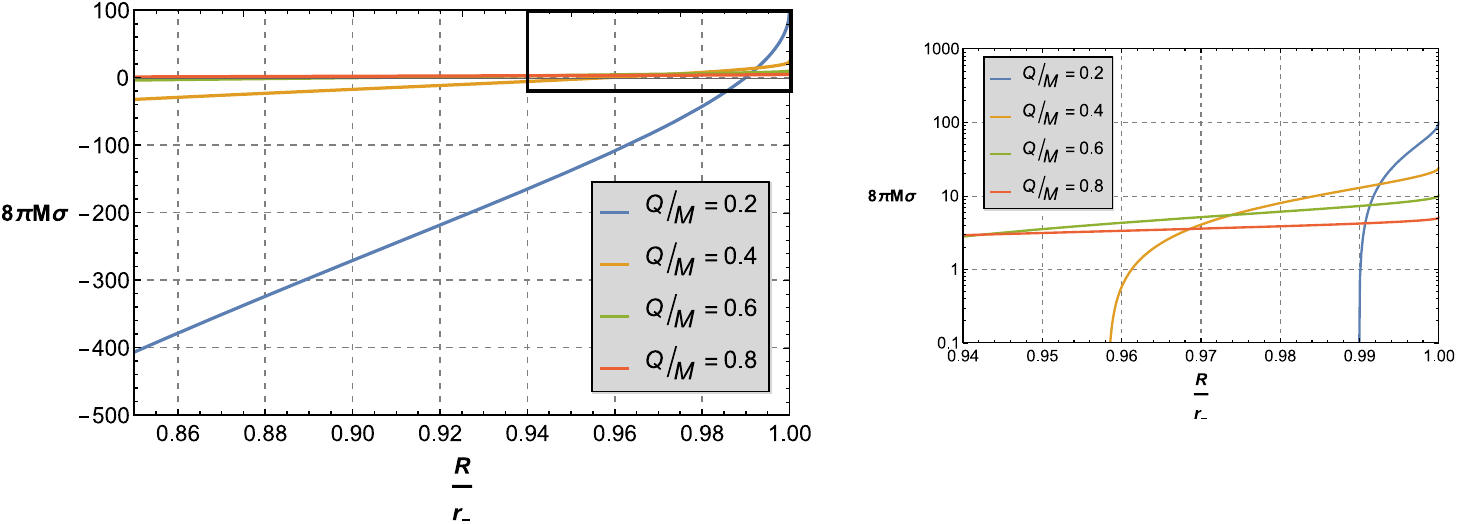}}\\
\subfloat[\label{Fig:Pressure_region_III_prime}]{
\includegraphics[scale=0.45]{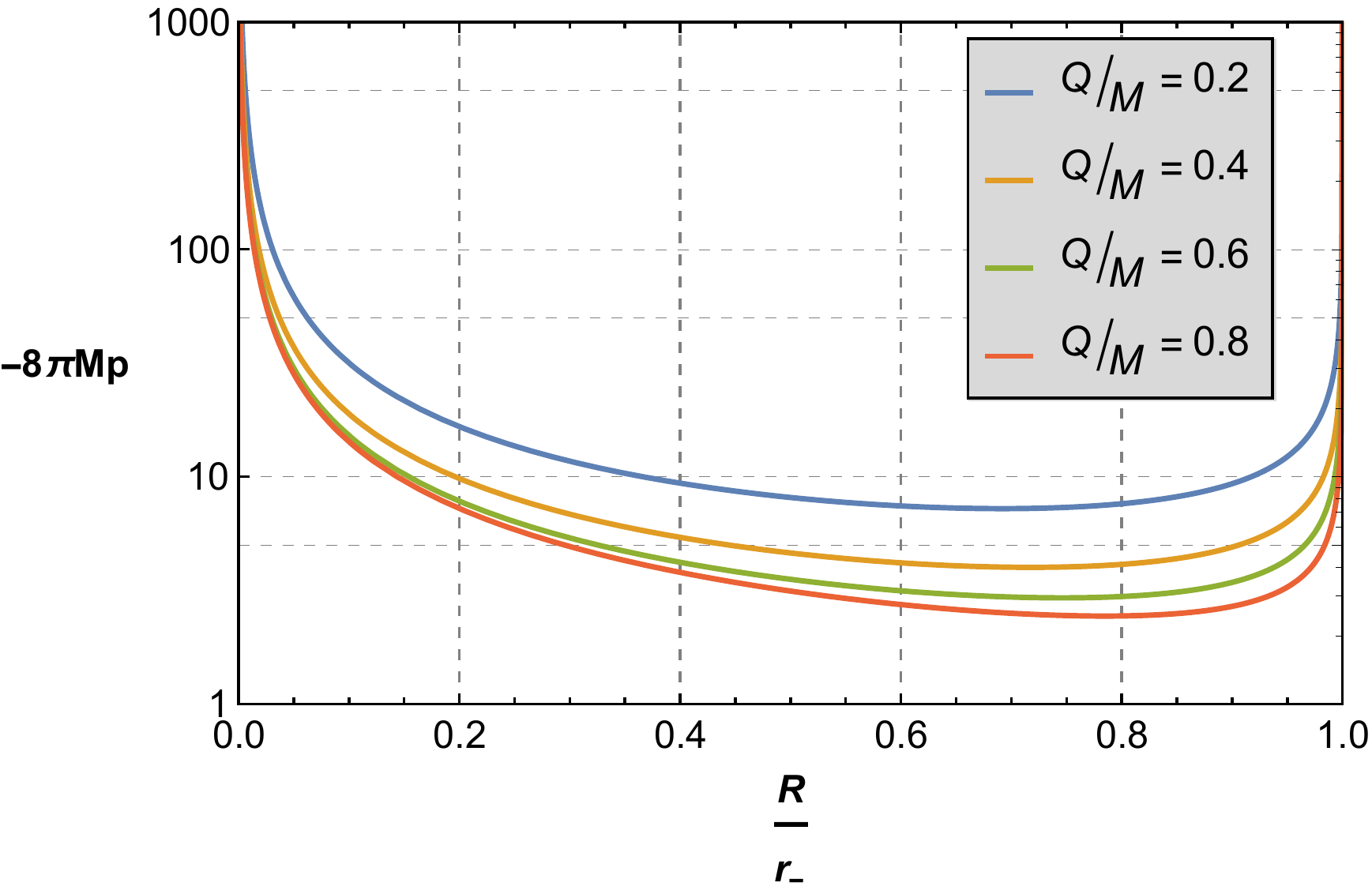}}
\caption{\label{Fig:Properties_region_III_prime}
Physical
properties of a nonextremal tension shell regular and
nonregular black hole, i.e.,
an electric perfect fluid
thin shell in a nonextremal Reissner-Nordstr\"om state, in the
location $R<r_-$, i.e., located inside the Cauchy radius,
and with orientation such that the normal points towards $r_-$.
The interior is Minkowski and the exterior is nonextremal
Reissner-Nordstr\"om spacetime.
Panel (a)
Energy density $\sigma$ of the shell as a function of the radius $R$
of the shell for various values of the $\frac{Q}{M}$ ratio. The energy
density is adimensionalized through the mass $M$, $8\pi M\sigma$, and
the radius is adimensionalized through the Cauchy radius $r_-$,
$\frac{R}{r_-}$. The marked zone on the top left is amplified on the
right.  Panel (b) Tension $-p$ on the shell as a function of the
radius $R$ of the shell for various values of the $\frac{Q}{M}$
ratio. The tension is adimensionalized through the mass $M$, $-8\pi
Mp$, and the radius is adimensionalized through the Cauchy
radius $r_-$, $\frac{R}{r_-}$.}
\end{figure}
Also, the electric charge density $\sigma_{e}$
is given in terms of $M$, $Q$, and $R$, by
\begin{equation}
8\pi \sigma_{e}=\frac{2Q}{R^{2}}\,.
\label{eq:chargedensity12cauchy}
\end{equation}
The behavior of $\sigma$ and $p$ as functions of the radial coordinate
$R$ of the shell for various values of the $\frac{Q}{M}$ ratio in this
case is shown in Figure~\ref{Fig:Properties_region_III_prime}.  We see
that, depending on the radial coordinate of the shell, the energy
density might take negative values. Indeed, from
Eq.~(\ref{eq:sigma_value_rminus_MQ_sign_plus}) we find that for
$R<\frac{Q^{2}}{2M}$ the energy density, $\sigma$, is negative. Also,
this kind of thin shell is always supported by tension, see also
Eq.~(\ref{eq:pressure_value_rminus_MQ_sign_plus}).  It is a tension
shell.  This is related to the fact that the Reissner-Nordstr\"om
singularity at $r=0$ is repulsive.  Moreover, we see that both the
energy density and the pressure of the shell diverge to negative
infinity as the shell gets closer to $R=0$.  On the other hand, in the
limit of $R\to r_-$ the pressure diverges to negative infinity, but
the energy density, $\sigma$, tends to $4\pi\sigma=\frac{1}{r_-}$.
When $Q=0$, i.e., $r_-=0$,
the solution is the vacuum
Schwarzschild solution, since as $R<r_-$, one
has in the limit $R=0$, which is singular.
In relation to the energy conditions of the shell we can
say that the null, the weak, the dominant, and the strong energy
conditions are never verified in this case, see a detailed
presentation ahead.

\newpage

\begin{figure}[h]
\subfloat[
\label{Fig:Penrose_diagram_Mink_RN_region_III_alt}]
{\includegraphics[height=0.31\paperheight]
{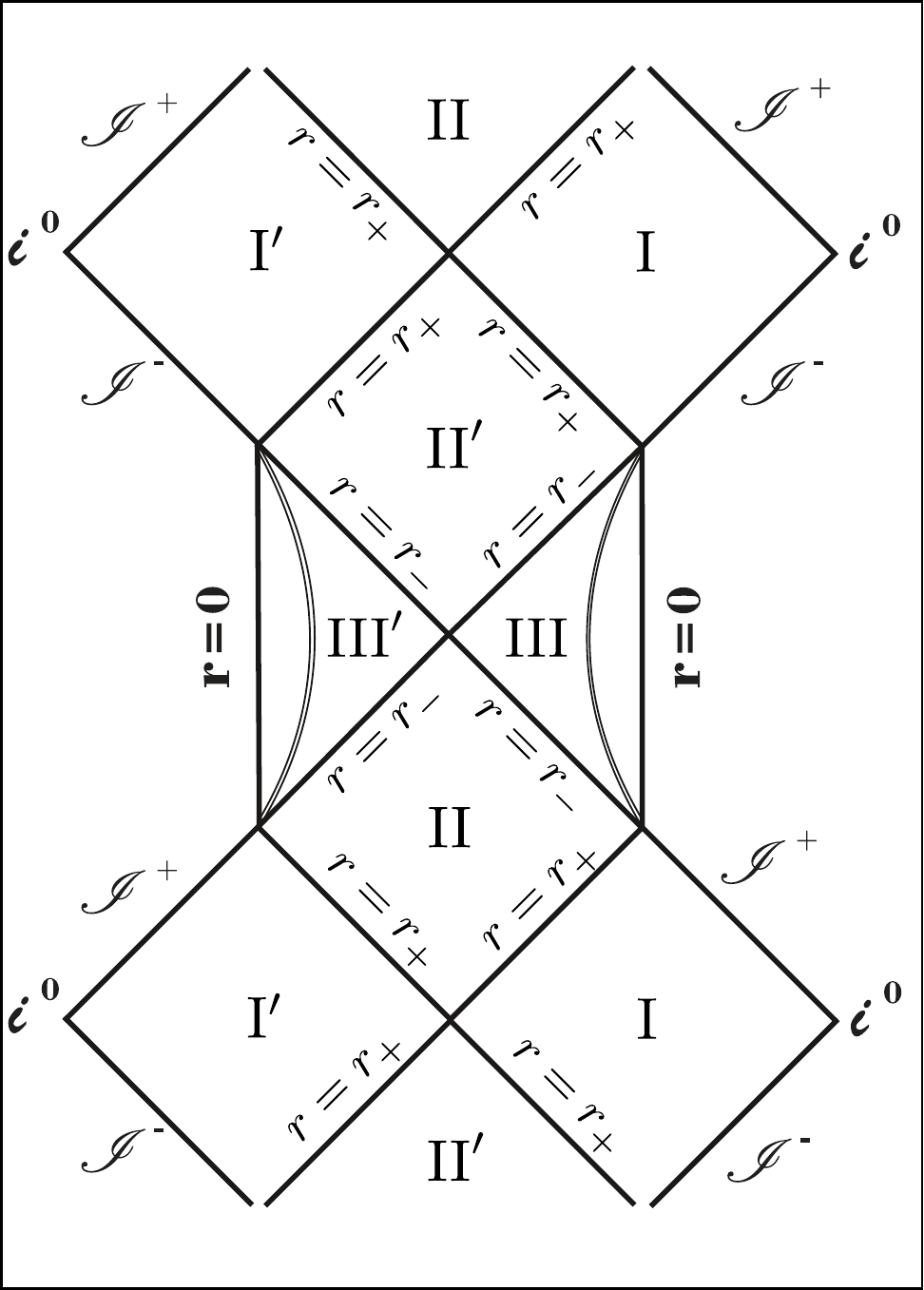}

}\hspace*{4cm}
\subfloat[
\label{Fig:Penrose_diagram_Mink_RN_region_III}]
{\includegraphics[height=0.31\paperheight]
{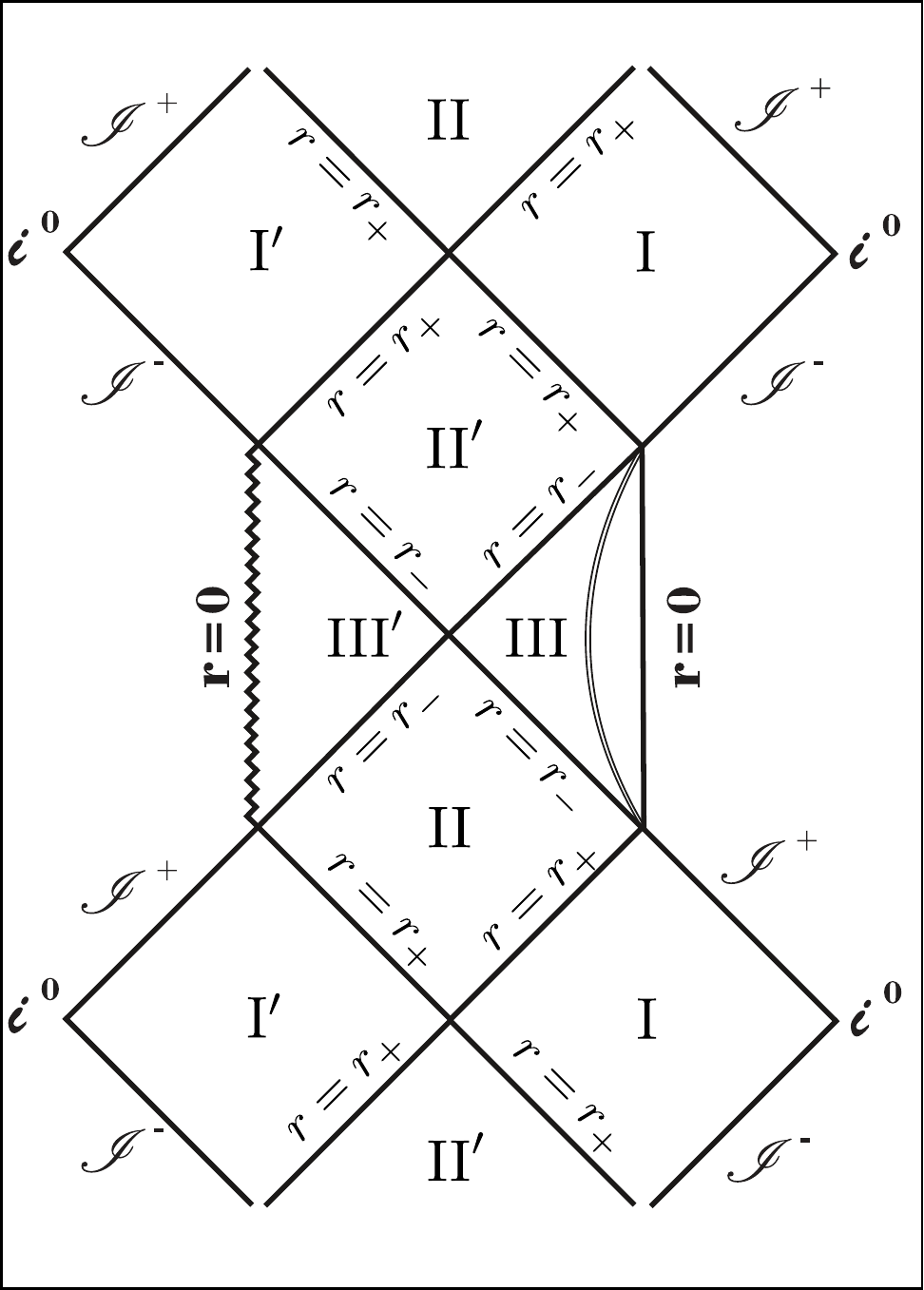}

}

\caption{
\label{Fig:Penrose_diagram_Mink_RN_region_III_towards_Cauchy_horizon}
Carter-Penrose diagrams of the tension shell black holes,
i.e., a thin shell spacetime
in a nonextremal Reissner-Nordstr\"om state, in the location $R<r_-$,
i.e., located inside the Cauchy radius, with orientation
such that the normal to the shell points towards $r_-$.
The interior is Minkowski, the exterior is Reissner-Nordstr\"om
spacetime.
Panel~(a) The Carter-Penrose diagram contains a shell in 
both regions $\mathrm{III}$ and $\mathrm{III'}$.  If this pattern is
repeated ad infinitum then it is a tension shell regular black hole.
Panel~(b) The Carter-Penrose diagram contains a shell in region
$\mathrm{III}$ and a singularity in region $\mathrm{III'}$. It is a
tension shell black hole, now not regular.
An infinite number of different Carter-Penrose diagrams can be drawn,
since there are an infinite number of combinations to locate the shell.
}
\end{figure}

The Carter-Penrose diagram for this case
can be drawn directly from the building
blocks of an interior Minkowski spacetime and the full
nonextremal Reissner-Nordstr\"om spacetime. In
Figure~\ref{Fig:Penrose_diagram_Mink_RN_region_III_towards_Cauchy_horizon}
two possible
Carter-Penrose diagrams of a
shell spacetime in a nonextremal Reissner-Nordstr\"om state,
in
the location $R<r_-$,  with orientation such that
the normal points towards $r_-$, i.e.,
$\text{sign}\left(X\right)=+1$, are shown.
It is a tension shell black hole spacetime. More specifically,
there is an infinitude of possible diagrams.
Indeed,
in the diagram (a) it is clear that the tension shell is inside the
Cauchy horizon in both regions  $\mathrm{III}$ and
$\mathrm{III'}$ of a Reissner-Nordstr\"om spacetime.  Admitting that
the portion shown of the diagram repeats itself ad infinitum then the
black hole is regular.  In the diagram (b) there is a shell in region
$\mathrm{III}$ and a singularity in region $\mathrm{III'}$, and so it
is not a regular black hole, it is a tension shell black hole with a
singularity. Since what one puts in the regions $\mathrm{III}$ and
$\mathrm{III'}$, either a shell or a singularity, is not decided by
the solution, an infinite number of different Carter-Penrose diagrams
can be drawn, as there are an infinite number of combinations to
locate a shell or a singularity when one goes upward or downward
through the diagram.  Note that $r_+$ and $r_-$ are the event horizon
and the Cauchy horizon radii, clearly, and the Einstein-Rosen bridge,
i.e., the dynamic wormhole, is there.
Regular black holes with shells that are
sandwiched between a de Sitter interior and a Reissner-Nordstr\"om
exterior were built in \cite{lemoszanchinregularbhs}.

\newpage

The physical interpretation of this case is of real interest. This
nonextremal thin shell solution provides a regular black hole
solution. The energy density and pressure never obey the energy
conditions for all shell radii, i.e., shell radii between zero and the
Cauchy horizon. The causal and global structure as displayed by the
Carter-Penrose diagram shows clearly that there is no singularity if
one adopts the simplest form of the diagram, meaning also that the
topology of the region inside the Cauchy horizons is a three-sphere,
as usual for regular black holes.  As in the Reissner-Nosdstr\"om
vacuum solution, this tension shell regular black holes possess Cauchy
horizons, and so they are subject to instabilities, which would lead
the solutions to an endpoint which can only be guessed. As regular
black holes these solutions join the other known regular black hole
solutions which are of interest in quantum gravitational settings that
presumably get rid of the singularities. So, this case falls into the
category of having the energy conditions never verified, and so in
this sense is odd, although of interest as regular black hole
matter solutions always are. As much as a regular black hole is
familiar so this shell solution is familiar.

\newpage

\subsection{Nonextremal electric thin shells inside the Cauchy radius:
Compact shell naked singularities}
\label{Subsec:nonextcompactshellnakedsingularity}

Here we study the case of a fundamental electric thin shell in the
nonextremal state, i.e., $r_+>r_-$ or $M>Q$, for which the shell's
location obeys $R<r_-$, and for which the orientation is
such that the normal to the shell points towards $r=0$.  In this case,
horizons do not exist and so, following the nomenclature, $r_+$ is
the gravitational radius, and $r_-$ is
the
Cauchy radius.  The normal to the shell
pointing towards $r=0$ means in the notation of the Kruskal
coordinate $X$ that we take $\text{sign}\left(X\right)=-1$, see the
end of this section and
Appendix~\ref{Appendix_sec:Kruskal-Szekeres_coordinates_RN} for
details.

\begin{figure}[h]
\subfloat[]
{\includegraphics[scale=0.45]{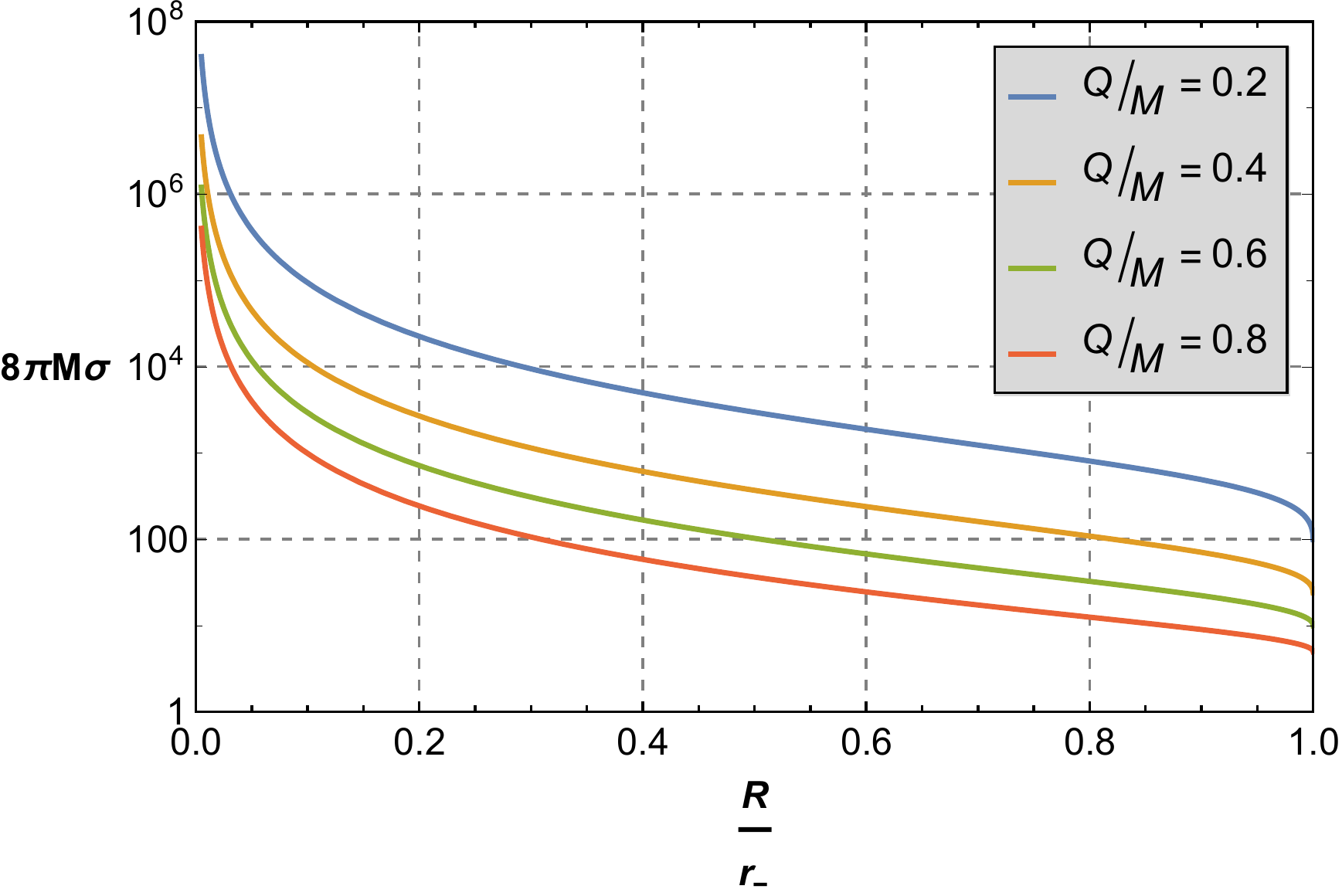}}
\hfill{}\subfloat[\label{Fig:Pressure_region_III}]
{\includegraphics[scale=0.45]{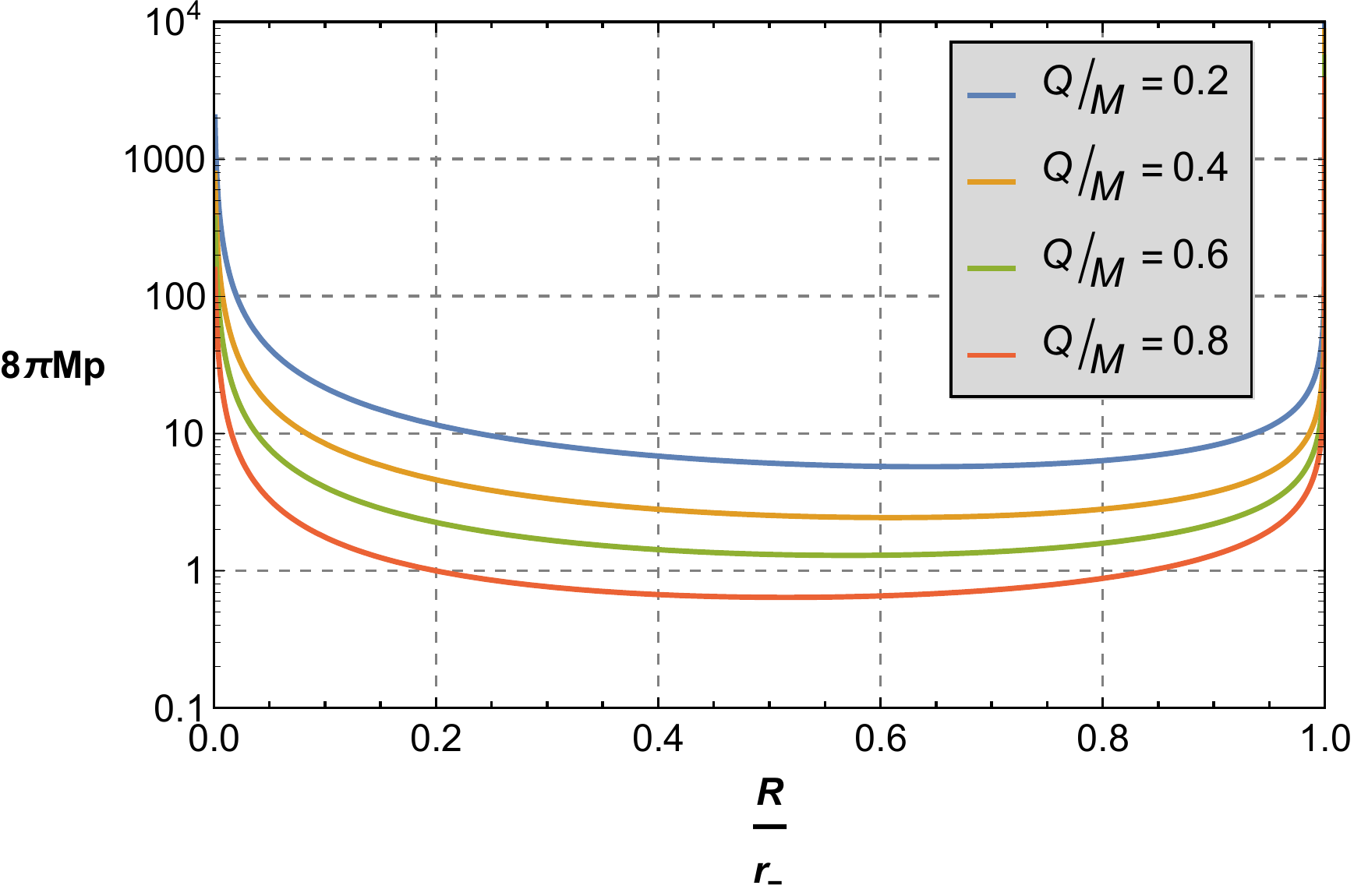}}
\caption{
\label{Fig:Properties_region_III}
Physical properties of a nonextremal
compact thin shell singularity, i.e., an
electric perfect fluid thin shell in a nonextremal
Reissner-Nordstr\"om state, in the location $R<r_-$, i.e., located
inside the Cauchy radius, and with orientation such that the normal
points towards $r=0$.  The interior is Minkowski and the exterior is
nonextremal Reissner-Nordstr\"om spacetime, although what it is
interior and what is exterior is blurred in this case.
Panel (a) Energy density $\sigma$ of the shell as a function of the
radius $R$ of the shell for various values of the $\frac{Q}{M}$
ratio. The energy density is adimensionalized through the mass $M$,
$8\pi M\sigma$, and the radius is adimensionalized through the Cauchy
radius $r_-$, $\frac{R}{r_-}$.
Panel (b) Pressure $p$ on the shell as a function of the radius $R$ of
the shell for various values of the $\frac{Q}{M}$ ratio. The pressure
is adimensionalized through the mass $M$, $8\pi Mp$, and the radius is
adimensionalizedthrough the Cauchy radius $r_-$, $\frac{R}{r_-}$.
}
\end{figure}

As functions of $M $, $Q$, and $R$, the shell's energy
density $\sigma$ and
pressure $p$, are, see the end of this section,
\begin{equation}
8\pi\sigma=\frac{2}{R}\left(1+k\right)\,,
\label{eq:sigma_value_rminus_MQ_sign_minus}
\end{equation}
\begin{equation}
8\pi p=-\frac{1}{2Rk}\left[\left(1+k\right)^{2}-
\frac{Q^{2}}{R^{2}}\right]\,,
\label{eq:pressure_value_rminus_MQ_sign_minus}
\end{equation}
respectively,
with $k=\sqrt{1-\frac{2M}{R}+\frac{Q^{2}}{R^{2}}}$.
The  electric charge density $\sigma_{e}$
is given in terms of $M$, $Q$, and $R$
by Eq.~(\ref{eq:chargedensity12cauchy}).
The behavior of $\sigma$ and $p$ as functions of the radial coordinate
$R$ of the shell for various values of the $\frac{Q}{M}$ ratio
in this case
is shown  in
Figure~\ref{Fig:Properties_region_III}.
We see that the energy density of the shell is always positive and the
shell is supported by pressure. As the radial coordinate of the shell,
$R$, goes to zero, both the energy density and pressure of the shell
diverge to infinity.  Moreover, as $R\to r_-$ the energy density tends
to $\frac1{4\pi\,r_-}$ and the pressure diverges to infinity.  When
$Q=0$ the solution does not exist.  In relation to the energy
conditions of the shell we can say that the null and the weak energy
conditions are verified for $0<R<r_-$, the dominant energy condition
is verified for $0<R<R_{\rm III}$, with $R_{\rm III}$ to be given
later, and the strong energy condition is verified for $0<R<r_-$, see
a detailed presentation ahead.

\newpage

\begin{figure}[h]
\includegraphics[height=0.25\paperheight]
{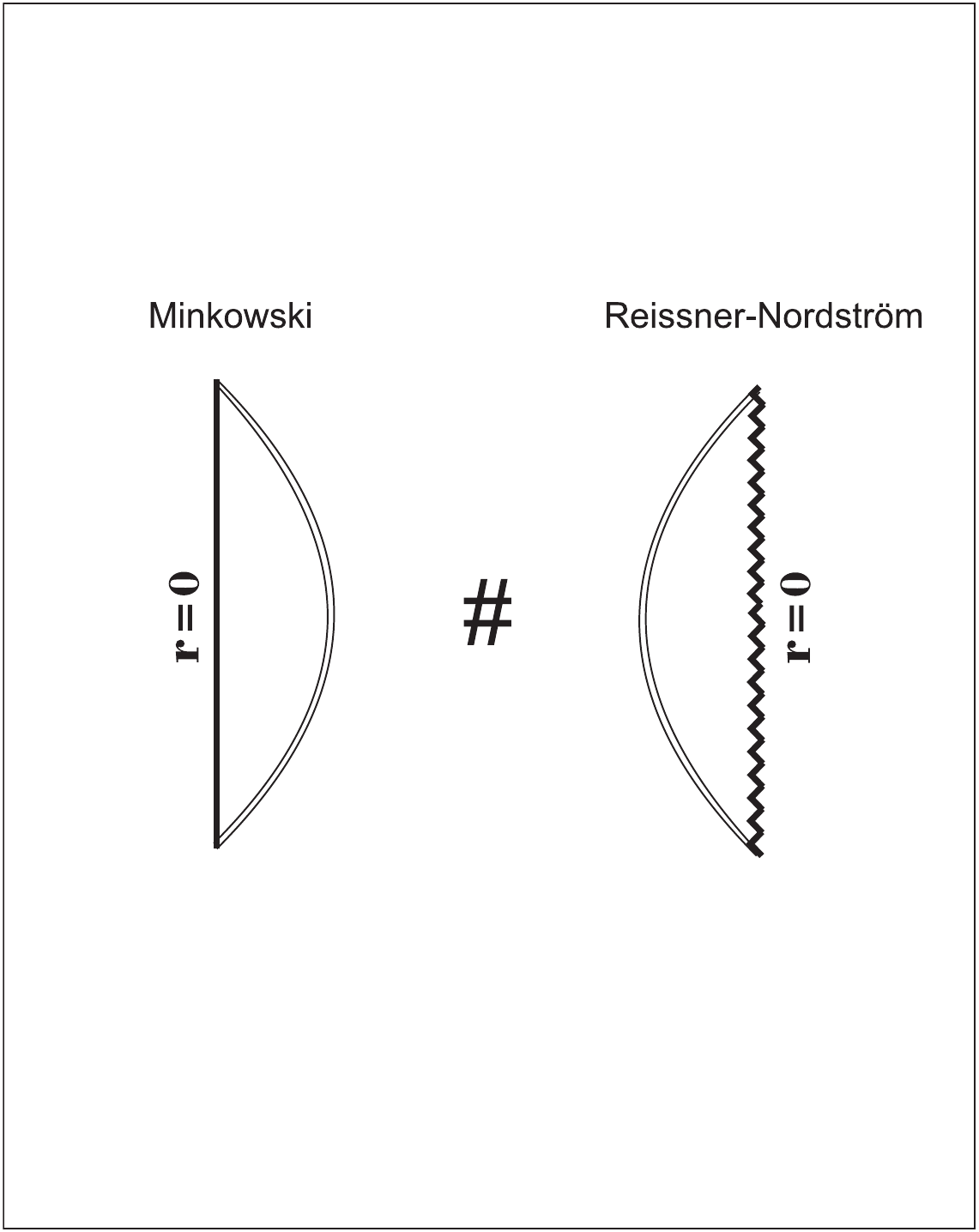}
\caption{
\label{Fig:Penrose_diagram_Mink_RN_region_III_towards_singularity}
Carter-Penrose diagram of the compact shell naked singularity, i.e., a
thin shell spacetime in a nonextremal Reissner-Nordstr\"om state, in
the location $R<r_-$, i.e., located inside the Cauchy radius, with
orientation such that the normal to the shell points towards $r=0$.
Part of the spacetime is Minkowski, part is Reissner-Nordstr\"om, in
this case there is no clear distinction what is interior from what is
exterior.  The hash symbol $\#$ represents the connected sum of the
two spacetimes.}
\end{figure}

The Carter-Penrose diagram for this case can be drawn directly from
the building blocks of an interior Minkowski spacetime and the full
nonextremal Reissner-Nordstr\"om spacetime. In
Figure~\ref{Fig:Penrose_diagram_Mink_RN_region_III_towards_singularity}
the Carter-Penrose diagram of a shell spacetime in a nonextremal
Reissner-Nordstr\"om state, in the location $R<r_-$, with orientation
such that the normal points towards $r=0$, i.e.,
$\text{sign}\left(X\right)=-1$, is shown.  It is a compact shell naked
singularity spacetime. It is clearly a compact space, $r$ goes from 0
to $R$ and then decreases back to 0 at the timelike singularity, such
that there is no clear distinction what is interior from what is
exterior. We use the hash symbol $\#$ to represent the connected sum
of the spacetime manifolds, in order to conserve the conformal
structure in the Carter-Penrose diagram of the total spacetime.  It is
difficult to understand if this solution can be achieved from a
physical phenomenon.  However, we expect the shell to be the source of
the singularity since the shell is the source of the exterior
spacetime, although it is very difficult to understand why the
singularity is formed away from the shell itself. Nonetheless, surely
the non-linearity of the theory just leads to this counterintuitive
behavior.

The physical interpretation of this case is most curious. This
nonextremal thin shell solution provides a closed spatial static
universe with a singularity at one pole. The energy density and
pressure obey the energy conditions for certain shell radii.  The
causal and global structure as displayed by the Carter-Penrose diagram
show the characteristics of this universe that has two sheets joined at
the shell with one sheet having a singularity at its pole and with no
horizons.  So, this case falls into the category of having the energy
conditions verified and the resulting spacetime being peculiar.

\newpage

\subsection{Formalism for nonextremal electric thin shells
inside the Cauchy radius}
\label{Subsec:Induced_MinkowskiandRN_inside_cauchy_horizon}

\subsubsection{Preliminaries}
\label{prel2}

We now make a careful study to derive the properties of the
fundamental electric thin shell used in the two previous subsections,
i.e., the thin shell in the nonextremal state, i.e., $r_+>r_-$ or
$M>Q$, for which the shell's location obeys $R<r_-$, and
for which the orientation is such that the normal to the shell points
towards $r_-$ or towards $r=0$.
It should be read as an appendix to the previous two
subsections.  We use the formalism developed in
Sec.~\ref{Sec:Junction_formalism} and
Appendix~\ref{Appendix_sec:Kruskal-Szekeres_coordinates_RN}.

\subsubsection{Induced metric and extrinsic curvature of
$\mathcal{S}$ as seen
from $\mathcal{M}_{\rm i}$}
\label{asseenfrom2}

Let us start by mentioning the interior
Minkowski spacetime, $\mathcal{M}_{\rm i}$. Since it is the same as
the analysis done in the previous section we only quote
the important equations.
They are the interior metric Eq.~(\ref{eq:Mink_metric_interior}),
the interior four-velocity of the shell
Eq.~(\ref{eq:Mink_vel_explicit}), the
metric for the shell at radius $R$ given in 
Eq.~(\ref{eq:induced_metric_Mink}), the normal to the shell
Eq.~(\ref{eq:normal_Mink}),
and the extrinsic curvature from the inside
Eq.~(\ref{eq:Extrinsic_curvature_Mink}).

\subsubsection{Induced metric, and extrinsic curvature of
$\mathcal{S}$ as seen
from $\mathcal{M}_{\rm e}$}
\label{Subsec:shells_inside_cauchy_horizon}

To proceed we have now to find the expressions for the induced metric
on $\mathcal{S}$ and the extrinsic curvature components as seen from
the exterior spacetime, $\mathcal{M}_{\rm e}$, in the nonextremal
state, i.e., $r_+>r_-$ or $M>Q$, see
Figure~\ref{Fig:Penrose_diagram_RN_non_extremal}, for which the
shell's location has radius $R$ obeying $R<r_-$, and for which the
orientation is such that the normal to the shell points towards
increasing $r$, i.e., towards $r_-$, or towards decreasing $r$, i.e.,
towards $r=0$, as seen from the exterior, as used in the two previous
subsections.

For a nonextremal shell with $R<r_-$ we work with the coordinate
patch that has no coordinate singularity at the gravitational radius
$r=r_-$.  Many of the previous results are also valid for the second
coordinate patch.  From the discussion in
Appendix~\ref{Appendix_subsec:Kruskal-Szekeres_coordinates0},
%
%
the line element for the Reissner-Nordstr\"om spacetime in
Kruskal-Szekeres coordinates in this patch is,
\begin{eqnarray}
ds_{\rm e}^{2}=4\left(\frac{r_++r_-}{r_+-r_-}\right)^{2}&&
\frac{r_-^{4}}{r^{2}}e^{\frac{r\left(r_+-r_-
\right)}{r_-^{2}}}\left(\frac{r_+-r}{r_++r_-}
\right)^{1+\left(\frac{r_+}{r_-}
\right)^{2}}\left(dX^{2}-dT^{2}\right)+r^{2}
\left(T,X\right)d\Omega^{2}\,,
\label{eq:metric_RN_rminus}\\
&&
X^{2}-T^{2}=e^{-\frac{r\left(r_+-r_-\right)}{r_-^{2}}}\left(
\frac{r_--r}{r_++r_-}\right)\left(\frac{r_+-r}{r_++r_-}
\right)^{-\left(\frac{r_+}{r_-}\right)^{2}}\,,
\nonumber
\end{eqnarray}
with $r\left(T,X\right)$ being given implicitly by the latter equation. 

The shell's radial coordinate when measured by an observer at
$\mathcal{M}_{\rm e}$ is constant since the shell is static, so from
the second of the equations in Eq.~(\ref{eq:metric_RN_rminus}) we take
that the $X$ and $T$ coordinates of the shell must verify
$X^{2}-T^{2}=\text{constant}$.  Now, as was argued in the previous
section, a static shell must be timelike as seen from both interior
and exterior spacetimes. The restriction $X^{2}-T^{2}=\text{constant}$
and the analysis performed in subsection~\ref{induceM+1}, imply that
the components of the 4-velocity $u_{\rm e}$ of an observer comoving
with the shell as seen from the exterior spacetime, are given by
\begin{equation}
u_{\rm e}^{\alpha}=-\sqrt{\frac{g^{^{XX}}}{X^{2}-T^{2}}}
\left(X,T,0,0\right)\,,
\label{eq:4velocity_value_rminus}
\end{equation}
where, in this case, $g^{^{XX}}$ is the $XX$ component of the inverse
of the metric in Eq.~(\ref{eq:metric_RN_rminus}).  We see that
Eq.~(\ref{eq:4velocity_value_rminus}) only makes sense physically, if
$X^{2}-T^{2}>0$, which, taking into account the second of the
equations in Eq.~(\ref{eq:metric_RN_rminus}), allows us to conclude
that the shell must then be located either at the region $\mathrm{III}$
or $\mathrm{III}'$, see
Figure~\ref{Fig:Penrose_diagram_RN_non_extremal}.  Let us remark that
the minus sign in Eq.~(\ref{eq:4velocity_value_rminus}) arises from
the convention that the 4-velocity points to the future for both
regions $\mathrm{III}$ and $\mathrm{III}'$. Making use of
Eqs.~(\ref{eq:metric_RN_rminus}) and (\ref{eq:4velocity_value_rminus})
to find the induced metric on $\mathcal{S}$ as seen by an observer at
$\mathcal{M}_{\rm e}$ and imposing the first junction condition,
Eq.~(\ref{eq:1st_junct_cond}), we deduce that the shell's radial
coordinate $R$ is the same as measured by an observer at
$\mathcal{M}_{\rm i}$ or $\mathcal{M}_{\rm e}$ and the induced metric
on $\mathcal{S}$ is given by Eq.~(\ref{eq:induced_metric_RN}), namely,
\begin{equation}
\left.ds^{2}\right|_{\mathcal{S}}=-d\tau^{2}+R^{2}d\Omega^{2}\,.
\label{eq:induced_metric_RNcauchy}
\end{equation}
Combining $n_{\rm e}^{\alpha}n_{{\rm e}\alpha}=1$, see
Eq.~(\ref{eq:normal_normalized}),
$n_{{\rm e}\alpha}u_{\rm e}^{\alpha}=0$, see
Eq.~(\ref{eq:normal_orthogonal}),
and Eq.~(\ref{eq:4velocity_value_rminus}), we find the expression
for the components of the unit normal to the hypersurface $\mathcal{S}$,
as seen from the exterior spacetime $\mathcal{M}_{\rm e}$, to be
$n_{{\rm e}\alpha}=\pm\sqrt{\frac{g_{_{XX}}}{X^{2}-T^{2}}}
\left(-T,X,0,0\right)$.
To specify the sign of the normal to $\mathcal{S}$ for each region
we  consider two orientations: the orientation
where the normal $n_{{\rm e}\alpha}$
points
towards the Cauchy radius at $r_-$ and the orientation
where the normal
points towards the singularity  $r=0$. These two orientations can be treated
in a concise way by assuming, for example, a shell located either in
the region $\mathrm{III}$ or $\mathrm{III}'$ and the normal pointing
in the direction of decreasing $X$ coordinate, such that 
\begin{equation}
n_{{\rm e}\alpha}=\text{sign}\left(X\right)\sqrt{
\frac{g_{_{XX}}}{X^{2}-T^{2}}}\left(T,-X,0,0\right)\,.
\label{eq:normal_value_rminus}
\end{equation}
Note the importance of the sign of the normal to
yield totally different physical and geometrical
properties to a shell in the same location, here
in the region $R<r_-$. 
Then, using the
results from
Appendix~\ref{Appendix_subsec:Extrinsic_curvature_inside_Cauchy_horizon},
we find the nonzero components of the extrinsic curvature of
$\mathcal{S}$
as seen from the exterior spacetime to be given by 
\begin{equation}
{K_{\rm e}}^{\tau}{}_{\tau}=
\frac{\text{sign}\left(X\right)}{2R^{2}k}\left[r_++r_--2
\frac{r_+r_-}{R}\right]\,,\quad\quad
{K_{\rm e}}^{\theta}{}_{\theta}=
{K_{\rm e}}^{\varphi}{}_{\varphi}
=\frac{\text{sign}\left(X\right)\left(r_+-r_-\right)}{2r_-^{2}R}
\sqrt{g_{_{XX}}\left(X^{2}-T^{2}\right)}\,,
\label{eq:Nonextremal_Extrinsic_RN_inside_cauchy_horizon}
\end{equation}
where $k$, here, is the redshift function given in
Eq.~(\ref{eq:redshift}),
evaluated at $R$, i.e., $k(R,r_+,r_-)=\sqrt{\left(1-
\frac{r_+}{R}\right)\left(1-\frac{r_-}{R}\right)}$.

A comment is in order here.  In our study of a
shell in a nonextremal
Reissner-Nordstr\"om state, we have worked with two coordinate patches
to describe the various regions of the Reissner-Nordstr\"om spacetime
exterior to the shell as was done in~\citep{Comer_Katz_1994}, see also
Appendix~\ref{Appendix_sec:Kruskal-Szekeres_coordinates_RN}.  It is
possible to find a coordinate system that covers the entire
Reissner-Nordstr\"om spacetime without coordinate singularities,
see~\citep{Graves_Brill_1960} and also \cite{Carter_1966_2} or
\cite{Hawking_Ellis_book,MTW_Book,felicebook},
but we have not followed this path, as it
is not the best one to our aims, and thus we
have separated the study of a
shell located in a region described by one coordinate patch and the
other.

\subsubsection{Shell's energy density and pressure
\label{subSubsec:shellsenergydensityandpressurecauchy}}

We are now in position to find the properties of a perfect fluid thin
shell in a nonextremal Reissner-Nordstr\"om state, 
located
inside the 
Cauchy horizon radius or Cauchy radius, depending on the case.
The shell's stress-energy tensor is
given in Eq.~(\ref{eq:perfect}), an expression containing the energy
per unit area $\sigma$, the tangential pressure of the fluid $p$, the
four-velocity $u_a$, and the induced metric $h_{ab}$.
From our choice
of coordinates on $\mathcal{S}$ we have that $\left\{ y^{a}\right\}
=\left(\tau,\theta,\varphi\right)$,
the four-velocity $u_a$ is given in
Eq.~(\ref{eq:4velocity_value_rminus}),
and the metric $h_{ab}$ is  given
in Eq.~(\ref{eq:induced_metric_RNcauchy}). Putting everything together
we find
$S_{\tau}^{\tau}=-\sigma$,
$S_{\theta}^{\theta}=S_{\varphi}^{\varphi}=p$.  Comparing these latter
equations with the second junction condition,
Eq.~(\ref{eq:2nd_junct_cond}), taking into account the components of
the induced metric, given through Eq.~(\ref{eq:induced_metric_RNcauchy}),
and the fact that
$\left[K_{\theta}^{\theta}\right]=\left[K_{\varphi}^{\varphi}\right]$,
we find $\sigma=-\frac{1}{4\pi}\left[K_{\theta}^{\theta}\right]$ and
$p=\frac{1}{8\pi}\left[K_{\tau}^{\tau}\right]-\frac{\sigma}{2}$.  With
the components of the extrinsic curvature found in
Eqs.~(\ref{eq:Extrinsic_curvature_Mink})
and~(\ref{eq:Nonextremal_Extrinsic_RN_inside_cauchy_horizon}) we
obtain the following properties of a perfect fluid thin shell
located inside of the Cauchy radius,
\begin{equation}
8\pi\sigma=\frac{2}{R}\left(1-\text{sign}\left(X\right)k\right)\,,
\label{eq:sigma_value_rminus}
\end{equation}
\begin{equation}
8\pi p=\frac{\text{sign}\left(X\right)}{2Rk}\left[\left(1-
\text{sign}\left(X\right)k\right)^{2}-\frac{r_+r_-}{R^{2}}
\right]\,,\label{eq:pressure_value_rminus}
\end{equation}
where $k$ here is the redshift function given
in Eq.~(\ref{eq:redshift})
evaluated at $R$, i.e., $k(R,r_+,r_-)
=\sqrt{\left(1-\frac{r_+}{R}\right)
\left(1-\frac{r_-}{R}\right)}$.
As the surface electric current density $s_a$ on the
thin shell is defined as $s_a=\sigma_{e}u_a$, where
$\sigma_{e}$ represents the
electric charge density and $u_a$ is the velocity
of the shell,
from
Eqs.~(\ref{eq:junct_cond_Faradayb})-(\ref{eq:junct_cond_Faraday2})
and~(\ref{eq:RN_FaradayMaxwell_value})
it follows that
\begin{equation}
8\pi\sigma_{e}=2\frac{\sqrt{r_+r_-}}{R^{2}}\,.
\label{eq:chargedensitycuachy}
\end{equation}
Now, the
expressions found for the energy density
and pressure for a shell locate inside the
Cauchy radius, Eqs.~(\ref{eq:sigma_value_rminus}) and
(\ref{eq:pressure_value_rminus}), are the same as
Eqs.~(\ref{eq:sigma_value_rplus}) and (\ref{eq:pressure_value_rplus})
found for for the energy density
and pressure for a shell locate outside the gravitational radius.
However, the behavior of the properties of the shell will be different
since, the radial coordinate of the shell, $R$, in this case ranges
between zero and $r_-$.  As before, we have to distinguish the two
possible orientations provided by the $\text{sign}\left(X\right)$.  In
Eqs.~(\ref{eq:sigma_value_rminus})
and~(\ref{eq:pressure_value_rminus}) it is clear that it is necessary
to pick the sign in $\text{sign} \left(X\right)$.  Let us start with
$\text{sign}\left(X\right)=+1$. It is useful to give the expressions
for the shell's energy density and pressure, $\sigma$ and $p$ in terms
of $M$ and $Q$.  Using Eq.~(\ref{eq:KS_horizons_radius0}) in
Eqs.~(\ref{eq:sigma_value_rplus}) and~(\ref{eq:pressure_value_rplus})
with $\text{sign} \left(X\right)=+1$ we have
$8\pi\sigma=\frac{2}{R}\left(1-k\right)$, $8\pi
p=\frac{1}{2Rk}\left[\left(1-k\right)^{2}-
\frac{Q^{2}}{R^{2}}\right]$, and also from
Eq.~(\ref{eq:chargedensity1}) we have $8\pi\sigma_{e}=\frac{2Q}{R^{2}}$.
Let us now take $\text{sign}\left(X\right)=-1$. It is useful
to give the expressions for the shell's energy density and pressure,
$\sigma$ and $p$ in terms of $M$ and $Q$.  Using
Eq.~(\ref{eq:KS_horizons_radius0}) in
Eqs.~(\ref{eq:sigma_value_rplus}) and~(\ref{eq:pressure_value_rplus})
with $\text{sign} \left(X\right)=-1$ we have
$8\pi\sigma=\frac{2}{R}\left(1+k\right)$, $8\pi
p=-\frac{1}{2Rk}\left[\left(1+k\right)^{2}-\frac{Q^{2}}{R^{2}}
\right]$,
with
$k=\sqrt{1-\frac{2M}{R}+\frac{Q^{2}}{R^{2}}}$,
and also from Eq.~(\ref{eq:chargedensity1}) we have
$8\pi\sigma_{e}=\frac{2Q}{R^{2}}$.  These are the expressions used in
the two previous subsections.
Note also that when $r_-=0$,
then $\sigma_{e}=0$ and the
electric charge $Q$ is zero, $Q=0$,
and since $R<r_-$ we obtain that
there either the solution is vacuum and singular
or there is no solution, in brief,
there is no shell solution.

\clearpage{}

\section{Extremal electric thin shells outside
the gravitational radius:
Majumdar-Papapetrou star shells and extremal tension shell singularities}
\label{Sec:Extremal-thin-shells-outside}

\subsection{Extremal electric thin shells outside the
gravitational radius: Majumdar-Papapetrou star shells}
\label{Subsec:extremalnormaloutside}

Here we study the case of a fundamental electric thin
shell in the
extremal state, i.e., $r_+=r_-$ or $M=Q$, and indeed, $r_+=r_-=M=Q$,
for which the shell's 
location obeys $R>r_+$, and for which the orientation is such that the
normal to the shell points towards spatial infinity.  In this case
horizons do not exist and so, following the nomenclature, $r_+$ is
the gravitational radius. Also, since
$r_+$ and $r_-$ have the same value we opt to use consistently
the gravitational radius $r_+$ rather than the Cauchy radius
$r_-$. In general we also opt to use $M$ rather than $Q$.
The normal to the shell
pointing towards spatial infinity means
that the new parameter $\xi$ we introduce for the extremal states
has value $\xi=+1$, see the
end of this section.

\begin{figure}[h]
\subfloat[]{
\includegraphics[scale=0.45]
{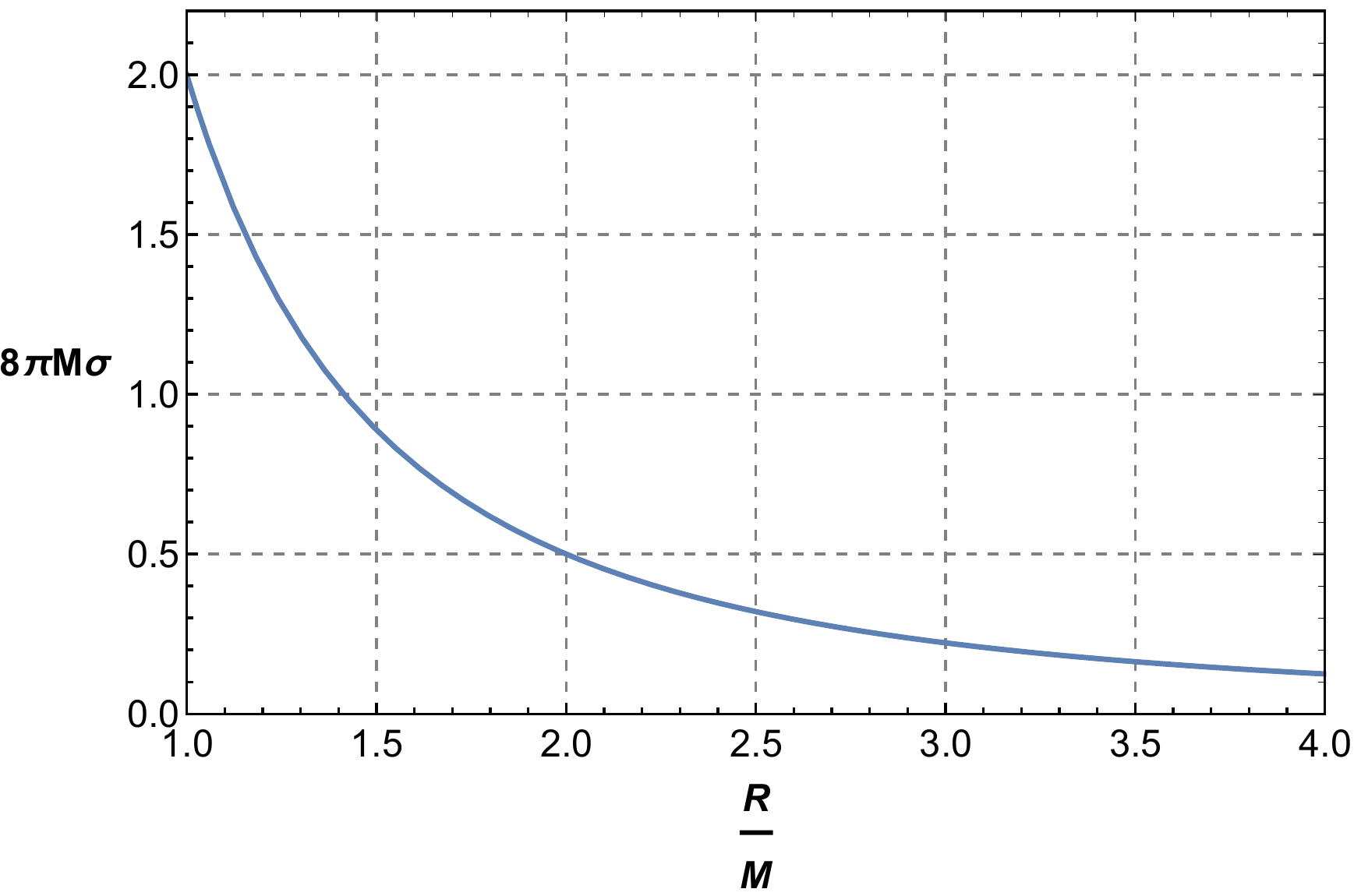}}
\hspace*{\fill}
\subfloat[]{
\includegraphics[scale=0.45]
{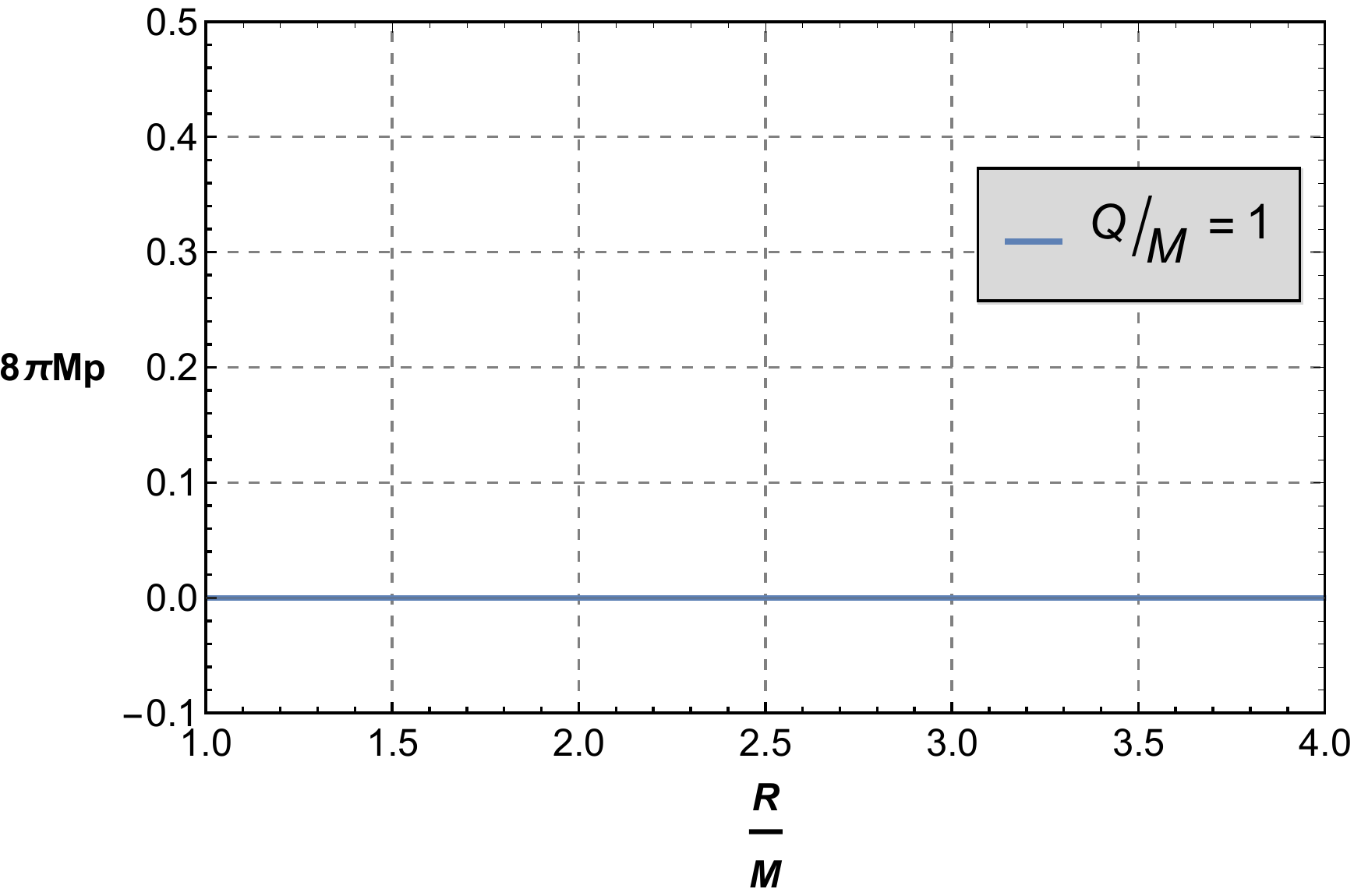}}
\caption{\label{Fig:Properties_extremal_outside_normal}
Physical properties of a Majumdar-Papapetrou star
shell, i.e., an electric
perfect fluid thin shell in an extremal Reissner-Nordstr\"om state, in
the location $R>r_+$, i.e., located outside the gravitational radius,
and with orientation such that the normal points towards spatial
infinity.
The interior is Minkowski and the exterior is extremal
Reissner-Nordstr\"om spacetime.  Extremal
means $\frac{Q}{M}=1$. Panel (a) Energy density $\sigma$ of
the shell as a function of the radius $R$ of the shell.
The energy density is adimensionalized through
the mass $M$, $8\pi M\sigma$, and the radius is adimensionalized
through the gravitational radius $r_+$, $\frac{R}{r_+}$.  Panel (b)
Pressure $p$ on the shell as a function of the radius $R$ of the
shell. 
The radius is adimensionalized through
the gravitational radius $r_+$, $\frac{R}{r_+}$.
The pressure is zero, the
shell is supported by electric repulsion alone, it is
Majumdar-Papapetrou matter.
}
\end{figure}

As functions of $M$ and $R$, the shell's energy density $\sigma$ and
pressure $p$, are, see the end of this section,
\begin{align}
8\pi\sigma= & \frac{2M}{ R^{2}}\,,
\label{eq:Extremal_sigma_value_outside_xi_positive}\\
8\pi p= & 0\,.\label{eq:Extremal_pressure_value_outside_xi_positive}
\end{align}
Also, the electric charge density $\sigma_{e}$
is given in terms of $M$ and $R$, by
\begin{equation}
8\pi\sigma_{e}=\frac{2M}{ R^{2}}\,,
\label{eq:chargedensity12extremal}
\end{equation}
The behavior of $\sigma$ and $p$, in
Eqs.~(\ref{eq:Extremal_sigma_value_outside_xi_positive})
and ~(\ref{eq:Extremal_pressure_value_outside_xi_positive}),
as functions of the radial coordinate
$R$ of the $\frac{Q}{M}=1$ extremal shell 
is shown in Figure~\ref{Fig:Properties_extremal_outside_normal}.
These shells are characterized by a positive energy density and
vanishing pressure support, and so the matter that composes this kind
of shells is Majumdar-Papapetrou matter, i.e., electric dust, there is
no need for matter pressure since there is an inbuilt equilibrium
between gravitational attraction and electrostatic repulsion.  These
are extremal star shells or Majumdar-Papapetrou star shells.
Majumdar-Papapetrou matter shells with a Minkowski interior matched to
an exterior extremal Reissner-Nordstr\"om spacetime, with the implicit
assumption that the outward unit normal to the matching surface points
towards spacial infinity, have been considered in
many works.  Notice that when $R\to\infty$, the energy density
$\sigma$ and the charge density $\sigma_{e}$, all tend to zero, i.e.,
the shell disperses away.  Notice also that when $R\to r_+$, the
energy density is finite, the pressure remains zero, and the charge
density $\sigma_{e}$ is also finite. Indeed, for $R= r_+$ one has a
quasiblack hole, discussed in detail ahead.  When
$Q=0$, and so $M=0$, there is no shell, only Minkowski spacetime.  In
relation to the energy conditions of the shell one can work out and
find that the null, the weak, the dominant, and the strong energy
conditions are verified for $R>r_+$, see a detailed presentation
ahead.

\begin{figure}[h]
\includegraphics[height=0.29\paperheight]
{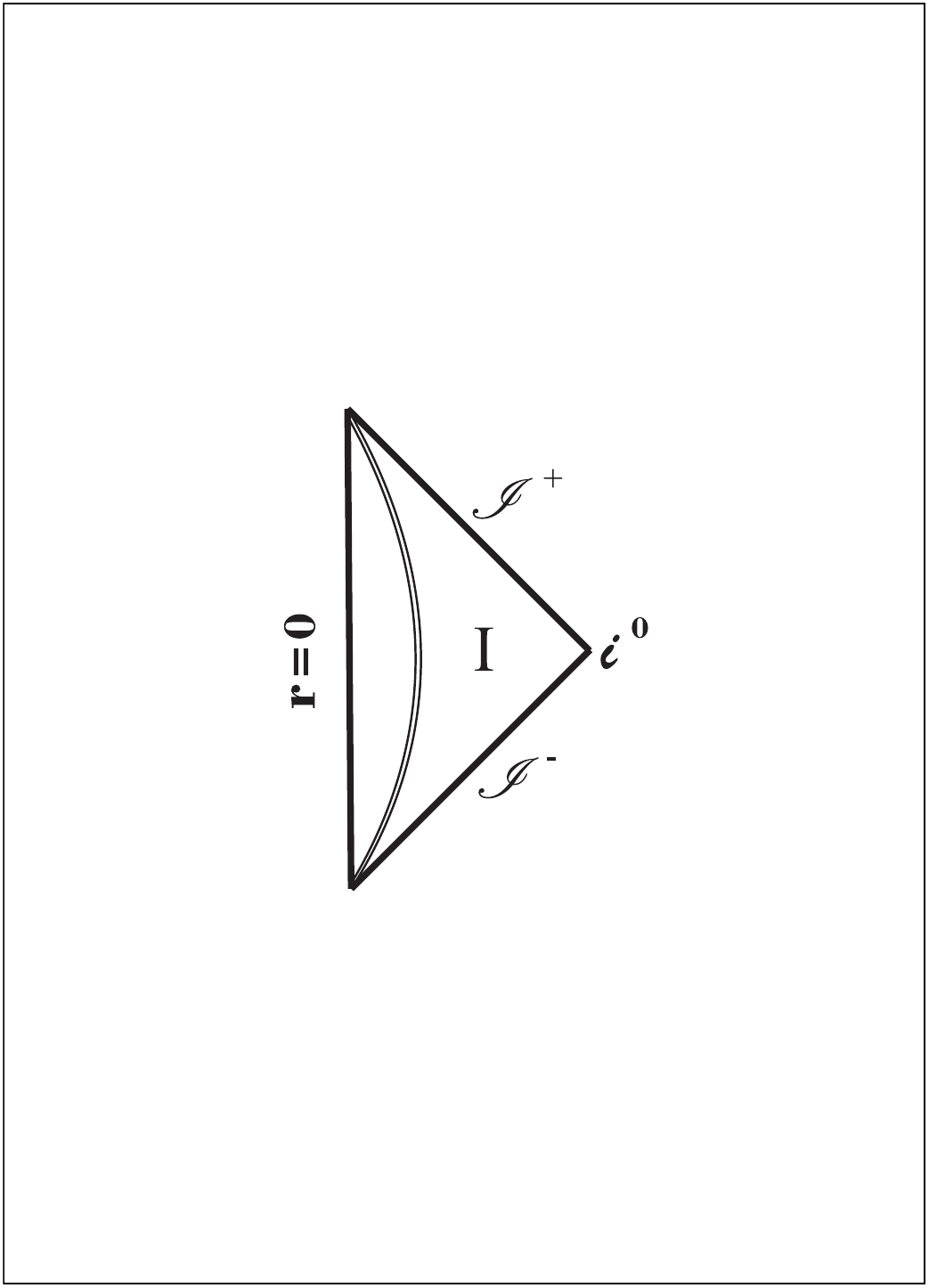}
\caption{\label{Fig:Penrose_diagram_Mink_extremal_RN_outside}
Carter-Penrose diagram of a Majumdar-Papapetrou star shell, i.e., a
thin shell spacetime in an extremal Reissner-Nordstr\"om state,
located at $R>r_+$, i.e., located outside the gravitational radius,
with orientation such that the normal points towards
spatial infinity.  The
interior is Minkowski, the exterior is extremal Reissner-Nordstr\"om.
This star shell is supported by electrical repulsion alone.
}
\end{figure}

The Carter-Penrose diagram can be drawn directly from the building
blocks of an interior Minkowski spacetime and the exterior asymptotic
region of an extremal Reissner-Nordstr\"om spacetime.  In
Figure~\ref{Fig:Penrose_diagram_Mink_extremal_RN_outside} the
Carter-Penrose diagram of an extremal Reissner-Nordstr\"om shell
spacetime for a junction surface with normal pointing towards spatial
infinity is shown.  It is clearly a star shell, a Majumdar-Papapetrou
star shell in an asymptotically flat spacetime.

The physical interpretation of this case is clear cut, and it is
similar to the corresponding nonextremal shell. This extremal thin
shell solution mimics an extremal star.  The energy density and
pressure obey the energy conditions for any radius, indeed the shell
is composed of Majumdar-Papapetrou matter. The causal and global
structure as displayed by the Carter-Penrose diagram are well behaved
and rather elementary.  So, this case falls into the category of having
the energy conditions verified and the geometrical setup is physically
reasonable.

\newpage

\subsection{Extremal electric thin shells outside the
event horizon: Extremal tension shell singularities}
\label{Subsec:extremalnormaloutsidenormaltoin}

Here we study the case of a fundamental electric thin shell in the
extremal state, i.e., $r_+=r_-$ or $M=Q$, and indeed, $r_+=r_-=M=Q$,
for which the shell's location obeys $R>r_+$, and for which the
orientation is such that the normal to the shell points towards $r_+$.
In this case horizons do exist and so, following the nomenclature,
$r_+$ is both the gravitational and the event horizon radius.  Also,
$r_+$ and $r_-$ have the same value and we opt to use the event
horizon radius $r_+$ rather than the Cauchy horizon radius $r_-$. We
also opt to use $M$ rather than $Q$.  The normal to the shell pointing
towards $r_+$ means in the notation we use that we take $\xi=-1$, see
the end of this section for details.

\begin{figure}[h]
\subfloat[]{\includegraphics[scale=0.45]
{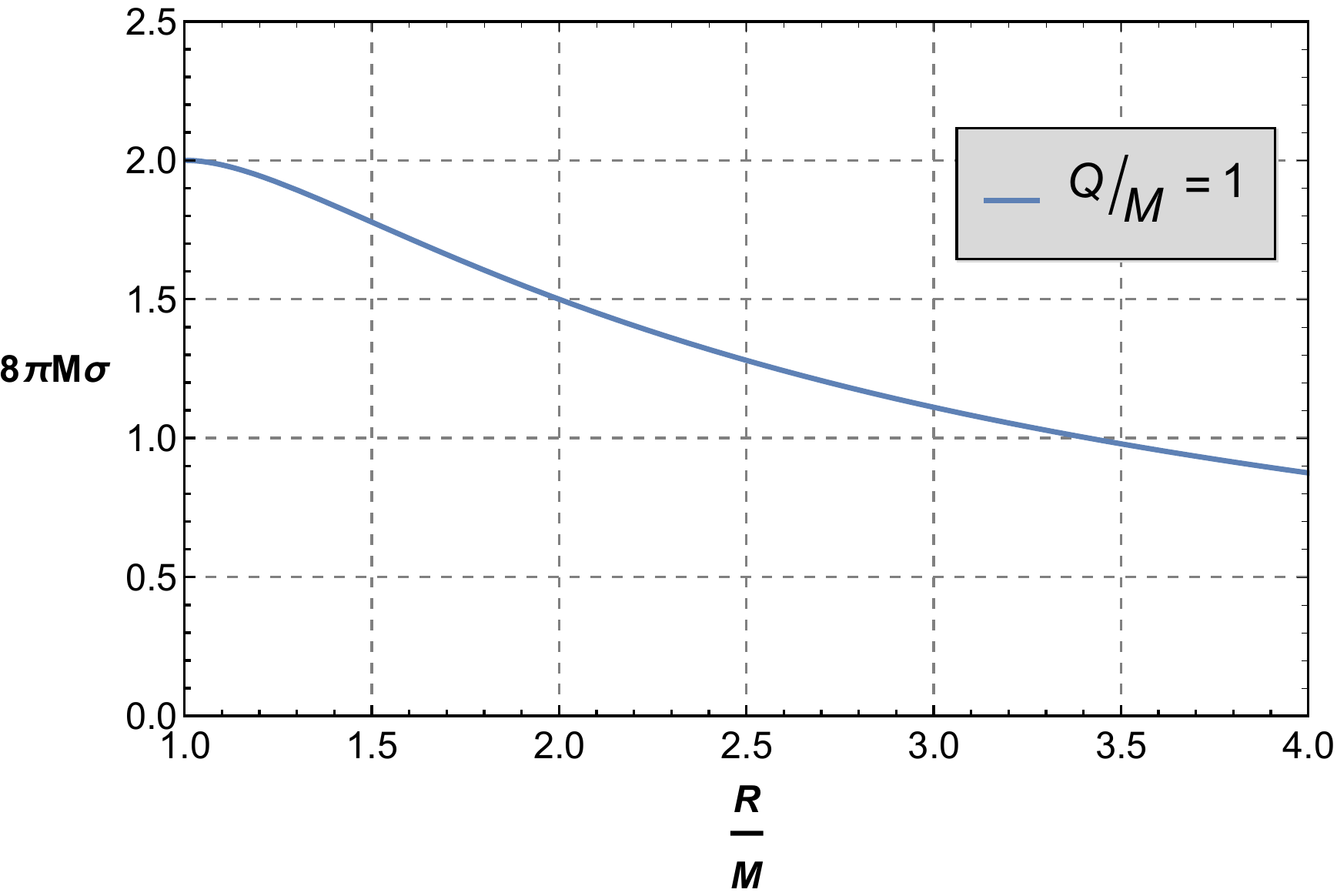}}
\hspace*{\fill}
\subfloat[]{\includegraphics[scale=0.45]
{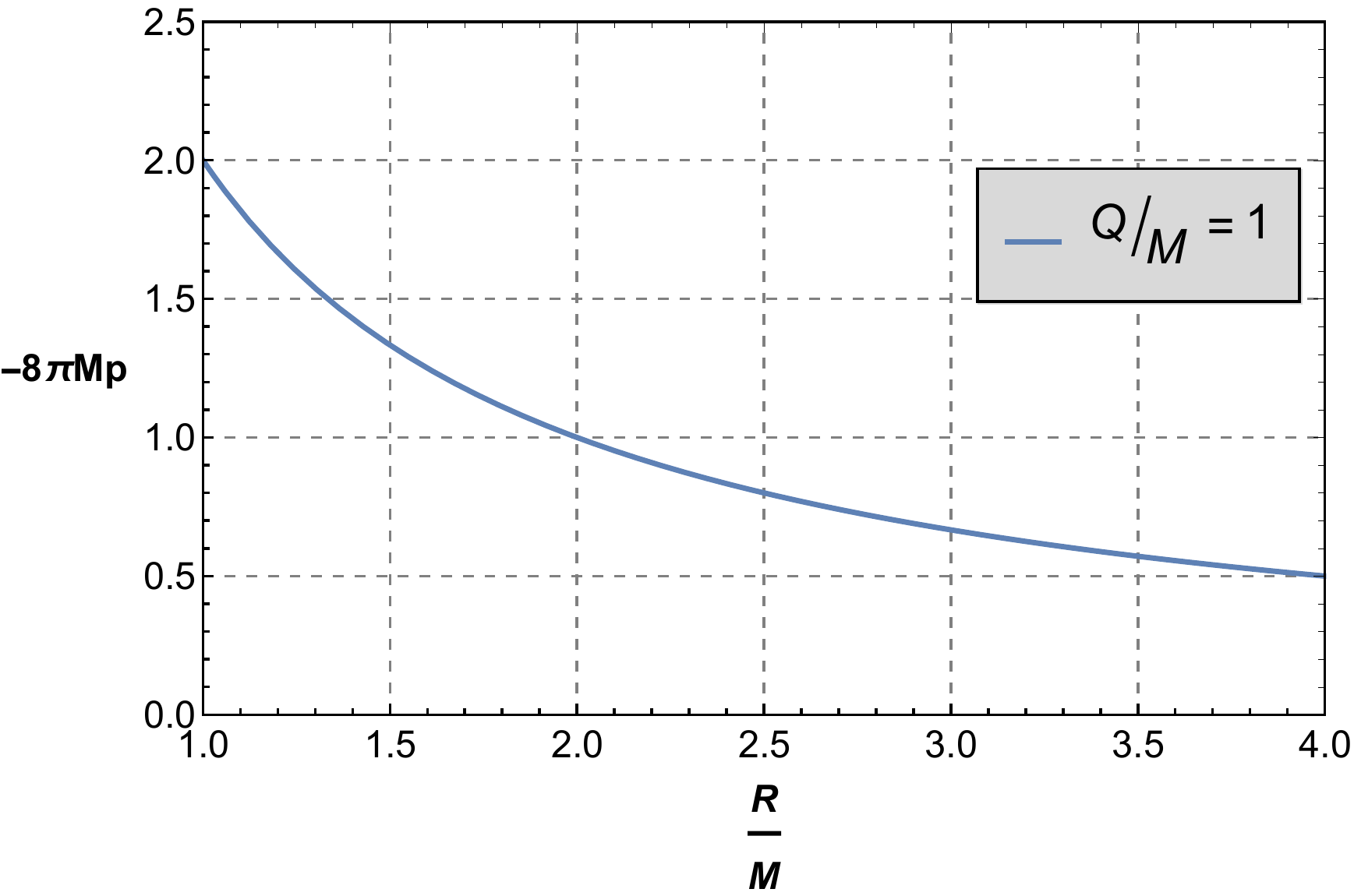}}
\caption{\label{Fig:Properties_extremal_outside_alternative}
Physical properties of an extremal tension shell singularity,
i.e.,  an electric
perfect fluid thin shell in an
extremal Reissner-Nordstr\"om state,
in the location
$R>r_+$, i.e., located outside the event horizon, with orientation
such that the normal points towards $r_+$.  The interior is Minkowski,
the exterior is extremal Reissner-Nordstr\"om spacetime.
Extremal means $\frac{Q}{M}=1$.
Panel (a)
Energy density $\sigma$ of the shell as a function of the radius $R$
of the shell.   The energy density is
adimensionalized through the mass $M$, $8\pi M\sigma$, and the radius
is adimensionalized through the gravitational radius $r_+$,
$\frac{R}{r_+}$.  Panel (b) Pressure $p$ on the shell as a function of
the radius $R$ of the shell.  The pressure is negative, so the shell
is supported by tension.  The radius is
adimensionalized through the gravitational radius $r_+$,
$\frac{R}{r_+}$.
}
\end{figure}

As functions of $M$ and $R$, the shell's energy density $\sigma$ and
pressure $p$, are, see the end of this section,
\begin{align}
8\pi\sigma= & \frac{2}{R}\left(2-\frac{M}{R}\right)\,,
\label{eq:Extremal_sigma_value_outside_xi_negative}\\
8\pi p= & -\frac{2}{R}\,.
\label{eq:Extremal_pressure_value_outside_xi_negative}
\end{align}
The electric charge density $\sigma_{e}$ is given in terms of
$M$ and $R$ by
$
8\pi\sigma_{e}=\frac{2M}{ R^{2}}$,
which is 
identical to Eq.~(\ref{eq:chargedensity12extremal}).
The
behavior of $\sigma$ and $p$, in
Eqs.~(\ref{eq:Extremal_sigma_value_outside_xi_negative})
and~(\ref{eq:Extremal_pressure_value_outside_xi_negative}),
as functions
of the radial coordinate $R$ of the $\frac{Q}{M}=1$ extremal shell is
shown in Figure~\ref{Fig:Properties_extremal_outside_alternative}.
The matter fluid that composes such shells is characterized by
positive energy density $\sigma$ and is supported by tension $-p$,
with both falling to zero when $R=\infty$.  In this case since $p$ is
not zero, the shell is not composed of Majumdar-Papapetrou matter.
Notwithstanding the exterior spacetime is extremal.  Examples of
spacetimes for which $M=Q$ globally whose interior is not made of
Majumdar-Papapetrou matter, as is the case here, are many.  However,
this case is of of particular interest since matter properties
provided by
Eqs.~(\ref{eq:Extremal_sigma_value_outside_xi_negative})-(\ref{eq:Extremal_pressure_value_outside_xi_negative})
and the electric charge density
$
8\pi\sigma_{e}=\frac{2M}{ R^{2}}$,
have specific relevant features.  Indeed, $\sigma$ has two terms,
namely, an intrinsic geometrical one given by $\frac4R$ and a
gravitational one which is negative given by
$-\frac{2M}{R^2}$. These two terms can be considered independent and
$\sigma$ is the sum of the two.
The first term of $\sigma$, $\frac4R$, is a geometrical term that also
gives rise to a geometrical tension given by $-\frac2R$ and ensures
that there is a shell for sure caused from the embedding of the shell
in the interior and exterior spacetimes, as the radial distance grows
up to a maximum at the shell with
radius $R$ nd then diminishes to $r_+$ and finally to zero at the
timelike singularity.  This geometric term exists independently on
whether there is spacetime mass $M$ or not, indeed, the spacetime mass
energy coming from this geometrical term is zero since $\frac4R+2p=0$.
The second term $-\frac{2M}{R^2}$ is negative and can be explained by
the fact that due to the electric charge density
$
8\pi\sigma_{e}=\frac{2M}{ R^{2}}$
on the shell, there is electric
repulsion, and on
the other hand, since positive gravity is on the direction of $r_+$
and $r=0$, to counterbalance the electric repulsion and the direction
of positive gravity, the shell has to have an anti repulsive negative
energy density, an anti gravity term or anti Majumdar-Papapetrou
energy density term, of value $-\frac{2M}{R^2}$ .  Note also that
$\sigma+2p+\sigma_e=0$.
When $Q=0$, and so $M=0$, there is still a shell of radius $R$, but
with a Minkowski spacetime on each side of it.
In relation to the energy conditions of the shell one can work out and
find that the null, the weak, and the dominant energy conditions are
verified for $R>r_+$, and the strong energy condition is never
verified, see a detailed presentation ahead.

\begin{figure}[h]
\subfloat[
\label{Fig:Penrose_diagram_Mink_extremal_RN_outside_alternative_2}]
{\includegraphics[height=0.29\paperheight]
{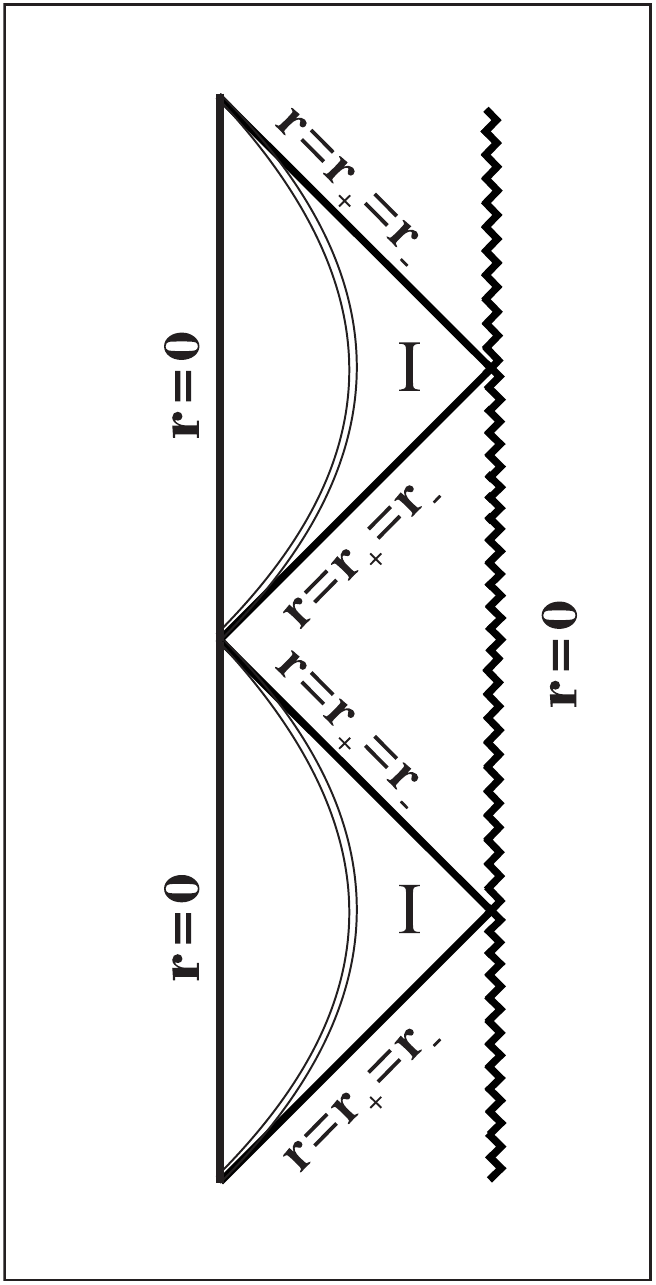}}
\hspace*{5cm}
\subfloat[
\label{Fig:Penrose_diagram_Mink_extremal_RN_outside_alternative_1}]
{\includegraphics[height=0.29\paperheight]
{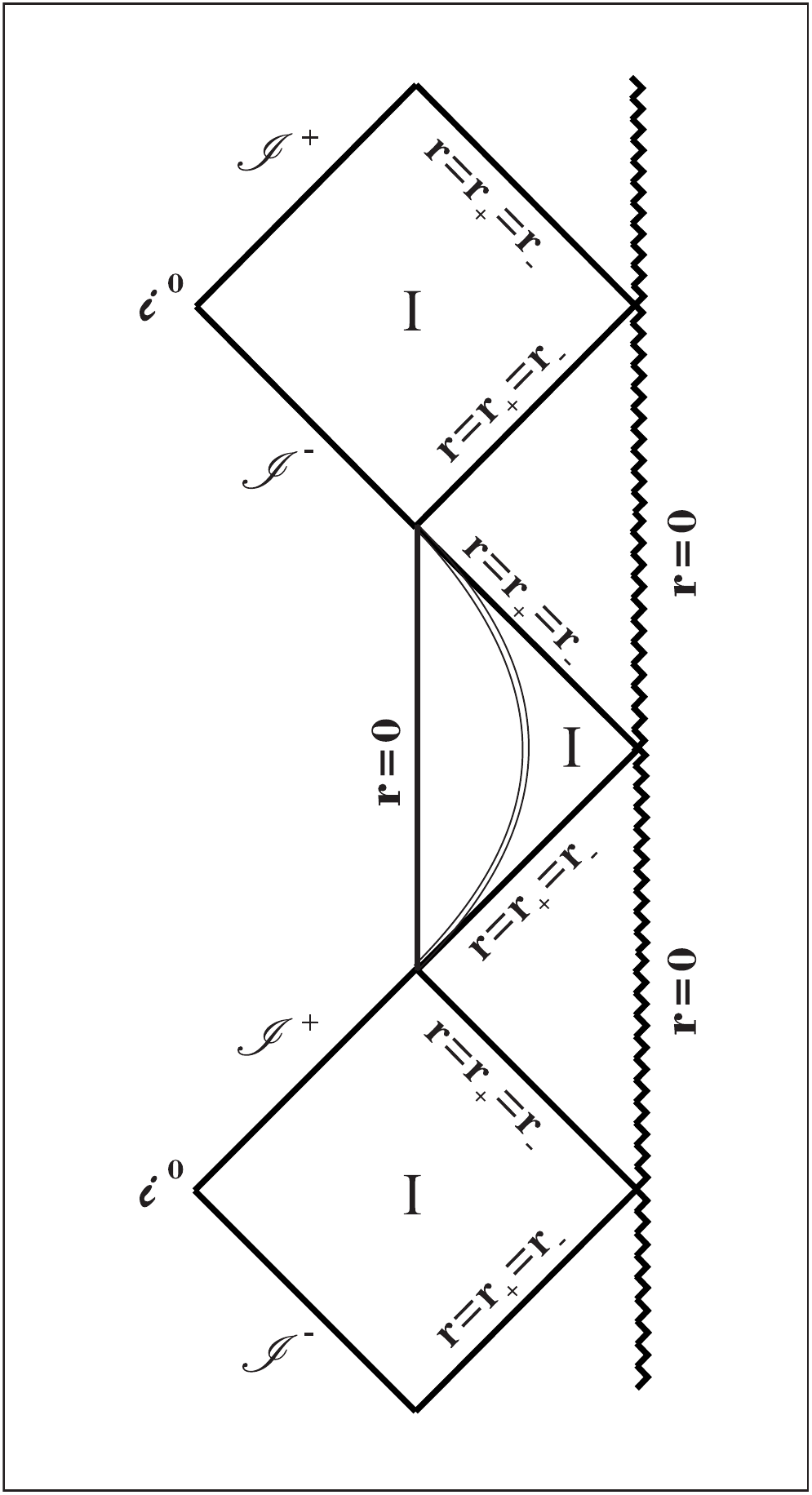}}
\caption{\label{Fig:Penrose_diagram_Mink_extremal_RN_outside_alternative}
Carter-Penrose diagrams of an extremal
tension shell singularity, i.e., a thin
shell spacetime in an extremal Reissner-Nordstr\"om state, in the
location $R>r_+$, i.e., located outside the event horizon, with
orientation such that the normal points towards $r_+$.  The interior
is Minkowski, the exterior is extremal Reissner-Nordstr\"om
spacetime.  
Panel~(a)
The Carter-Penrose diagram contains a shell in region
$\mathrm{I}$ and repeats itself upwards.
Panel~(b)
The Carter-Penrose diagram contains a shell in the regions
$\mathrm{I}$ shown that passes into asymptotically flat regions.
An infinite number of different Carter-Penrose diagrams can be drawn,
since there are an infinite number of combinations to place the shell
and infinity.
}
\end{figure}

The Carter-Penrose diagram for this case
can be drawn directly from the building
blocks of an interior Minkowski spacetime and the full extremal
Reissner-Nordstr\"om spacetime.
In
Figure~\ref{Fig:Penrose_diagram_Mink_extremal_RN_outside_alternative}
the Carter-Penrose diagram of a shell spacetime in an extremal
Reissner-Nordstr\"om state, in the location $R>r_+$, with orientation
such that the normal points towards $r_+$, i.e.,
$\xi=-1$, is shown.
It has a horizon,
but the existence of the singularity is more
striking, i.e., it is
an extremal  tension shell singularity.
There is an infinitude of possible diagrams as the maximal analytical
extension of the resulting spacetime can always contain a thin matter
shell outside the event horizon or only at a discrete number of these
regions.  In the diagram (a) the tension shell is outside the
event horizon in region
$\mathrm{I}$. Then, the tension shell repeats itself
in the next portion of the diagram. It is a compact tension shell that
repeats itself.  In the diagram (b) of the figure the tension shell is
outside the event horizon in region $\mathrm{I}$. Then, an asymptotic
infinity takes over in the next portion of the diagram.  Since what
one puts in the regions $\mathrm{I}$, either a shell or infinity, is
not decided by the solution, indeed an infinite number of different
Carter-Penrose diagrams can be drawn, as there are an infinite number
of combinations to locate a shell or infinity when one goes upward or
downward through the diagram.  This is a tension shell, but since it
is extremal there is no Einstein-Rosen bridge, no dynamic wormhole.

The physical interpretation of this case is somewhat simple, with the
case itself being unusual. This extremal thin shell solution, in its
simplest form, turns the space around up to a horizon and then opens
up to another universe with another shell, or to a singularity, and so
on.  The energy density and pressure have special features as has
been just pointed out, and obey some of the energy conditions.  The
causal and global structures as displayed by the Carter-Penrose
diagram show the unique features of this spacetime.  So, this case
falls into the category of having some of
the energy conditions verified and
the geometrical setup is strange.

\newpage

\subsection{Formalism for extremal electric thin
shells outside the gravitational radius
\label{Subsec:Extremal_induced_Minkowski_RN_ouside_event_horizon}}

\subsubsection{Preliminaries}
\label{prel3}

We now make a careful study to derive the properties of the
fundamental electric thin shell used in the two previous subsections,
i.e., the thin shell in an extremal state, i.e., $r_+=r_-$ or
$M=Q$, and indeed 
 $r_+=r_-=M=Q$,
for which the shell's radius $R$ location obeys $R>r_+$, and
for which the orientation is such that the normal to the shell points
towards infinity or towards $r_+$.
It should be read as an appendix to the previous two
subsections.  We use the formalism developed in
Sec.~\ref{Sec:Junction_formalism}.

\subsubsection{Induced metric, and extrinsic curvature of
$\mathcal{S}$ as seen
from $\mathcal{M}_{\rm i}$}
\label{Subsec:Extremal_induced_Mink_outside}

Let us start by analyzing the interior
Minkowski spacetime, $\mathcal{M}_{\rm i}$. Since it is the same as
the analysis done previously we only quote
the important equations.
They are the interior metric Eq.~(\ref{eq:Mink_metric_interior}),
the interior four-velocity of the shell
Eq.~(\ref{eq:Mink_vel_explicit}), the
metric for the shell at radius $R$
Eq.~(\ref{eq:induced_metric_Mink}), the normal to the shell
Eq.~(\ref{eq:normal_Mink}),
and the extrinsic curvature from the inside
Eq.~(\ref{eq:Extrinsic_curvature_Mink}).

\subsubsection{Induced metric, and extrinsic curvature
of $\mathcal{S}$ as seen from $\mathcal{M}_{\rm e}$}
\label{Subsec:Extremal_induced_RN_outside}

To proceed we have now to find the expressions for the induced metric
on $\mathcal{S}$ and the extrinsic curvature components as seen from
the exterior spacetime, $\mathcal{M}_{\rm e}$, in the extremal
state, i.e., $r_+=r_-$ or $M=Q$, see
Figure~\ref{Fig:Penrose_diagram_RN_extremal}, for which the
shell's radius $R$ location obeys $R>r_+$, and for which the
orientation is such that the normal to the shell points towards
increasing $r$, i.e., towards infinity,
or towards decreasing $r$ i.e., towards $r_+$,
as seen from the exterior, as
used in the two previous subsections.

For an extremal shell located at $R>r_+$
one has also to be concerned
about the normal vector to the shell. 
In the extremal Reissner-Nordstr\"om spacetime
there is no Einstein-Rosen-bridge and so 
there is no
ambiguity in the definition of the radial coordinate as the value
of the circumferential radius. Thus, there is no need
for the Kruskal-Skekeres $\left( T,X,\theta,\varphi\right)$
coordinates and we can resort
in this analysis of the induced
metric and extrinsic curvature of the matching surface
using simply the Schwarzschild
coordinates $\left(t,r,\theta,\varphi\right)$. The
Reissner-Nordstr\"om line element for the exterior extremal solution is
\begin{equation}
ds_{\rm e}^{2}=-\left(1-\frac{r_+}{r}\right)^2dt^{2}
+\frac{dr^{2}}{\left(1-\frac{r_+}{r}\right)^2}
+r^{2}d\Omega^{2}\,.
\label{RNextremal2}
\end{equation}

Assuming the circumferential radius of the matching surface
$\mathcal{S}$ to be described by a function ${R}
\left(\tau\right)$, where $\tau$ is the proper time of an observer
comoving with $\mathcal{S}$ and imposing the shell to be static implies
that $\frac{d{R}}{d\tau}=0$. Then, the 4-velocity of an observer
comoving with $\mathcal{S}$, as seen from $\mathcal{M}_{\rm e}$, is
given by
\begin{equation}
u_{\rm e}^{\alpha}=\left(\frac{1}{k},0,0,0\right)\,,
\label{eq:Extremal_4velocity_exterior_outside_horizon}
\end{equation}
where, in this situation the redshift function  $k$
at $\mathcal{S}$ is given in
Eq.~(\ref{eq:redshift}),
evaluated at $R$, i.e., $k(R,r_+=r_-)\equiv k(R,r_+)=1-
\frac{r_+}{R}$.
Equation~(\ref{eq:Extremal_4velocity_exterior_outside_horizon})
can now be used to compute the induced metric on $\mathcal{S}$ by
$\mathcal{M}_{\rm e}$, and we find
$
\left.ds^{2}_{\rm e}\right|_{\mathcal{S}}
=-d\tau^{2}+{R}^{2}d\Omega^{2}$.
Imposing the the first junction condition
Eq.~(\ref{eq:1st_junct_cond})
and
Eq.~(\ref{eq:induced_metric_Mink}) we find
that the shell's radial functions at each
sice of $\mathcal{S}$ are the same, and so the
matching surface $\mathcal{S}$
is characterized by the line element 
\begin{equation}
\left.ds^{2}\right|_{\mathcal{S}}=-d\tau^{2}+R^{2}d\Omega^{2}\,.
\label{eq:Extremal_induced_metric_outside_horizon}
\end{equation}
Using the normalization and orthogonality
relations~(\ref{eq:normal_normalized})
and (\ref{eq:normal_orthogonal}) allows us to find the following
expression for the normal
\begin{equation}
n_{{\rm e}\alpha}=\xi\left(0,\frac{1}{k},0,0\right)\,,
\label{eq:Extremal_normal_exterior_outside_horizon}
\end{equation}
where the parameter $\xi=\left\{ -1,1\right\} $ is defined as $\xi=+1$
if the outside unit normal to the shell points in the direction of
increasing radial coordinate $r$, measured by an observer in the
exterior $\mathcal{M}_{\rm e}$ spacetime, and $\xi=-1$ if the outside
unit normal to the shell points in the direction of decreasing radial
coordinate $r$, again, measured by an observer in the exterior
$\mathcal{M}_{\rm e}$
spacetime. In the extremal case
the parameter $\xi$ takes the place of the $\text{sign}\left(X\right)$
used in the nonextremal case.
Taking into account
Eqs.~(\ref{eq:normal_orthogonal}),
(\ref{eq:Extremal_4velocity_exterior_outside_horizon}),
and~(\ref{eq:Extremal_normal_exterior_outside_horizon}) we find that
the nonzero components of the extrinsic curvature of the matching
surface, see Eq.~(\ref{eq:extrinsic1}), are given by
\begin{equation}
{K_{\rm e}}^{\tau}{}_{\tau}=
\xi\frac{r_+}{R^{2}}
\,,\quad\quad
{K_{\rm e}}^{\theta}{}_{\theta}=
{K_{\rm e}}^{\varphi}{}_{\varphi}=
\xi\frac{k}{R}\,.
\label{eq:Extremal_extrinsic_curvature_RN_outside}
\end{equation}

\subsubsection{Shell's energy density and pressure
\label{subSubsec:shellsenergydensityandpressureextremalouts}}

Having determined the components of the extrinsic curvature of the
matching surface $\mathcal{S}$ as seen from the interior and exterior
spacetimes we are now in position to use the second junction
condition given in Eq.~(\ref{eq:2nd_junct_cond})
to find the expressions for the energy density and pressure support
of the extremal thin shell in these cases.
The shell's stress-energy tensor is given in Eq.~(\ref{eq:perfect}),
so
Eqs.~(\ref{eq:Extrinsic_curvature_Mink})
and (\ref{eq:Extremal_extrinsic_curvature_RN_outside}) yield
\begin{align}
8\pi\sigma= & \frac{2}{R}
\left[1-\xi\left(1-\frac{r_+}{R}\right)\right]\,,
\label{eq:Extremal_sigma_value_outside}\\
8\pi p= & \frac{1}{R}\left(\xi-1\right)\,,
\label{eq:Extremal_pressure_value_outside}
\end{align}
where again
here $k=1-\frac{r_+}{R}$. Note that $p$ in
Eq.~(\ref{eq:Extremal_pressure_value_outside})
is independent of $M$, it only depends on
$R$ and thus on the geometry of the shell as embedded in the ambient
spacetime.
Moreover, since the surface electric current density $s_a$ on
the thin shell is $s_a=\sigma_{e}u_a$, where $\sigma_{e}$
represents the electric charge density, and since the Minkowski
spacetime has zero electric charge,
from Eqs.~(\ref{eq:junct_cond_Faradayb}), (\ref{eq:junct_cond_Faraday2})
and~(\ref{eq:RN_FaradayMaxwell_value})
it follows that
\begin{equation}
8\pi \sigma_{e}=2\frac{r_+}{ R^{2}}\,.
\label{eq:chargedensity1extremalout}
\end{equation}
The radial coordinate of the shell is in the range $r_+<R<\infty$.

Equations~(\ref{eq:Extremal_sigma_value_outside})
and (\ref{eq:Extremal_pressure_value_outside}),
together with 
(\ref{eq:chargedensity1extremalout}),
can now be used to study the properties of the thin matter shells
separating a Minkowski spacetime from an exterior
extremal Reissner-Nordstr\"om
spacetime, located outside the extremal
gravitational radius $r_+$.
In Eqs.~(\ref{eq:Extremal_sigma_value_outside})
and~(\ref{eq:Extremal_pressure_value_outside})
it is clear that it is necessary to pick the sign of $\xi$.
Let us start with $\xi=+1$. It is useful to
give the expressions for the shell's energy density and pressure,
$\sigma$ and $p$. in terms of $M=Q$, where we opt for $M$.
Using Eq.~(\ref{eq:KS_horizons_radius0}), i.e., $r_+=M$, in
Eqs.~(\ref{eq:Extremal_sigma_value_outside})
and~(\ref{eq:Extremal_pressure_value_outside}) with $\xi=+1$ we have
$8\pi\sigma=  \frac{M}{4\pi R^{2}}$,
$8\pi p=  0$, and 
also from Eq.~(\ref{eq:chargedensity1extremalout}) we have
$8\pi \sigma_{e}=\frac{2M}{ R^{2}}$.
Let us now take $\xi=-1$.
It is useful to
give the expressions for the shell's energy density and pressure,
$\sigma$ and $p$, in terms of $M=Q$, where as usual we opt for $M$.
Using Eqs.~(\ref{eq:Extremal_sigma_value_outside})
and~(\ref{eq:Extremal_pressure_value_outside})
with $\xi=-1$ we have
$8\pi\sigma=  \frac{2}{R}\left(2-\frac{M}{R}\right)$,
$8\pi p=  -\frac{2}{R}$, and 
also from Eq.~(\ref{eq:chargedensity1extremalout}) we have
again $8\pi \sigma_{e}=\frac{2M}{ R^{2}}$.
These are the expressions used in
the two previous subsections.

\clearpage{}

\section{Extremal electric thin shells inside the gravitational
radius:
Extremal tension shell regular and nonregular black holes and
Majumdar-Papapetrou compact naked singularities}
\label{Sec:Extremal-thin-shells-inside12}

\subsection{Extremal electric thin shells inside the event horizon:
Extremal tension shell regular and nonregular black holes}
\label{Subsec:extremalnormalinsideplus}

Here we study the case of a fundamental electric thin shell in the
extremal state, i.e., $r_+=r_-$ or $M=Q$, and indeed, $r_+=r_-=M=Q$,
for which the shell's location obeys $R<r_+$, and so also $R<r_-$, and
for which the orientation is such that the normal to the shell points
towards $r_+$, i.e., we choose the quantity $\xi$ which gives the
direction of the normal as $\xi=+1$, see the end of this section for
details..  In this case horizons do exist and so, following the
nomenclature, $r_+$ is both the gravitational and the event horizon
radius, and since $r_+=r_-$ it is also the Cauchy horizon radius and
the Cauchy radius. We opt to use $r_+$ and $M$.

\begin{figure}[h]
\subfloat[]
{\includegraphics[scale=0.45]
{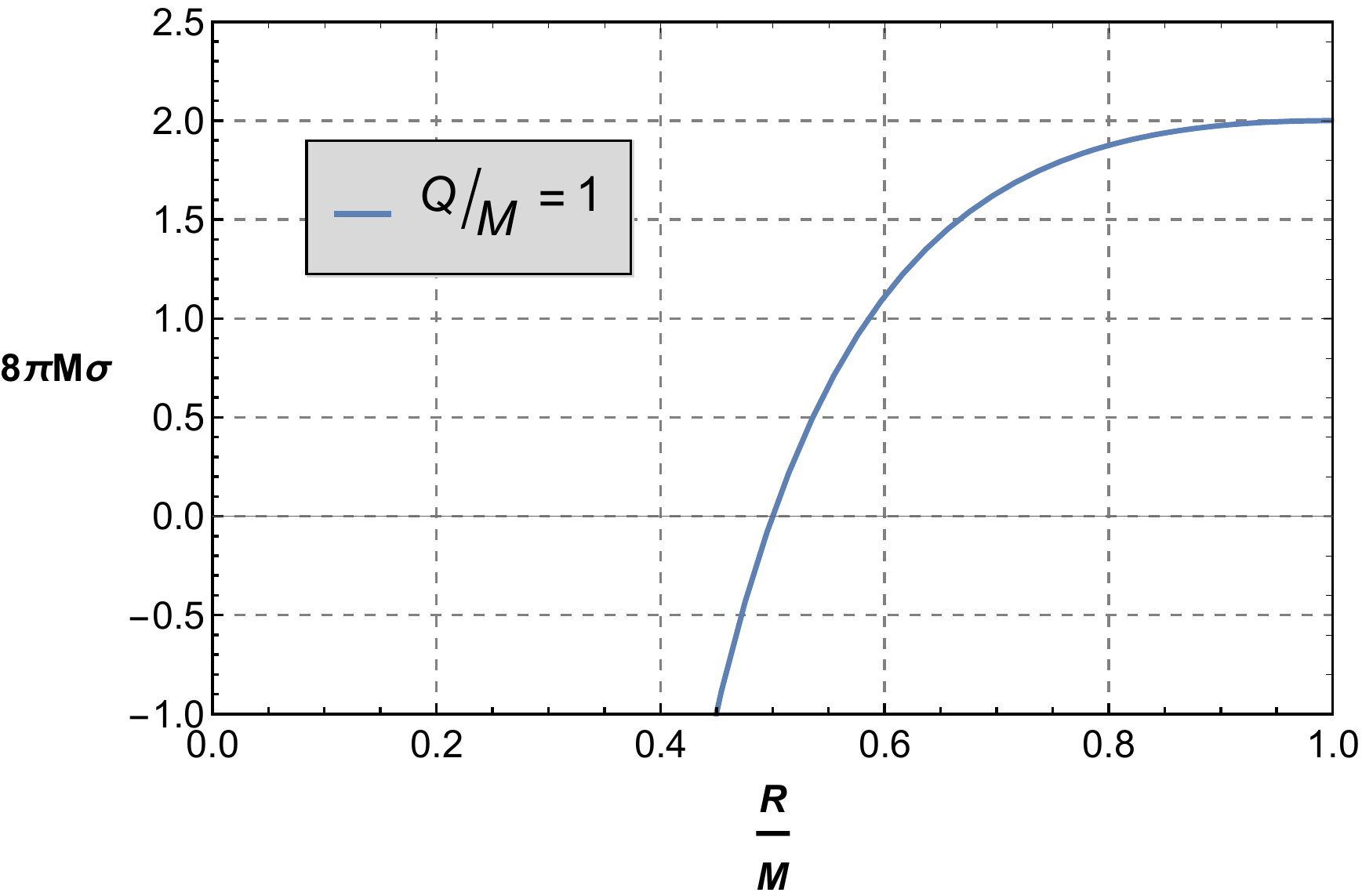}}
\hspace*{\fill}
\subfloat[]
{\includegraphics[scale=0.45]
{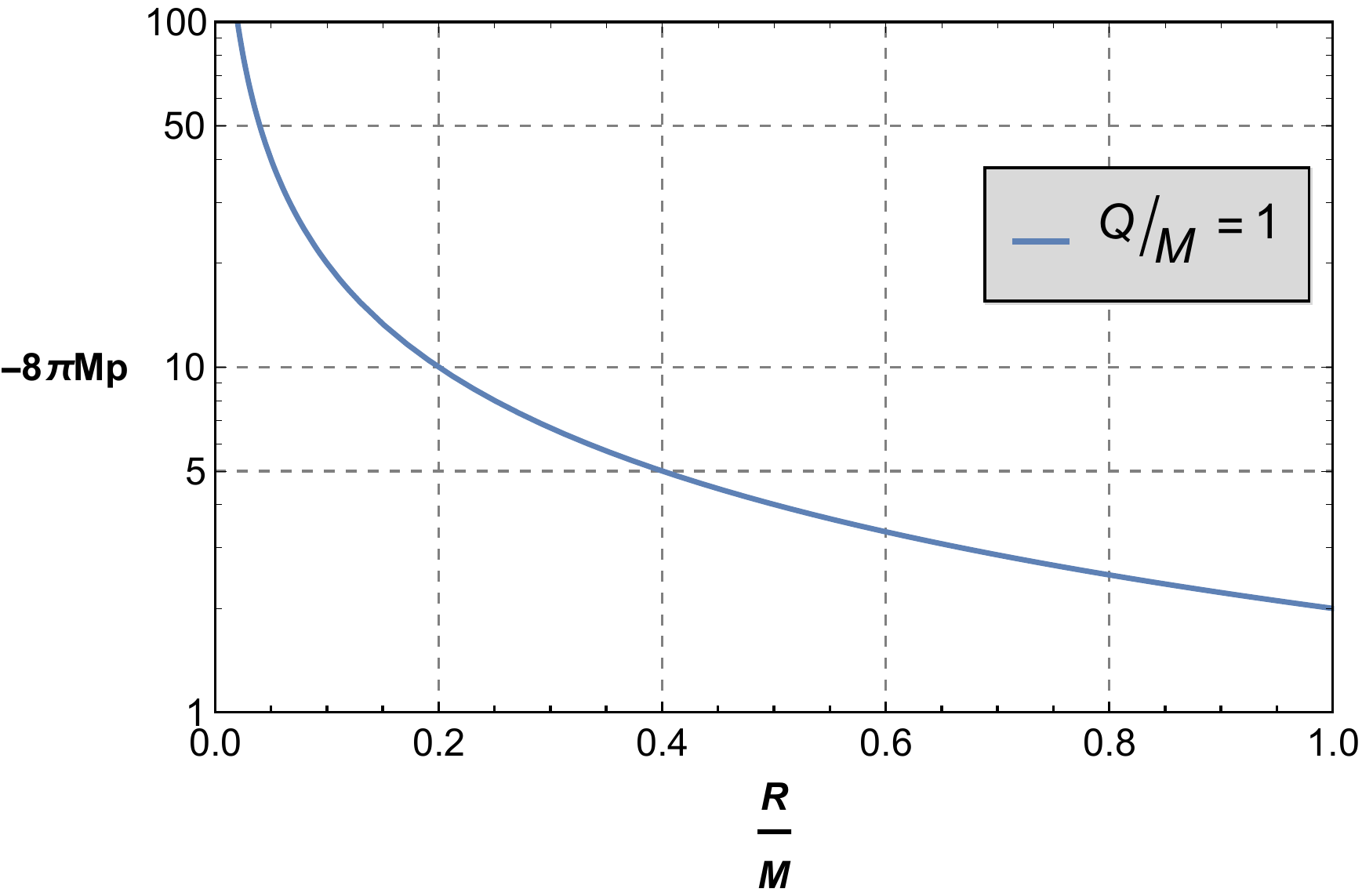}}
\caption{\label{Fig:Properties_extremal_inside_normal}
Physical
properties of an
extremal tension shell regular and nonregular black hole,
i.e., an electric
perfect fluid thin shell in an extremal Reissner-Nordstr\"om state, in
the location $R<r_+=r_-$, i.e., located inside the even hoizon, and
with orientation such that the normal points towards $r_+$.  The
interior is Minkowski and the exterior is extremal
Reissner-Nordstr\"om spacetime.
Extremal means $\frac{Q}{M}=1$.
Panel (a) Energy density $\sigma$ of
the shell as a function of the radius $R$ of the shell.
The energy density is adimensionalized through
the mass $M$, $8\pi M\sigma$, and the radius is adimensionalized
through the gravitational radius $r_+$, $\frac{R}{r_+}$.  Panel (b)
Tension $-p$ on the shell as a function of the radius $R$ of the
shell.   The tension is
adimensionalized through the mass $M$, $-8\pi Mp$, and the radius is
adimensionalized through the event horizon radius $r_+$,
$\frac{R}{r_+}$.
}
\end{figure}

As functions of $M$ and $R$, the shell's energy density $\sigma$ and
pressure $p$, are, see the end of this section,
\begin{align}
8\pi\sigma= & \frac{2}{R}\left(2-\frac{M}{R}\right)\,,
\label{eq:Extremal_sigma_value_inside_xi_positive}\\
8\pi p= & -\frac{2}{R}\,.
\label{eq:Extremal_pressure_value_inside_xi_positive}
\end{align}
Also, the electric charge density $\sigma_{e}$
is given in terms of $M$ and $R$, by
\begin{equation}
8\pi\sigma_{e}=\frac{2M}{R^{2}}\,,
\label{eq:chargedensity12insextremal}
\end{equation}
The behavior of $\sigma$ and $p$, in
Eqs.~(\ref{eq:Extremal_sigma_value_inside_xi_positive})
and ~(\ref{eq:Extremal_pressure_value_inside_xi_positive}),
as functions of the radial coordinate
$R$ of the $\frac{Q}{M}=1$ extremal shell 
is shown in Figure~\ref{Fig:Properties_extremal_inside_normal}.
These shells are characterized by a positive energy density for $R$
near $r_+$ that changes sign from positive to negative values when the
radius of the shell $R$ obeys $R=\frac{M}2$ up to minus infinity when
$R=0$.  The exterior spacetime is extremal although $p$ is not zero
and so the shell is not made of Majumdar-Papapetrou matter, this case
providing thus another instance,
of the many instances found
in the literature, for which $M=Q$ globally but with an
interior that is not made of Majumdar-Papapetrou matter.
Equation~(\ref{eq:Extremal_sigma_value_inside_xi_positive})
shows that
$\sigma$ is the sum of a geometrical term given by $\frac4R$ and
a gravitational term which is negative given by 
$-\frac{2M}{R^2}$, wit the two terms being independent.
The first term of $\sigma$, $\frac4R$, is a geometrical term that also
gives rise to a geometrical tension given by $-\frac2R$ and ensures
that there is a shell for sure with radius
$R$ inside the Cauchy horizon
$r_+=r_-$.
This geometric term exists independently on
whether there is spacetime mass $M$ or not, indeed, the spacetime mass
energy coming from this geometrical term is zero since $\frac4R+2p=0$.
The second term $-\frac{2M}{R^2}$ is negative and can be explained by
the fact that inside a Cauchy horizon $r_+=r_-$ gravity is repulsive,
here manifested by $\sigma_e=\frac{2M}{R^2}$, and since the shell is
indeed inside $r_+=r_-$ the shell tends naturally to $r_+$, so to
counterbalance this effect and produce a static shell, the shell has
to have an anti repulsive negative energy density, an anti
Majumdar-Papapetrou energy density, of value $-\frac{2M}{R^2}$ .  Note
also that $\sigma+2p+\sigma_e=0$.
When $Q=0$, and so $M=0$, and since $R<M$, in the
limiting case one has $R=0$, and we are left with
a singular massless null shell at $R=0$
with $\sigma+2p=0$
surrounded by a massless spacetime, i.e., a Minkowski spacetime.
This  Minkowski spacetime
with a well defined singularity
at its center is a new and interesting solution of
Einstein equation. 
In relation to the energy conditions of the shell one can work out and
find that the null, the weak, the dominant, and the strong energy
conditions are never verified, see a detailed presentation ahead.

\begin{figure}[h]
\subfloat[\label{Fig:Penrose_diagram_Mink_extremal_RN_inside_normal_1}]
{\includegraphics[height=0.26\paperheight]
{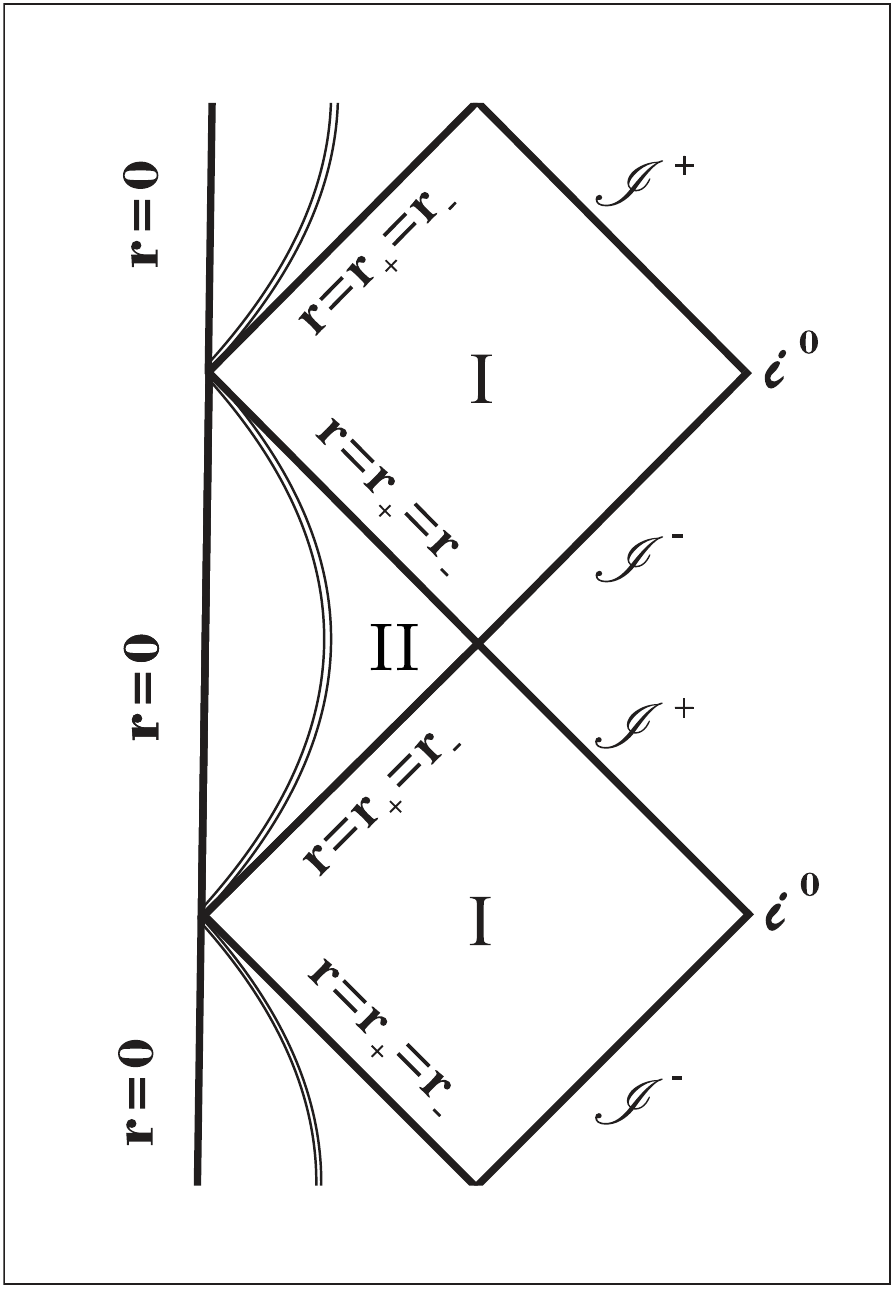}}
\hspace*{5cm}
\subfloat[\label{Fig:Penrose_diagram_Mink_extremal_RN_inside_normal_2}]
{\includegraphics[height=0.26\paperheight]
{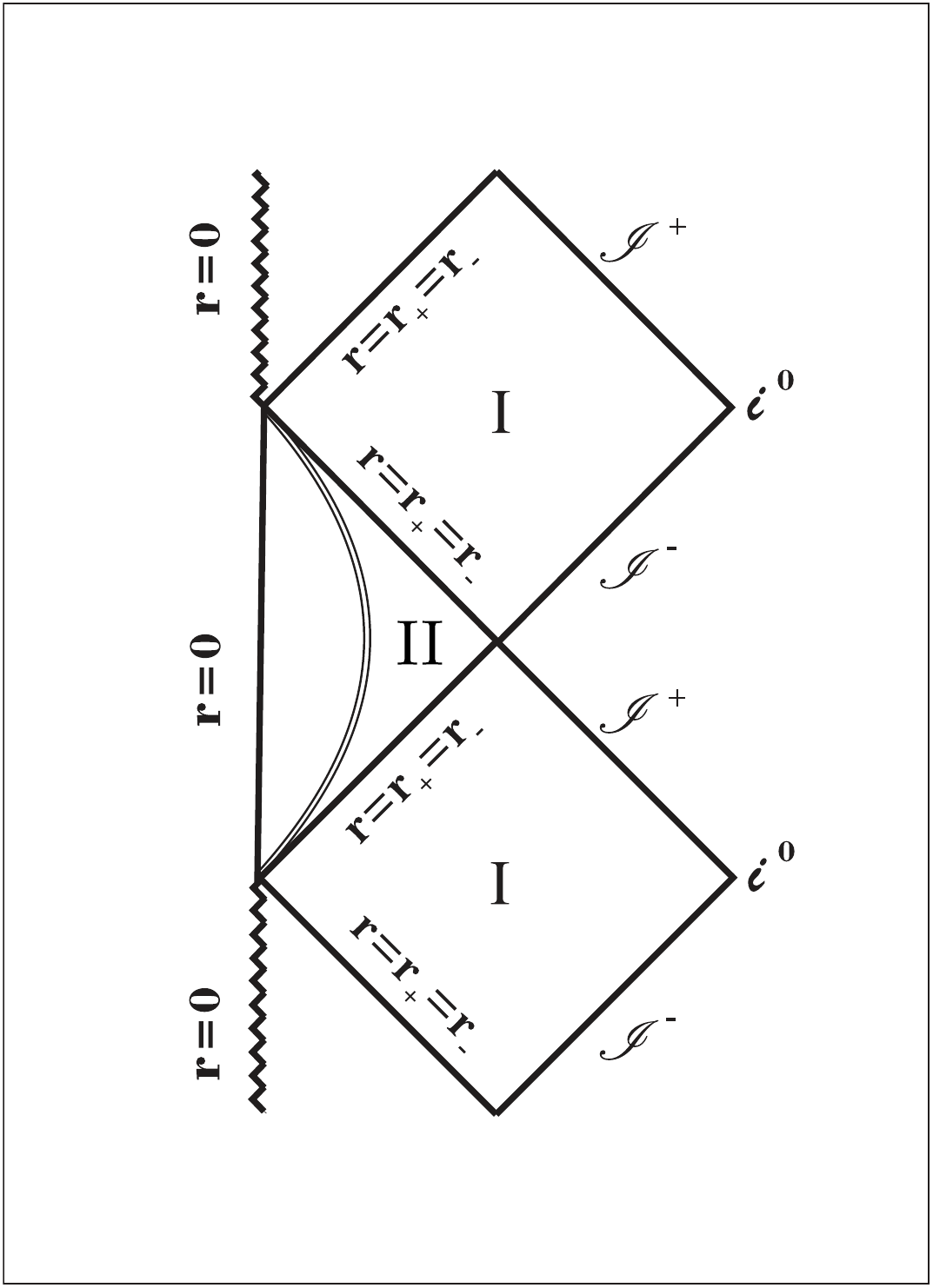}}
\caption{\label{Fig:Penrose_diagram_Mink_extremal_RN_inside}
Carter-Penrose diagrams of the extremal tension shell black holes,
i.e., a thin shell spacetime in an extremal Reissner-Nordstr\"om
state, in the location $R<r_+=r_-$, i.e., located inside the event
horizon radius, with orientation such that the normal to the shell
points towards $r_+$.  The interior is Minkowski, the exterior is
extremal Reissner-Nordstr\"om spacetime.  Panel~(a) The Carter-Penrose
diagram contains a shell in the region $\mathrm{II}$.  If this pattern
is repeated ad infinitum then it is an extremal tension shell regular
black hole.  Panel~(b) The Carter-Penrose diagram contains a shell in
region $\mathrm{II}$ and a singularity in regions $\mathrm{II}$ above
and below.  It is a tension shell black hole, now not regular.
An infinite number of different Carter-Penrose diagrams can be drawn,
since there are an infinite number of combinations to place the shell
and the singularity.
}
\end{figure}

The Carter-Penrose diagram can be drawn directly from the building
blocks of an interior Minkowski spacetime and the full extremal
Reissner-Nordstr\"om spacetime.  In
Figure~\ref{Fig:Penrose_diagram_Mink_extremal_RN_inside} two possible
Carter-Penrose diagrams of a shell spacetime in an extremal
Reissner-Nordstr\"om state, in the location $R<r_+=r_-$, with
orientation such that the normal points towards $r_+$, are shown.
It is clearly a black hole, more specifically, a tension shell black
hole.  In the diagram (a) the tension shell is inside the event
horizon in region $\mathrm{II}$.  Then,
in the next portion of the diagram there is another shell and so
onwards.  So, this realization it is a regular tension black hole.  In
the diagram (b) the tension shell is also inside the event horizon
in region $\mathrm{II}$. Then, the tension shell is replaced by the
timelike singularity at $r=0$. So, in this realization it is a
nonregular tension black hole.  Since what one puts in the regions
$\mathrm{II}$, either a shell or a singularity, is not decided by the
solution, an infinite number of different Carter-Penrose diagrams can
be drawn, as there are an infinite number of combinations to locate a
shell or a singularity when one goes upward or downward through the
diagram.  So, similarly to the
previous subsection, in the case of shells whose unit normal points
towards the event horizon, the maximal analytical extension of the
spacetime may always contain a thin shell inside the event horizon or
only at some regions.

The physical interpretation of this case is of some interest. This
extremal thin shell solution provides an extremal regular black hole
solution. The energy density and pressure never obey the energy
conditions for all shell radii, i.e., shell radii between zero and the
horizon. The causal and global structure as displayed by the
Carter-Penrose diagram shows clearly that there is no singularity if
one adopts the simplest form of the diagram. As regular extremal black
holes these solutions join the other known regular black hole
solutions which are of interest in quantum gravitational settings that
presumably get rid of the singularities. So, this case falls into the
category of having the energy conditions never verified, and in this
sense is odd, although of interest as regular black hole matter
solutions always are.


\subsection{Extremal electric thin shells inside the
gravitational radius:
Majumdar-Papapetrou compact shell naked singularities}
\label{Subsec:extremalnormalinsideminus}

Here we study the case of a fundamental electric thin shell in the
extremal state, i.e., $r_+=r_-$ or $M=Q$, and indeed,
$r_+=r_-=M=Q$, for which the shell's location obeys $R<r_+=r_-$,
and for which the orientation is such that the normal to
the shell points towards $r=0$, i.e., we choose the quantity $\xi$
which gives the direction of the normal as $\xi=-1$, see the end of
this section for details..  In this case horizons do not exist and so,
following the nomenclature, $r_+$ is both the gravitational radius,
and since $r_+=r_-$ it is also the Cauchy radius. We opt to use $r_+$
and $M$.

\begin{figure}[h]
\subfloat[]{
\includegraphics[scale=0.45]
{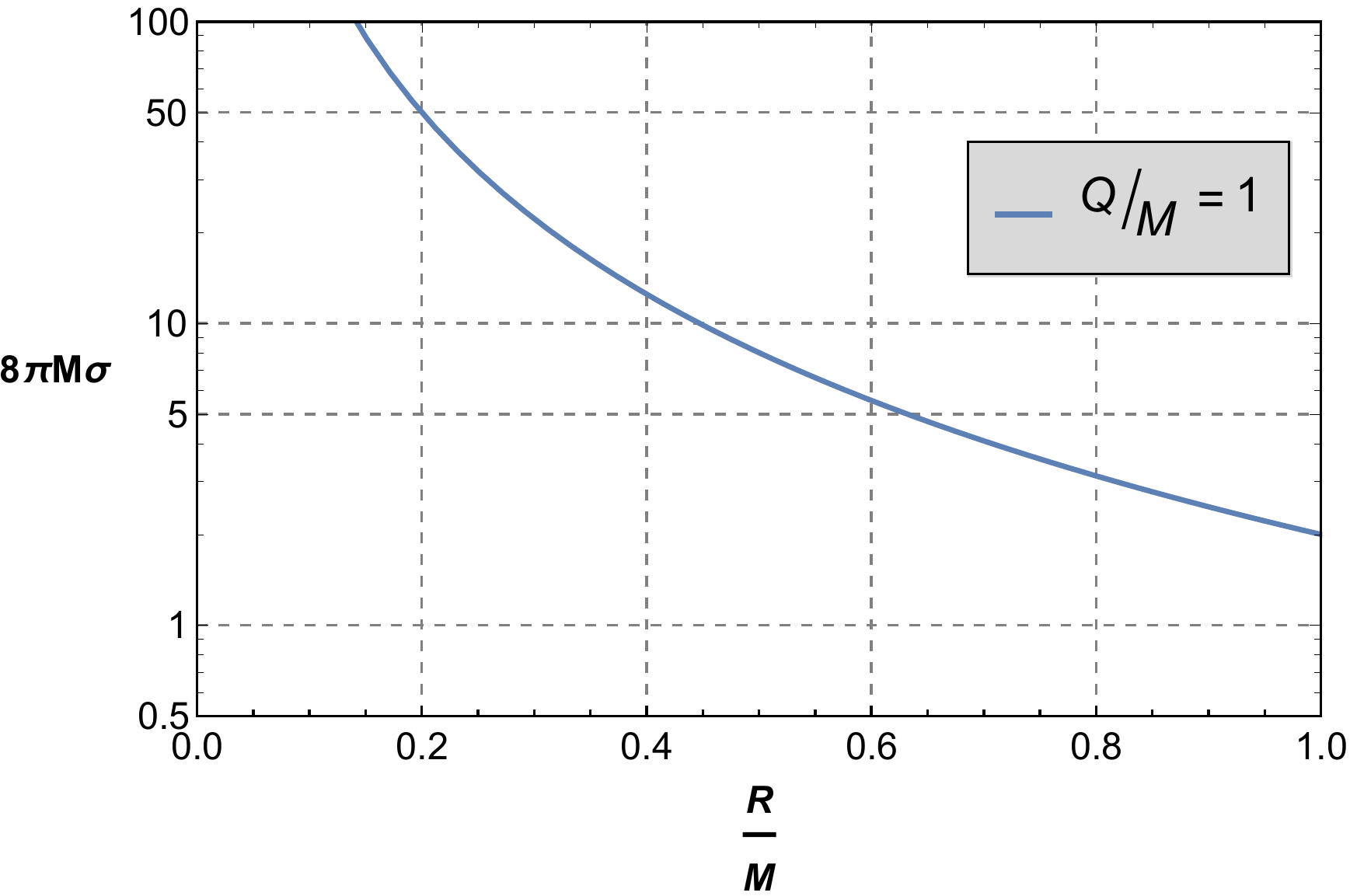}}
\hspace*{\fill}
\subfloat[]{
\includegraphics[scale=0.45]
{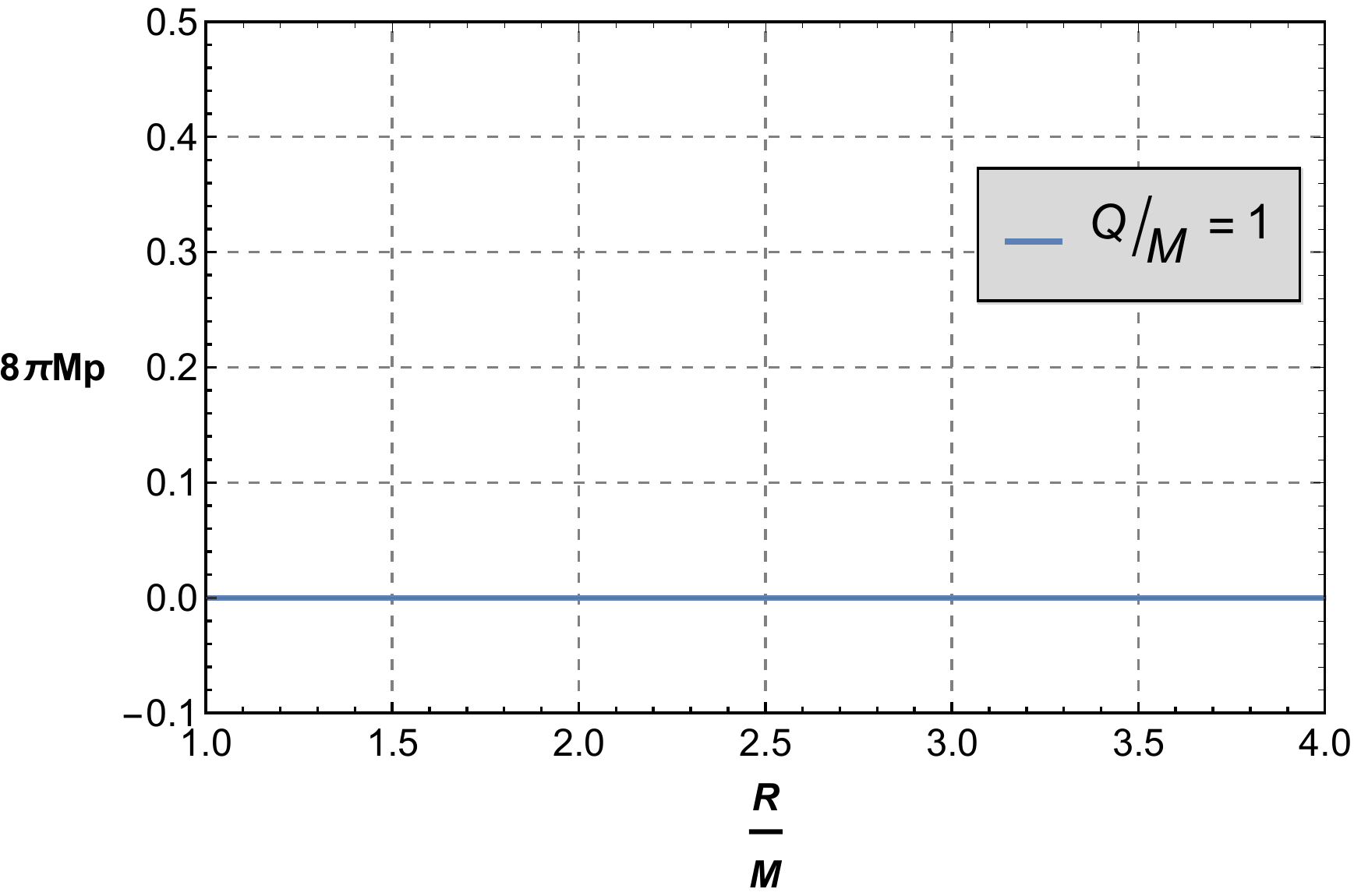}}
\caption{\label{Fig:Properties_extremal_inside_alternative}
Physical properties of a Majumdar-Papapetrou compact shell naked
singularity, i.e., an electric perfect fluid thin shell in an extremal
Reissner-Nordstr\"om state, in the location $R<r_+=r_-$, and with
orientation such that the normal points towards $r=0$.  The interior
is Minkowski and the exterior is extremal Reissner-Nordstr\"om
spacetime.   Extremal means $\frac{Q}{M}=1$.
Panel (a) Energy density $\sigma$ of the shell as a
function of the radius $R$ of the shell.
The energy density is adimensionalized through the
mass $M$, $8\pi M\sigma$, and the radius is adimensionalized through
the gravitational radius $r_+$, $\frac{R}{r_+}$.  Panel (b) Pressure
on the shell as a function of the radius $R$ of the shell.  The
pressure is zero, the shell is supported by electric repulsion, it is
Majumdar-Papapetrou matter.  The radius is adimensionalized through
the gravitational radius $r_+$, $\frac{R}{r_+}$.
}
\end{figure}

As functions of $M$ and $R$, the shell's energy density $\sigma$ and
pressure $p$, are, see the end of this section,
\begin{align}
8\pi\sigma= & \frac{2M}{R^2}\,,
\label{eq:Extremal_sigma_value_insideother}\\
8\pi p= & 0\,.\label{eq:Extremal_pressure_value_insideother}
\end{align}
Also, the electric charge density $\sigma_{e}$
is given in terms of $M$ and $R$ by
Eq.~(\ref{eq:chargedensity12insextremal}).
The behavior of $\sigma$ and $p$, in
Eqs.~(\ref{eq:Extremal_sigma_value_insideother})
and ~(\ref{eq:Extremal_pressure_value_insideother}),
as functions of the radial coordinate
$R$ of the $\frac{Q}{M}=1$ extremal shell 
is shown in Figure~\ref{Fig:Properties_extremal_inside_alternative}.
These shells are characterized by a positive energy density for all
shell's radii.  The pressure is zero, and so the matter is
Majumdar-Papapetrou matter.  When $Q=0$, and so $M=0$, there is no
shell spacetime.  In relation to the
energy conditions of the shell one can work out and find that the
null, the weak, the dominant, and the strong energy conditions are
verified for $0<R<r_+$, see a detailed presentation ahead.

\begin{figure}[h]
\includegraphics[height=0.26\paperheight]
{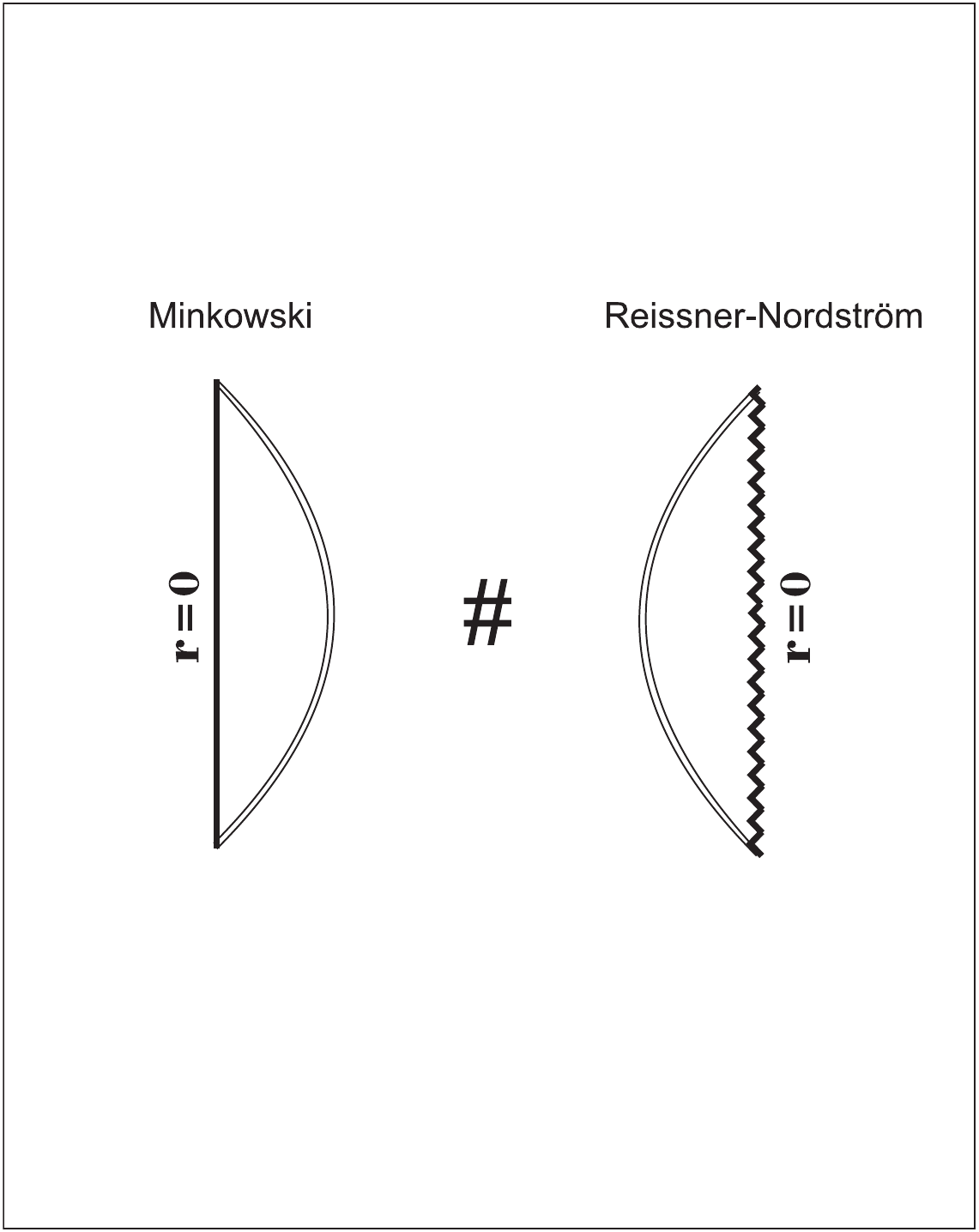}
\caption{
\label{Fig:Penrose_diagram_Mink_extremal_RN_inside_alternative}
Carter-Penrose diagram of the Majumdar-Papapetrou compact shell naked
singularity spacetime, i.e., a shell in an extremal
Reissner-Nordstr\"om
state, in the location $R<r_+=r_-$, i.e., located inside the event
horizon radius, with orientation such that the normal to the shell
points towards $r=0$. The interior is Minkowski, the exterior is
extremal
Reissner-Nordstr\"om.
There is no clear distinction
of what is outside from what is inside.
The hash symbol $\#$ represents the
connected sum of the two spacetimes.
}
\end{figure}

The Carter-Penrose diagram can be drawn directly from the building
blocks of an interior Minkowski spacetime and the full extremal
Reissner-Nordstr\"om spacetime.  In
Figure~\ref{Fig:Penrose_diagram_Mink_extremal_RN_inside_alternative}
the Carter-Penrose diagram of a shell spacetime in an extremal
Reissner-Nordstr\"om state, in the location $R<r_+=r_-$, with
orientation such that the normal points towards $r=0$, is shown.  It
is a Majumdar-Papapetrou, i.e., extremal, compact shell naked
singularity spacetime. It is clearly a compact space, the coordinate
$r$ goes from 0 to $R$ and then decreases back to 0 at the timelike
singularity, such that there is no clear distinction
of what is outside from what is inside.
We use the hash symbol $\#$ to represent the connected
sum of the spacetime manifolds, in order to conserve the conformal
structure in the Carter-Penrose diagram of the total spacetime.

The physical interpretation of this case is noteworthy, and it is
similar to the corresponding nonextremal shell. This
extremal thin shell solution provides a closed spatial static universe
with a singularity at one pole. There are no horizons.  The energy
density and pressure obey the energy conditions for all shell radii,
indeed the shell is composed of Majumdar-Papapetrou matter.  The
causal and global structure as displayed by the Carter-Penrose diagram
show the characteristics of this universe that has two sheets joined at
the shell with one sheet having a singularity at its pole and with no
horizons. The singularity is avoidable to timelike curves. So, this
case falls into the category of having the energy conditions verified
and the resulting spacetime being peculiar.

\subsection{Formalism for extremal electric thin
shells inside the gravitational radius
\label{Subsec:Extremal_induced_Minkowski_RN_inside_event_horizon}}

\subsubsection{Preliminaries}
\label{prel4}

We now make a careful study to derive the properties of the
fundamental electric thin shell used in the two previous subsections,
i.e., the thin shell in an extremal state, i.e., $r_+=r_-$ or
$M=Q$,
for which the shell's radius $R$ location obeys $R<r_+=r_-$, and
for which the orientation is such that the normal to the shell points
towards $r_+$ or towards $r=0$.
It should be read as an appendix to the previous two
subsections.  We use the formalism developed in
Sec.~\ref{Sec:Junction_formalism}.

\subsubsection{Induced metric, and extrinsic curvature of
$\mathcal{S}$ as seen from $\mathcal{M}_{\rm i}$}
\label{Subsec:Extremal_induced_Mink_inside2}

Let us start by analyzing the interior
Minkowski spacetime, $\mathcal{M}_{\rm i}$. Since it is the same as
the analysis done previously we only quote
the important equations.
They are the interior metric Eq.~(\ref{eq:Mink_metric_interior}),
the interior four-velocity of the shell
Eq.~(\ref{eq:Mink_vel_explicit}), the
metric for the shell at radius $R$
Eq.~(\ref{eq:induced_metric_Mink}), the normal to the shell
Eq.~(\ref{eq:normal_Mink}),
and the extrinsic curvature from the inside
Eq.~(\ref{eq:Extrinsic_curvature_Mink}).

\subsubsection{Induced metric, and extrinsic curvature of
$\mathcal{S}$ as seen
from $\mathcal{M}_{\rm e}$}
\label{Subsec:Extremal_induced_RN_inside}

To proceed we have now to find the expressions for the induced metric
on $\mathcal{S}$ and the extrinsic curvature components as seen from
the exterior spacetime, $\mathcal{M}_{\rm e}$, in the extremal
state, i.e., $r_+=r_-$ or $M=Q$, see
Figure~\ref{Fig:Penrose_diagram_RN_extremal}, for which the
shell's obeys $R<r_+=r_-$, and for which the
orientation is such that the normal to the shell points towards
increasing $r$, i.e., towards $r_+$,
or towards decreasing $r$ i.e., towards $r=0$,
as seen from the exterior, as
used in the two previous subsections.

Most of the analysis and results of
Section~\ref{Sec:Extremal-thin-shells-outside} are still verified,
namely,
the extremal 
Reissner-Nordstr\"om line element $ds_{\rm e}^{2}$
given in Eq.~(\ref{RNextremal2}),
the four-velocity
$u_{\rm e}^\alpha$
given in Eq.~(\ref{eq:Extremal_4velocity_exterior_outside_horizon}),
the line element on $\mathcal{S}$, $\left.ds^{2}\right|_{\mathcal{S}}$
given in Eq.~(\ref{eq:Extremal_induced_metric_outside_horizon}),
and the normal to the surface $\mathcal{S}$
given in Eq.~(\ref{eq:Extremal_normal_exterior_outside_horizon}).
Then, taking
into account that here we are considering that, $R$, the radial
coordinate of $\mathcal{S}$ as seen from $\mathcal{M}_{\rm e}$,
verifies $R<r_+=r_-$, we find the following expressions for the nonzero
components of the extrinsic curvature of the matching hypersurface
\begin{equation}
{K_{\rm e}}^{\tau}{}_{\tau}
=-\xi\frac{r_+}{R^{2}}
\,,\quad\quad
{K_{\rm e}}^{\theta}{}_{\theta}=
{K_{\rm e}}^{\varphi}{}_{\varphi}
=\xi\frac{k}{R}
\,,\label{eq:eq:Extremal_extrinsic_curvature_RN_inside}
\end{equation}
where, as before, the parameter $\xi$ is defined
as $\xi=+1$ if the orientation is such
that the outside unit normal to the shell points in the
direction of increasing radial coordinate $r$, measured by an observer
in the exterior $\mathcal{M}_{\rm e}$ spacetime, and $\xi=-1$ if the
the orientation is such
that the outside unit normal to the shell points in the
direction of decreasing
radial coordinate $r$, and the redshift function $k$ at the
shell is given by
$k=|1-\frac{r_+}{R}|$, i.e., since $R<r_+$
one has $k=\frac{r_+}{R}-1$.

\subsubsection{Shell's energy density and pressure}
\label{subSubsec:shellsenergydensityandpressureextremalins}

Having determined the components of the extrinsic curvature of the
matching surface $\mathcal{S}$ as seen from the interior and exterior
spacetimes we are now in position to use the second junction
condition
given in Eq.~(\ref{eq:2nd_junct_cond}) to find the expressions for the
energy density and pressure support of the thin shell.
The shell's stress-energy tensor is given in Eq.~(\ref{eq:perfect}),
and 
Eqs.~(\ref{eq:Extrinsic_curvature_Mink})
and~(\ref{eq:eq:Extremal_extrinsic_curvature_RN_inside})
then yield
\begin{align}
8\pi\sigma= & \frac{2}{R}\left[1+\xi
\left(1-\frac{r_+}{R}\right)\right]\,,
\label{eq:Extremal_sigma_value_inside}\\
8\pi p= & -\frac{1}{R}\left(1+\xi\right)\,,
\label{eq:Extremal_pressure_value_inside}
\end{align}
where we used
$k=\frac{r_+}{R}-1=\frac{M}{R}-1$. Note that $p$ in
Eq.~(\ref{eq:Extremal_pressure_value_outside})
is independent of $M$, it only depends on
$R$ and thus on the geometry of the shell as embedded in the ambient
spacetime.
Moreover, defining the surface electric current density
$s_a$ on the
thin shell as $s_a=\sigma_{e}u_a$, where $\sigma_{e}$
represents the
electric charge density, and since the Minkowski spacetime
has zero electric
charge,
from Eqs.~(\ref{eq:junct_cond_Faradayb})-(\ref{eq:junct_cond_Faraday2})
and~(\ref{eq:RN_FaradayMaxwell_value})
it follows that
\begin{equation}
8\pi \sigma_{e}=2\frac{r_+}{R^{2}}\,.
\label{eq:chargedensity1extremalin}
\end{equation}
The radial coordinate of the shell is in the range $0<R<r_+$.

Equations~(\ref{eq:Extremal_sigma_value_inside})
and (\ref{eq:Extremal_pressure_value_inside}),
together with 
(\ref{eq:chargedensity1extremalin}),
can now be used to study the properties of the thin matter shells
separating a Minkowski spacetime from an exterior
extremal Reissner-Nordstr\"om
spacetime, located inside the event horizon $r_+$.
In Eqs.~(\ref{eq:Extremal_sigma_value_inside})
and (\ref{eq:Extremal_pressure_value_inside})
it is clear that it is necessary to pick the sign  $\xi$.
Let us start with $\xi=+1$. It is useful to
give the expressions for the shell's energy density and pressure,
$\sigma$ and $p$ in terms of $M=Q$, we opt for $M$.
Using Eq.~(\ref{eq:KS_horizons_radius0}), i.e., $r_+=M$, in
Eqs.~(\ref{eq:Extremal_sigma_value_inside})
and (\ref{eq:Extremal_pressure_value_inside})
with $\xi=+1$ we have
$8\pi\sigma=  \frac{2}{R}\left(2-\frac{M}{R}\right)$,
$8\pi p=  -\frac{2}{ R}$, and 
also from Eq.~(\ref{eq:chargedensity1extremalout}) we have
$8\pi\sigma_{e}=\frac{2M}{R^{2}}$.
Let us now take $\xi=-1$.
It is useful to
give the expressions for the shell's energy density and pressure,
$\sigma$ and $p$ in terms of $M=Q$, we opt for $M$.
Using Eqs.~(\ref{eq:Extremal_sigma_value_inside})
and (\ref{eq:Extremal_pressure_value_inside})
with $\xi=-1$ we have
$8\pi\sigma=  \frac{2M}{R^2}$,
$8\pi p= 0$, and 
also from Eq.~(\ref{eq:chargedensity1extremalout}) we have again
$8\pi\sigma_{e}=\frac{2M}{R^{2}}$. These are the expressions used in
the two previous subsections.

\newpage

\section{Extremal electric thin shells at the gravitational radius:
Majumdar-Papapetrou shell quasiblack holes,
extremal null shell quasinonblack holes, extremal null shell
singularities, 
and Majumdar-Papapetrou null shell singularities
\label{Sec:Horizon_thin_shells}}

\subsection{Extremal electric thin shells at the event horizon:
Majumdar-Papapetrou shell quasiblack holes
and extremal null shell quasinonblack holes }
\label{Subsec:extremalqbhdnormalinsideplus}

\subsubsection{Majumdar-Papapetrou shell quasiblack holes}
\label{majumdarpapapetroushellquasiblackholes}

Here we study the case of a fundamental electric thin shell in the
extremal state, i.e., $r_+=r_-$ or $M=Q$, and indeed, $r_+=r_-=M=Q$,
for which the shell's location obeys $R=r_+$, and for which the
orientation is such that the normal to the shell points towards
spatial infinity.  Moreover, there is an additional characterization
for shells at the horizon. This case comes from the limit of $R\to
r_+$ from above and so is the limiting case of the case studied in
Sec.\ref{Subsec:extremalnormaloutside}.  In this case a horizon is
barely formed, namely, we have a
quasihorizon, and so, following the nomenclature, $r_+$ is both the
gravitational radius and the quasihorizon radius.  This is an extremal
quasiblack hole~\citep{lemoszasla2020}. Also $r_+$ and $r_-$ have the
same value.  In general we also opt to use $M$ rather than $Q$.  The
normal to the shell pointing towards spatial infinity means in the
notation for the extremal states that the new parameter $\xi$ has
value $\xi=+1$, see the end of this section for details.

As functions of $M$ and $R$, the shell's energy density $\sigma$ and
pressure $p$, are, see the end of this section,
\begin{align}
8\pi\sigma= & \frac{2}{M}\,,
\label{eq:qbh_sigma}\\
8\pi p= & 0\,.\label{eq:qbhpressure}
\end{align}
Also, the electric charge density $\sigma_{e}$
is given in terms of $M$
by 
\begin{equation}
8\pi\sigma_{e}=  \frac{2}{M}\,.
\label{qbhchargedensity}
\end{equation}
Since it is one point in a plot of $\sigma$ or $p$ as functions of
$\frac{R}{M}$, there is no need to draw a figure.  The shell is
characterized by a positive energy density.  The pressure is zero, and
so the matter is Majumdar-Papapetrou matter, i.e., $\sigma_e=\sigma$,
and therefore is fully supported by electric repulsion.  This is an
interesting system to consider, this case when the shell's radius is
taken to the event horizon radius. It is a quasiblack hole
configuration. The Majumdar-Papapetrou shell
quasiblack hole is regular in that all
curvature scalars are finite everywhere.  When $Q=0$,  so
$M=0$ and $r_+=0$, the shel is at $R=0$,
and the spacetime is singular being  
Minkowski in the exterior.
In relation to the energy conditions of the shell one can
work out and find that the null, the weak, the dominant, and the
strong energy conditions are always verified, see a detailed
presentation ahead.

\begin{figure}[h]
\includegraphics[height=0.26\paperheight]
{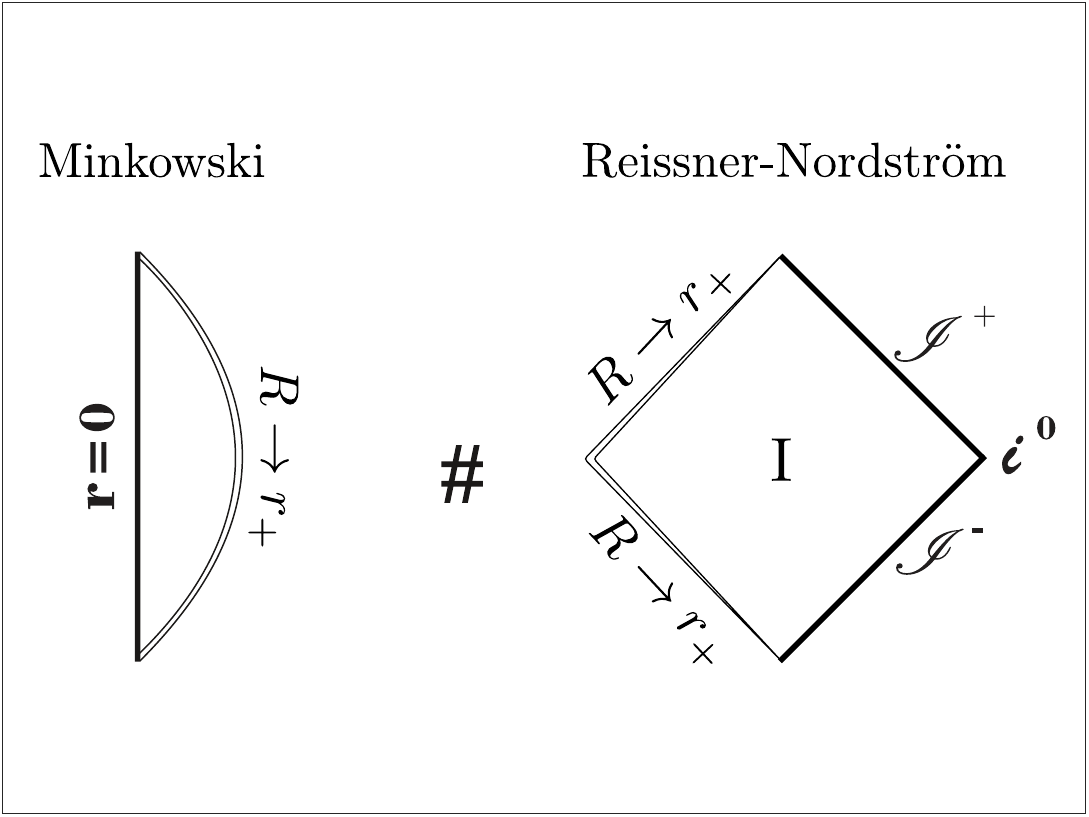}
\caption{\label{Fig:Penrose_diagram_Mink_horizonshell_RN1}
Carter-Penrose diagram of a Majumdar-Papapetrou
shell quasiblack hole, i.e., a
thin shell spacetime in an extremal Reissner-Nordstr\"om state, with
the shell located at $R=r_+$, i.e., located at the gravitational
radius or quasihorizon, with orientation such that the normal points
towards infinity, and such that $R\to r_+$ from $R> r_+$.  The
interior is Minkowski, the exterior is extremal Reissner-Nordstr\"om.
This quasiblack hole shell is supported by electrical
repulsion alone.      
}       
\end{figure}

The Carter-Penrose diagram can be drawn with some care
from the building
blocks of an interior Minkowski spacetime and the
exterior asymptotic region of an
extremal Reissner-Nordstr\"om spacetime, see
\cite{qbh4lemoszaslacp} and for more details see
\cite{lemoszasla2020}.
In Figure~\ref{Fig:Penrose_diagram_Mink_horizonshell_RN1}
the Carter-Penrose
diagram
of a Majumdar-Papapetrou shell quasiblack hole, i.e.,
for $R=r_+$ and a junction surface
with orientation such that the outside normal points towards
spacial infinity is shown. We use the hash symbol $\#$
to represent the connected sum
of the spacetime manifolds, in order to conserve the conformal
structure in the Carter-Penrose diagram of the total spacetime.
We see that when the shell is at $R=r_+$, i.e., the shell
is at a null
surface, the two regions contain incomplete
geodesics with ending points at the matching surface, so that,
observers at each spacetime are disconnected and the manifold is
composed by two separate regions. 

\newpage

Some remarks on quasiblack holes should be made.  In this section we
treated an extremal quasiblack hole,
namely, a  Majumdar-Papapetrou
shell quasiblack hole.  Since it is Majumdar-Papapetrou the
pressure on the shell is zero, $p=0$, and so the extremal quasiblack
hole is regular in this sense.  On the other hand, in
Section~\ref{Subsec:nonextremalnormaloutside} on nonextremal shells
that are located outside $r_+$, $R>r_+$, with orientation such that
the normal points towards spatial infinity, one has that
Eqs.~(\ref{eq:sigma_value_rplus_MQ_sign_plus}),
(\ref{eq:pressure_value_rplus_MQ_sign_plus}), and
(\ref{eq:chargedensity12}), in the limit that the shell is located
at the gravitational radius, $R=r_+$, yield that the surface density
$\sigma$ is finite, the pressure support $p$ of the thin matter shell
diverges to infinity, and the electric charge density $\sigma_e$ is
finite.  This case defines a nonextremal quasiblack hole.  Since the
pressure diverges the spacetime of nonextremal shells at $R=r_+$
presents some type of singularity.  This singularity is mild however,
with entropy and the mass formulas being derived in this limiting case,
see~\citep{lemoszasla2020}.  The Carter-Penrose diagram of a
nonextremal quasiblack hole is similar to the Carter-Penrose diagram
for a Majumdar-Papapetrou one, i.e., the one showed in
Figure~\ref{Fig:Penrose_diagram_Mink_horizonshell_RN1}.
Since nonextremal quasiblack holes are somewhat singular and
extremal ones are not, we have treated these
within the extremal state and mentioned the nonextremal here.

The physical interpretation of this case is known and it is
remarkable. The extremal thin shell solution with its radius at the
horizon radius, is inherited from the extremal thin shell star, and
provides a typical extremal quasiblack hole. A quasiblack hole is an
object on the verge of becoming a black hole, but cannot turn into
such one.
The energy density and pressure shows that the matter is
Majumdar-Papapetrou and obey the energy conditions. The causal and
global structure as displayed by the Carter-Penrose diagram show the
quasiblack hole characteristics. These quasiblack holes have no
curvature singularities, although at the quasihorizon there is some
form of singular degeneracy that disconnects the interior from the
exterior. It can form in a limiting process of quasistatic collapse.
Quasiblack holes are of great interest because they reveal new black
hole properties or black hole properties in a new perspective. So, this
case falls into the category of having some of
the energy conditions verified
and the geometrical setup is interesting and peculiar.

\subsubsection{Extremal null shell quasinonblack holes}
\label{extremalnullshellblackholes}

Here we study the case of a fundamental electric thin shell in the
extremal state, i.e., $r_+=r_-$ or $M=Q$, and indeed, $r_+=r_-=M=Q$,
for which the shell's location obeys $R=r_+$, and for which the
orientation is such that the normal to the shell points towards
spatial infinity. Moreover, there is an additional characterization
for shells at the horizon. This case here comes from the limit of $R\to
r_+$ from below and so is the limiting case of the case studied in
Sec.\ref{Subsec:extremalnormalinsideplus}.
In this case $r_+$ is timelike on
one side and lightlike on the other side.
Thus, a horizon, or rather a quasinonhorizon, does exist and
so, following the nomenclature, $r_+$ is both the gravitational radius
and the quasinonhorizon radius.
Also $r_+$ and $r_-$ have the same value.
In general, we also opt here
to use $M$ rather than $Q$.  The normal to the
shell pointing towards spatial infinity means in the notation for the
extremal states that the new parameter $\xi$ has value $\xi=+1$, see
the end of this section for details.

As functions of $M$ and $R$, the shell's energy density $\sigma$ and
pressure $p$, are, see the end of this section,
\begin{align}
8\pi\sigma&= \;\; \frac{2}{M}\,,
\label{eq:qbh_sigma2new2}\\ 8\pi p&= -\frac{2}{M}\,.
\label{eq:qbhpressure2new2}
\end{align}
Also, the electric
charge density $\sigma_{e}$ is given in terms of $M$ by
Eq.~(\ref{qbhchargedensity}).  Since it is one point in a plot of
$\sigma$ or $p$ as functions of $\frac{R}{M}$, there is no need to
draw a figure. The shell is characterized by a positive energy
density.  The pressure is negative, so it is a tension.  The equation
of state is $\sigma+2p+\sigma_e=0$, inherited from
the extremal $R<r_+$ shell.
When $Q=0$, and so $M=0$, there is a singular null shell at $R=0$,
and a Minkowski spacetime in the exterior.
In relation to the energy conditions of the
shell one can work out and find that the null, the weak and the
dominant energy conditions are always verified, and the strong energy
condition is always violated, see a detailed presentation ahead.

\begin{figure}[h]
\includegraphics[height=0.26\paperheight]
{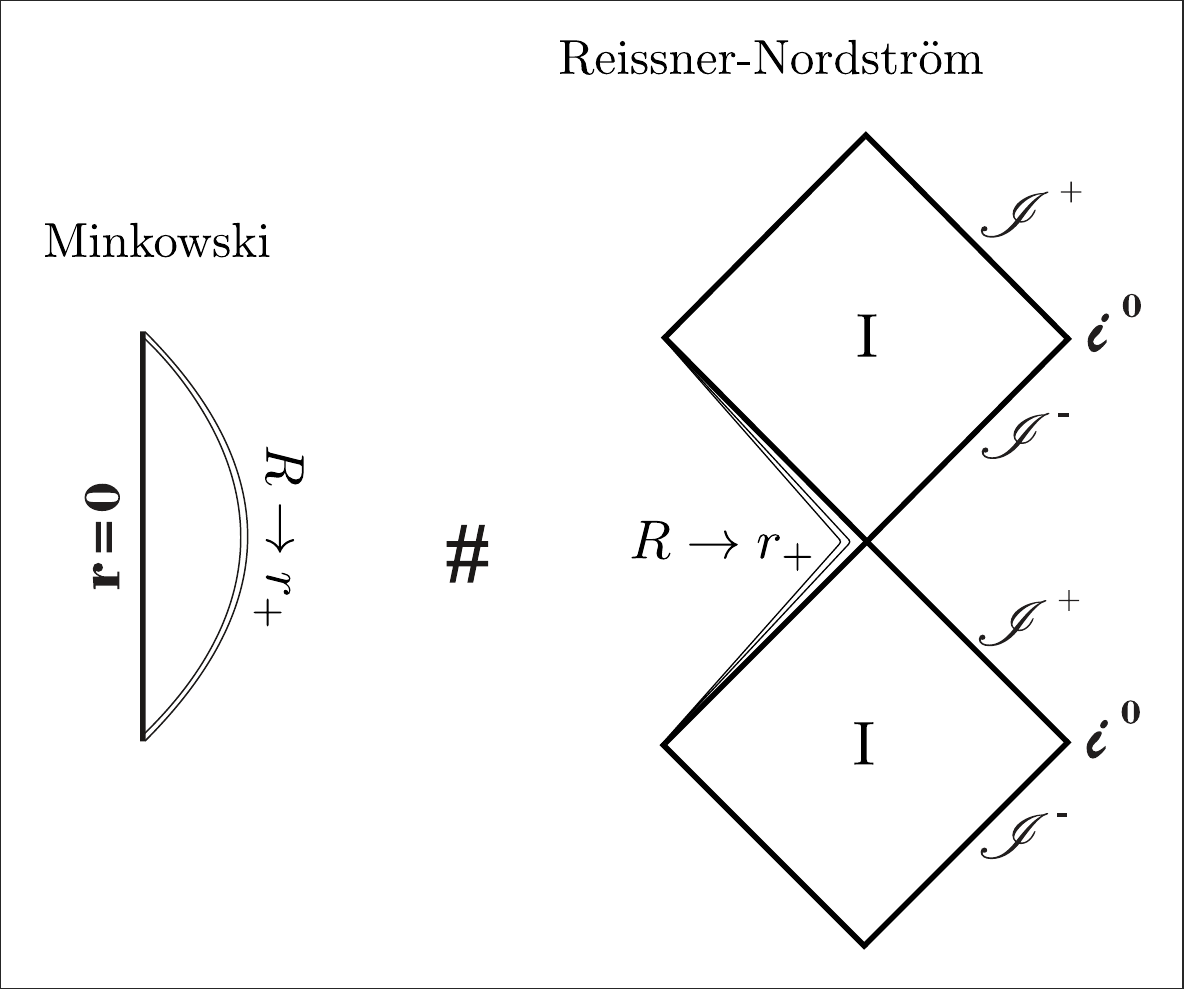}
\caption{\label{Fig:Penrose_diagram_Mink_horizonshell_RN2}
Carter-Penrose diagram of an extremal null shell quasinonblack hole,
i.e., a
thin shell spacetime in an extremal Reissner-Nordstr\"om state, with
the shell located at $R=r_+$, i.e., located at the gravitational
radius or horizon, with orientation such that the normal points
towards $r=0$, and such that
$R\to r_+$ from $R< r_+$. The interior
is Minkowski, the exterior is extremal Reissner-Nordstr\"om.
}
\end{figure}

The Carter-Penrose diagram can be drawn with some care from the
building blocks of an interior Minkowski spacetime and the exterior
asymptotic region of an extremal Reissner-Nordstr\"om spacetime.  In
Figure~\ref{Fig:Penrose_diagram_Mink_horizonshell_RN2}, the
Carter-Penrose diagram of an extremal null shell quasinonblack hole,
i.e., for
$R=r_+$ and a junction surface with orientation such that the outside
normal points towards spatial infinity, is shown.  
We use the hash symbol $\#$ to represent the connected sum of the
spacetime manifolds, in order to conserve the conformal structure in
the Carter-Penrose diagram of the total spacetime. This setup is very
different from the quasiblack hole limit of the last section leading
to a new Carter-Penrose diagram.  Nonetheless, we see that as in the
previous case, when the shell is at $R=r_+$, i.e., the shell,
for one of the regions, is at a
null surface, the two regions contain incomplete geodesics with ending
points at the matching surface, so that, observers at each spacetime
are disconnected and the manifold is composed by two separate regions.

The physical interpretation of this case is also remarkable. The
extremal thin shell solution with its radius at the horizon radius, is
inherited from the extremal regular black hole, and provides an
example of an extremal quasinonblack hole It is an object that is on
the verge of becoming a star solution, but cannot turn into one.  The
energy density and pressure shows that the matter obeys some of the
energy conditions.  These quasinonblack holes have no curvature
singularities, although at the quasinonhorizon there is some form of
singular degeneracy that disconnects the interior from the exterior.
The causal and global structure as displayed by the Carter-Penrose
diagram show the characteristics pertaining to quasinonblack hole.
These quasinonblack hole solutions are new, they have showed up here
for the first time.  So, this case falls into the category of having
the energy conditions verified and the geometrical setup is new,
very interesting, and peculiar.

\subsection{Extremal electric thin shells at the gravitational radius:
Extremal null shell singularities, and Majumdar-Papapetrou null shell
singularities}
\label{Subsec:extremalqbgnormalinsideminus}

\subsubsection{Extremal null shell singularities}
\label{extremalnullshellsingularities}

Here we study the case of a fundamental electric thin shell in the
extremal state, i.e., $r_+=r_-$ or $M=Q$, and indeed, $r_+=r_-=M=Q$,
for which the shell's location obeys $R=r_+$, and for which the
orientation is such that the normal to the shell points towards
the singularity at $r=0$.  Moreover,
as we have seen above, there is an additional characterization
for shells at the horizon, this case comes from the limit of $R\to
r_+$ from above and so is the limiting case of the case studied in
Sec.\ref{Subsec:extremalnormaloutsidenormaltoin}.
In this case the shell
is at the horizon, thus in a sense
a quasihorizon does exist, and
so, following the nomenclature, $r_+$ is both the gravitational radius
and the quasihorizon radius.
Also $r_+$ and $r_-$ have the same value.
In general we also opt to use $M$ rather than $Q$. 
This is an extremal null shell singularity.
The normal to the
shell pointing towards the singularity at $r=0$
means in the notation for the
extremal states that the new parameter $\xi$ has value $\xi=-1$, see
the end of this section for details.

As functions of $M$ and $R$, the shell's energy density $\sigma$ and
pressure $p$, are, see the end of this section,
\begin{align}
8\pi\sigma&= \;\; \frac{2}{M}\,,
\label{eq:qbh_sigma2new}\\ 8\pi p&= 
-\frac{2}{M}\,,
\label{eq:qbhpressure2new}
\end{align}
so the matter is not Majumdar-Papapetrou.
Also, the electric
charge density $\sigma_{e}$ is given in terms of $M$ by
Eq.~(\ref{qbhchargedensity}). The equation
of state is $\sigma+2p+\sigma_e=0$, inherited from
the extremal $R<r_+$ shell.
In relation to the energy conditions of the shell one can work out and
find that the null, the weak and the dominant energy conditions are
always verified whereas, the strong energy condition is never
verified, see a detailed presentation ahead.

\begin{figure}[h]
\includegraphics[height=0.26\paperheight]
{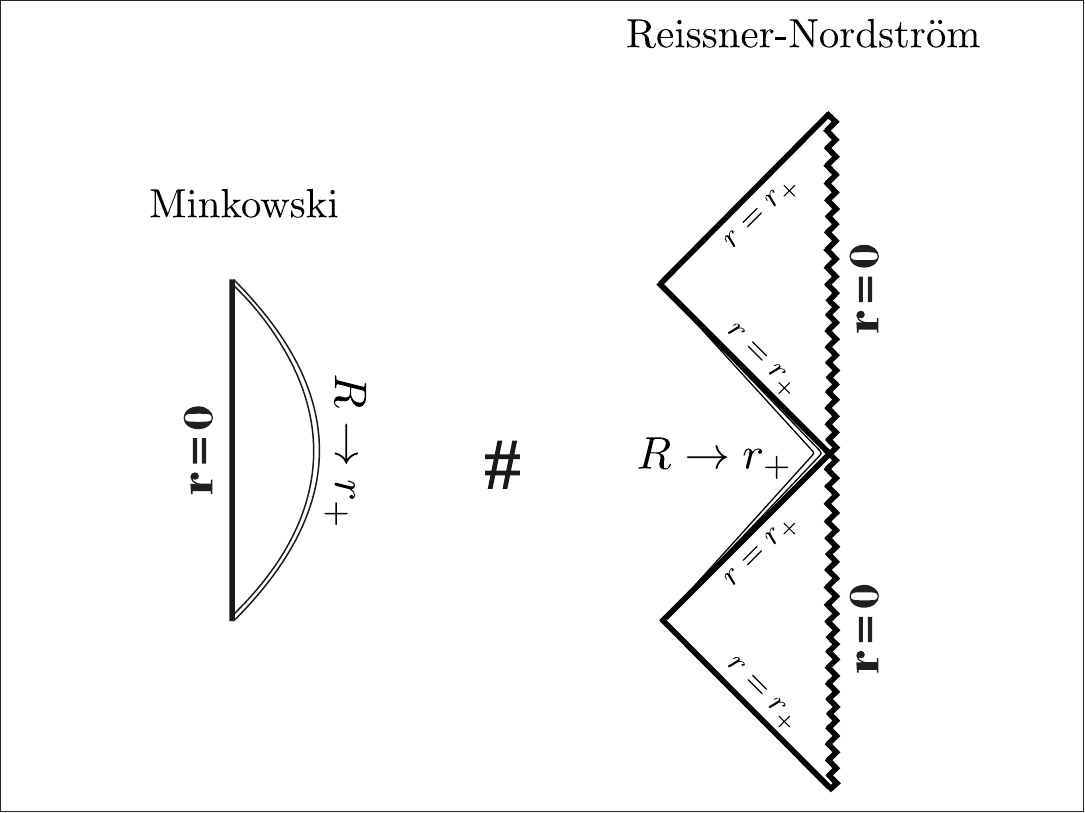}
\caption{\label{Fig:Penrose_diagram_Mink_horizonnullshell_RN2}
Carter-Penrose diagram of an extremal null shell singularity, i.e., a
thin shell spacetime in an extremal Reissner-Nordstr\"om state, with
the shell located at $R=r_+$, i.e., located at the gravitational
radius from above, with orientation such that the normal points
towards the singularity at $r=0$, and such that
$R\to r_+$ from $R> r_+$.  The interior is Minkowski,
the exterior is extremal
Reissner-Nordstr\"om.
}       
\end{figure}

The Carter-Penrose diagram can be drawn with some care
from the building
blocks of an interior Minkowski spacetime and the
exterior asymptotic region of an
extremal Reissner-Nordstr\"om spacetime.
In Figure~\ref{Fig:Penrose_diagram_Mink_horizonnullshell_RN2}
the Carter-Penrose
diagram of an extremal null shell singularity, i.e.,
for $R=r_+$ from above and a junction surface
with orientation such that the outside normal points towards the
singularity at $r=0$ is shown. We see that when the shell is at
$R=r_+$, that is the shell is at a null
surface, the two regions contain incomplete
geodesics with ending points at the matching surface, so that,
observers at each spacetime are disconnected and the manifold is
composed by two separate regions. 

The physical interpretation of this case follows from the
corresponding extremal shell outside the gravitational radius.  This
extremal thin shell solution, with the shell itself at the horizon, or
more properly, at the quasinonhorizon, turns the space around at the
quasinonhorizon and then ends in a singularity.  The energy density and
pressure obey some of the energy conditions.  The causal and global
structures as displayed by the Carter-Penrose diagram are interesting
and the two parts up to the shell and from the shell to the
singularity are disjoint, with the quasinonhorizon presenting some form
of degeneracy, although there are no curvature singularities there.
So, this case falls into the category of having some of the energy
conditions verified and the geometrical setup is rather strange.

\newpage

\subsubsection{Majumdar-Papapetrou null shell singularities}
\label{extremalmajumdarpapapetroushellsingularities}

Here we study the case of a fundamental electric thin shell in the
extremal state, i.e., $r_+=r_-$ or $M=Q$, and indeed, $r_+=r_-=M=Q$,
for which the shell's location obeys $R=r_+$, and for which the
orientation is such that the normal to the shell points towards
the singularity at $r=0$.  Moreover, there is an additional
characterization
for shells at the horizon, this case comes from the limit of $R\to
r_+$ from below and so is the limiting case of the case studied in
Sec.\ref{Subsec:extremalnormalinsideminus}.
In this case there is a null shell, which is not a horizon,
and
so, following the nomenclature, $r_+$ is the gravitational radius.
Also $r_+$ and $r_-$ have the same value.
In general we also opt to use $M$ rather than $Q$.  The normal to the
shell pointing towards $r=0$ means in the notation for the
extremal states that the new parameter $\xi$ has value $\xi=-1$, see
the end of this section for details.

As functions of $M$ and $R$, the shell's energy density $\sigma$ and
pressure $p$, are, see the end of this section,
\begin{align}
8\pi\sigma&= \;\; \frac{2}{M}\,,
\label{eq:qbh_sigma2}\\ 8\pi p&= 0\,.
\label{eq:qbhpressure2}
\end{align}
Also, the electric
charge density $\sigma_{e}$ is given in terms of $M$ by
Eq.~(\ref{qbhchargedensity}).  Since it is one point in a plot of
$\sigma$ or $p$ as functions of $\frac{R}{M}$, there is
no need to draw a
figure.
The shell is
characterized by a positive energy density.  The pressure is zero, and
so the matter is Majumdar-Papapetrou matter, i.e., $\sigma_e=\sigma$,
and therefore is fully supported by electric repulsion.
When $Q=0$, and so
$M=0$, there is a singularity at $R=0$ and
Minkowski in the exterior.
In relation to the energy conditions of the shell one can
work out and find that the null, the weak, the dominant, and the
strong energy conditions are always verified, see a detailed
presentation ahead.

\begin{figure}[h]
\includegraphics[height=0.26\paperheight]
{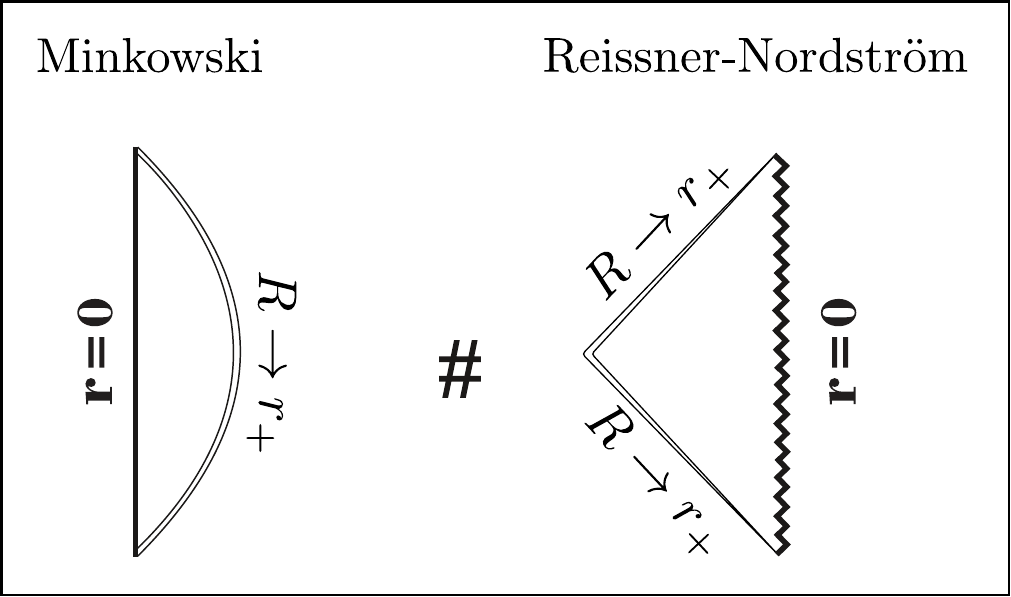}
\caption{\label{Fig:Penrose_diagram_Mink_horizonnullshell2_RN2}
Carter-Penrose diagram of an extremal Majumdar-Papapetrou
null shell
singularity, i.e., a thin shell spacetime in an extremal
Reissner-Nordstr\"om state, with the shell located at $R=r_+$, i.e.,
located at the gravitational radius, with orientation such that the
normal points towards $r=0$, and such that
$R\to r_+$ from $R<r_+$.  The interior is Minkowski, the exterior
is extremal Reissner-Nordstr\"om.
}       
\end{figure}

The Carter-Penrose diagram can be drawn with some care from the
building blocks of an interior Minkowski spacetime and the exterior
asymptotic region of an extremal Reissner-Nordstr\"om spacetime.  In
Figure~\ref{Fig:Penrose_diagram_Mink_horizonnullshell2_RN2} the
Carter-Penrose diagram of an extremal Majumdar-Papapetrou shell
singularity, i.e., for $R=r_+$ from below and a junction surface with
orientation such that the outside normal points towards $r=0$ is
shown.  The two regions contain complete geodesics so that the
manifold is composed by two connected regions, where in the interior
there is Minkowski spacetime, and on the exterior extremal
Reissner-Nordstr\"om spacetime.

\newpage

The physical interpretation of this case  follows from the
corresponding extremal shell inside the gravitational radius.  This
extremal thin shell solution provides a closed spatial static universe
with a singularity at one pole. There are quasihorizons. The energy
density and pressure obey the energy conditions for all shell radii,
indeed the shell is composed of Majumdar-Papapetrou matter. The causal
and global structure as displayed by the Carter-Penrose diagram show
the characteristics of this universe that has two sheets joined at the
shell. For one sheet, i.e., for one side of the universe, the shell is
timelike, for the other sheet, the shell is null, and possesses a
timelike singularity.  So, this case falls into the category of having
the energy conditions verified and the resulting spacetime being
strange.

\subsection{Formalism for extremal electric thin
shells at the gravitational radius
\label{Subsec:Extremal_induced_Minkowski_RN_qbhinsidenormalboth}}

\subsubsection{Preliminaries}
\label{prel5}

We now make a careful study to derive the properties of the
fundamental electric thin shell used in the two previous subsections,
i.e., the thin shell in an extremal state, i.e., $r_+=r_-$ or
$M=Q$, for which the shell's radius $R$ location obeys $R=r_+=r_-$, and
for which the orientation is such that the normal to the shell points
towards infinity or towards $r=0$.
It should be read as an appendix to the previous two
subsections.  We use the formalism developed in
Sec.~\ref{Sec:Junction_formalism}.

\subsubsection{Induced metric, and extrinsic curvature of
$\mathcal{S}$ as seen from $\mathcal{M}_{\rm i}$}
\label{Subsec:Extremal_induced_Minkowski_RN_qbhinsidenormalsubsub}

Let us start by analyzing the interior
Minkowski spacetime, $\mathcal{M}_{\rm i}$. Since it is the same as
the analysis done previously we only quote
the important equations.
They are the interior metric Eq.~(\ref{eq:Mink_metric_interior}),
the interior four-velocity of the shell
Eq.~(\ref{eq:Mink_vel_explicit}), the
metric for the shell at radius $R$
Eq.~(\ref{eq:induced_metric_Mink}), the normal to the shell
Eq.~(\ref{eq:normal_Mink}),
and the extrinsic curvature from the inside
Eq.~(\ref{eq:Extrinsic_curvature_Mink}).

\subsubsection{Induced metric, and extrinsic curvature of
$\mathcal{S}$ as seen
from $\mathcal{M}_{\rm e}$}
\label{Subsec:Extremal_induced_RN_insideqbhinsidenormalsubsub2}

To proceed we have now to find the expressions for the induced metric
on $\mathcal{S}$ and the extrinsic curvature components as seen from
the exterior spacetime, $\mathcal{M}_{\rm e}$, in the extremal state,
i.e., $r_+=r_-$ or $M=Q$, see
Figure~\ref{Fig:Penrose_diagram_RN_extremal}, for which the radius of
the shell $R$ tends towards $r_+=r_-$, and for which the orientation is
such that the normal to the shell points towards increasing $r$, i.e.,
towards spatial infinity, or towards decreasing $r$, i.e., towards
$r=0$, as seen from the exterior, as we considered in the two previous
subsections.
Moreover, besides the direction of the
normal as seen from the exterior spacetime, we also have to
differentiate between the cases when the shell is located outside or
inside the event horizon, i.e., $R>r_+$, or
$R<r_+$,
see
Sec.~\ref{Sec:Extremal-thin-shells-outside} and
\ref{Sec:Extremal-thin-shells-inside12},
respectively.

The direction of the normal is taken into account by the
parameter $\xi$ as previously used. 
In order to account for the two possibilities
$R>r_+$ and 
$R<r_+$ when $R$ tends to $r_+$,
we 
introduce a new sign parameter $\chi$ defined by
$\chi=\text{sign}\left(R-r_{+}\right)$. Then, we can take directly
from Eqs.~(\ref{eq:Extremal_extrinsic_curvature_RN_outside}) and
(\ref{eq:eq:Extremal_extrinsic_curvature_RN_inside}) the expressions
for the extrinsic curvature
\begin{equation}
{K_{\rm e}}^{\tau}{}_{\tau}
=\chi\,\xi\frac{1}{r_+}
\,,\quad\quad
{K_{\rm e}}^{\theta}{}_{\theta}=
{K_{\rm e}}^{\varphi}{}_{\varphi}
=\xi\frac{k}{R}\,,
\label{eq:eq:qbhExtremal_extrinsic_curvature_RN_inside}
\end{equation}
where again $\xi$ is defined
as $\xi=+1$ if the outside unit normal to the shell points in the
direction of increasing radial coordinate $r$, measured by an observer
in the exterior $\mathcal{M}_{\rm e}$ spacetime, and $\xi=-1$ if the
outside unit normal to the shell points in the direction of decreasing
radial coordinate $r$, and $k=\left|1-\frac{r_+}{R}\right|$.

\subsubsection{Shell's energy density and pressure}
\label{subSubsec:shellsenergydensityandpressureextremalqbhins}

Having determined the components of the extrinsic curvature of the
matching surface $\mathcal{S}$ as seen from the interior and exterior
spacetimes we are now in position to use the second junction
condition~(\ref{eq:2nd_junct_cond}) to find the expressions for the
energy density and pressure support of the thin shell.  The relations
$\sigma=-\frac{1}{4\pi}\left[K_{\theta}^{\theta}\right]$ and
$p=\frac{1}{8\pi}\left[K_{\tau}^{\tau}\right]-\frac{\sigma}{2}$ then
yield
\begin{align}
8\pi\sigma &=  \frac{2}{r_+}\,,
\label{eq:qbhExtremal_sigma_value_inside}\\
8\pi p &= \frac{\xi}{r_+}\left(\chi-\xi\right)\,.
\label{eq:qbhExtremal_pressure_value_inside}
\end{align}
Moreover, defining the surface electric current density $s_a$ on
the thin shell as $s_a=\sigma_{e}u_a$, where $\sigma_{e}$
represents the electric charge density, and since the Minkowski
spacetime has zero electric charge, from
Eqs.~(\ref{eq:junct_cond_Faradayb})-(\ref{eq:junct_cond_Faraday2})
and~(\ref{eq:RN_FaradayMaxwell_value}) it follows that
\begin{equation}
8\pi\sigma_{e}=\frac{2}{r_+}\,.
\label{eq:chargedensity1extremalqbh}
\end{equation}
The radial coordinate of the shell is $R=r_+$.

In Eq.~(\ref{eq:qbhExtremal_pressure_value_inside}) it is clear that
it is necessary to pick the signs of $\xi$ and $\chi$.
It is useful to give the expressions for
the shell's energy density and pressure, $\sigma$ and $p$ in terms of
$M=Q$, where as usual we opt for $M$.  Using
Eq.~(\ref{eq:KS_horizons_radius0}), i.e., $r_+=M$, in
Eqs.~(\ref{eq:qbhExtremal_sigma_value_inside})-(\ref{eq:chargedensity1extremalqbh})
with $\xi=+1$ and $\chi=+1$ we have $8\pi\sigma=\frac{2}{M}$, $8\pi p=0$,
and $8\pi\sigma_{e}=\frac{2}{M}$.  Choosing now $\xi=+1$ and
$\chi=-1$, with $r_+=M$, in
Eqs.~(\ref{eq:qbhExtremal_sigma_value_inside})-(\ref{eq:chargedensity1extremalqbh})
we have $8\pi\sigma=\frac{2}{M}$, $8\pi p=-\frac{2}{M}$, and
$8\pi\sigma_{e}=\frac{2}{M}$.  Choosing then $\xi=-1$ and $\chi=+1$,
with $r_+=M$, in
Eqs.~(\ref{eq:qbhExtremal_sigma_value_inside})-(\ref{eq:chargedensity1extremalqbh})
we have $8\pi\sigma=\frac{2}{M}$, $8\pi p=-\frac{2}{M}$, and
$8\pi\sigma_{e}=\frac{2}{M}$.
Choosing finally
$\xi=-1$ and $\chi=-1$, with $r_+=M$, in
Eqs.~(\ref{eq:qbhExtremal_sigma_value_inside})-(\ref{eq:chargedensity1extremalqbh})
we have $8\pi\sigma=\frac{2}{M}$, $8\pi p=0$, and
$8\pi\sigma_{e}=\frac{2}{M}$.  These are the expressions used in the
two previous subsections to study the properties of the thin matter
shells located at the event horizon $r_+$ separating a Minkowski
spacetime from an exterior extremal Reissner-Nordstr\"om spacetime.

\clearpage{}

\clearpage{}

\section{Overcharged electric thin shells:
Overcharged star shells and compact overcharged
shell naked singularities
\label{Sec:Overcharged_thin_shells}}

\subsection{Overcharged electric thin shells:
Overcharged star like shells}
\label{Subsec:overchargednormalplus}

Here we study the case of a fundamental electric thin
shell in the overcharged state, i.e.,
$r_+$ and $r_-$ are not real, or $M<Q$,
for which the shell's 
location is anywhere, i.e.,  $0<R<\infty$,
and for which the orientation is such that the
normal to the shell points towards spatial infinity.  In this case
horizons do not exist and moreover $r_+$
and $r_-$ do not exist, and so there is
neither gravitational radius nor Cauchy radius.
The normal to the shell
pointing towards spatial infinity means in the notation
we have been using that $\xi=+1$, see the
end of this section for
details.

\begin{figure}[h]
\subfloat[]
{\includegraphics[scale=0.45]{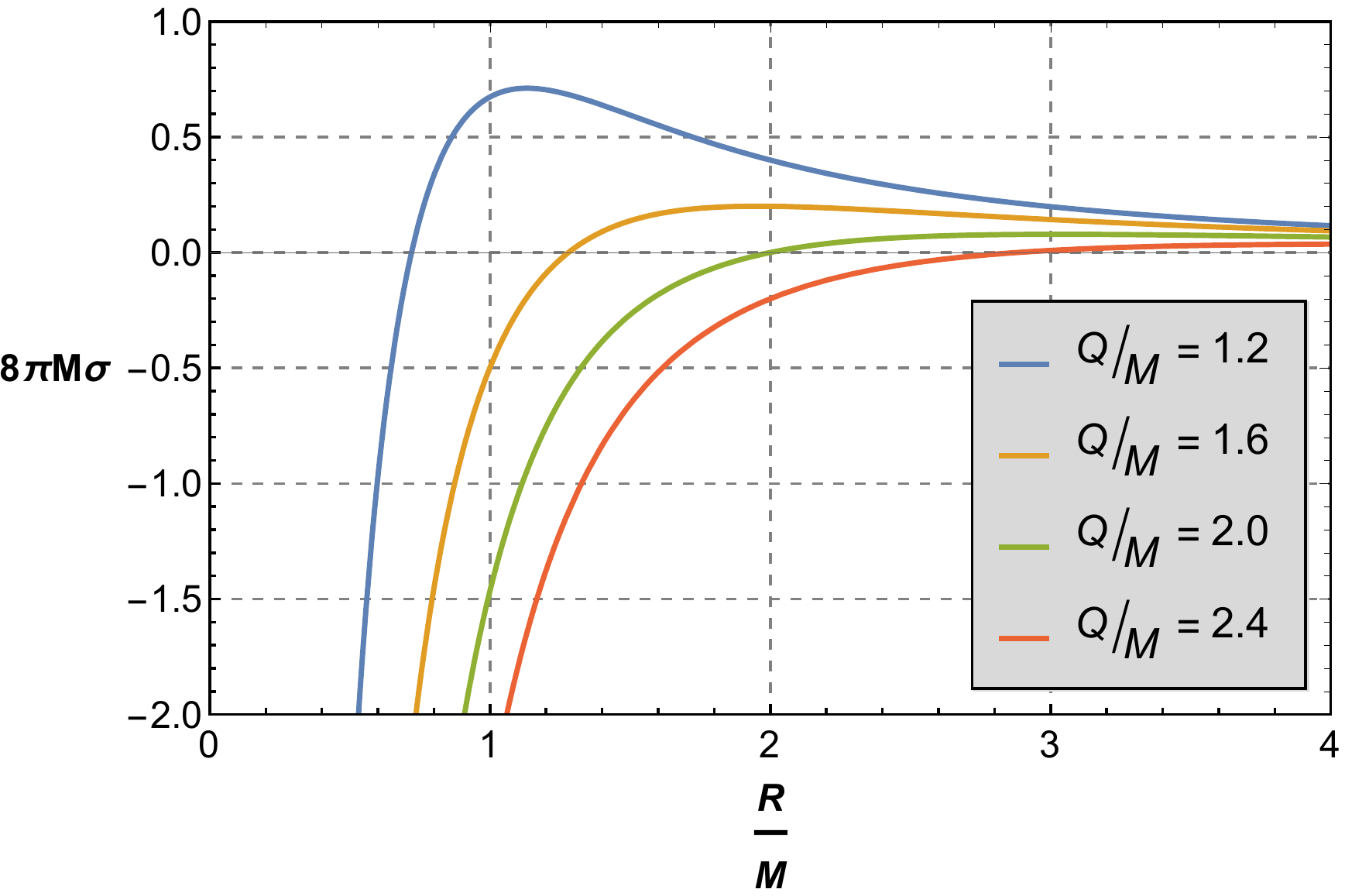}}
\hspace*{\fill}
\subfloat[]
{\includegraphics[scale=0.45]{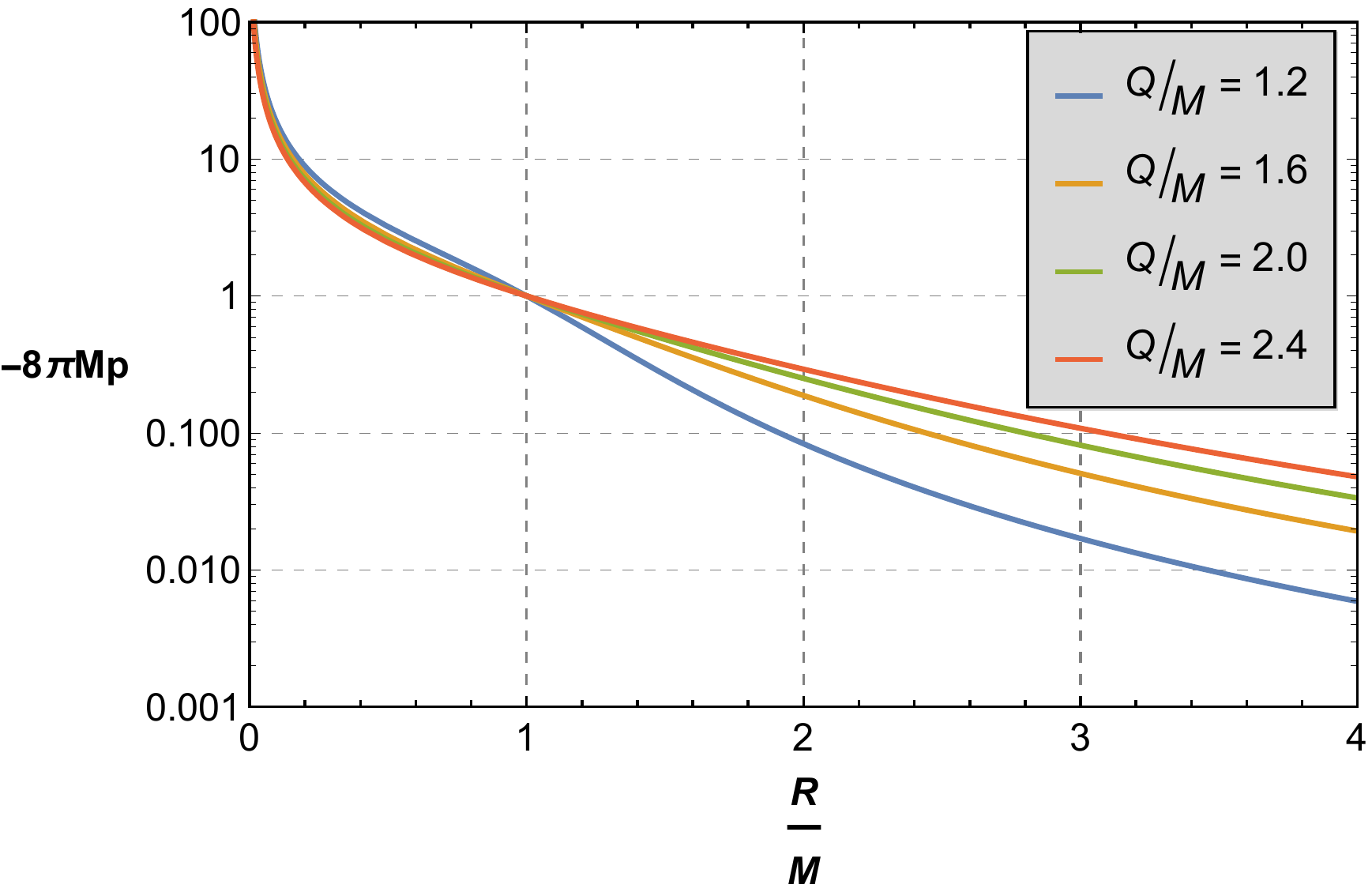}}
\caption{
\label{Fig:Properties_overcharged_normal}
Physical properties of an overcharged star shell i.e., an electric
perfect fluid thin shell in an overcharged Reissner-Nordstr\"om state,
in any location, i.e., $0<R<\infty$,
and with orientation such that the normal points towards spatial infinity.
Panel (a)
Energy density $\sigma$ of the shell as a function of the radius $R$
of the shell for various values of the $\frac{Q}{M}$ ratio. The energy
density is adimensionalized through the mass $M$, $8\pi M\sigma$, and
the radius is adimensionalized through the mass $M$, $\frac{R}{M}$.
Panel (b) Tension $-p$ on the shell as a function of the radius $R$ of
the shell for various values of the $\frac{Q}{M}$ ratio. The tension
is adimensionalized through the mass $M$, $-8\pi Mp$, and the radius
is adimensionalized through the mass $M$, $\frac{R}{M}$.
}
\end{figure}

As functions of $M $, $Q$, and $R$,  the shell's energy
density $\sigma$ and
pressure $p$, are, see the end of this section,
\begin{equation}
8\pi\sigma=\frac{2}{R}\left(1-k\right)\,,
\label{eq:sigma_value_overc}
\end{equation}
\begin{equation}
8\pi p=\frac{1}{2Rk}\left[\left(1-k\right)^{2}-\frac{Q^{2}}{R^{2}}
\right]\,,
\label{eq:pressure_value_overc}
\end{equation}
respectively, with $k=\sqrt{1-\frac{2M}{R}+\frac{Q^{2}}{R^{2}}}$.
Also, the electric charge density $\sigma_{e}$
is given in terms of $M$, $Q$, and $R$, by
\begin{equation}
8\pi \sigma_{e}=\frac{2Q}{R^{2}}\,.
\label{eq:chargedensityoverc}
\end{equation}
The behavior of $\sigma$ and $p$ as functions of the
radial coordinate $R$ of the shell for various values of the
$\frac{Q}{M}$ ratio in this case is shown in
Figure~\ref{Fig:Properties_overcharged_normal}.
From Figure~\ref{Fig:Properties_overcharged_normal} we see that,
depending on the radial coordinate of the shell, the energy density
might take negative values. Indeed, from
Eq.~(\ref{eq:sigma_value_overc}) we find that for $R<\frac{Q^{2}}{2M}$
the energy density $\sigma$ is negative. Also, this
case of thin shell is always supported by negative pressure, i.e.,
tension, see 
Eq.~(\ref{eq:pressure_value_overc}).  It is a tension shell and can also
be a negative energy density shell.  The fact that it is supported by
negative energy density sometimes and by tension
translates the well known fact that the Reissner-Nordstr\"om
singularity at $r=0$ is repulsive.  Moreover, we see that both the
energy density and the pressure of the shell diverge to negative
infinity as the shell gets closer to $R=0$.  On the other hand, in the
limit of $R\to\infty$ both energy density and the pressure go to zero.
When $Q=0$ there are no shells,
since then $M=0$ as we are not considering negative
$M$.
In relation to the energy conditions of the shell we can say that the
null, the weak, and the dominant, energy conditions are verified when
$R\geq R_{{\mathrm I}'}$, and the strong energy condition is verified
when $R\geq\frac{Q^2}{M}$, see a detailed presentation
ahead.

\newpage

\begin{figure}[h]
\label{Fig:Penrose_diagram_Mink_overcharged_RN_normal}
{\includegraphics[height=0.27\paperheight]
{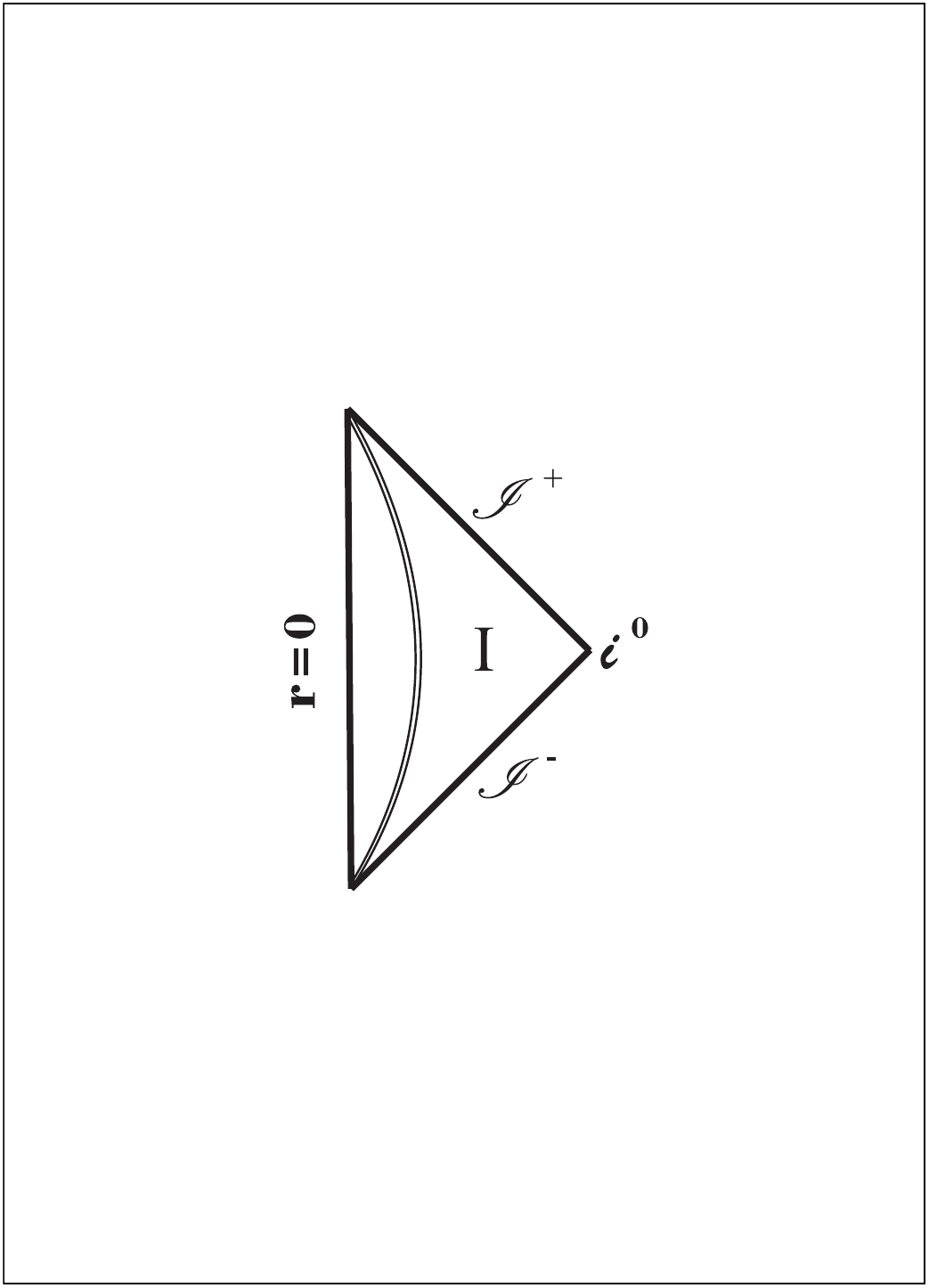}}
\caption{\label{Fig:Penrose_diagram_Mink_overcharged_RN}
Carter-Penrose diagram of an overcharged star thin shell, i.e., a
thin shell spacetime in an overcharged
Reissner-Nordstr\"om state,
with a thin shell located at any radius $R$,
with orientation such that the normal points towards spatial infinity. 
The interior is Minkowski, the exterior
is overcharged Reissner-Nordstr\"om. The shell is a star
shell supported by tension and for sufficiently small $R$ also by
negative energy density.
}
\end{figure}

The Carter-Penrose diagram can be drawn directly from the building
blocks of an interior Minkowski spacetime and the exterior asymptotic
infinite region of the overcharged
Reissner-Nordstr\"om spacetime. In
Figure~\ref{Fig:Penrose_diagram_Mink_overcharged_RN}
the
Carter-Penrose diagram of an overcharged
Reissner-Nordstr\"om star shell
spacetime for a junction surface with normal pointing towards spatial
infinity is shown. It is clearly a star shell, a
star in an asymptotically flat spacetime.  

The physical interpretation of this case is clear cut, and it is
similar to the corresponding nonextremal and extremal
shells. This overcharged thin
shell solution mimics an overcharged star.  The energy density and
pressure obey the energy conditions for certain radii.
The causal and global
structure as displayed by the Carter-Penrose diagram are well behaved
and rather elementary.  So, this case falls into the category of having
the energy conditions verified and the geometrical setup is physically
reasonable.

\subsection{Overcharged electric thin shells: Overcharged compact 
shell naked singularities}
\label{Sec:Overcharged_thin_shells2}

Here we study the case of a fundamental electric thin
shell in the overcharged state, i.e.,
$r_+$ and $r_-$ are not real, or $M<Q$,
for which the shell's 
location is anywhere, i.e.,  $0<R<\infty$,
and for which the orientation is such that the
normal to the shell points towards $r=0$.  In this case
horizons do not exist and moreover $r_+$
and $r_-$ do not exist, and so there is
neither gravitational radius nor Cauchy radius.
The normal to the shell
pointing towards $r=0$ means in the notation
we have been using that $\xi=-1$, see the
end of this section for
details.

\begin{figure}[h]
\subfloat[]
{\includegraphics[scale=0.45]
{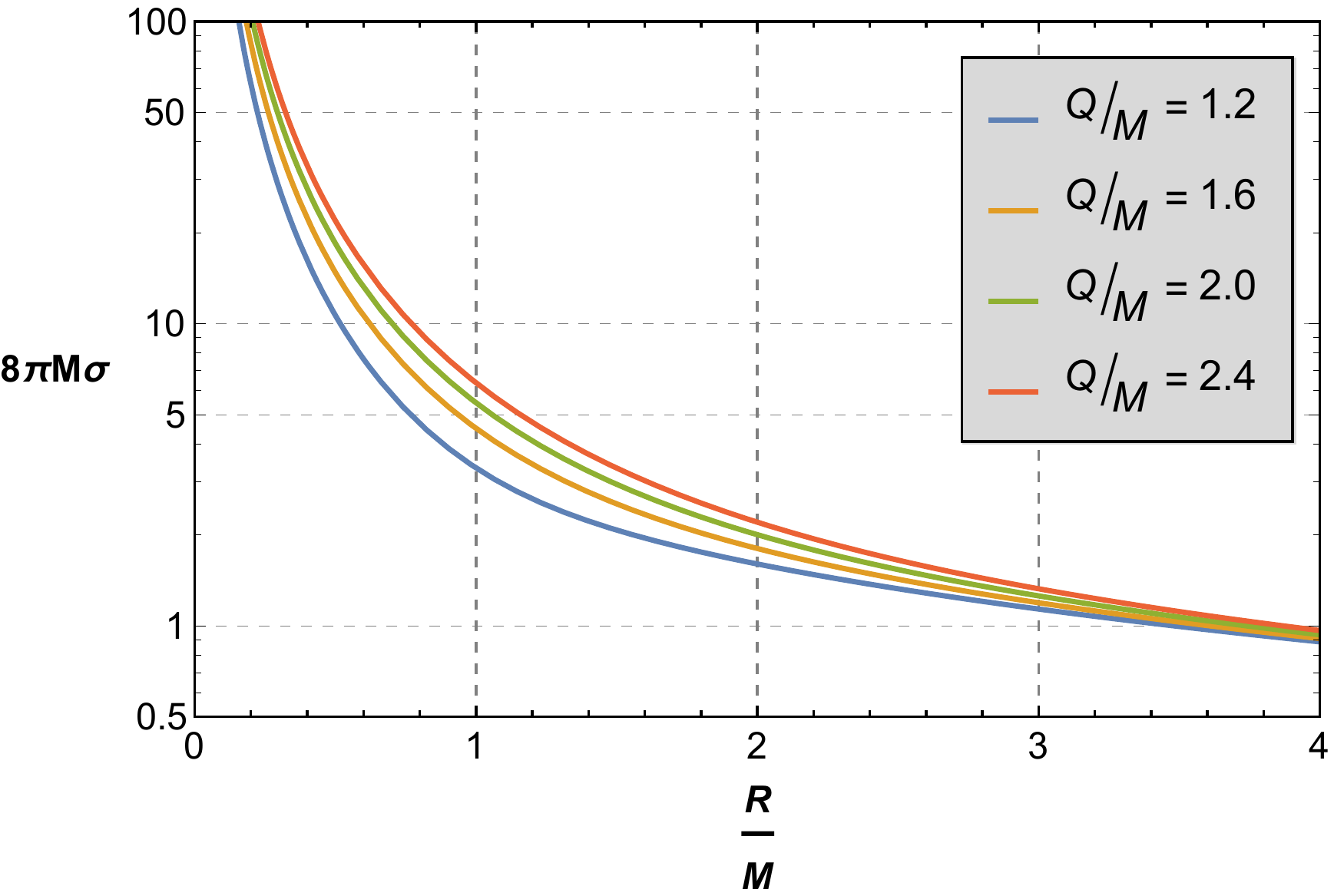}}
\hspace*{\fill}\subfloat[]
{\includegraphics[scale=0.45]
{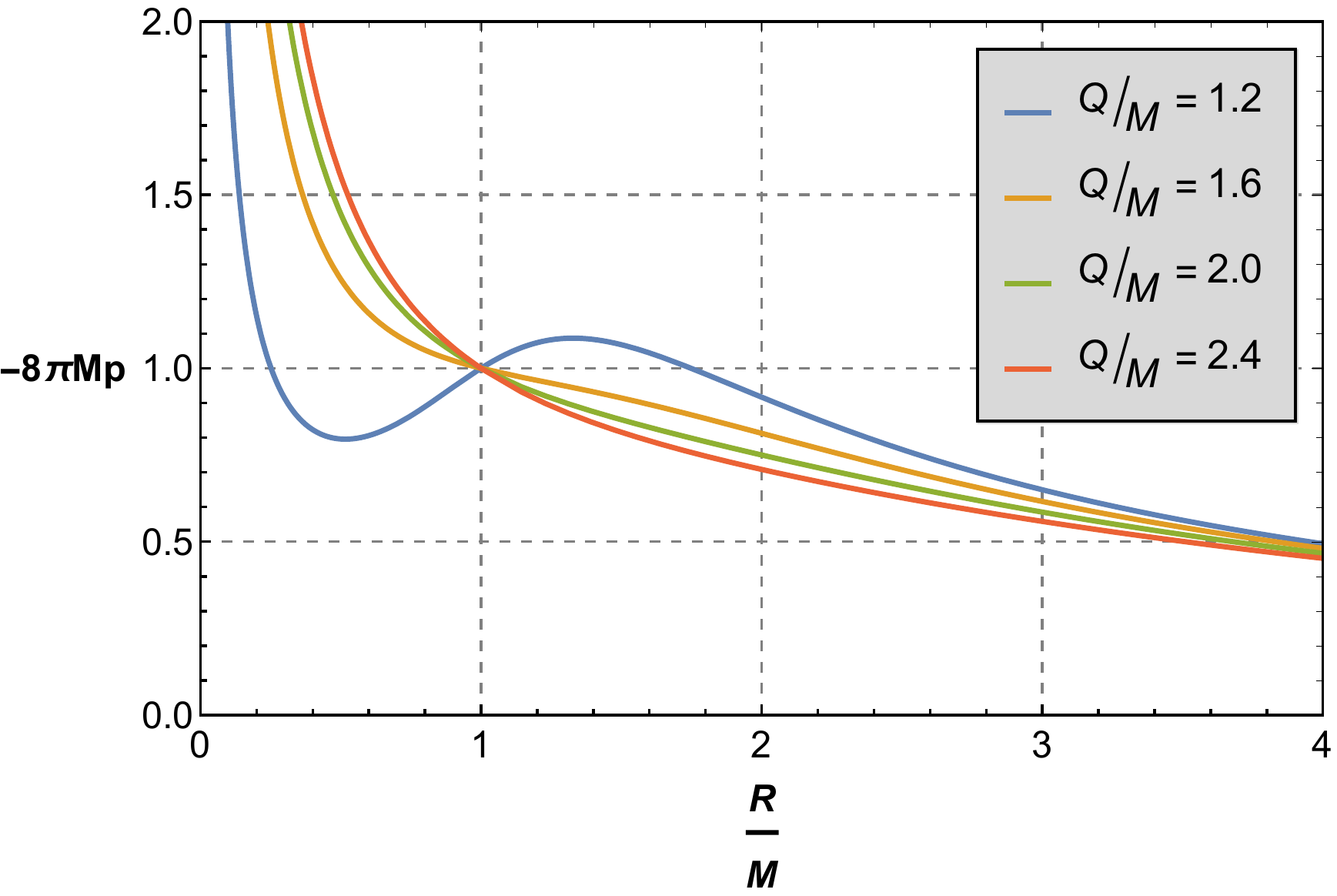}}
\caption{\label{Fig:Properties_overcharged_alternative}
Physical properties of an overcharged
compact shell naked singularity, i.e., an electric
perfect fluid thin shell in an overcharged Reissner-Nordstr\"om state,
in any location, i.e., $0<R<\infty$,
and with orientation such that the normal points towards $r=0$.
Panel (a) Energy density $\sigma$ of the shell as a
function of the radius $R$ of the shell for various values of the
$\frac{Q}{M}$ ratio. The energy density is adimensionalized through
the mass $M$, $8\pi M\sigma$, and the radius is adimensionalized
through the mass $M$, $\frac{R}{M}$.  Panel (b) Tension $-p$ on the
shell as a function of the radius $R$ of the shell for various values
of the $\frac{Q}{M}$ ratio. The tension is adimensionalized through
the mass $M$, $-8\pi Mp$, and the radius is adimensionalized through
the mass $M$, $\frac{R}{M}$.
}
\end{figure}

As functions of $M $, $Q$, and $R$,  the shell's energy
density $\sigma$ and
pressure $p$, are, see the end of this section,
\begin{equation}
8\pi\sigma=\frac{2}{R}\left(1+k\right)\,,
\label{eq:sigma_value_overcnormalto0}
\end{equation}
\begin{equation}
8\pi p=-\frac{1}{2Rk}\left[\left(1+k\right)^{2}-\frac{Q^{2}}{R^{2}}
\right]\,,
\label{eq:pressure_value_overcnormalto0}
\end{equation}
respectively, with $k=\sqrt{1-\frac{2M}{R}+\frac{Q^{2}}{R^{2}}}$.
Also, the electric charge density $\sigma_{e}$
is given in terms of $M$, $Q$, and $R$, by
Eq.~(\ref{eq:chargedensityoverc}).
The behavior of $\sigma$ and $p$ as functions of the
radial coordinate $R$ of the shell for various values of the
$\frac{Q}{M}$ ratio in this case is shown in
Figure~\ref{Fig:Properties_overcharged_alternative}.
From Figure~\ref{Fig:Properties_overcharged_alternative}
we see that
the energy density
is positive for all shells.
Also, this
kind of thin shell is always supported by tension.
It is a tension shell.  The fact that it is supported by
tension translates the well known fact that the Reissner-Nordstr\"om
singularity at $r=0$ is repulsive.  Moreover, we see that both the
energy density and the tension of the shell diverge to 
infinity as the shell gets closer to $R=0$.  On the other hand, in the
limit of $R\to\infty$ both go to zero.
An interesting feature
of this kind of shells is the change in the behavior of $-p$
for $R>M$, where the tension needed to support such shells is smaller
as the $\frac{Q}{M}$ ratio increases. Moreover, if the ratio
$\frac{Q}{M}$ is in the range 
$1<\frac{Q}{M}<\sqrt{2}$
we find that the tension support of the matter fluid that composes
this type of shells is an increasing function at $R=M$ and this function
contains a local minimum in the region $0<R<M$. Notwithstanding,
the minimum value is zero only in the extremal case, $\frac{Q}{M}=1$.
When $Q = 0$ there are no shells since we
are not considering negative $M$.
In relation to the energy
conditions of the shell we can say that the null, the weak, and the
dominant, energy conditions are verified when $R>0$, and the
strong energy condition is verified when $R\leq\frac{Q^2}{M}$ in this case,
see a detailed presentation ahead.

\newpage

\begin{figure}[h]
\label{Fig:Penrose_diagram_Mink_overcharged_RN_alternative}
{\includegraphics[height=0.27\paperheight]
{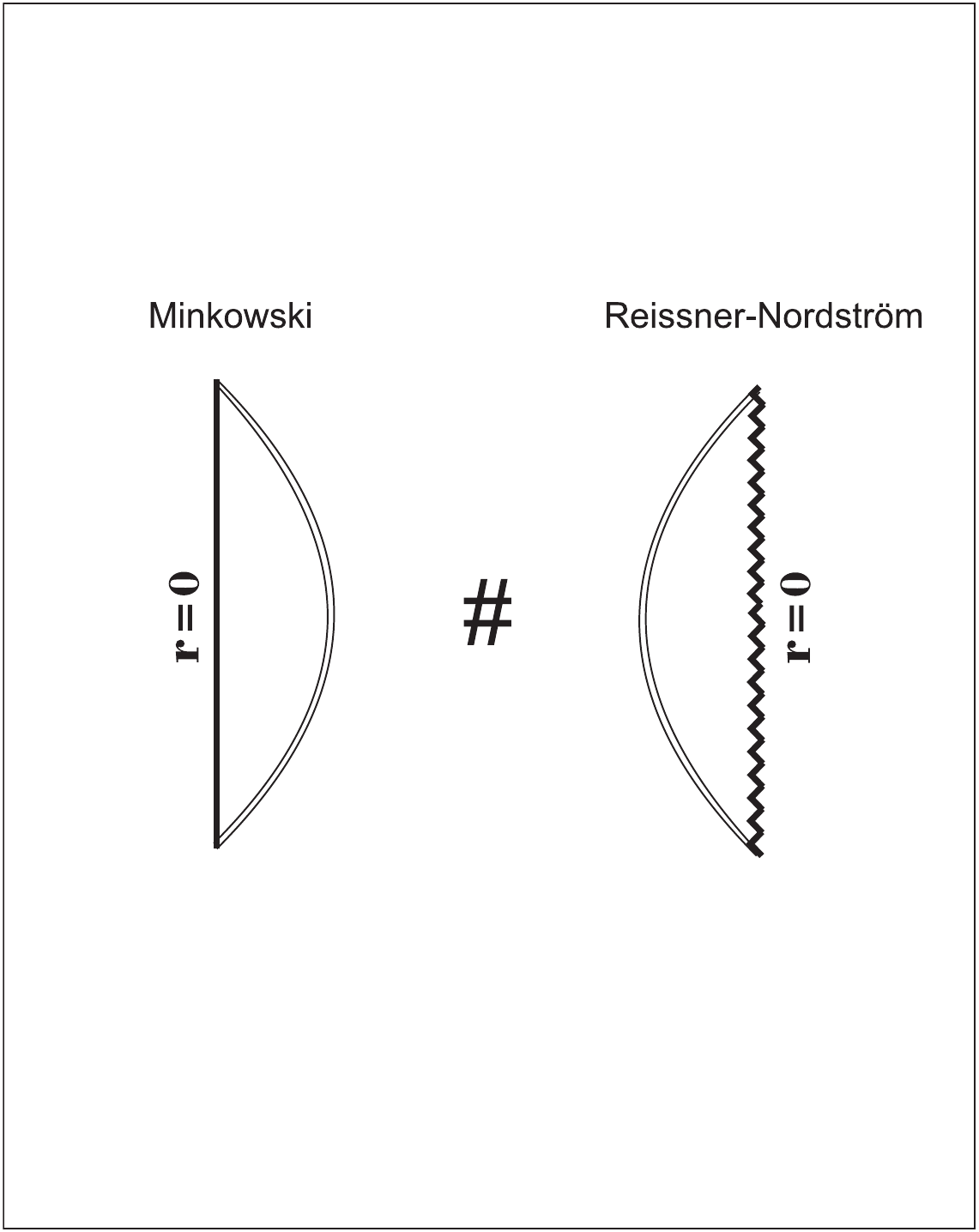}}
\caption{\label{Fig:Penrose_diagram_Mink_overcharged_RN2}
Carter-Penrose diagram of an overcharged compact
shell naked singularity, i.e., a thin
shell spacetime in an overcharged Reissner-Nordstr\"om state, located
at any radius $R$, with orientation such that the normal points
towards $r=0$. There is no clear distinction
of what is outside from what is inside.
The interior is Minkowski, the exterior is overcharged
Reissner-Nordstr\"om. This is a compact shell naked singularity
spacetime.
}
\end{figure}

The Carter-Penrose diagram can be drawn directly from the building
blocks of an interior Minkowski spacetime and the exterior
region neighbor to $r=0$ of the overcharged Reissner-Nordstr\"om
spacetime. In
Figure~\ref{Fig:Penrose_diagram_Mink_overcharged_RN2} the
Carter-Penrose diagram of an overcharged Reissner-Nordstr\"om star
shell spacetime for a junction surface with normal pointing towards
the $r=0$ singularity is shown. It is clearly a compact shell naked
singularity, such that there is no clear distinction
of what is outside from what is inside.

\newpage

The physical interpretation of this case is understood by now, it is
similar to the corresponding nonextremal and extremal shells. This
overcharged thin shell solution provides a closed spatial static
universe with a singularity at one pole. There are no horizons.  The
energy density and pressure obey the energy conditions for certain
shell radii.  The causal and global structure as displayed by the
Carter-Penrose diagram show the characteristics of this universe that
has two sheets joined at the shell with one sheet having a singularity
at its pole and with no horizons. The singularity is avoidable to
timelike curves. So, this case falls into the category of having the
energy conditions verified and the resulting spacetime being peculiar.

\subsection{Formalism for overcharged shells
\label{Subsec:Overcharged_induced_Minkowski_RN}}

\subsubsection{Preliminaries}
\label{prel6}

We now make a careful study to derive the properties of the
fundamental electric thin shell used in the two previous subsections,
i.e., the thin shell in an overcharged
state, i.e., $r_+$ and $r_-$
do not exist or
$M<Q$, for which the shell's radius $R$ location
obeys $0<R<\infty$, and
for which the orientation is such that the normal to the shell points
towards spatial infinity or towards $r=0$.
It should be read as an appendix to the previous two
subsections.  We use the formalism developed in
Sec.~\ref{Sec:Junction_formalism}.

\subsubsection{Induced metric, and extrinsic curvature of
$\mathcal{S}$ as seen from $\mathcal{M}_{\rm i}$}
\label{Subsec:Overcharged_induced_Minkowski_RN2}

Let us start by analyzing the interior
Minkowski spacetime, $\mathcal{M}_{\rm i}$. Since it is the same as
the analysis done previously we only quote
the important equations.
They are the interior metric Eq.~(\ref{eq:Mink_metric_interior}),
the interior four-velocity of the shell
Eq.~(\ref{eq:Mink_vel_explicit}), the
metric for the shell at radius $R$
Eq.~(\ref{eq:induced_metric_Mink}), the normal to the shell
Eq.~(\ref{eq:normal_Mink}),
and the extrinsic curvature from the inside
Eq.~(\ref{eq:Extrinsic_curvature_Mink}).

\subsubsection{Induced metric, and extrinsic curvature of
$\mathcal{S}$ as seen
from $\mathcal{M}_{\rm e}$}
\label{Subsec:Overcharged_induced_RN}

To proceed we have now to find the expressions for the induced metric
on $\mathcal{S}$ and the extrinsic curvature components as seen from
the exterior spacetime, $\mathcal{M}_{\rm e}$, in the overcharged
state, i.e., $r_+$ and $r_-$ do not exist or $M<Q$, see
Figure~\ref{Fig:Penrose_diagram_RN_extremal}, for which the
shell's obeys $0<R<\infty$, and for which the
orientation is such that the normal to the shell points towards
increasing $r$, i.e., towards spatial infinity,
or towards decreasing $r$ i.e., towards $r=0$,
as seen from the exterior, as
used in the two previous subsections.

The line element for the overcharged Reissner-Nordstr\"om spacetime,
now in the quantities
$M$ and $Q$, since $r_+$ and $r_-$ do not exist, is
\begin{equation}
ds_{\rm e}^{2}=
-\left(1-\frac{2M}{r}+\frac{Q^2}{r^2}\right)dt^2+
\frac{dr^2}{1-\frac{2M}{r}+\frac{Q^2}{r^2}}
+r^{2}
d\Omega^{2}\,,
\label{eq:metricovercharged}
\end{equation}
where $M<Q$.

Considering a static shell
as we have been doing,
the components of the 4-velocity $u^\alpha$ of an observer comoving
with the shell as seen from the exterior spacetime, are given by 
\begin{equation}
u_{\rm e}^{\alpha}=-
\left(\frac{1}{k},0,0,0\right)\,,
\label{eq:overc4velocity_value_rminus}
\end{equation}
where 
$k=\sqrt{1-
\frac{2M}{R}+
\frac{Q^2}{R^2}}$.
To find the induced metric
on $\mathcal{S}$ as seen by an observer at $\mathcal{M}_{\rm e}$ and
imposing the first junction condition, Eq.~(\ref{eq:1st_junct_cond}),
we find that the shell's radial coordinate $R$ is the same as measured
by an observer at $\mathcal{M}_{\rm i}$ or $\mathcal{M}_{\rm e}$ and the
induced metric on $\mathcal{S}$ is given by
Eq.~(\ref{eq:induced_metric_RN}),
namely,
\begin{equation}
\left.ds^{2}\right|_{\mathcal{S}}=-d\tau^{2}+R^{2}d\Omega^{2}\,.
\label{eq:induced_metric_RNoverc}
\end{equation}
Combining $n_{\rm e}^{\alpha}n_{{\rm e}\alpha}=1$, see
Eq.~(\ref{eq:normal_normalized}),
$n_{{\rm e}\alpha}u_{\rm e}^{\alpha}=0$, see
Eq.~(\ref{eq:normal_orthogonal}),
and Eq.~(\ref{eq:overc4velocity_value_rminus}), we find the expression
for the components of the unit normal to the hypersurface $\mathcal{S}$,
as seen from the exterior spacetime $\mathcal{M}_{\rm e}$, to be
$n_{{\rm e}\alpha}=\pm\left(0,\frac{1}{k},0,0\right)$.
To specify the sign of the normal to $\mathcal{S}$ for each region
we  consider two orientations: the orientation where the normal $n$ points
towards spatial infinity and the orientation where the normal
points towards the singularity  $r=0$. These two orientations can be treated
in a concise way by using $\xi=\pm1$, such that
\begin{equation}
n_{{\rm e}\alpha}=\xi
\left(0,\frac{1}{k},0,0\right)\,.
\label{eq:overcnormal_value_rminus}
\end{equation}
Then, we find the nonzero components of the extrinsic curvature of
$\mathcal{S}$ as seen from the exterior spacetime to be given by
\begin{equation}
{K_{\rm e}}^{\tau}{}_{\tau}=
\frac{\xi}{R^{2}k}\left( M-
\frac{Q^2}{R}\right)\,,
\quad\quad
{K_{\rm e}}^{\theta}{}_{\theta}=
{K_{\rm e}}^{\varphi}{}_{\varphi}
=
\xi\frac{k}{R}\,,
\label{eq:overc_Extrinsic_RN_inside_cauchy_horizon}
\end{equation}
where again $k$ is the redshift function given by  
$k=\sqrt{1-
\frac{2M}{R}+
\frac{Q^2}{R^2}}$.

\subsubsection{Shell's energy density and pressure}
\label{subSubsec:shellsenergydensityandpressureextremalovercharged}

Having determined the components of the extrinsic curvature of the
matching surface $\mathcal{S}$ as seen from the interior and exterior
spacetimes we are now in position to use the second junction
condition~(\ref{eq:2nd_junct_cond}) to find the expressions for the
energy density and pressure support  of a perfect fluid thin shell
in an overcharged state.
Using the shell's stress-energy tensor
given in Eq.~(\ref{eq:perfect})
we find
\begin{equation}
8\pi\sigma=\frac{2}{R}\left(1-\xi k\right)\,,
\label{eq:overcsigma}
\end{equation}
\begin{equation}
8\pi p=\frac{\xi}{2Rk}\left[\left(1-\xi k\right)^{2}-\frac{Q^2}{R^{2}}
\right]\,,
\label{eq:overcpressure}
\end{equation}
where the redshift function of the shell at $r=R$ is given by
$k=\sqrt{1-\frac{2M}{R}+\frac{Q^2}{R^2}}$.
Moreover, defining the surface electric current density $s_a$ on
the thin shell as $s_a=\sigma_{e}u_a$, where $\sigma_{e}$
represents the electric charge density, and since the Minkowski
spacetime has zero electric charge,
from Eqs.~(\ref{eq:junct_cond_Faradayb})-(\ref{eq:junct_cond_Faraday2})
and~(\ref{eq:RN_FaradayMaxwell_value})
it follows that
\begin{equation}
8\pi \sigma_{e}=\frac{2Q}{R^{2}}\,.
\label{eq:overcchargedensity}
\end{equation}
As before, we have to distinguish the two possible orientations provided by
$\xi$.  In
Eqs.~(\ref{eq:overcsigma})
and~(\ref{eq:overcpressure}) it is clear that it is
necessary to pick the sign in $\xi$.  Let us
start with $\xi=+1$.  Eqs.~(\ref{eq:overcsigma})
and~(\ref{eq:overcpressure}) with $\xi=+1$
yield $8\pi\sigma=\frac{2}{R}\left(1-k\right)$, $8\pi
p=\frac{1}{2Rk}\left[\left(1-k\right)^{2}-
\frac{Q^{2}}{R^{2}}\right]$, and also from
Eq.~(\ref{eq:overcchargedensity}) we have
$8\pi \sigma_{e}=\frac{2Q}{R^{2}}$. Let us now take $\xi=-1$.
Eqs.~(\ref{eq:overcsigma})
and~(\ref{eq:overcpressure}) with $\xi=-1$
yield $8\pi\sigma=\frac{2}{R}\left(1+k\right)$, $8\pi
p=-\frac{1}{2Rk}\left[\left(1+k\right)^{2}-\frac{Q^{2}}{R^{2}}
\right]$, and also from Eq.~(\ref{eq:overcchargedensity}) we have
$8\pi \sigma_{e}=\frac{2Q}{R^{2}}$.  These are the expressions used in
the two previous subsections. Note that for the overcharged
case $0<R<\infty$.

\newpage

\section{A synopsis to all
the fundamental  electric thin shells: Energy conditions and the
bewildering variety of Carter-Penrose
diagrams}
\label{Sec:SinopECandCP}

\subsection{Energy conditions for the fundamental electric thin shells}

\subsubsection{Energy-conditions}

The analysis of the properties of the fundamental electric
shells, i.e., timelike, static, perfect fluid thin
shells with a Minkowski interior and a Reissner-Nordstr\"om exterior
showed that both the energy density and pressure support depend on the
state of the shell, on the location of the shell and on the
orientation of the shell, i.e., on the direction of the outside
pointing normal. Moreover, we saw that in some situations the energy
density and pressure may take negative values, and this feature can
also depend on the value of the radial coordinate of the shell.
Here, we address the question of which shells and in what
conditions do they verify the various energy conditions.

The energy conditions are a set of restrictions on the stress-energy
tensor. In the case of a perfect fluid they lead to specific
constraints on the energy density and pressure, see,
e.g.,~\citep{andreasson2009} for energy conditions on shells, see
also~\citep{visserbook,Lemos_Lobo_2008} for energy conditions
on shells and 
\citep{Hawking_Ellis_book} for the original setting
of energy conditions. Here we will
study the null, weak, dominant, and strong energy conditions
for the fundamental electric thin shells. Now,
each energy condition may be considered to hold at any point of the
spacetime or along a flowline, where the specific energy condition is
only verified on average, allowing for pointwise violations. We
consider the pointwise version of the energy conditions.  Let us
first briefly explain the physical motivation for each energy
condition and their implications on the properties of a perfect fluid
thin shell.



The null energy condition, or NEC, represents the restriction that the
energy density of any matter distribution in spacetime experienced
by a ligh-ray is nonnegative. For a generic
stress-energy tensor ${T}_{\alpha\beta}$,
this is represented by 
${T}_{\alpha\beta}k^{\alpha}k^{\beta}\geq0$
for any future pointing null vector field $k^\alpha$. For a perfect fluid
thin shell with stress-energy tensor ${S}_{ab}$
given by Eq.~(\ref{eq:perfect})
this implies 
\begin{equation}
\hskip -1.6cm
\sigma+p\geq0\hfill
\,.
\label{eq:NEC_perfect_fluid}
\end{equation}


The weak energy condition, or WEC, is a more restrictive version of
the NEC where it is imposed that the energy density of any matter
distribution in spacetime measured by any timelike observer must be
nonnegative, then ${T}_{\alpha\beta}v^{\alpha}v^{\beta}\geq0$ for any
future pointing, timelike vector field $v^\alpha$.   For a perfect fluid
thin shell with stress-energy tensor ${S}_{ab}$
given by Eq.~(\ref{eq:perfect}) this leads to the following restrictions
\begin{equation}
\sigma\geq0\,,\quad\quad
\sigma+p\geq0
\,.
\end{equation}


The dominant energy condition, or DEC, represents the
statement that in addition
to the WEC being verified, the flow of energy can never be observed
to be faster than light, that is, in addition to
${T}_{\alpha\beta}v^{\alpha}v^{\beta}\geq0$,
the vector field $Y^\alpha$ with components given by
$Y^{\alpha}=-{T}_{\beta}{}^{\alpha}v^{\beta}$,
verifies $Y^{\alpha}Y_{\alpha}\leq0$, for any timelike future pointing
vector field $v^\alpha$.   For a perfect fluid
thin shell with stress-energy tensor ${S}_{ab}$
given by Eq.~(\ref{eq:perfect}) this 
implies 
\begin{equation}
\sigma\geq0\,,\quad\quad
\sigma-|p|\geq0\,.
\label{eq:DEC_perfect_fluid}
\end{equation}


The strong energy condition, or SEC, represents the restriction that nearby
timelike geodesics are always focused towards each other,
essentially
guaranteeing that gravity is always perceived to be attractive by
any timelike observer. In the case of general relativity,
this is found by guaranteeing
$\left({T}_{\alpha\beta}-\frac{1}{2}
g_{\alpha\beta}\mathcal{T}_{\gamma}{}^{\gamma}\right)
v^{\alpha}v^{\beta}\geq0$
for any timelike vector field $v^\alpha$. For a perfect fluid
thin shell with stress-energy tensor ${S}_{ab}$
given by Eq.~(\ref{eq:perfect})
we find, 
\begin{equation}
\sigma+p\geq0
\,,\quad\quad
\sigma+2p\geq0
\,.
\label{eq:SEC}
\end{equation}

\subsubsection{Limiting radii from an
analysis of the energy conditions on fundamental
electric thin shells}



\begin{figure}[h]
\subfloat[\label{Fig:Energy_radii_undercharged}]
{\includegraphics[height=0.2\paperheight]
{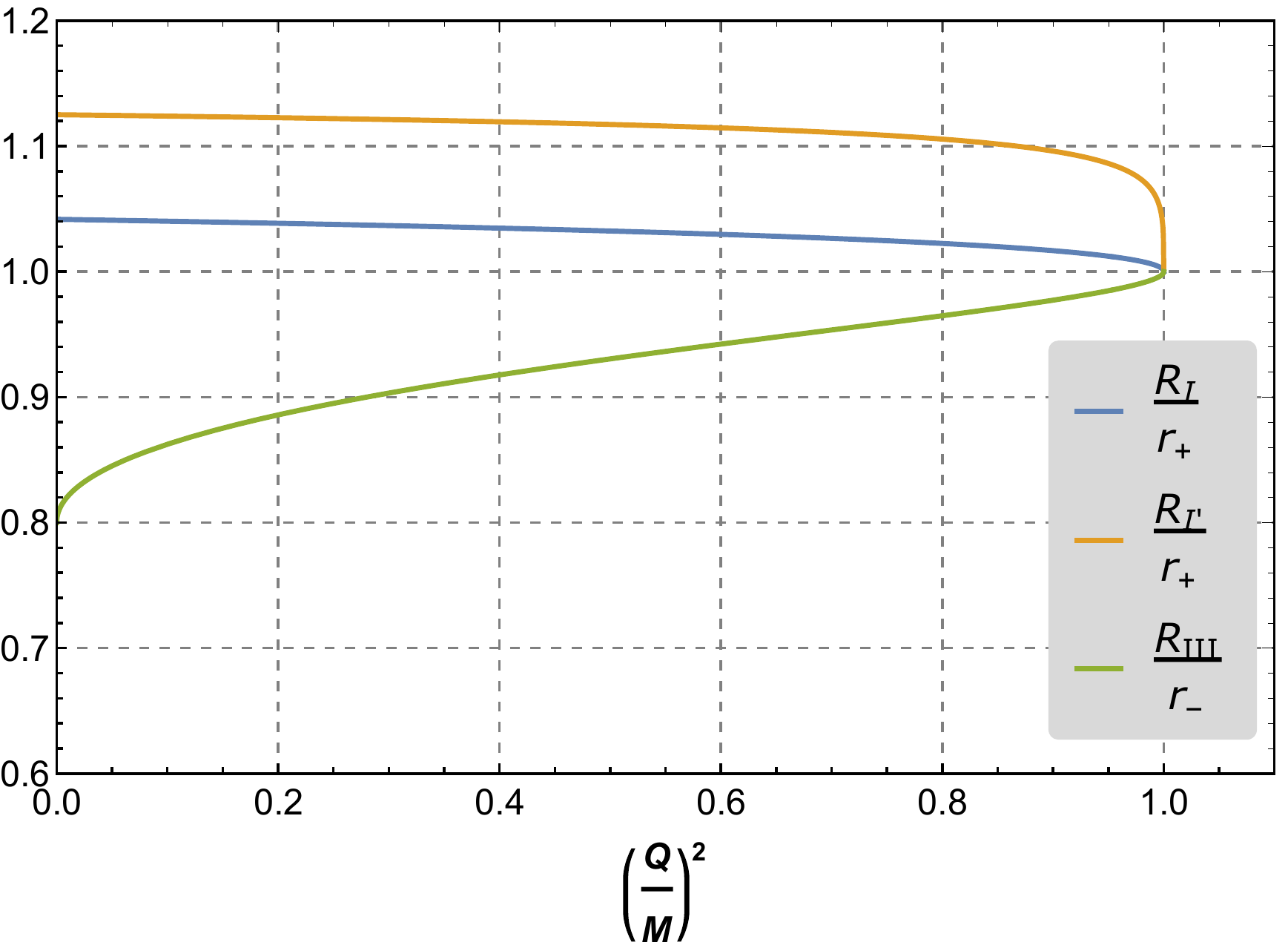}}
\hfill{}
\subfloat[\label{Fig:Energy_radii_overcharged}]
{\includegraphics[height=0.2\paperheight]
{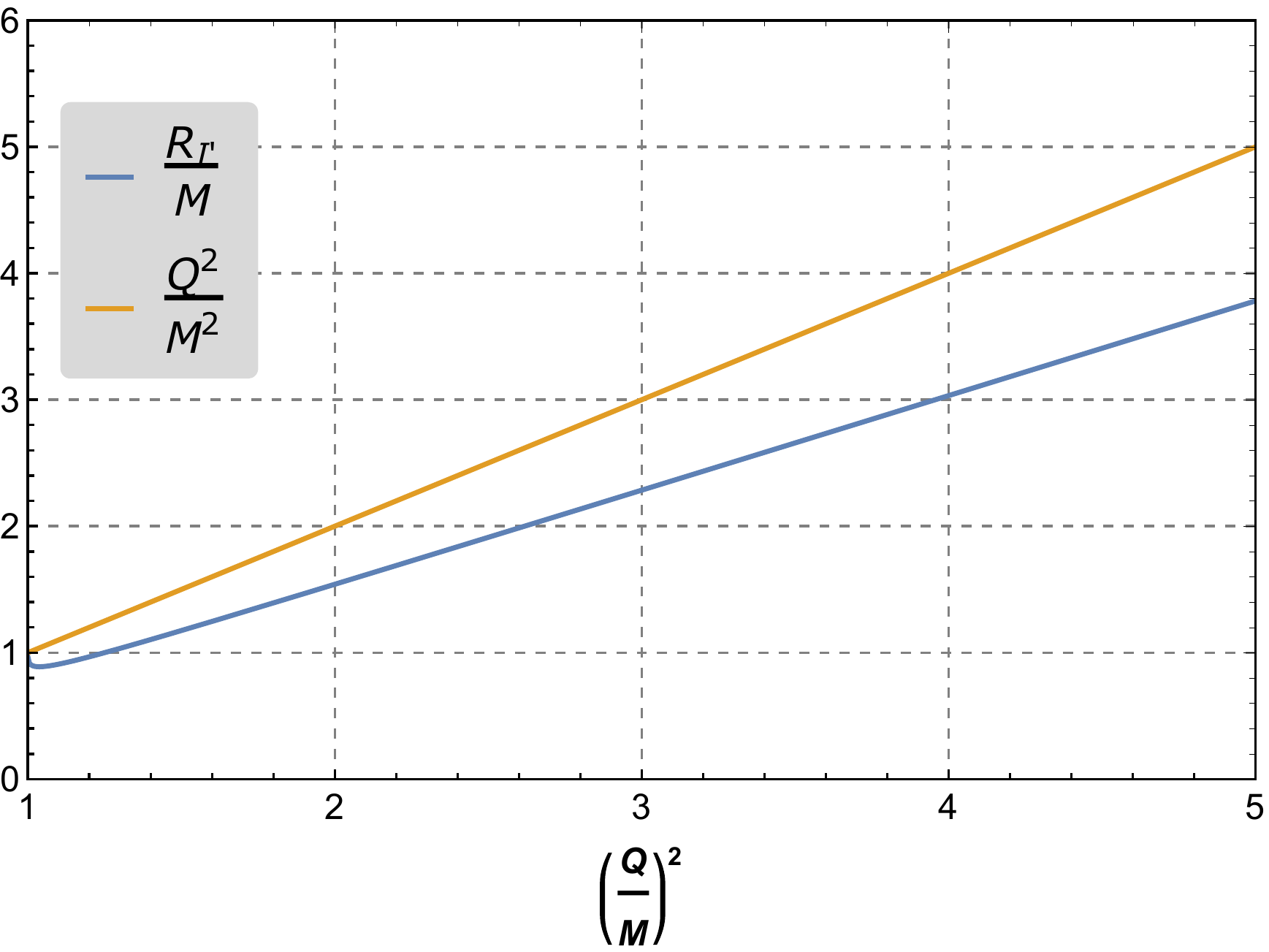}}
\caption{\label{Fig:Energy_radii}Behavior of the various radii, whose
expressions are given by Eqs.~(\ref{eq:energy_cond_RI}) -
(\ref{eq:energy_cond_RIII}), found by imposing the null, weak,
dominant and strong energy conditions to the thin shells present at
the matching surface of the various junction spacetimes.}
\end{figure}

From Eqs.~(\ref{eq:NEC_perfect_fluid})-(\ref{eq:SEC}) we see
that the energy conditions imply various restrictions on the energy
density and pressure of a perfect fluid. In the considered setup,
we have found that the properties of the perfect fluid
fundamental electric thin shells are
functions essentially
of the radius $R$ of the shell. Hence, the constrains
imposed by the energy conditions on the thin shell for the various
possible spacetimes will lead to restrictions on $R$.
Anticipating what follows, we present
the expressions for the limiting radii  $R_{\mathrm{I}}$,
$R_{\mathrm{I}'}$,
and $R_{\mathrm{III}}$, that arise from solving the
inequalities (\ref{eq:NEC_perfect_fluid})-(\ref{eq:SEC}) for
the various junction spacetimes, i.e.,
\begin{align}
R_{\mathrm{I}}= & \frac{M}{36}\left[25+3\left(\frac{Q}{M}\right)^{2}+
\frac{9\left(\frac{Q}{M}\right)^{4}-570\left(\frac{Q}{M}\right)^{2}+
625}{\Delta_{\mathrm{I}}}+\Delta_{\mathrm{I}}\right]\,,
\label{eq:energy_cond_RI}\\
R_{\mathrm{I}'}= & \frac{M}{4}\left[3+\left(\frac{Q}{M}\right)^{2}+
\frac{\left(\frac{Q}{M}\right)^{4}-10\left(\frac{Q}{M}\right)^{2}+
9}{\Delta_{\mathrm{I}'}}+\Delta_{\mathrm{I}'}\right]\,,
\label{eq:energy_cond_RIp}\\
R_{\mathrm{III}}= & \frac{M}{72}\left[50+6\left(\frac{Q}{M}\right)^{2}-
\frac{\left(1-i\sqrt{3}\right)\left[9\left(\frac{Q}{M}\right)^{4}-570
\left(\frac{Q}{M}\right)^{2}+625\right]}{\Delta_{\mathrm{I}}}-
\left(1+i\sqrt{3}\right)\Delta_{\mathrm{I}}\right]\,,
\label{eq:energy_cond_RIII}
\end{align}
with 
\begin{equation}
\begin{aligned}\Delta_{\mathrm{I}} & =\sqrt[3]{27
\left(\frac{Q}{M}\right)^{6}+216\left(\frac{Q}{M}
\right)^{3}\sqrt{9\left(\frac{Q}{M}\right)^{4}+366
\left(\frac{Q}{M}\right)^{2}-375}+5211\left(\frac{Q}{M}
\right)^{4}-21375\left(\frac{Q}{M}\right)^{2}+15625}\,,\\
\\
\Delta_{\mathrm{I}'} & =\sqrt[3]{8\left(\frac{Q}{M}
\right)^{3}\,\,\sqrt{\left(\left(\frac{Q}{M}\right)^{2}-1
\right)^{2}}+\left(\frac{Q}{M}\right)^{6}+17\left(\frac{Q}{M}
\right)^{4}-45\left(\frac{Q}{M}\right)^{2}+27}\,.
\end{aligned}
\end{equation}
The expressions for
$R_{\mathrm{I}}$ and $R_{\mathrm{I}'}$ can be read directly,
the expression for $R_{\mathrm{III}}$
is written in terms of the imaginary unit $i$, but for the range of values
of the ratio $\frac{Q}{M}$ of interest, this function takes purely real values.
Moreover, although it is not clear from the expressions, the values
of the radii $R_{\mathrm{I}}$, $R_{\mathrm{I}'}$, and $R_{\mathrm{III}}$
are independent of the sign of $Q$, as expected.
For completeness, in Figure~\ref{Fig:Energy_radii} we present the
behavior of the various limiting radii defined in
Eqs.~(\ref{eq:energy_cond_RI})-(\ref{eq:energy_cond_RIII})
as functions of the ratio $\frac{Q}{M}$.

\newpage


\subsubsection{Table of the energy conditions on fundamental
electric thin shells}

\begin{figure}[h]
{\includegraphics[height=0.27\paperheight]
{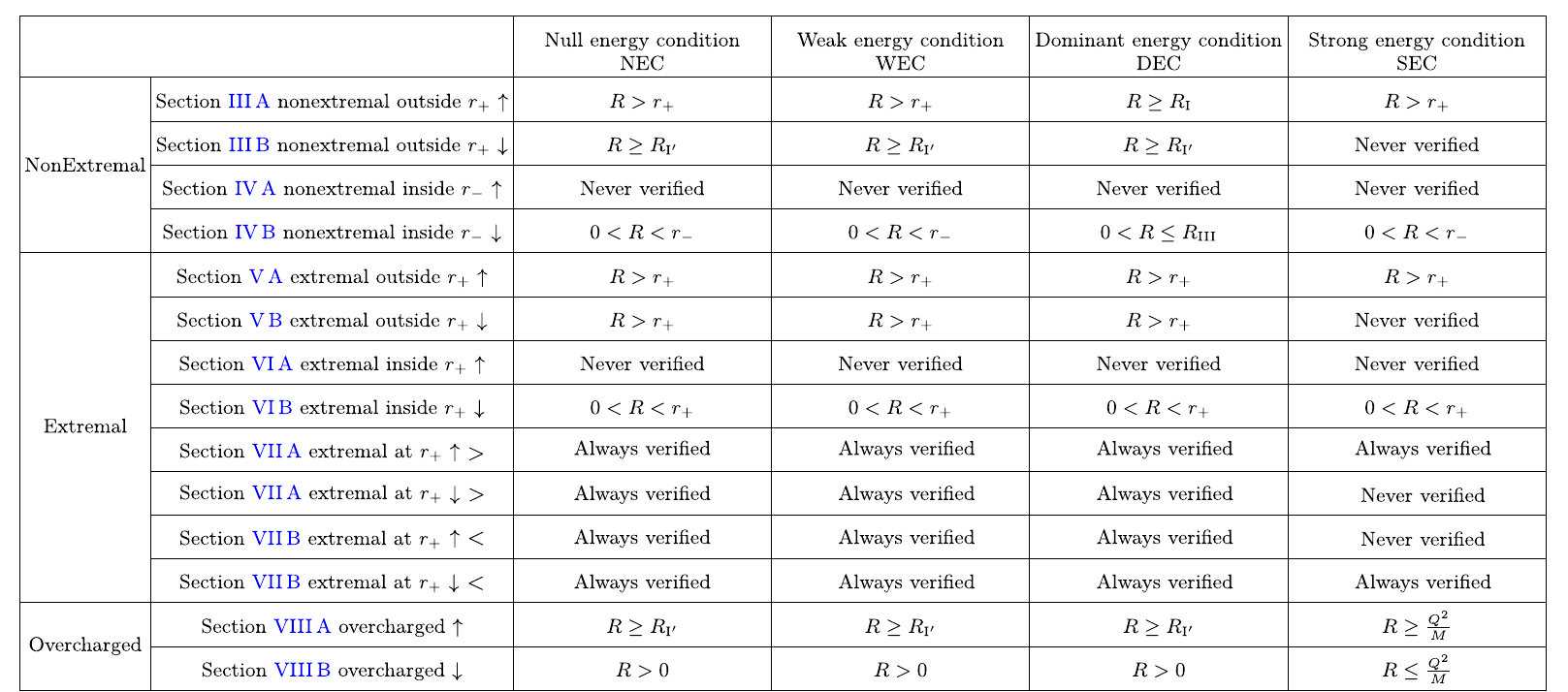}}
\caption{\label{Table:Energy_cond_table}
Range of values of the radius $R$ of the fundamental electric thin
shell, in the various allowed locations of the exterior
Reissner-Nordstr\"om spacetime
for which, the null, weak, dominant and
strong energy conditions are verified. The symbols $\uparrow$ and
$\downarrow$ denote the orientation of the shell, i.e., outward normal
pointing to increasing radius and to decreasing radius from the shell,
respectively. The symbols $>$ and $<$ for Sections VII A and VIIB
in the table denote whether the approach to $r_+$ is done
through $R>r_+$ or
$R<r_+$, respectively.
}
\end{figure}

Using the expressions for the energy density and pressure support for
the thin matter shell for each resulting junction spacetime in the
inequalities (\ref{eq:NEC_perfect_fluid})-(\ref{eq:SEC}),
allows us to find the constraints on the
shell's location so that each of the tested energy conditions is
verified. In the table of Figure~\ref{Table:Energy_cond_table}
we summarize the results.

\subsubsection{Detailed description}

For the fundamental electric shells in a nonextremal state, located
outside the gravitational radius $r_+$, $R>r_+$, we find that when
their orientation is such that the outward normal points to spatial
infinity, Section~\ref{Subsec:nonextremalnormaloutside}, i.e., the
star shells, they always verify the NEC and WEC, they verify the DEC
for $R>R_{\mathrm{I}}$, and also always verify the SEC, and when their
orientation is such that the outward normal points to the gravitational
radius $r_+$, Section~\ref{Subsec:nonextremaltensionoutside}, i.e.,
the tension shell black holes, they verify the NEC, WEC, and DEC for
$R>R_{\mathrm{I}'}$, and the SEC is always violated.  Moreover, the
limiting radius $R_{\mathrm{I}'}$ of Eq.~(\ref{eq:energy_cond_RIp})
also determines the value of the circumferential radius of the shell
for which its energy density is maximum and thus it is connected to
the bumps in de energy density $\sigma$ of
Figure~\ref{Fig:Properties_region_I_prime}.
For the fundamental electric shells in a nonextremal state, located
inside the Cauchy radius $r_-$, $R<r_-$, we find that when their
orientation is such that the outward normal points to $r_-$,
Section~\ref{Subsec:nonextremalnormalcauchy}, the tension shell
regular and nonregular black holes, none of the energy conditions are
verified, and when their orientation is such that the outward normal
points to the $r=0$ singularity,
Section~\ref{Subsec:nonextcompactshellnakedsingularity}, the compact
shell naked singularities, the shells always verify the NEC and WEC,
verify the DEC in the domain $0<R\leq R_{\mathrm{III}}$, and always
verify the SEC.

For the fundamental electric shells in an extremal state, $r_+=r_-$,
located outside the gravitational radius $r_+$, $R>r_+$, we find that
when their orientation is such that the outward normal points to
spatial infinity, Section~\ref{Subsec:extremalnormaloutside}, i.e.,
the Majumdar-Papapetrou star shells, they always verify the NEC, WEC,
DEC, and SEC, and when their orientation is such that the outward
normal points to the event horizon,
Section~\ref{Subsec:extremalnormaloutsidenormaltoin}, the extremal
tension shell black holes, they always verify the NEC, WEC, and DEC,
and the SEC is always violated.
For the fundamental electric shells in an extremal state, $r_+=r_-$,
located inside the gravitational radius, $R<r_+$, we find that when
their orientation is such that the outward normal points to spatial
infinity, Section~\ref{Subsec:extremalnormalinsideplus}, the extremal
tension shell regular and nonregular black holes, none of the energy
conditions are verified by the shells, and when their orientation is
such that the outward normal points to the $r=0$ singularity,
Section~\ref{Subsec:extremalnormalinsideminus}, the
Majumdar-Papapetrou compact shell naked singularities, the shells
always verify the NEC, WEC, DEC, and SEC.
For the fundamental electric shells in an extremal state, $r_+=r_-$,
located in the limit at the gravitational radius, $R=r_+$, we find
that when their orientation is such that the outward normal points to
spatial infinity and the limit of $R\to r_+$ comes from above,
Section~\ref{majumdarpapapetroushellquasiblackholes}, the
Majumdar-Papapetrou shell quasiblack holes, the shells always verify
the NEC, WEC, DEC, and SEC, the matter is Majumdar-Papapetrou matter,
whereas when the limit of $R\to r_+$ comes from below,
Section~\ref{extremalnullshellblackholes}, the extremal null shell
black holes, the shells verify the NEC, WEC, DEC, and never verify the
SEC,
and when their orientation is such that the outward normal points to
the $r=0$ singularity and the limit of $R\to r_+$ comes from above,
Section~\ref{extremalnullshellsingularities}, the extremal tension
shell null singularities, one has that these shells always verify the
NEC, WEC and DEC, and never verify the SEC,
whereas when the limit of $R\to r_+$ comes from below,
Section~\ref{extremalmajumdarpapapetroushellsingularities}, the
extremal Majumdar-Papapetrou null shell singularities, the
shells verify
the NEC, WEC, DEC, and SEC, the matter is Majumdar-Papapetrou matter.

For the fundamental electric shells in an overcharged state, $r_+$ and
$r_-$ do not exist and $M<Q$, located at any radius $R$, we find that
when their orientation is such that the outward normal points to
spatial infinity, Section~\ref{Subsec:overchargednormalplus}, the
overcharged star shells, the shells verify the NEC, WEC, DEC for
$R\geq R_{\mathrm{I}'}$, and the SEC when $R\geq \frac{Q^{2}}{M}$, and
when their orientation is such that the outward normal points to the
$r=0$ singularity, Section~\ref{Sec:Overcharged_thin_shells2}, the
overcharged compact shell naked singularities, the NEC, WEC, DEC are
always satisfied and the SEC when $R\leq \frac{Q^{2}}{M}$.  The
results for the strong energy condition
of an overcharged shell 
indicate that in the
overcharged Reissner-Nordstr\"om spacetime the singularity is
repulsive in a core region within $r<\frac{Q^{2}}{M}$. Our result
extends that of~\citep{Graves_Brill_1960,Carter_1966_2} where it was
found that the nonextremal and extremal Reissner-Nordstr\"om
solutions are characterized by a
repulsive region delimited, respectively, by the Cauchy or event
horizons. Here, although there are no horizons, we see that the same
conclusion holds, and confirm the result given
in, e.g.,~\citep{felicebook} that there is a repulsive region in the
overcharged Reissner-Nordstr\"om spacetime near the singularity, and,
in addition, find the limiting radius of this repulsive region.

\newpage

\subsection{The bewildering variety
of the Carter-Penrose diagrams for
the fundamental electric thin shells}
\label{Sec:Bewild}

\begin{figure}[h]
{\includegraphics[height=0.60\paperheight]
{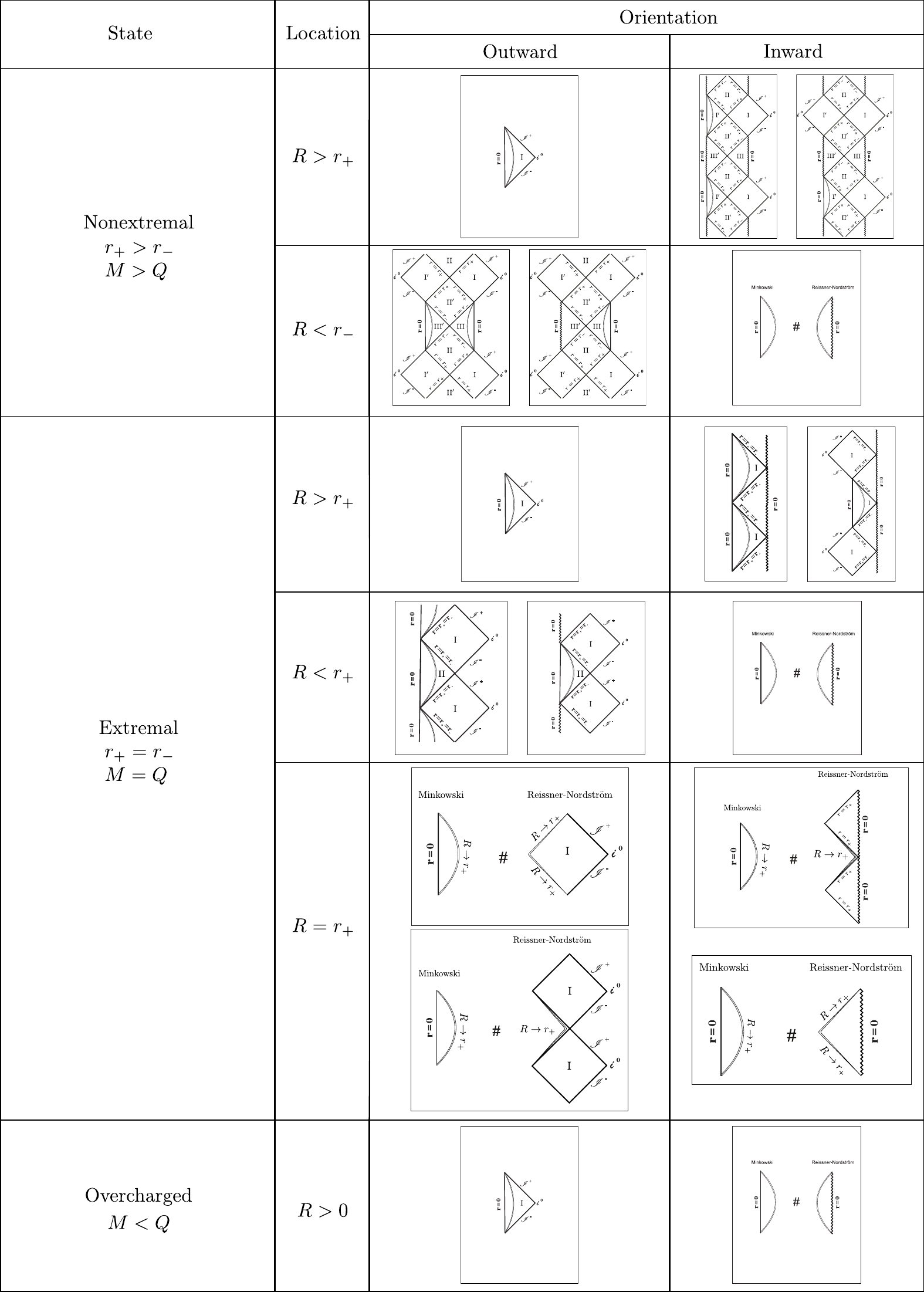}}
\caption{\label{alldiagrams}
A chart with all the fourteen
different Carter-Penrose diagrams for the fundamental electric
charged shells, i.e., static shells with a Minkowski interior and a
Reissner-Nordstr\"om exterior.}
\end{figure}

In addition to performing an analysis
on the physical properties of the shells, i.e.,
their energy density $\sigma$,  pressure $p$,
and the corresponding energy conditions, we have
drawn the
Carter-Penrose diagram in each of the fourteen
cases.
These diagrams
for
the fundamental electric thin shells are summarized in the chart of 
Figure~\ref{alldiagrams} which displays
clearly their bewildering variety.
\noindent There were cases that the
solution does not tell precisely how to continue the
Carter-Penrose diagram, one can
either repeat the shell, or draw horizons and infinities at will, in
any possible combination.

\newpage
\section{Conclusions \label{Sec:Conclusions}}

We have classified and studied
the spacetimes generated by a fundamental
electric thin shell, i.e., a spherical static electrical thin shell
with a Minkowski interior and a Reissner-Nordstr\"om exterior.
All three
main states a shell with
a Reissner-Nordstr\"om exterior can have were considered,
namely, nonextremal, extremal, and overcharged.  In the
nonextremal state there are still two possible locations
for the shell, namely, the
shell is located outside the gravitational radius or the shell is
located inside the Cauchy radius. In the extremal state there are
three possibilities, namely, the shell is located outside the
gravitational radius, the shell is located inside the gravitational
radius, or the shell is located at the gravitational radius.  In the
overcharged state there is only one possibility, the shell can be
located anywhere.  We have seen, in the wake of the work of
Lynden-Bell and Katz for non-electrical thin shells with a
Schwarzschild exterior, that each of the locations has still two
possibilities, either the outward normal to the shell points toward
increasing radius or it points toward decreasing radius.
For extremal shells at the gravitational radius there is still
a subdivision, either the shell approaches the gravitational
radius from above, or it approaches the gravitational
radius from below.
In all there
are fourteen different cases.

For each of the fourteen different shells we have worked out the energy
density $\sigma$ and the pressure $p$ and analyzed the energy
conditions of the matter on the shell.  In addition we have drawn the
Carter-Penrose diagrams in all the fourteen cases.  There were cases
that the solution does not tell precisely how to continue the diagram,
one can either repeat the shell, or draw horizons and infinities at
will, in any possible combination. In addition,
in some cases the distinction between what is 
interior and what is exterior is blurred.
The maximum analytical extension of the fundamental electric
shells and consequent Carter-Penrose diagrams, 
showed that there is a plethora of
solutions that encompass
nonextremal star shells,
nonextremal tension shell black holes,
nonextremal tension shell regular and
    nonregular black holes,
nonextremal compact shell naked singularities,
Majumdar-Papapetrou star
   shells,
extremal tension shell singularities,
extremal tension shell regular and nonregular black holes,
Majumdar-Papapetrou compact shell naked singularities,
Majumdar-Papapetrou shell quasiblack
     holes,
extremal null shell quasinonblack holes,
extremal null shell singularities,
Majumdar-Papapetrou null shell singularities,
overcharged star shells,
and overcharged compact shell naked singularities.

In some of the cases it was found that the energy conditions are
verified and the geometrical setup is physically reasonable, in other
cases it was found that the energy conditions are verified but the
resulting geometry is rather peculiar, or even strange, or that the
energy conditions are violated but the resulting geometry seems
physically reasonable.  Therefore, the set of solutions might be
greatly reduced if we only choose solutions which indeed obey the
energy conditions and are physically reasonable or, regard only
solutions that verify the energy conditions, independently of the
geometrical setup, or, maintain only solutions whose geometry seems
reasonable. Here we choose to maintain everything as good and
interesting solutions, to be tested a posteriori.

\section*{Acknowledgments}
JPSL acknowledges Funda\c c\~ao para a
Ci\^encia e Tecnologia - FCT, Portugal, for financial
support through Project No.~UIDB/00099/2020.
PL acknowledges IDPASC and FCT, Portugal, for financial support
through Grant No.~PD/BD/114074/2015, and thanks 
Centro de Matem\'{a}tica, Universidade do Minho,
where part of this work has been performed, for
the hospitality.

\newpage

\appendix
\renewcommand\thefigure{\thesection\arabic{figure}}
\setcounter{figure}{0}

\section{Kruskal-Szekeres coordinates of the
nonextremal Reissner-Nordstr\"om spacetime}
\label{Appendix_sec:Kruskal-Szekeres_coordinates_RN}

\subsection{General formalism for a static spherically symmetric
spacetime}
\label{Appendix_subsec:Kruskal-Szekeres_coordinates}

In this appendix we will construct two coordinate systems for the
nonextremal
Reissner-Nordstr\"om spacetime, one well behaved in a neighborhood
of the event horizon, $r=r_+$, and the other
in a neighborhood of the Cauchy
horizon, $r=r_-$, which, together, cover the full nonextremal
Reissner-Nordstr\"om
spacetime. To find these new coordinate systems we will use the formalism
introduced in~\citep{Graves_Brill_1960} which we now
present briefly.

Given a static, spherically symmetric spacetime whose line element
can be written in the form 
\begin{equation}
ds^{2}=-\Phi\left(r\right)dt^{2}+\Phi^{-1}
\left(r\right)dr^{2}+r^{2}d\Omega^{2}\,,
\label{Appendix_eq:KS_Non_regular_metric}
\end{equation}
where the function $\Phi\left(r\right)$ is assumed to have zeros
or poles representing coordinate
singularities, which can be removed by
a change of coordinates. Let us determine a simultaneous transformation
of the coordinates $r$ and $t$ to new coordinates $X\left(r,t\right)$
and $T\left(r,t\right)$ such that the line element can be written
as 
\begin{equation}
ds^{2}=f^{2}\left(X,T\right)\left(dX^{2}-dT^{2}\right)+r^{2}
\left(X,T\right)d\Omega^{2}\,,\label{Appendix_eq:KS_Regular_metric}
\end{equation}
where $f^{2}\left(X,T\right)$ is to be regular in a sub-region covered
by the coordinates $X$ and $T$.
Comparing Eqs.(\ref{Appendix_eq:KS_Non_regular_metric}) and
(\ref{Appendix_eq:KS_Regular_metric})
it is found that~\citep{Graves_Brill_1960} 
\begin{equation}
\begin{gathered}X=h\left(r^{*}+t\right)+g\left(r^{*}-t\right)\,,\\
T=h\left(r^{*}+t\right)-g\left(r^{*}-t\right)\,,
\end{gathered}
\label{Appendix_eq:KS_General_u_v}
\end{equation}
with $dr^{*}=\Phi^{-1}\left(r\right)dr$, i.e.,
\begin{equation}
r^{*}=\int \Phi^{-1}\left(r\right)dr\,,
\label{Appendix_eq:KS_r_asterisk_def}
\end{equation}
$h$ and $g$ are arbitrary functions of one variable, and 
\begin{equation}
f^{2}=\frac{\Phi\left(r\right)}{4h'\left(r^{*}+t
\right)g'\left(r^{*}-t\right)}\,,\label{Appendix_eq:KS_General_f2}
\end{equation}
where prime denotes differentiation with respect to the functions
variable, and $r^{*}=r^{*}(r)$
as given in Eq.~(\ref{Appendix_eq:KS_r_asterisk_def}).
In order to $f^{2}$ given in Eq.~(\ref{Appendix_eq:KS_General_f2})
be non singular, any singularity in the numerator $\Phi\left(r\right)$
must be canceled by the denominator, for all $t$. Assuming $\Phi$
to have only poles of order 1, and setting 
\begin{equation}
\begin{gathered}h\left(r^{*}+t\right)=A\,e^{\gamma\left(r^{*}+t\right)}\,,\\
g\left(r^{*}-t\right)=B\,e^{\gamma\left(r^{*}-t\right)}\,,
\end{gathered}
\label{Appendix_eq:KS_h_and_g_explicit}
\end{equation}
where the scale factors $A$ and $B$
are complex numbers, it is possible to choose
a value for the constant $\gamma$ such that $f^{2}$ is regular and
positive throughout the region covered by the coordinate patch. Substituting
Eq.~(\ref{Appendix_eq:KS_h_and_g_explicit}) into
Eqs.~(\ref{Appendix_eq:KS_General_u_v})
and (\ref{Appendix_eq:KS_General_f2}) we find 
\begin{equation}
f^{2}=\frac{\Phi\left(r\right)\,e^{-2\gamma r^{*}}}{4AB\gamma^{2}}\,,
\label{Appendix_eq:KS_f2_explicit}
\end{equation}
and 
\begin{equation}
\begin{gathered}X\left(r,t\right)=A\,
e^{\gamma\left(r^{*}+t\right)}+B\,e^{\gamma\left(r^{*}-t\right)}\,,\\
T\left(r,t\right)=A\,e^{\gamma\left(r^{*}+t\right)}-B\,
e^{\gamma\left(r^{*}-t\right)}\,,
\end{gathered}
\label{Appendix_eq:KS_u_v_explicit}
\end{equation}
in terms of the coordinates $r$ and $t$. From
Eq.~(\ref{Appendix_eq:KS_u_v_explicit})
we can find the inverse transformation and define the coordinate $r$,
implicitly, in terms of the coordinates $X$ and $T$, such that 
\begin{equation}
X^{2}-T^{2}=4AB\,e^{2\gamma r^{*}}\,.\label{Appendix_eq:KS_u2_v2}
\end{equation}
Lastly, since $f^{2}$ in Eq.~(\ref{Appendix_eq:KS_f2_explicit})
depends on the values of $A$ and $B$, $A$ and $B$ themselves
have to
be chosen in such a way that $f^{2}$ is positive. Moreover, given
that the transformation between the coordinates $\left\{ r,t\right\} $
and $\left\{ X,T\right\} $ depends on $A$ and $B$, these must be
chosen such that the coordinates $X$ and $T$ take only real values.

\subsection{The general formalism applied
specifically to the nonextremal Reissner-Nordstr\"om spacetime}
\label{Appendix_subsec:Kruskal-Szekeres_coordinates0}

\subsubsection{The Reissner-Nordstr\"om spacetime}
\label{Appendix_subsec:Kruskal-Szekeres_coordinates00}

Having introduced the general formalism, we can apply it specifically to
the the nonextremal
Reissner-Nordstr\"om spacetime. For this spacetime,
the line element in terms of the coordinates $\left\{ r,t\right\}$ is
given by Eq.~(\ref{Appendix_eq:KS_Non_regular_metric})
with 
\begin{equation}
\Phi\left(r\right)=
\frac{\left(r-r_+\right)\left(r-r_-\right)}{r^{2}}\,,
\label{Appendix_eq:KS_RN_phi}
\end{equation}
where $r_+$ is the gravitational or event
horizon  radius, and $r_-$ is the Cauchy horizon radius.
In terms of the mass $M$ and the electric charge
$Q$, $r_+$ and $r_-$ are given by
\begin{equation}
r_{\pm}=M\pm\sqrt{M^{2}-Q^{2}}\,,
\label{Appendix_eq:KS_horizons_radius}
\end{equation}
such that 
$\Phi\left(r\right)$ in Eq.~(\ref{Appendix_eq:KS_RN_phi})
can also be written as
$\Phi\left(r\right)=1-
\frac{2M}{r}+\frac{Q^2}{r^2}$.
Inverting Eq.~(\ref{Appendix_eq:KS_horizons_radius})
one has
$2M=r_++r_-$ and $Q=\sqrt{r_+r_-}$.
From Eq.~(\ref{Appendix_eq:KS_RN_phi}), in these coordinates we
see that the line element for the nonextremal
Reissner-Nordstr\"om
spacetime, i.e., $\left(M^{2}>Q^{2}\right)$ contains two coordinate
singularities
at $r=r_+$ and at $r=r_-$. Then, using the formalism of
subsection~\ref{Appendix_subsec:Kruskal-Szekeres_coordinates},
two coordinate patches need to be found, each well defined in the
neighborhood of each of the coordinate
singularities. Notice, however,
that there is a common region where both coordinate
patches overlap.

\subsubsection{Removal of the coordinate singularity at the event horizon
$r_+$}
\label{Appendix_subsec:Kruskal-Szekeres_coordinates_event_horizon}

Let us first find a coordinate patch that covers a neighborhood of
the coordinate singularity at $r=r_+$. Using
Eq.~(\ref{Appendix_eq:KS_r_asterisk_def}),
for the nonextremal Reissner-Nordstr\"om spacetime, $r^{*}$ is given
by
\begin{equation}
r^{*}=r+\frac{r_+^{2}}{r_+-r_-}
\log\left(\frac{r-r_+}{{r_++r_-}}\right)-\frac{r_-^{2}}{r_+-r_-}
\log\left(\frac{r-r_-}{{r_++r_-}}\right)\,,
\label{Appendix_eq:KS_r_asterisk_rplus}
\end{equation}
where we have set the value of the integration constant to ${r_++r_-}$.
To remove the coordinate singularity at $r=r_+$ we will impose
the constant $\gamma$ that appears in
Eqs.~(\ref{Appendix_eq:KS_h_and_g_explicit})-(\ref{Appendix_eq:KS_u2_v2})
to take the following value 
\begin{equation}
\gamma=\frac{r_+-r_-}{2r_+^{2}}\,.\label{Appendix_eq:KS_gamma_rplus}
\end{equation}
Substituting Eqs. (\ref{Appendix_eq:KS_r_asterisk_rplus}) and
(\ref{Appendix_eq:KS_gamma_rplus})
into Eq. (\ref{Appendix_eq:KS_f2_explicit}) we find 
\begin{equation}
f^{2}=\frac{\left( {r_++r_-}\right)^2}{AB\,r^{2}}
\left( \frac{r_+^{2}}{r_+-r_-}\right)^2
e^
{
-\frac{r(r_+-r_-)}{r_+^{2}}
}
\left(\frac{r-r_-}{{r_++r_-}}\right)^{1+
\left(\frac{r_-}{r_+}\right)^{2}}\,,
\label{Appendix_eq:KS_unscaled_metric_rplus}
\end{equation}
which is well behaved near the event horizon at $r=r_+$.  As was
stated in the previous section, the choice of the scale-factors $A$
and $B$ is quite arbitrary and we impose their values to be such that
in the limit when the electric charge $Q$ goes to zero we recover the
Kruskal-Szekeres coordinates defined in~\citep{MTW_Book} for the
Schwarzschild spacetime. So, the values for the scale-factors for the
various regions are
\begin{equation}
\begin{aligned}\mathrm{I\,} & \begin{cases}
A= & \;\;\,\frac{1}{2}\\
B= & \;\;\,\frac{1}{2}
\end{cases}\,, & \qquad & \mathrm{I}' & \begin{cases}
A= & -\frac{1}{2}\\
B= & -\frac{1}{2}
\end{cases}\,,\\
\mathrm{II\,} & \begin{cases}
A= & -\frac{i}{2}\\
B= & \;\;\,\frac{i}{2}
\end{cases}\,, & \qquad & \mathrm{II'} & \begin{cases}
A= &  \;\;\,\frac{i}{2}\\
B= & -\frac{i}{2}
\end{cases}\,.
\end{aligned}
\label{Appendix_eq:KS_scale_factors_rplus}
\end{equation}
We see that our choice for the scale-factors differs for each region.
This is a consequence of the behavior of the coordinates $\left\{
r,t\right\} $. Nonetheless, obviously, the geometry of the spacetime
is unaltered since different choices for the scale factor that obey
the restrictions imposed in the previous section give the same
expression for the metric, aside a conformal constant factor. Our
choice, though, leaves the metric completely unaltered, hence,
substituting the various values for the scale-factors listed in
Eq.~(\ref{Appendix_eq:KS_scale_factors_rplus}) into
Eq.~(\ref{Appendix_eq:KS_unscaled_metric_rplus}) we get, for every
region covered by the coordinate patch,
\begin{equation}
f^{2}=\frac{4\left( {r_++r_-}\right)^2}{r^{2}}
\left(\frac{r_+^{2}}{r_+-r_-}\right)^2
e^{-\frac{r(r_+-r_-)}{r_+^2}}
\left(\frac{r-r_-}{{r_++r_-}}\right)^{1+\left(
\frac{r_-}{r_+}\right)^{2}}\,.
\label{Appendix_eq:KS_g_uu_rplus}
\end{equation}
Substituting Eq.~(\ref{Appendix_eq:KS_scale_factors_rplus}) into
Eq.~(\ref{Appendix_eq:KS_u2_v2})
allow us to write the inverse transformation for the coordinate $r$
in terms of the coordinates $X$ and $T$ as 
\begin{equation}
X^{2}-T^{2}=e^{\frac{r(r_+-r_-)}{r_+^2}}
\left(\frac{r-r_+}{{r_++r_-}}\right)
\left(\frac{r-r_-}{{r_++r_-}}\right)^{-
\left(\frac{r_-}{r_+}
\right)^{2}}\,.\label{Appendix_eq:KS_u2_v2_rplus}
\end{equation}

For completeness, we also define the transformations for the
coordinates $\left\{ T,X\right\} $ in terms of
the coordinates $\left\{ r,t\right\} $
for the various regions. Substituting
Eqs.~(\ref{Appendix_eq:KS_gamma_rplus})
and (\ref{Appendix_eq:KS_scale_factors_rplus}) into
Eq.~(\ref{Appendix_eq:KS_u_v_explicit})
we find
\begin{equation}
\begin{array}{cc}
\mathrm{I\,} & \begin{cases}
X= & e^{\frac{r(r_+-r_-)}{2r_+^2}}\left(\frac{r-r_+}{{r_++r_-}}
\right)^{\frac{1}{2}}\left(\frac{r-r_-}{{r_++r_-}}
\right)^{-\frac{r_-^2}{2r_+^2}}
\cosh\left(\frac{t(r_+-r_-)}{2r_+^2}\right)\\
T= & e^{\frac{r(r_+-r_-)}{2r_+^2}}
\left(\frac{r-r_+}{{r_++r_-}}\right)^{
\frac{1}{2}}\left(\frac{r-r_-}{{r_++r_-}}\right)^{-
\frac{r_-^2}{2r_+^2}}\sinh\left(\frac{t(r_+-r_-)}{2r_+^2}\right)
\end{cases}\,,\\
\mathrm{I}' & \begin{cases}
X= & -e^{\frac{r(r_+-r_-)}{2r_+^2}}\left(\frac{r-r_+}{{r_++r_-}}
\right)^{\frac{1}{2}}\left(\frac{r-r_-}{{r_++r_-}}
\right)^{-\frac{r_-^2}{2r_+^2}}\cosh\left(
\frac{t(r_+-r_-)}{2r_+^2}\right)\\
T= & -e^{\frac{r(r_+-r_-)}{2r_+^2}}\left(\frac{r-r_+}{{r_++r_-}}
\right)^{\frac{1}{2}}\left(\frac{r-r_-}{{r_++r_-}}\right)^{-
\frac{r_-^2}{2r_+^2}}\sinh\left(\frac{t(r_+-r_-)}{2r_+^2}\right)
\end{cases}\,,\\
\mathrm{II\,} & \begin{cases}
X= & e^{\frac{r(r_+-r_-)}{2r_+^2}}
\left(\frac{r_+-r}{{r_++r_-}}\right)^{
\frac{1}{2}}\left(\frac{r-r_-}{{r_++r_-}}\right)^{-
\frac{r_-^2}{2r_+^2}}\sinh\left(\frac{t(r_+-r_-)}{2r_+^2}\right)\\
T= & e^{\frac{r(r_+-r_-)}{2r_+^2}}\left(\frac{r_+-r}{{r_++r_-}}
\right)^{\frac{1}{2}}\left(\frac{r-r_-}{{r_++r_-}}
\right)^{-\frac{r_-^2}{2r_+^2}}\cosh\left(
\frac{t(r_+-r_-)}{2r_+^2}\right)
\end{cases}\,,\\
\mathrm{II'} & \begin{cases}
X= & -e^{\frac{r(r_+-r_-)}{2r_+^2}}\left(\frac{r_+-r}{{r_++r_-}}
\right)^{\frac{1}{2}}\left(\frac{r-r_-}{{r_++r_-}}\right)^{-
\frac{r_-^2}{2r_+^2}}\sinh\left(\frac{t(r_+-r_-)}{2r_+^2}
\right)\\
T= & -e^{\frac{r(r_+-r_-)}{2r_+^2}}\left(\frac{r_+-r}{{r_++r_-}}
\right)^{\frac{1}{2}}\left(\frac{r-r_-}{{r_++r_-}}\right)^{-
\frac{r_-^2}{2r_+^2}}\cosh\left(\frac{t(r_+-r_-)}{2r_+^2}\right)
\end{cases}\,.
\end{array}\label{Appendix_eq:KS_u_v_regions_rplus}
\end{equation}
This relations for $X$ and $T$ can be used to find the coordinate
$t$ as a function of these coordinates, such that
\begin{equation}
\begin{aligned}t=
\frac{2r_+^{2}}{r_+-r_-}
\text{ arctanh}\left(\frac{T}{X}\right) &
\qquad\text{, in regions I and I\ensuremath{\mathrm{'}},}\\
t=\frac{2r_+^{2}}{r_+-r_-}\text{ arctanh}\left(\frac{X}{T}\right) &
\text{\qquad, in regions II and II\ensuremath{\mathrm{'}}.}
\end{aligned}
\label{Appendix_eq:KS_Schwarzschild_time_KS_relation}
\end{equation}

\begin{figure}
\centering\includegraphics[scale=0.45]
{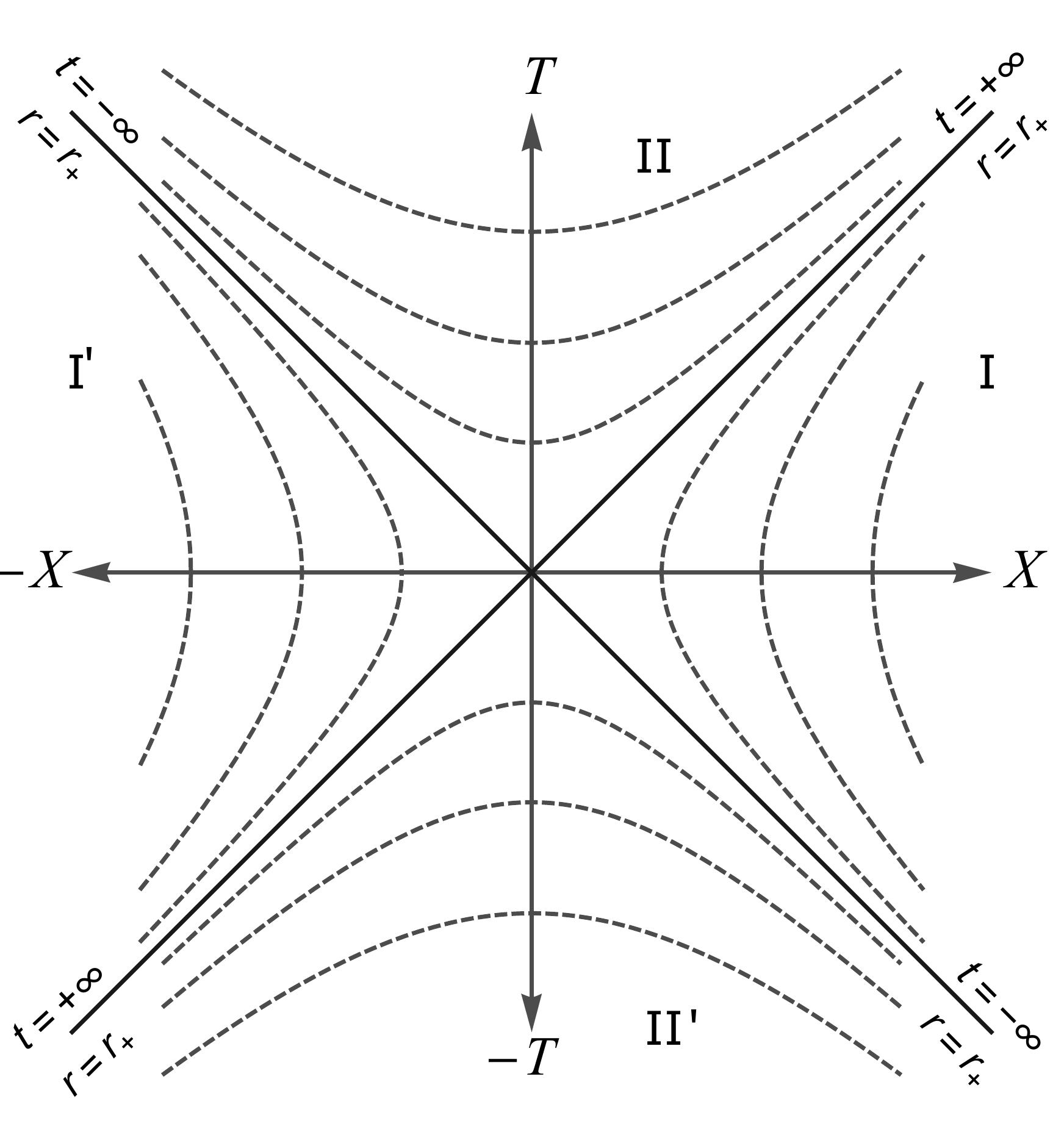}
\caption{\label{Appendix_fig:Coordinate_patch_1}
Relation between the Kruskal-Szekeres coordinates $\left\{ X,T\right\}$
that cover a neighborhood of the event horizon and the Schwarzschild
coordinates $\left\{ r,t\right\} $.  The hyperbolas represent curves
of constant $r$ coordinate while curves of constant $t$ are straight
lines through the origin.}
\end{figure}

The spacetime diagram for this coordinate patch 
together with the relevant
coordinate transformations are exhibited
graphically in Figure~\ref{Appendix_fig:Coordinate_patch_1}.
We further note that in-between 
region I and region  I' there is an Einstein-Rosen bridge,
i.e., a nontraversable dynamical wormhole.

\newpage

\subsubsection{Removal of the coordinate singularity at the Cauchy horizon
$r_-$}
\label{Appendix_subsec:Kruskal-Szekeres_coordinates_Cauchy_horizon}

Let us now define a second coordinate patch where
the coordinate singularity
at the Cauchy horizon, $r=r_-$, is removed. Here the function
$r^{*}$ is now given
by 
\begin{equation}
r^{*}=r+\frac{r_+^{2}}{r_+-r_-}
\log\left(\frac{r_+-r}{{r_++r_-}}\right)-
\frac{r_-^{2}}{r_+-r_-}\log
\left(\frac{r_--r}{{r_++r_-}}\right)\,.
\label{Appendix_eq:KS_r_asterisk_rminus}
\end{equation}
This change in the expression for the function $r^{*}$, compared
with Eq.~(\ref{Appendix_eq:KS_r_asterisk_rplus}), should not
come with surprise since, as was stated in the previous subsection,
the function $r^{*}$ is defined up to an integration constant which
will be chosen such that the metric is real for each coordinate patch.

Now, since we want to remove the coordinate singularity at $r=r_-$,
we impose 
\begin{equation}
\gamma=-\frac{r_+-r_-}{2r_-^2}\,.
\label{Appendix_eq:KS_gamma_rminus}
\end{equation}
Substituting Eqs.~(\ref{Appendix_eq:KS_r_asterisk_rminus})
and (\ref{Appendix_eq:KS_gamma_rminus})
into Eq.~(\ref{Appendix_eq:KS_f2_explicit}) we find 
\begin{equation}
f^{2}=\frac{\left( {r_++r_-}\right)^2}{AB\,r^{2}}
\left(\frac{r_-^{2}}{r_+-r_-}
\right)^2e^{\frac{r(r_+-r_-)}{r_-^2}}
\left(\frac{r_+-r}{{r_++r_-}}\right)^{1+
\left(\frac{r_+}{r_-}
\right)^{2}}\,.
\end{equation}
As in the previous subsection, we have now to choose the scale-factors
$A$ and $B$ for the various sub-regions covered by the second coordinate
patch hence, 
\begin{equation}
\begin{aligned}\mathrm{III} & \begin{cases}
A= & \;\;\,\frac{1}{2}\\
B= & \;\;\,\frac{1}{2}
\end{cases}\,, & \qquad & \mathrm{III'} & \begin{cases}
A= & -\frac{1}{2}\\
B= & -\frac{1}{2}
\end{cases}\,,\\
\mathrm{II} & \begin{cases}
A= & \;\;\,\frac{i}{2}\\
B= & -\frac{i}{2}
\end{cases}\,, & \qquad & \mathrm{II}' & \begin{cases}
A= & -\frac{i}{2}\\
B= & \;\;\,\frac{i}{2}
\end{cases}\,.
\end{aligned}
\label{Appendix_eq:KS_scale_factors_rminus}
\end{equation}
This choice for the scale factors leaves the expression for the metric
unaltered for the various sub-regions covered by the coordinate patch,
such that,
\begin{equation}
f^{2}=\frac{4\left( {r_++r_-}\right)^2}{r^{2}}
\left(\frac{r_-^{2}}{r_+-r_-}
\right)^2e^{\frac{r(r_+-r_-)}{r_-^2}}
\left(\frac{r_+-r}{{r_++r_-}}\right)^{1+\left(
\frac{r_+}{r_-}\right)^{2}}\,.
\label{Appendix_eq:KS_g_uu_rminus}
\end{equation}
Substituting Eq.~(\ref{Appendix_eq:KS_g_uu_rminus}) in
Eq.~(\ref{Appendix_eq:KS_u2_v2}) we find 
\begin{equation}
X^{2}-T^{2}=e^{-\frac{r(r_+-r_-)}{r_-^2}}
\left(\frac{r_--r}{{r_++r_-}}
\right)\left(\frac{r_+-r}{{r_++r_-}}\right)^{-\left(
\frac{r_+}{r_-}\right)^{2}}\,,
\label{Appendix_eq:KS_u2_v2_rminus}
\end{equation}
which defines, implicitly, the coordinate $r$ in terms of the
coordinates $X$ and $T$. For
completeness we define the transformations for the various sub-regions
covered by the coordinate patch that relate the coordinates $\left\{
X,T\right\} $ with the coordinates $\left\{ r,t\right\}
$. Substituting Eqs.~(\ref{Appendix_eq:KS_gamma_rminus}) and
(\ref{Appendix_eq:KS_scale_factors_rminus}) in
Eq.~(\ref{Appendix_eq:KS_u_v_explicit})
we find 
\begin{equation}
\begin{array}{cl}
\mathrm{III} & \begin{cases}
X= & e^{-\frac{r(r_+-r_-)}{2r_-^2}}
\left(\frac{r_--r}{{r_++r_-}}
\right)^{\frac{1}{2}}\left(\frac{r_+-r}{{r_++r_-}}
\right)^{-
\frac{ r_+^{2} } { 2r_-^{2} }
}
\cosh\left(-
\frac{t(r_+-r_-)}{2r_-^2}\right)\\
T= & e^{-\frac{r(r_+-r_-)}{2r_-^2}}\left(
\frac{r_--r}{{r_++r_-}}
\right)^{\frac{1}{2}}\left(\frac{r_+-r}{{r_++r_-}}
\right)^{-\frac{ r_+^{2} }{ 2r_-^{2} } }\sinh\left(-
\frac{t(r_+-r_-)}{2r_-^2}\right)
\end{cases}\,,\\
\mathrm{III'} & \begin{cases}
X= & -e^{-\frac{r(r_+-r_-)}{2r_-^2}}\left(
\frac{r_--r}{{r_++r_-}}\right)^{\frac{1}{2}}
\left(\frac{r_+-r}{{r_++r_-}}\right)^{-
\frac{r_+^{2}}{2r_-^{2}}}
\cosh\left(-\frac{t(r_+-r_-)}{2r_-^2}\right)\\
T= & -e^{-\frac{r(r_+-r_-)}{2r_-^2}}\left(
\frac{r_--r}{{r_++r_-}}
\right)^{\frac{1}{2}}\left(\frac{r_+-r}{{r_++r_-}}
\right)^{-\frac{r_+^{2}}{2r_-^{2}}}\sinh\left(-
\frac{t(r_+-r_-)}{2r_-^2}\right)
\end{cases}\,,\\
\mathrm{II'} & \begin{cases}
X= & e^{-\frac{r(r_+-r_-)}{2r_-^2}}\left(
\frac{r-r_-}{{r_++r_-}}
\right)^{\frac{1}{2}}\left(\frac{r_+-r}{{r_++r_-}}
\right)^{-\frac{r_+^{2}}{2r_-^{2}}}\sinh\left(-
\frac{t(r_+-r_-)}{2r_-^2}\right)\\
T= & e^{-\frac{r(r_+-r_-)}{2r_-^2}}\left(
\frac{r-r_-}{{r_++r_-}}
\right)^{\frac{1}{2}}\left(\frac{r_+-r}{{r_++r_-}}
\right)^{-\frac{r_+^{2}}{2r_-^{2}}}\cosh\left(-
\frac{t(r_+-r_-)}{2r_-^2}\right)
\end{cases}\,,\\
\mathrm{II}\, & \begin{cases}
X= & -e^{-\frac{r(r_+-r_-)}{2r_-^2}}\left(
\frac{r-r_-}{{r_++r_-}}\right)^{\frac{1}{2}}
\left(\frac{r_+-r}{{r_++r_-}}\right)^{-
\frac{r_+^{2}}{2r_-^{2}}}
\sinh\left(-
\frac{t(r_+-r_-)}{2r_-^2}\right)\\
T= & -e^{-\frac{r(r_+-r_-)}{2r_-^2}}\left(
\frac{r-r_-}{{r_++r_-}}\right)^{\frac{1}{2}}
\left(\frac{r_+-r}{{r_++r_-}}\right)^{-
\frac{r_+^{2}}{2r_-^{2}}}
\cosh\left(-\frac{t(r_+-r_-)}{2r_-^2}\right)
\end{cases}\,.
\end{array}\label{Appendix_eq:KS_u_v_regions_rminus}
\end{equation}
These relations can then be used to find the transformation that gives
the coordinate $t$ in terms of the coordinates $X$ and $T$, such
that
\begin{equation}
\begin{aligned}
t=-\frac{2r_-^{2}}{r_+-r_-}\text{ arctanh}
\left(\frac{T}{X}\right) &
\qquad\text{, in regions III and III\ensuremath{\mathrm{'}},}\\
t=-\frac{2r_-^{2}}{r_+-r_-}\text{ arctanh}
\left(\frac{X}{T}\right)
& \qquad\text{, in regions II and II\ensuremath{\mathrm{'}}.}
\end{aligned}
\end{equation}

\begin{figure}[h]
\centering
\includegraphics[scale=0.45]{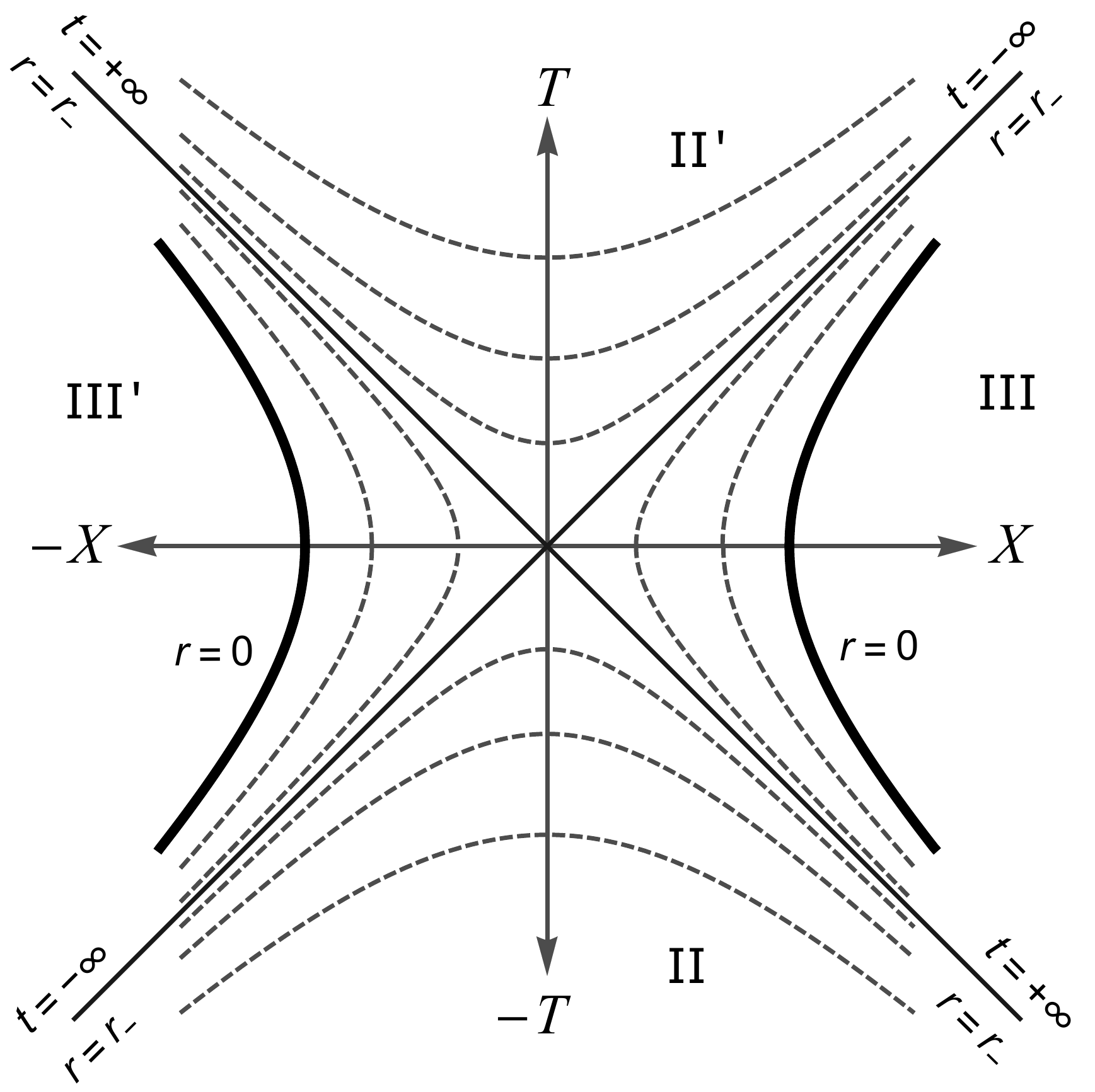}
\caption{\label{Appendix_fig:Coordinate_patch_2}Relation between the
Kruskal-Szekeres coordinates $\left\{ X,T\right\} $ that cover a
neighborhood of the Cauchy horizon and the Schwarzschild coordinates
$\left\{ r,t\right\} $.  The hyperbolas represent curves of constant
$r$ coordinate while curves of constant $t$ are straight lines through
the origin. The thick black line represent the singularity at $r=0$.}
\end{figure}

The spacetime diagram for this coordinate patch 
together with the relevant
coordinate transformations are exhibited
graphically in Figure~\ref{Appendix_fig:Coordinate_patch_2}.

\newpage

\section{Extrinsic curvature as seen
from $\mathcal{M}_{\rm e}$ in a nonextremal
Reissner-Nordstr\"om spacetime}
\label{Appendix_sec:Extrinsic_curvature}

\subsection{Boundary surface outside the event horizon}
\label{Appendix_subsec:Extrinsic_curvature_outside_event_horizon}

We want to calculate the
extrinsic curvature
of a shell $\mathcal{S}$
in a nonextremal Reissner-Nordstr\"om spacetime,
see also the nomenclature and some details in
Section~\ref{Sec:Junction_formalismproper}.  Assuming the matching
surface to be timelike, static and spherically symmetric, one finds
that the nontrivial components of the extrinsic curvature of the
matching hypersurface $\mathcal{S}$ are given by
\begin{equation}
K_{\tau\tau}=-a^{\alpha}n_{\alpha}\,,\quad
K_{\theta\theta}=\nabla_{\theta}n_{\theta}\,,\quad
K_{\varphi\varphi}=\nabla_{\varphi}n_{\varphi}\,,
\label{ktautau}
\end{equation}
where
$a^{\alpha}\equiv u^{\beta}\nabla_{\beta}u^{\alpha}$
is the acceleration of an observer
with four-velocity $u^{\alpha}$ comoving with $\mathcal{S}$, 
$n^{\alpha}$ is the normal to $\mathcal{S}$, and 
$\nabla_\alpha$ is the covariant derivative
using the Levi-Civita connection.
In our study we will allow the shell to be located  at any region of
the nonextremal Reissner-Nordstr\"om spacetime.
It is possible to
find a coordinate system that covers the entire Reissner-Nordstr\"om
spacetime without coordinate singularities.
For our analysis it is simpler to define, instead and as was done
in Appendix~\ref{Appendix_sec:Kruskal-Szekeres_coordinates_RN},
two coordinate
patches to describe the various regions of the
Reissner-Nordstr\"om spacetime exterior to the shell. 
Hence,
we will separate the study of a shell located in a region described by
one coordinate patch and the other.

Here we will make the derivation of the expressions for the induced
metric on $\mathcal{S}$ and the extrinsic curvature components as seen
from the exterior spacetime, $\mathcal{M}_{\rm e}$, in the nonextremal
state, $r_+>r_-$, i.e., $M>Q$, and with $\mathcal{S}$
having radius $R$ obeying $R>r_+$, i.e., 
$\mathcal{S}$ is located 
outside the sphere defined by the
gravitational radius or outside
the event horizon, depending on the
orientation of $\mathcal{S}$ itself.

We start by studying the properties of a static shell located in a
region described by the coordinate patch defined in
Appendix~\ref{Appendix_sec:Kruskal-Szekeres_coordinates_RN}, the
coordinate patch without the coordinate singularity at the event
horizon $r=r_+$. In this region, the line element for the
Reissner-Nordstr\"om spacetime in Kruskal-Szekeres coordinates is
given by, see
Section~\ref{Appendix_subsec:Kruskal-Szekeres_coordinates0}
for details,
\begin{eqnarray}
ds_{\rm e}^{2}=4\left(\frac{r_++r_-}{r_+-r_-}\right)^{2}&&
\frac{r_+^{4}}{r^{2}}e^{-\frac{r\,\left(r_+-r_-\right)}{r_+^{2}}}
\left(\frac{r-r_-}{r_++r_-}
\right)^{1+\left(\frac{r_-}{r_+}\right)^{2}}\left(dX^{2}-dT^{2}
\right)+r^{2}\left(X,T\right)d\Omega^{2}\,,
\label{eq:metric_RN_rplusAppendixA}\\
&&X^{2}-T^{2}=e^{\frac{r\,\left(r_+-r_-\right)}{r_+^{2}}}
\left(\frac{r-r_+}{r_++r_-}\right)\left(
\frac{r-r_-}{r_++r_-}\right)^{-\left(
\frac{r_-}{r_+}\right)^{2}}\,.
\nonumber
\end{eqnarray}
The components of the 4-velocity of an
observer comoving with $\mathcal{S}$
as seen from $\mathcal{M}_{\rm e}$ are,
\begin{equation}
u_{\rm e}^{\alpha}=\sqrt{\frac{g^{^{XX}}}{X^{2}-T^{2}}}
\left(X,T,0,0\right)\,,\label{eq:4velocity_value_rplusAppendixA}
\end{equation}
where the sign was chosen so that $u$ points to the future and
$g^{^{XX}}$ is the $XX$ component of the inverse metric associated
with Eq.~(\ref{eq:metric_RN_rplusAppendixA}).
We  use $R$ to describe the radial
coordinate of the shell
and so, the intrinsic line element of $\mathcal{S}$ 
is
\begin{equation}
\left.ds^{2}\right|_{\mathcal{S}}=-d\tau^{2}+R^{2}d\Omega^{2}\,.
\label{eq:induced_metric_RNAppendixA}
\end{equation}
Imposing that 
the normal points in the direction of increasing $X$ coordinate
implies that the choice of the sign is different if we consider
the shell to be in the region $\mathrm{I}$ or $\mathrm{I}'$, see
Figure~\ref{Fig:Penrose_diagram_RN_non_extremal} and 
also Figure~\ref{Appendix_fig:Coordinate_patch_1}.
Then the normal $n_{{\rm e}\alpha}$ is
\begin{equation}
n_{{\rm e}\alpha}=\text{sign}\left(X\right)\sqrt{
\frac{g_{_{XX}}}{X^{2}-T^{2}}}\left(T,-X,0,0\right)\,.
\label{eq:normal_value_rplusAppendixA}
\end{equation}
where the quantities on the right-hand side are to be evaluated at
$r=R$ and $\text{sign}\left(X\right)$ is the signum function of the
coordinate $X$ of the shell.  Having found the normal to the
hypersurface $\mathcal{S}$ as seen from the exterior nonextremal
Reissner-Nordstr\"om spacetime, we can now compute the nonzero
components of the extrinsic curvature.

Anticipating some of the intermediate
results, we first find the derivative of the radial coordinate
$r$ in order to the coordinates $\left\{ X,T\right\} $. Taking the
derivative of the second of the equations
given in Eq.~(\ref{eq:metric_RN_rplusAppendixA}) in order to $X$ and $T$
independently, we find that
$\frac{\partial r}{\partial X}=\frac{g_{_{XX}}}{2}
\frac{r_+-r_-}{r_+^{2}}X$ and 
$\frac{\partial r}{\partial T}=-\frac{g_{_{XX}}}{2}
\frac{r_+-r_-}{r_+^{2}}T$,
as well as
$\frac{\partial r}{\partial T}=-\frac{T}{X}\frac{\partial r}{\partial X}$.
The Christoffel symbols
are given by
$\Gamma_{\alpha\beta}^{\gamma}=
\frac12 g^{\gamma\sigma}
\left( g_{\alpha\sigma,\beta}+
g_{\beta\sigma,\alpha}-
g_{\alpha\beta,\sigma}
\right)
$.
So, to compute the Christoffel symbols associated with the
metric~(\ref{eq:metric_RN_rplusAppendixA})
we need to find the derivatives of the metric components.
Noting that
$g_{_{TT}}=-g_{_{XX}}$,
and using the three previous equations we find
$\partial_{_{X}}g_{_{XX}}  =  \frac{\left(g_{_{XX}}\right)^{2}}{r-r_-}
\frac{r_+-r_-}{2r_+^{4}}\left[2\frac{r_+^{2}r_-}{r}-
\left(r_+-r_-\right)\left(r_++r\right)\right]X$,
$\partial_{_{X}}g_{_{TT}}  =  -\partial_{_{X}}g_{_{XX}}$,
$\partial_{_{T}}g_{_{XX}}  =  -\frac{T}{X}\partial_{_{X}}
g_{_{XX}}$,
$\partial_{_{T}}g_{_{TT}}  =  \frac{T}{X}\partial_{_{X}}
g_{_{XX}}$. One can 
find the Christoffel symbols needed
to compute
the component $K_{\tau\tau}$ of the extrinsic curvature. They are
$\Gamma_{_{XX}}^{^{X}}  =  \frac{g_{_{XX}}}{4r_+^{4}}
\frac{r_+-r_-}{r-r_-}\left[2\frac{r_+^{2}r_-}{r}-
\left(r_+-r_-\right)\left(r_++r\right)\right]X$,
$\Gamma_{_{TT}}^{^{X}}  =  \Gamma_{_{XX}}^{^{X}}$,
$\Gamma_{_{XT}}^{^{X}}  =  \Gamma_{_{TX}}^{^{X}}=
-\frac{T}{X}\Gamma_{_{XX}}^{^{X}}$,
$\Gamma_{_{TT}}^{^{T}}  =  -\frac{T}{X}\Gamma_{_{XX}}^{^{X}}$,
$\Gamma_{_{XX}}^{^{T}}  =  -\frac{T}{X}\Gamma_{_{XX}}^{^{X}}$,
$\Gamma_{_{TX}}^{^{T}}  =  \Gamma_{_{XT}}^{^{T}}=
\Gamma_{_{XX}}^{^{X}}$.
Substituting these into the acceleration vector $a^\alpha$, 
we find that components $a^{^{X}}$ and $a^{^{T}}$ of the acceleration
vector field are given by
$a^{^{X}}  =  \frac{dU^{^{X}}}{d\tau}+
\frac{X^{2}-T^{2}}{T^{2}}\Gamma_{_{XX}}^{^{X}}U^{^{X}}U^{^{X}}$, 
$a^{T}  =  \frac{dU^{^{T}}}{d\tau}+\frac{X^{2}-
T^{2}}{X\,T}\Gamma_{_{XX}}^{^{X}}U^{^{X}}U^{^{X}}$, 
where the repetition of the indices does not mean summation but actual
products of the components.
Substituting Eq.~(\ref{eq:4velocity_value_rplusAppendixA})
in these two equations we find,
$a^{X}  =  \frac{dU^{^{X}}}{d\tau}+g^{^{XX}}\Gamma_{_{XX}}^{^{X}}$
and
$a^{T}  =  \frac{dU^{^{T}}}{d\tau}+g^{^{XX}}
\Gamma_{_{XX}}^{^{X}}\frac{T}{X}$.
We are now in position to compute the $K_{{\rm e}\, {\tau\tau}}$
component of
the extrinsic curvature of $\mathcal{S}$
given in Eq.~(\ref{ktautau})
embedded in  the exterior nonextremal
Reissner-Nordstr\"om spacetime.
Indeed, with the components $a^{^{X}}$ and $a^{T}$,
Eq.~(\ref{eq:normal_value_rplusAppendixA}),
 and using
$\frac{\partial X}{\partial \tau}=\frac{T}{X}
\frac{\partial T}{\partial \tau}$ that we encountered before, in
Eq.~(\ref{ktautau})
yields,
$
K_{{\rm e}\, {\tau\tau}}
=-\text{sign}\left(X\right)
\sqrt{\frac{g_{_{XX}}}{X^{2}-T^{2}}}\left(g^{^{XX}}+g^{^{XX}}
\Gamma_{_{XX}}^{^{X}}\frac{X^{2}-T^{2}}{X}\right)$.
Then, using Eqs.~(\ref{eq:metric_RN_rplusAppendixA}), and 
$\Gamma_{_{XX}}^{^{X}}$ above, yields
${K_{\rm e}}^{\tau}{}_{\tau}
=\frac{\text{sign}\left(X
\right)}{2R^{2}k}\left(r_++r_--2\frac{r_-r_+}{R}\right)$,
where the induced metric in Eq.~(\ref{eq:induced_metric_RNAppendixA}) was
used to raise the indices and
$k=\sqrt{\left(1-\frac{r_+}{R}\right)\left(1-\frac{r_-}{R}\right)}$.
To find the other nonzero components of the extrinsic curvature of
$\mathcal{S}$
embedded in  the exterior nonextremal
Reissner-Nordstr\"om spacetime,
$K_{{\rm e}\, {\theta\theta}}$ and
$K_{{\rm e}\, {\varphi\varphi}}$, we
have to compute the remaining entries of the Christoffel symbols.
Equation~(\ref{eq:metric_RN_rplusAppendixA}) and
$\frac{\partial r}{\partial T}=-\frac{T}{X}\frac{\partial r}
{\partial X}$,
yield 
$\Gamma_{\theta\theta}^{X}  =  -rg^{^{XX}}\frac{\partial r}{\partial X}$,
$\Gamma_{\theta\theta}^{T}  =  -rg^{^{XX}}\frac{T}{X}
\frac{\partial r}{\partial X}$,
$\Gamma_{\varphi\varphi}^{X}  =  -r\sin^{2}\theta\,g^{^{XX}}$,
$\Gamma_{\varphi\varphi}^{T}  =  -r\sin^{2}\theta\,g^{^{XX}}
\frac{T}{X}\frac{\partial r}{\partial X}$.
Substituting these
Christoffel symbols and previous found equations
into Eq.~(\ref{ktautau})
we find
${K_{\rm e}}^{\theta}{}_{\theta}=
{K_{\rm e}}^{\varphi}{}_{\varphi}
=  \frac{\text{sign}\left(X\right)}{2}
\frac{r_+-r_-}{r_+^{2}}
\frac{\sqrt{g_{_{XX}}\left(X^{2}-T^{2}\right)}}{R}$.
Thus, in brief
\begin{equation}
{K_{\rm e}}^{\tau}{}_{\tau}=
\frac{\text{sign}
\left(X\right)}{2R^{2}k}\left(r_++r_--2
\frac{r_-r_+}{R}\right)\,,\quad
{K_{\rm e}}^{\theta}{}_{\theta}=
{K_{\rm e}}^{\varphi}{}_{\varphi}=  \frac{\text{sign}
\left(X\right)(r_+-r_-)}{2r_+^{2}R}
\sqrt{g_{_{XX}}\left(X^{2}-T^{2}\right)}\,,
\label{Appendix_eq:ExtCurv_ktautau_value_rplus}
\end{equation}
where $k=\sqrt{\left(1-\frac{r_+}{R}\right)\left(1-\frac{r_-}{R}\right)}$.
With these
geometrical
quantities one can compute the physical quantities
of a thin shell, such as its energy density and tangential pressure,
assuming it is made of a perfect fluid, at the boundary surface
outside the event horizon,
as we did in the text.

\subsection{Boundary surface  inside the Cauchy horizon
\label{Appendix_subsec:Extrinsic_curvature_inside_Cauchy_horizon}}

Here we will make the derivation of the expressions for the induced
metric on $\mathcal{S}$ and the extrinsic curvature components as seen
from the exterior spacetime, $\mathcal{M}_{\rm e}$, in the nonextremal
state, $r_+>r_-$, i.e., $M>Q$, and with $\mathcal{S}$
having radius $R$ obeying $R<r_-$, i.e. 
$\mathcal{S}$ is located inside
 the sphere defined by the Cauchy radius or inside
the Cauchy horizon depending on
the orientation of $\mathcal{S}$ itself.

We start by studying the properties of a static shell located in
a region described by the coordinate patch defined
in 
Appendix~\ref{Appendix_sec:Kruskal-Szekeres_coordinates_RN},
the coordinate patch without the coordinate singularity at
the Cauchy horizon $r=r_-$.
In this region,
the line element for the Reissner-Nordstr\"om
spacetime in Kruskal-Szekeres
coordinates is given by, see
Section~\ref{Appendix_subsec:Kruskal-Szekeres_coordinates0}
for details,
\begin{eqnarray}
ds_{\rm e}^{2}=4\left(\frac{r_++r_-}{r_+-r_-}\right)^{2}&&
\frac{r_-^{4}}{r^{2}}{\rm e}^{\frac{r\left(r_+-r_-
\right)}{r_-^{2}}}\left(\frac{r_+-r}{r_++r_-}
\right)^{1+\left(\frac{r_+}{r_-}
\right)^{2}}\left(dX^{2}-dT^{2}\right)+r^{2}
\left(X,T\right)d\Omega^{2}\,,
\label{eq:metric_RN_rminusAppendixB}\\
&&
X^{2}-T^{2}={\rm e}^{-\frac{r\left(r_+-r_-\right)}{r_-^{2}}}\left(
\frac{r_--r}{r_++r_-}\right)\left(\frac{r_+-r}{r_++r_-}
\right)^{-\left(\frac{r_+}{r_-}\right)^{2}}\,.
\nonumber
\end{eqnarray}
The components of the 4-velocity of an observer comoving with the as
seen from $\mathcal{M}_{\rm e}$ are,
\begin{equation}
u_{\rm e}^{\alpha}=-\sqrt{\frac{g^{^{XX}}}{X^{2}-T^{2}}}
\left(X,T,0,0\right)\,,
\label{eq:4velocity_value_rminusAppendixB}
\end{equation}
where the sign was chosen so that $u$ points to the future and
$g^{^{XX}}$ is the $XX$ component of the inverse metric associated
with Eq.~(\ref{eq:metric_RN_rminusAppendixB}).
We  use $R$ to describe the radial
coordinate of the shell
and so, the intrinsic line element of $\mathcal{S}$,
is
\begin{equation}
\left.ds^{2}\right|_{\mathcal{S}}=-d\tau^{2}+R^{2}d\Omega^{2}\,.
\label{eq:induced_metric_RNAppendixB}
\end{equation}
Imposing that 
the normal points in the direction of increasing $X$ coordinate
implies that the choice of the sign is different if we consider
the shell to be in the region $\mathrm{III}$ or $\mathrm{III}'$, see
Figure~\ref{Fig:Penrose_diagram_RN_non_extremal} and 
also Figure~\ref{Appendix_fig:Coordinate_patch_2}.
Then the normal $n_{{\rm e}\alpha}$ is
\begin{equation}
n_{{\rm e}\alpha}=\text{sign}\left(X\right)\sqrt{
\frac{g_{_{XX}}}{X^{2}-T^{2}}}\left(-T,X,0,0\right)\,,
\label{eq:normal_value_rplusAppendixB}
\end{equation}
where the quantities on the right-hand side are to be evaluated at
$r=R$ and $\text{sign}\left(X\right)$ is the signum function of the
coordinate $X$ of the shell.  Having found the normal to the
hypersurface $\mathcal{S}$ as seen from the exterior nonextremal
Reissner-Nordstr\"om spacetime, we can now compute the nonzero
components of the extrinsic curvature.

Let us now compute the nonzero components of the extrinsic curvature
of $\mathcal{S}$ for a thin shell inside the Cauchy horizon. Similarly
to the previous subsection, we have first to compute some intermediate
quantities. Taking the derivative to $X$ and $T$, independently,
of the second of the equations
given in Eq.~(\ref{eq:metric_RN_rminusAppendixB}) we find
$\frac{\partial r}{\partial
X}=-\frac{g_{_{XX}}}{2}\frac{r_+-r_-}{r_-^{2}}X$ and
$\frac{\partial r}{\partial
T}=\frac{g_{_{XX}}}{2}\frac{r_+-r_-}{r_-^{2}}T$.
These latter expressions are then related by
$\frac{\partial r}{\partial T}=-\frac{T}{X}
\frac{\partial r}{\partial X}$.
Equation~(\ref{eq:metric_RN_rminusAppendixB}) and
$\frac{\partial r}{\partial X}=-
\frac{g_{_{XX}}}{2}\frac{r_+-r_-}{r_-^{2}}X$,
yield 
$\partial_{_{X}}g_{_{XX}}  = 
\frac{\left(g_{_{XX}}\right)^{2}}{2r_-^{4}}
\frac{r_+-r_-}{r_+-r}\left[2\frac{r_-^{2}r_+}{r}+
\left(r_+-r_-\right)\left(r+r_-\right)\right]X$.
Now, in this coordinate patch one still has 
$\frac{\partial r}{\partial T}=-\frac{T}{X}
\frac{\partial r}{\partial X}$ and 
$g_{_{TT}}=-g_{_{XX}}$, the other derivatives of the metric
are also 
$\partial_{_{X}}g_{_{TT}}  =  -\partial_{_{X}}g_{_{XX}}$,
$\partial_{_{T}}g_{_{XX}}  =  -\frac{T}{X}\partial_{_{X}}
g_{_{XX}}$,
$\partial_{_{T}}g_{_{TT}}  =  \frac{T}{X}\partial_{_{X}}
g_{_{XX}}$, where $g_{_{XX}}$ refers here to the $XX$ component of
the metric in Eq.~(\ref{eq:metric_RN_rminusAppendixB}).
From Eq.~(\ref{eq:metric_RN_rminusAppendixB})
and the one just found for $\partial_{_{X}}g_{_{XX}}$ 
we find 
$\Gamma_{_{XX}}^{^{X}}=\frac{g_{_{XX}}}{4r_-^{4}}
\frac{r_+-r_-}{r_+-r}\left[2\frac{r_-^{2}r_+}{r}+
\left(r_+-r_-\right)\left(r_+r_-\right)\right]X$.
Unsurprisingly, the other entries of the Christoffel symbols
are
given by
$\Gamma_{_{TT}}^{^{X}}  =  \Gamma_{_{XX}}^{^{X}}$,
$\Gamma_{_{XT}}^{^{X}}  =  \Gamma_{_{TX}}^{^{X}}=
-\frac{T}{X}\Gamma_{_{XX}}^{^{X}}$,
$\Gamma_{_{TT}}^{^{T}}  =  -\frac{T}{X}\Gamma_{_{XX}}^{^{X}}$,
$\Gamma_{_{XX}}^{^{T}}  =  -\frac{T}{X}\Gamma_{_{XX}}^{^{X}}$,
$\Gamma_{_{TX}}^{^{T}}  =  \Gamma_{_{XT}}^{^{T}}=
\Gamma_{_{XX}}^{^{X}}$,
which imply that the $a^{^{X}}$ and $a^{^{T}}$ components of the
acceleration are also given by
$a^{^{X}}  =  \frac{dU^{^{X}}}{d\tau}+
\frac{X^{2}-T^{2}}{T^{2}}\Gamma_{_{XX}}^{^{X}}U^{^{X}}U^{^{X}}$ and
$a^{T}  =  \frac{dU^{^{T}}}{d\tau}+\frac{X^{2}-
T^{2}}{X\,T}\Gamma_{_{XX}}^{^{X}}U^{^{X}}U^{^{X}}$, 
where the repetition of the indices does not mean summation but actual
products of the components.
Then, using
Eq.~(\ref{eq:4velocity_value_rminusAppendixB}) we find 
$a^{^{X}}  =  \frac{dU^{^{X}}}{d\tau}+g^{^{XX}}\Gamma_{_{XX}}^{^{X}}$
and
$a^{^{T}}  =  \frac{dU^{^{T}}}{d\tau}+g^{^{XX}}
\Gamma_{_{XX}}^{^{X}}\frac{T}{X}$.
Finally, substituting Eq.~(\ref{eq:normal_value_rplusAppendixB}),
and these
two latter equations
for $a^{^{X}}$ and $a^{^{T}}$,
into Eq.~(\ref{ktautau})
gives
$K_{{\rm e}\, \tau\tau}
=\text{sign}
\left(X\right)\sqrt{\frac{g_{_{XX}}}{X^{2}-T^{2}}}
\left[g^{^{XX}}+g^{^{XX}}\Gamma_{_{XX}}^{^{X}}\frac{X^{2}-
T^{2}}{X}\right]$.
Using then Eq.~(\ref{eq:metric_RN_rminusAppendixB})
and the equation for $\Gamma_{_{XX}}^{^{X}}$ found above,
leads to
${K_{\rm e}}^{\tau}{}_{\tau}
=
\frac{\text{sign}\left(X\right)}{2R^{2}k}
\left[r_++r_--2\frac{r_-r_+}{R}\right]$,
where the induced metric on $\mathcal{S}$,
Eq.~(\ref{eq:induced_metric_RNAppendixB}),
was used to raise the indices and $k$ is given by
$k=\sqrt{\left(1-\frac{r_+}{R}\right)
\left(1-\frac{r_-}{R}\right)}$.
All is left now is to find the Christoffel symbols necessary to compute
the components
${K_{\rm e}}^{\theta}{}_{\theta}$
and 
${K_{\rm e}}^{\phi}{}_{\phi}$
of the extrinsic curvature of $\mathcal{S}$.
However, since the angular
part of the metric is the same for both
coordinate patches, the Christoffel
symbols are also given by
$\Gamma_{\theta\theta}^{X}  =  -rg^{^{XX}}\frac{\partial r}{\partial X}$,
$\Gamma_{\theta\theta}^{T}  =  -rg^{^{XX}}\frac{T}{X}
\frac{\partial r}{\partial X}$,
$\Gamma_{\varphi\varphi}^{X}  =  -r\sin^{2}\theta\,g^{^{XX}}$,
$\Gamma_{\varphi\varphi}^{T}  =  -r\sin^{2}\theta\,g^{^{XX}}
\frac{T}{X}\frac{\partial r}{\partial X}$.
These, in conjunction with Eq.~(\ref{eq:normal_value_rplusAppendixB}),
and the equations $\frac{\partial r}{\partial
X}=-\frac{g_{_{XX}}}{2}\frac{r_+-r_-}{r_-^{2}}X$ and
$\frac{\partial r}{\partial
T}=\frac{g_{_{XX}}}{2}\frac{r_+-r_-}{r_-^{2}}T$
found above 
yield
${K_{\rm e}}^{\theta}{}_{\theta}=
{K_{\rm e}}^{\varphi}{}_{\varphi}=
\frac{\text{sign}\left(X\right)}{2}
\frac{r_+-r_-}{r_-^{2}}
\frac{\sqrt{g_{_{XX}}\left(X^{2}-T^{2}\right)}}{R}$.
Thus, in brief
\begin{equation}
{K_{\rm e}}^{\tau}{}_{\tau}
=
\frac{\text{sign}\left(X\right)}{2R^{2}k}
\left[r_++r_--2\frac{r_-r_+}{R}\right]
\,,\quad
{K_{\rm e}}^{\theta}{}_{\theta}=
{K_{\rm e}}^{\varphi}{}_{\varphi}
=\frac{\text{sign}
\left(X\right)(r_+-r_-)}{2r_-^{2}R}
\sqrt{g_{_{XX}}\left(X^{2}-T^{2}\right)}\,.
\label{Appendix_eq:ExtCurv_ktautau_value_rminus2}
\end{equation}
where again
$k=\sqrt{\left(1-\frac{r_+}{R}\right)\left(1-\frac{r_-}{R}\right)}$.
With these 
geometrical
quantities one can compute the physical quantities
of a thin shell, such as its energy density and tangential pressure,
assuming it is made of a perfect fluid, at the boundary surface
inside the Cauchy horizon,
as we did in the text.


\newpage






\begin{thebibliography}{10}

\bibitem{Israel_1966}
W. Israel, ``Singular hypersurfaces and thin shells in general
relativity'', Il Nouvo Cimento {\bf 44}, 10 (1966).

\bibitem{papapetrouhamoui}
A. Papapetrou and A. Hamoui, ``Couches simples de mati\`ere en
relativit\'e g\'en\'erale'', Annales de l'Institute Henri Poincar\'e
A {\bf 9}, 179 (1968).

\bibitem{taub}
A. H. Taub, ``Space-times with distribution valued curvature
tensors'', Journal of Mathematical Physics {\bf 21}, 1423 (1980).

\bibitem{Barrabes_Israel_1991}
C. Barrab\`{e}s and W. Israel, ``Thin shells in general relativity and
cosmology: The lightlike limit'', Phys. Rev. D {\bf 43}, 1129 (1991).

\bibitem{israel2}
W. Israel, ``Gravitational collapse and causality'',
Phys. Rev. {\bf 153}, 1388 (1967).

\bibitem{evans}
A. B. Evans, ``Relativistic dynamics of spherical counter-rotating
dust bodies'', Gen. Relativ. Gravit.  {\bf 8}, 155 (1977).

\bibitem{papapetrouhamoui2}
A. Papapetrou and A. Hamoui, ``Spherically symmetric surface
layers in general relativity - correction of an error'',
Gen. Relativ. Gravit. {\bf 10}, 253 (1979).

\bibitem{draythooft}
G. 't Hooft and T. Dray, ``The effect of spherical shells of matter on
the Schwarzschild black hole'', Communications in Mathematical Physics
{\bf 99}, 613 (1985).

\bibitem{blauguendelmanguth}
S. K. Blau, E. I. Guendelman, and A. H. Guth, ``Dynamics of
false-vacuum bubbles'', Phys.  Rev. D {\bf 35}, 1747 (1987).

\bibitem{frauen}
J. Frauendiener, C. Hoenselaers, and W. Konrad, ``A shell around a
black hole'', Classical Quantum Gravity {\bf 7}, 585 (1990).

\bibitem{brad}
P. R. Brady, J. Louko, and E. Poisson, ``Stability of a
shell around a black hole'', Phys.  Rev. D {\bf 44}, 1891 (1991).

\bibitem{Katz_Lynden-Bell_1991}
J. Katz and D. Lynden-Bell, ``Tension shells and tension stars'',
Classical Quantum Gravity {\bf 8}, 2231 (1991).

\bibitem{Comer_Katz_1994}
G. L. Comer and J. Katz, ``Some conditions for the existence of
tension stars'', Mon. Not. R. Astron. Soc. {\bf 267}, 51 (1994).

\bibitem{visserbook}
M. Visser, {\it Lorentzian Wormholes. From Einstein to Hawking}
(Springer, New York, 1996).

\bibitem{Lemos_Lobo_2008}
J. P. S. Lemos, F. S. N. Lobo,  Sergio Q. Oliveira,
``Morris-Thorne wormholes with a cosmological constant'', Phys.
Rev. D {\bf 68}, 064004 (2003); arXiv:gr-qc/0302049.

\bibitem{l5}
R. Andr\'e, J. P. S. Lemos, and G. M. Quinta, ``Thermodynamics and
entropy of self-gravitating matter shells and black holes in $d$
dimensions'', Phys. Rev. D {\bf 99}, 125013 (2019); arXiv:1905.05239
[hep-th].

\bibitem{bergli}
S. E. P. Bergliaffa, M. Chiapparini, L. M. Reyes, ``Thermodynamical
and dynamical equilibrium of a self-gravitating uncharged thin
shell'', Eur. Phys. J. C {\bf 80}, 719 (2020); arXiv:2006.06766
[gr-qc].

\bibitem{cruzisrael}
V. de la Cruz and W. Israel, `Gravitational bounce'', Il Nuovo Cimento
A {\bf 51}, 744 (1967).

\bibitem{Kuchar_1968}
K. Kucha\v{r}, ``Charged shells in general relativity and their
gravitational collapse'', Czechoslovak J. Phys.  B {\bf 18}, 435
(1968).

\bibitem{chase}
J. E. Chase, ``Gravitational instability and collapse of charged fluid
shells'', Il Nuovo Cimento B {\bf 67}, 136 (1970).

\bibitem{boulware}
D. G. Boulware, ``Naked singularities, thin shells, and the
Reissner-Nordstr\"om metric'', Phys. Rev. D {\bf 8}, 2363 (1973).

\bibitem{vilenkinfomin}
A. V. Vilenkin and P. I. Fomin, ``Schwarzschild sphere and classical
electron self-energy problem'', Il Nuovo Cimento A {\bf 45} 59 (1978).

\bibitem{hiscock}
W. A. Hiscock, 
``On the topology of charged spherical collapse'',
Journal of Mathematical Physics {\bf 22}, 215 (1981).

\bibitem{lemoszanchin2006}
J. P. S. Lemos and V. T. Zanchin,
``Gravitational magnetic monopoles and Majumdar-Papapetrou stars'',
J. Math. Phys. {\bf 47} 042504 (2006); arXiv:gr-qc/0603101.

\bibitem{lz2}
G. A. S. Dias, S. Gao, and J. P. S. Lemos, ``Charged shells in
Lovelock gravity: Hamiltonian treatment and physical implications'',
Phys. Rev. D {\bf 75}, 024030 (2007); arXiv:gr-qc/0612072.

\bibitem{gaolemos}
S. Gao and J. P. S. Lemos, ``Collapsing and static thin massive
charged dust shells in a Reissner-Nordstr{\"o}m black hole background
in higher dimensions'', Int. J. Mod. Phys. A {\bf 23}, 2943 (2008);
arXiv:0804.0295 [hep-th].

\bibitem{andreasson2009}
H. Andr\'easson, ``Sharp bounds on the critical stability radius for
relativistic charged spheres'', Communications in Mathematical Physics
{\bf 288}, 715 (2009); arXiv:0804.1882 [gr-qc].

\bibitem{andreasson2}
H. Andr\'easson, M. Eklund and G. Rein, ``A numerical investigation
of the steady states of the spherically symmetric
Einstein-Vlasov-Maxwell system'',
Classical Quantum Gravity {\bf 26}, 145003
(2009); arXiv:0903.4092 [gr-qc].

\bibitem{diaslemosworm2010}
G. A. S. Dias and J. P. S. Lemos, ``Thin-shell wormholes in
$d$-dimensional general relativity: Solutions, properties, and
stability'', Phys. Rev. D {\bf 82}, 084023 (2010); arXiv:1008.3376
[gr-qc].

\bibitem{lemoszanchinregularbhs}
J. P. S. Lemos and V. T. Zanchin, ``Regular black holes: Electrically
charged solutions, Reissner-Nordstr\"om outside a de Sitter core'',
Phys. Rev. D {\bf 83} 124005 (2011); arXiv:1104.4790 [gr-qc].

\bibitem{qbh4lemoszaslacp}
J. P. S. Lemos and O. B. Zaslavskii, ``Quasiblack holes: Properties
and Carter-Penrose diagrams'', in {\it Proceedings of the 13th Marcel
Grossmann Meeting} (Stockholm, July 2012), edited by K. Rosquist et al
(World Scientific, Singapore 2015), p. 1195.

\bibitem{berezin1}
V. A. Berezin and V. I. Dokuchaev, {\it Global Geometry of Space-Time
with the Charged Shell} (Nova Science Publishers, New York 2014);
arXiv:1404.2726 [gr-qc].

\bibitem{l4}
J. P. S. Lemos, G. M. Quinta, and O. B. Zaslavskii, ``Entropy of a
self-gravitating electrically charged thin shell and the black hole
limit'', Phys. Rev. D {\bf 91}, 104027 (2015); arXiv:1503.00018
[hep-th].

\bibitem{l6}
J. P. S. Lemos, G. M. Quinta, and O. B. Zaslavskii, ``Entropy of an
extremal electrically charged thin shell and the extremal black
hole'', Phys. Lett. B {\bf 750}, 306 (2015); arXiv:1505.05875
[hep-th].



\bibitem{luzlemos}
P. Luz and J. P. S. Lemos, ``Electrically charged tension shells'',
in {\it Proceedings of the Fourteenth Marcel Grossmann Meeting},
edited by M. Bianchi et al
(World Scientific, Singapore, 2017), p.~1641.

\bibitem{thaller}
M. Thaller, ``Existence of static solutions of the
Einstein-Vlasov-Maxwell system and the thin shell limit'',
SIAM J. Math. Anal. {\bf 51},
2231 (2019); arXiv:1801.05664 [gr-qc].

\bibitem{lemoszasla2020}
J. P. S. Lemos and O. B. Zaslavskii, ``Compact objects in
general relativity: From Buchdahl stars to quasiblack holes'',
Int. J. Mod. Phys. D {\bf 29},  2041019 (2020);
arXiv:2007.1094 [gr-qc].

\bibitem{Graves_Brill_1960}
J. C. Graves and D. R. Brill, ``Oscillatory character of
Reissner-Nordstr\"om metric for an ideal charged wormhole'',
Phys. Rev. {\bf 120}, 1507 (1960).

\bibitem{Penrose1964}
R. Penrose, ``Structure of space-time'', in {\it Battelle Rencontres:
1967 Lectures in Mathematics and Physics}, edited by C. M. DeWitt and
J. A. Wheeler (Benjamin, New York 1968), p. 121.

\bibitem{Carter_1966_2}
B. Carter, ``The complete analytic extension of the
Reissner-Nordstr\"om metric in the special case $e^{2}=m^{2}$'',
Phys. Lett. {\bf 21}, 423 (1966).

\bibitem{Hawking_Ellis_book}
S. W. Hawking and G. F. R. Ellis, {\it The Large Scale Structure of
Spacetime} (Cambridge University Press, Cambridge 1973).

\bibitem{MTW_Book}
C. W. Misner, K. S. Thorne, and J. A. Wheeler, {\it Gravitation}
(W. H. Freeman, San Francisco 1973). 

\bibitem{felicebook}
F. de Felice and C. J. S. Clarke,
{\it Relativity on Curved Manifolds}
(Cambridge University Press, Cambridge, 1990). 

\end{thebibliography}
\end{document}